\documentclass{lmcs} 
\pdfoutput=1

\usepackage[utf8]{inputenc}

\usepackage{lastpage}
\lmcsdoi{20}{1}{23}
\lmcsheading{}{\pageref{LastPage}}{}{}%
{Feb.~17,~2022}{Mar.~22,~2024}{}

\keywords{Interval Temporal Logic, Star-Free Regular Languages,
Satisfiability,
Complexity}

\usepackage[caption=false]{subfig}
\usepackage{marginnote}
\usepackage{transparent}
\usepackage{stmaryrd}
\usepackage{xspace}
\usepackage{wrapfig}
\usepackage{pgf,tikz}
\usepackage{hyperref}
\usepackage{underscore}
\usepackage{amssymb}
\usepackage{refcount}
\usepackage{enumitem}
\usepackage{fdsymbol}
\usepackage{amsthm}
\usepackage{sidenotes}
\usepackage{xcolor}
\usepackage{refcount}
\usepackage{enumitem}
\usepackage{transparent}
\usepackage{tcolorbox}
\usepackage[makeroom]{cancel}
\usepackage{contour}
\usepackage{ulem}
\usepackage{sidenotes}
\usepackage{transparent}
\usepackage{lineno}
\usepackage{thmtools, thm-restate}

\newlist{compactitem}{itemize}{2}      
\setlist[compactitem,1]{label=\textbullet,labelindent=*,leftmargin=*,nosep}
\setlist[compactitem,2]{label=$-$,labelindent=*,leftmargin=*,nosep}

\newlist{compactenum}{enumerate}{2}
\setlist[compactenum,1]{font=\normalfont,label=(\arabic*),labelindent=*,leftmargin=*,start=1,nosep}
\setlist[compactenum,2]{font=\normalfont,label=(\alph*),labelindent=*,leftmargin=*,start=1,nosep}

\contourlength{0.8pt}
\newcommand{\myUnderline}[1]{%
  \uline{\phantom{#1}}%
  \llap{\contour{white}{#1}}%
}

\usepackage{adjustbox}

\newcommand{\red}[1]{{\color{red} #1}}

\usetikzlibrary{calc}
\usetikzlibrary{patterns}
\usetikzlibrary{backgrounds}
\usetikzlibrary{fit}
\usetikzlibrary{math}
\usetikzlibrary{shapes.geometric}

\usetikzlibrary{decorations.pathreplacing}
\pgfdeclarelayer{background}
\pgfdeclarelayer{foreground}
\pgfsetlayers{background,main,foreground}
\usetikzlibrary{positioning}

\newcommand{\mathset}[1]{\{#1\}}

\usepackage{hyperref}

\theoremstyle{plain}

\newcommand{\vpair}[1]{$\begin{array}{c}
   #1    
\end{array}$}
\newcommand{\vpairmath}[1]{\!\!\!
\begin{array}{c}
   #1    
\end{array}
\!\!\!}

\newcommand{\expspacecomplete}{\textsc{ExpSpace-complete}}
\newcommand{\pspacecomplete}{\textsc{PSpace-complete}}

\newcommand{\expspace}{\textsc{ExpSpace}}
\newcommand{\expspacehard}{\textsc{ExpSpace-hard}}
\newcommand{\expspacehardness}{\textsc{ExpSpace-hardness}}
\newcommand{\logspace}{\textsc{LogSpace}}
\newcommand{\pspace}{\textsc{PSpace}}

\newcommand{\hsER}{\langle R\rangle}
\newcommand{\hsEB}{\langle B\rangle}
\newcommand{\hsEA}{\langle A\rangle}

\newcommand{\hsED}{\langle D\rangle}
\newcommand{\hsEE}{\langle E\rangle}

\newcommand{\hsEL}{\langle L\rangle}
\newcommand{\hsEO}{\langle O\rangle}

\newcommand{\hsEBbar}{\langle \overline{B}\rangle}
\newcommand{\hsEAbar}{\langle \overline{A}\rangle}

\newcommand{\hsEDbar}{\langle \overline{D}\rangle}
\newcommand{\hsEEbar}{\langle \overline{E}\rangle}

\newcommand{\hsELbar}{\langle \overline{L}\rangle}
\newcommand{\hsEObar}{\langle \overline{O}\rangle}

\newcommand{\Rcon}{\overrightarrow{\scalebox{0.8}{$Con$}}}

\newcommand{\escapebackslash}{\scalebox{0.9}{$\backslash$}}
\newcommand{\respace}{\tikz{\node[circle, fill, inner sep=2pt, opacity=0.2]{};} }

\newcommand{\hsAR}{[R]}
\newcommand{\hsAA}{[A]}
\newcommand{\hsAB}{[B]}
\newcommand{\hsAD}{[D]}

\newcommand{\allenScale}[1]{\scalebox{0.5}{$#1$}}
\newcommand{\allenA}{\ \allenScale{\mathrm{MEETS}}\ }
\newcommand{\allenB}{\ \allenScale{\mathrm{STARTED\mbox{-}BY}}\ }
\newcommand{\allenD}{\ \allenScale{\mathrm{CONTAINS}}\ }
\newcommand{\allenE}{\ \allenScale{\mathrm{FINISHED\mbox{-}BY}}\ }
\newcommand{\allenL}{\ \allenScale{\mathrm{BEFORE}}\ }
\newcommand{\allenO}{\ \allenScale{\mathrm{OVERLAPS}}\ }

\newcommand{\allenScaleText}[1]{\scalebox{0.75}{$#1$}}
\newcommand{\allenTextA}{\ \allenScaleText{\mathrm{MEETS}}\ }
\newcommand{\allenTextB}{\ \allenScaleText{\mathrm{STARTED\mbox{-}BY}}\ }
\newcommand{\allenTextD}{\ \allenScaleText{\mathrm{CONTAINS}}\ }
\newcommand{\allenTextE}{\ \allenScaleText{\mathrm{FINISHED\mbox{-}BY}}\ }
\newcommand{\allenTextL}{\ \allenScaleText{\mathrm{BEFORE}}\ }
\newcommand{\allenTextO}{\ \allenScaleText{\mathrm{OVERLAPS}}\ }

\newcommand{\Rlang}{\scalebox{0.9}{$\overrightarrow{\Lang}$}}

\newcommand{\equivmodA}{\equiv_{\neg A}}
\newcommand{\nequivmodA}{\nequiv_{\neg A}}

\renewcommand{\min}{min}

\newcommand{\LogicABDhom}{\ensuremath{\mathsf{BDA}_{hom}}\xspace}
\newcommand{\LogicBDAhom}{\ensuremath{\mathsf{BDA}_{hom}}\xspace}

\newcommand{\LogicABhom}{\ensuremath{\mathsf{BA}_{hom}}\xspace}

\newcommand{\LogicCDT}{\ensuremath{\mathsf{CDT}}\ }

\newcommand{\cdtC}{\ \mathbf{C}\ }
\newcommand{\cdtD}{\ \mathbf{D}\ }
\newcommand{\cdtT}{\ \mathbf{T}\ }

\newcommand{\LogicHS}{\ensuremath{\mathsf{HS}}}

\newcommand{\Prop}{\ensuremath{\mathrm{Prop}}}

\newcommand{\TFA}{TF^\varphi_A}

\newcommand{\Lang}{\mathrm{Lang}}

\newcommand{\bfM}{\mathbf{M}}

\newcommand{\ob}{\overline{b}}

\newcommand{\hF}{\hat{F}}
\newcommand{\hx}{\hat{x}}
\newcommand{\hy}{\hat{y}}

\newcommand{\bbN}{\mathbb{N}}

\newcommand{\oF}{\overline{F}}
\newcommand{\oG}{\overline{G}}

\newcommand{\witnesses}{\mathrm{Wit}_{\cG}}

\newcommand{\bN}{\mathbb{N}}

\newcommand{\Pre}{\mathrm{Pre}}
\newcommand{\Sub}{\mathrm{Inf}}

\newcommand{\bS}{\mathbb{S}}

\newcommand{\cG}{\mathcal{G}}
\newcommand{\cL}{\mathcal{L}}
\newcommand{\cO}{\mathcal{O}}

\newcommand{\cP}{\mathcal{P}}

\newcommand{\cV}{\mathcal{V}}

\newcommand{\bI}{\mathbb{I}}
\newcommand{\bG}{\mathbb{G}}

\newcommand{\rowG}{\mathrm{Row}_{\cG}}

\renewcommand{\closure}{\mathrm{Cl}(\varphi)}
\newcommand{\atoms}{\mathrm{At}(\varphi)}

\newcommand{\ox}{\overline{x}}
\newcommand{\oy}{\overline{y}}

\newcommand{\thenB}{\rightarrow_B}
\newcommand{\thenNB}{\xcancel{\rightarrow_B}}

\newcommand{\thenD}{\rightarrow_D}
\newcommand{\thenND}{\xcancel{\rightarrow_D}}

\newcommand{\shadingB}{\mathrm{Sh}_{B}}
\newcommand{\hshadingB}{\hat{\mathrm{Sh}}_{B}}

\newcommand{\shadingG}{\mathrm{Sh}^{\cG}}
\newcommand{\shadingGB}{\shadingG_B}
\newcommand{\shadingGN}{\shadingG_{\bbN}}

\newcommand{\Closest}{\mathrm{Closest}_{\mathrm{wit}}}

\newcommand{\obsR}{\mathrm{Obs}_R}
\newcommand{\obsB}{\mathrm{Obs}_B}
\newcommand{\obsD}{\mathrm{Obs}_D}
\newcommand{\obsA}{\mathrm{Obs}_A}

\newcommand{\reqR}{\mathrm{Req}_R}
\newcommand{\reqB}{\mathrm{Req}_B}
\newcommand{\reqD}{\mathrm{Req}_D}
\newcommand{\reqA}{\mathrm{Req}_A}

\newcommand{\boxR}{\mathrm{Box}_R}
\newcommand{\boxB}{\mathrm{Box}_B}
\newcommand{\boxD}{\mathrm{Box}_D}
\newcommand{\boxA}{\mathrm{Box}_A}

\newcommand{\Deltareq}{\Delta_{\uparrow}}
\newcommand{\future}{\bS_{\rightarrow}}

\newcommand{\asat}{\scalebox{0.7}{$\checkmark$}}
\newcommand{\areq}{\scalebox{0.7}{$\uparrow$}}
\newcommand{\abox}{{\scalebox{0.7}{$\notin$}}}

\newcommand{\ltlUntil}{\ \mathrm{U}\ }
\newcommand{\ltlNext}{\varbigcirc}

\newcommand{\gray}[1]{{\color{gray} #1}}
\newcommand{\gzero}{{\color{gray} 0}}

\tcbset{
  top=0pt,bottom=0pt,left=0mm,right=0mm,
  leftrule=0pt,rightrule=0pt,toprule=0mm,
  bottomrule=0mm,boxsep=0mm, colback=black!20!white
}

\begin{document}

\title[The logic of prefixes, sub-intervals, and temporal neighborhood]{The addition of temporal neighborhood makes the logic of prefixes and sub-intervals EXPSPACE-complete\ {\lsuper*} }

\titlecomment{{\lsuper*}This paper is a revised and extended version of 
\cite{DBLP:conf/mfcs/BozzelliMPS20} and \cite{DBLP:journals/corr/abs-2109-08320}}

\author[L.~Bozzelli]{Laura Bozzelli\lmcsorcid{0000-0003-0963-8169}}[a]

\author[A.~Montanari]{Angelo 
Montanari\lmcsorcid{0000-0002-4322-
769X}}[b]	
 
\author[A.~Peron]{Adriano 
Peron\lmcsorcid{0000-0002-7111-3171}}[c]	

\author[P.~Sala]{Pietro 
Sala\lmcsorcid{0000-0002-2612-1519}}[d]	

\address{University of Napoli 
``Federico II'', Napoli, Italy}	
\email{lr.bozzelli@gmail.com} 

\address{University of Udine, Udine, Italy}	
\email{angelo.montanari@uniud.it} 

\address{University of Trieste, 
Trieste, Italy}	
\email{adriano.peron@units.it}  

\address{University of Verona, Verona, 
Italy}	
\email{pietro.sala@univr.it}

\begin{abstract}
 

A classic result by Stockmeyer \cite{stockmeyer1974complexity} 
gives a non-elementary lower bound to the emptiness problem for 
generalized $*$-free regular expressions. 
This result is intimately connected to the satisfiability problem
for the interval temporal logic of the chop modality under 
the homogeneity assumption \cite{10.1007/BFb0036915}. The chop modality can indeed be viewed as the inverse of the concatenation operator of regular languages,
and such a correspondence enables reductions between the two problems.

In this paper, we study the complexity of the satisfiability
problem for suitable weakenings of the chop interval temporal logic,
that can be equivalently viewed as fragments of Halpern and Shoham 
interval logic. We first introduce the logic $\mathsf{BD}_{hom}$
featuring modalities $B$ (for \emph{begins}), 
corresponding to the prefix relation on pairs of intervals, and $D$ (for \emph{during}), corresponding to the infix relation,
whose satisfiability problem, under 
the homogeneity assumption, has been recently shown to be 
\pspacecomplete\ \cite{BMPS21}.
The homogeneous models of $\mathsf{BD}_{hom}$ naturally correspond  
to languages defined by restricted forms of generalized $*$-free regular expressions, that feature operators
for union, complementation, and the inverses of the 
prefix and infix relations. 
Then, we study the extension of $\mathsf{BD}_{hom}$ with the temporal neighborhood 
modality $A$, corresponding to the Allen  relation \emph{Meets}, and prove that such an addition
increases both the expressiveness and the complexity of the logic. In particular, we
show that the resulting logic $\mathsf{BDA}_{hom}$ is \expspacecomplete.


\end{abstract}


\maketitle

\newcommand{\tempcut}[1]{#1}

\tempcut{


\section{Introduction}\label{sec:intro}

Interval temporal logics (ITLs for short) are versatile and expressive formalisms to specify properties of sequences of states and their duration. When it comes to fundamental problems like satisfiability, their high expressive power is often obtained at the price of undecidability. As an example, the satisfiability problem of the most widely known ITLs, namely, Halpern and Shoham's $\mathsf{HS}$ \cite{DBLP:journals/jacm/HalpernS91} and Venema's $\mathsf{CDT}$  \cite{10.1093/logcom/1.4.453}, turn out to be highly undecidable. Despite these negative results, a number of decidable ITLs have been identified by suitably weakening $\mathsf{HS}$ (see \cite{DBLP:journals/tcs/BresolinMMSS14} for a complete classification of $\mathsf{HS}$ fragments). 
Here the term ``weakening'' is intended as a set of syntactic and/or semantics restrictions imposed on the formulas of the logic
and/or the temporal structures on which such formulas are interpreted, respectively. 
Among the plethora of possible weakenings, here
we focus on (the combination of) the following two natural and well-studied restrictions:

\smallskip

\begin{compactitem}

\item \emph{Restrict the set of interval relations}. 

\noindent A number of decidable fragments of $\mathsf{HS}$ and $\mathsf{CDT}$ have been obtained by considering a restricted set of Allen's relations over pairs of intervals. This approach naturally induces fragments of $\mathsf{HS}$ with modalities corresponding to the selected subset of interval relations. As an example, the logic of temporal neighborhood ($\mathsf{PNL}$ for short)  features only two modalities, corresponding to two interval relations among the  thirteen possible ones, namely, $A$ (\emph{adjacent to the right}) and its inverse $\bar A$ (\emph{adjacent to the left}) \cite{DBLP:conf/compos/ChaochenH97}. $\mathsf{PNL}$ has been shown to be decidable over all meaningful classes of linear orders \cite{DBLP:conf/tableaux/BresolinMSS11,DBLP:conf/time/MontanariS12}.

\smallskip

\item \emph{Restrict the class of models}.

\noindent As an alternative, it is possible to tame the complexity of ITLs by restricting to classes of models that satisfy some specific assumptions. An example of such an approach can be found in a recent series of papers that study the model checking problem for ITLs (see, e.g., the seminal paper \cite{DBLP:journals/acta/MolinariMMPP16}),
as well as their expressiveness compared to that of classical point-based temporal logics, like $\mathsf{LTL}$, $\mathsf{CTL}$, and $\mathsf{CTL^*}$ \cite{DBLP:journals/tocl/BozzelliMMPS19}. 
In this setting, models are represented as Kripke structures, and are inherently point-based rather than interval-based. The very same models can be obtained from interval temporal structures by making the so-called \emph{homogeneity assumption}, i.e., by assuming that every proposition letter holds over an interval if and only if it holds at all its points \cite{roe80}. Under such an assumption, full $\mathsf{HS}$ has a decidable satisfiability problem (as a matter of fact, the model checking procedures proposed in the aforementioned series of papers can be easily turned into satisfiability procedures, often retaining the same complexity) \cite{DBLP:journals/acta/MolinariMMPP16}. In the light of the above, the focus in studying $\mathsf{HS}$ fragments under the homogeneity assumption was shifted from decidability to complexity. 
\end{compactitem}

\smallskip

Under the homogeneity assumption, a natural connection to generalized $*$-free regular languages emerges from the analysis of the complexity of ITLs over finite linear orders.
A classic result 
by Stockmeyer states that the emptiness problem for  generalized $*$-free
regular expressions is non-elementarily decidable (tower-complete) 
for unbounded nesting of negation~\cite{Schmitz:2016,stockmeyer1974complexity} 
(it is \textsc{(K-1)-}\expspacecomplete\ for expressions where the nesting 
of negation is at most $\textsc{K}\in\mathbb{N}^+$). 
Such a problem can be easily turned into the satisfiability problem for the logic $\mathsf{C}$ of the chop modality, over finite linear orders, under the homogeneity assumption~\cite{DBLP:conf/csl/HodkinsonMS08,digitalcircuitsthesis,
\detokenize{choppingintervals}}, and vice versa. $\mathsf{C}$ is a proper fragment of $\mathsf{CDT}$ with a single binary modality, the so-called \emph{chop} operator, that allows one to split the current interval in two parts and to state what is true over the first part and what over the second one. It can be easily shown that there is a  reduction, that operates in logarithmic space, of the emptiness problem for generalized $*$-free regular expressions to the satisfiability problem for $\mathsf{C}$ with unbounded nesting of the chop operator, and vice versa.

The close relationships between formal languages and ITLs have been already pointed out in \cite{DBLP:conf/lics/MontanariS13,DBLP:conf/lata/MontanariS13,DBLP:journals/tcs/MonicaMS23}, where the ITL counterparts of regular languages, $\omega$-regular languages, and extensions of them ($\omega  B$-, $\omega S$-, and $\omega T$-regular languages) have been provided. Here, we focus on some meaningful fragments of $\mathsf{C}$ under the homogeneity assumption. Hereafter, for any ITL $\mathsf{X}$, we  write $\mathsf{X}_{hom}$ to point out that we are considering $\mathsf{X}$ under the homogemeity assumption. Modalities for the prefix,  suffix, and  infix relations over (finite) intervals can be easily defined in $\mathsf{C}$. It holds that a formula holds over a prefix of the current interval if and only if it is possible to split the interval in such a way that the formula holds over the first part and the second part contains at least two points. The case of suffixes is completely symmetric. Infixes can be defined in terms of prefixes and suffixes: a proper sub-interval of the current interval is a suffix of one of its prefixes or, equivalently, a prefix of one of its suffixes. 

The satisfiability problem for the logic $\mathsf{D}_{hom}$ of the infix relation has been shown to be \pspacecomplete\ by a suitable contraction method~\cite{DBLP:conf/icalp/BozzelliMMPS17,DBLP:journals/lmcs/BozzelliMMPS22}. \pspace\ completeness has been recently proved also for the logic $\mathsf{BD}_{hom}$ (resp., $\mathsf{DE}_{hom}$) that extends $\mathsf{D}_{hom}$ with modality $B$ (resp., $E$) for the prefix (resp., suffix) relation \cite{BMPS21,DBLP:journals/iandc/BozzelliMPS23}. 

A lot of work has been done on the logic $\mathsf{BE}_{hom}$ of prefixes and suffixes. 
In \cite{DBLP:journals/tcs/BozzelliMMPS19}, 
its satisfiability problem has been shown to be \expspacehard\ by a polynomial-time reduction from a domino-tiling problem for grids with rows of single exponential length. Moreover, all the other  $\mathsf{HS}_{hom}$ fragments whose satisfiability problem is known to be \expspacehard\ feature $\mathsf{BE}_{hom}$ as a proper fragment. As for the upper bound, a trivial one is given by the non-elementary decision procedure for full $\mathsf{HS}_{hom}$ devised in \cite{DBLP:journals/acta/MolinariMMPP16} ($\mathsf{BE}_{hom}$ is a small fragment of $\mathsf{HS}_{hom}$). In \cite{DBLP:conf/time/BozzelliMP19}, Bozzelli et al.\ showed that it is not possible to improve such a bound by tailoring the proof techniques exploited for $\mathsf{HS}_{hom}$ in \cite{DBLP:journals/acta/MolinariMMPP16}, which are based on the notion of $\mathsf{BE}$-descriptor, to $\mathsf{BE}_{hom}$, as it is not possible to give an elementary upper bound on the size of $\mathsf{BE}$-descriptors for $\mathsf{BE}_{hom}$ \cite{DBLP:conf/time/BozzelliMP19}. \expspace\ membership (from which \expspace\ completeness immediately follows) of the satisfiability problem for $\mathsf{BE}_{hom}$ has been very recently shown by devising an equi-satisfiable normal form with boundedly many nested modalities \cite{DBLP:conf/lics/MonicaMPS23}. The normalization technique somehow resembles Scott’s quantifier elimination, but it turns out to be much more involved due to the limitations enforced by the homogeneity assumption.




In this paper, we focus on the logic $\mathsf{BDA}_{hom}$, that extends $\mathsf{BD}_{hom}$ with the \emph{meet} modality $\mathsf{A}$, and prove that its satisfiability problem is \expspacecomplete\ by a model-theoretic argument. As a matter of fact, $\mathsf{BDA}_{hom}$ turns out to be the first \expspacecomplete\
fragment of $\mathsf{HS}_{hom}$  that does not feature $\mathsf{BE}_{hom}$ as a proper fragment. 
As a preparatory step, we apply the proposed model-theoretic proof technique to the simpler fragment  $\mathsf{BD}_{hom}$, and then we show how to extend it to $\mathsf{BDA}_{hom}$ without any increase in complexity.

\smallskip

The paper is organized as follows. In Section~\ref{sec:itlintro}, we provide a gentle introduction to ITLs. We first give an informal account of the two main propositional ITLs, namely $\mathsf{CDT}$ and $\mathsf{HS}$, interpreted over finite linear orders. Then, by making use of a simple example, we compare their expressive power with that of Linear Temporal Logic ($\mathsf{LTL}$). In Section \ref{sec:logic}, we introduce the logic 
$\mathsf{BD}_{hom}$. We specify its syntax and semantics, and we point out some interesting connections between its formulas and restricted forms of generalized $*$-free regular expressions. In the following three sections, we prove a small model theorem for the satisfiability of $\mathsf{BD}_{hom}$ formulas over finite linear orders, which provides a doubly exponential bound (in the size of the formula) on their models, and exploit it to devise a satisfiability checking procedure that works in exponential space with respect to the size of the input $\mathsf{BD}_{hom}$ formula. More precisely, in Section \ref{sec:compass}, we introduce and discuss a spatial representation of the models of $\mathsf{BD}_{hom}$ formulas, called \emph{compass structure}. Then, in Section \ref{sec:properties}, we prove a series of meaningful spatial properties of these structures.
%
%
Finally, in Section \ref{sec:expspace}, 
we prove a small model theorem for $\mathsf{BD}_{hom}$, that allows us to prove the  \expspace\ membership of the problem of checking the satisfiability of $\mathsf{BD}_{hom}$ formulas over finite linear orders\footnote{It is worth noticing that, in view of the results in \cite{BMPS21}, where \pspace\ membership is shown, the proposed decision procedure is sub-optimal. However, it plays a fundamental  instrumental role in the proof of the main result of the paper about $\mathsf{BDA}_{hom}$.}
%
%
In Section~\ref{sec:abdexpspace},  we define the logic $\mathsf{BDA}_{hom}$, that extends $\mathsf{BD}_{hom}$ with modality $\mathsf{A}$; then, we show that the properties stated for $\mathsf{BD}_{hom}$ in Section \ref{sec:properties} holds for $\mathsf{BDA}_{hom}$ as well, and, building on them, we prove the \expspace\ membership of the problem of checking the satisfiability of $\mathsf{BDA}_{hom}$ formulas over finite linear orders.
The decision procedure for $\mathsf{BDA}_{hom}$ can be obtained from that for $\mathsf{BD}_{hom}$ by making a few small adjustments. \expspacehardness\ of the satisfiability problem for $\mathsf{BDA}_{hom}$, over finite linear orders, is proved in Section \ref{sec:hardness}, by a reduction from the acceptance problem for  (non-deterministic) Turing Machines working in exponential space, thus allowing 
us to conclude that the complexity bound for $\mathsf{BDA}_{hom}$ finite satisfiability is tight. 
Section~\ref{sec:conclusions} concludes the paper with an assessment of the work done and an outline of future research directions. Supplementary material, including additional examples and some of the most technical proofs, that will be referenced throughout the paper, is provided in the appendices. In particular, proofs are reported in Appendix \ref{appendix:proofs}.

%


\section{A gentle introduction to Interval Temporal Logics} 
\label{sec:itlintro}

In this section, we provide a gentle introduction to Interval Temporal Logics (ITLs), focusing on the features that distinguish them from point-based ones. As a natural term of comparison, we choose $\mathsf{LTL}$, and,
for the sake of simplicity, we restrict our attention to totally ordered finite temporal structures, that is, finite prefixes of $\bbN$
(finite \emph{traces} or models). With a little abuse of notation, we denote the ordered set $\{0, 1, 2, \ldots, N\}$ by $N$. 
$\mathsf{LTL}$ over finite traces is often referred to as $\mathsf{LTL}_f$ in the literature \cite{DBLP:conf/ijcai/GiacomoV13, DBLP:conf/aaai/GiacomoMM14}.

Let $\Prop$ be a set of proposition letters. The first, crucial difference between ITLs and $\mathsf{LTL}_f$ is the way in which $\Prop$ is interpreted over models. Let $\bI_N = \{[x,y] : 0 \leq x
\leq y \leq N\}$ be the set of all and only the intervals on $N$. In the case of $\mathsf{LTL}_f$, the valuation function is $\pi : N \rightarrow 2^\Prop$, while, in the case of ITLs, it is $\cV: \bI_N \rightarrow 2^\Prop$. It is immediate to see that $\cV$ is a generalization of $\pi$, as the point-based semantics $\pi$ can be embedded into the interval-based one $\cV$ by assuming $\pi(x) = \cV([x,x])$, for all $x \in N$. From now on, we will refer to intervals of the forms $[x,x]$ and 
$[x,y]$, with $x < y$, as \emph{point-intervals} and \emph{strict-intervals}, respectively. Whenever we will not need to distinguish between point- and strict-intervals, we will simply refer to them as \emph{intervals}.


\begin{figure}
\begin{tikzpicture}[scale=1, node distance=2cm]


\node[draw, circle, inner sep=1, label={[]270:$\scalebox{0.7}{$ 0 $}$},
      label={[]90:$\scalebox{0.7}{$ p $}$}](0-0) {};
      
\node[draw, circle, inner sep=1, label={[]270:$\scalebox{0.7}{$ 1 $}$},
      label={[]90:$\scalebox{0.7}{$ p $}$}](1-1) [right of=0-0] {};
      
\node[draw, circle, inner sep=1, label={[]270:$\scalebox{0.7}{$ 2 $}$},
      label={[]90:$\scalebox{0.7}{$ p,q $}$}](2-2) [right of=1-1] {};
      
\node[draw, circle, inner sep=1, label={[]270:$\scalebox{0.7}{$ 3 $}$},
      label={[]90:$\scalebox{0.7}{$  $}$}](3-3) [right of=2-2] {};
      
\node[draw, circle, inner sep=1, label={[]270:$\scalebox{0.7}{$ 4 $}$},
      label={[]90:$\scalebox{0.7}{$ q $}$}](4-4) [right of=3-3] {};
      
\draw[|-|, opacity=0.5] ($(0-0.center) + (0,0.75)$) -- ($(1-1.center)+ (0,0.75)$)
node[pos=0.5, above](47) {$\scalebox{0.7}{$  $}$};

\draw[|-|, opacity=1] ($(0-0.center) + (0,1.25)$) -- ($(2-2.center)+ (0,1.25)$)
node[pos=0.5, above](47) {$\scalebox{0.7}{$ p,q $}$};

\draw[|-|, opacity=1] ($(0-0.center) + (0,2.25)$) -- ($(3-3.center)+ (0,2.25)$)
node[pos=0.5, above](47) {$\scalebox{0.7}{$ q $}$};

\draw[|-|, opacity=0.5] ($(0-0.center) + (0,3.4)$) -- ($(4-4.center)+ (0,3.4)$)
node[pos=0.5, above](47) {$\scalebox{0.7}{$  $}$};

\draw[|-|, opacity=0.5] ($(1-1.center) + (0,0.75)$) -- ($(2-2.center)+ (0,0.75)$)
node[pos=0.5, above](47) {$\scalebox{0.7}{$  $}$};

\draw[|-|, opacity=1] ($(1-1.center) + (0,1.75)$) -- ($(3-3.center)+ (0,1.75)$)
node[pos=0.5, above](47) {$\scalebox{0.7}{$ p $}$};

\draw[|-|, opacity=1] ($(1-1.center) + (0,2.75)$) -- ($(4-4.center)+ (0,2.75)$)
node[pos=0.5, above](47) {$\scalebox{0.7}{$ p,q $}$};

\draw[|-|, opacity=0.5] ($(2-2.center) + (0,0.75)$) -- ($(3-3.center)+ (0,0.75)$)
node[pos=0.5, above](47) {$\scalebox{0.7}{$  $}$};

\draw[|-|, opacity=1] ($(2-2.center) + (0,1.25)$) -- ($(4-4.center)+ (0,1.25)$)
node[pos=0.5, above](47) {$\scalebox{0.7}{$ p,q $}$};

\draw[|-|, opacity=0.5] ($(3-3.center) + (0,0.75)$) -- ($(4-4.center)+ (0,0.75)$)
node[pos=0.5, above](47) {$\scalebox{0.7}{$  $}$};

\node(PI) at (11,0.25) {\scalebox{0.75}{
    $
    \begin{array}{c||c|c|c|c|c}
      \pi(x) & 0 & 1 & 2 & 3 & 4 \\
      \hline\hline
      p & 1 & 1 & 1 & \gzero & \gzero \\ 
      q & \gzero & \gzero & 1 & \gzero & 1
    \end{array}  
    $
    }
};  

\pgftransformshift{\pgfpoint{6cm}{-2cm}}

\node[inner sep=0](V) {\scalebox{0.74}{
    $
    \begin{array}{c||c|c|c|c|c|c|c|c|c|c|c|c|c|c|c}
      \cV([x,y]) & [0,0] & [0,1] & [0,2] & [0,3] & [0,4] & [1,1] & [1,2] & [1,3] & [1,4] & [2,2] & [2,3] & [2,4] & [3,3] & [3,4] & [4,4] \\
      \hline\hline
      p & 1 & \gzero & 1 & \gzero & \gzero & 1 & \gzero & 1 & 1 & 1 & \gzero & 1 & \gzero & \gzero & \gzero \\ 
q & \gzero & \gzero & 1 & 1 & \gzero & \gzero & \gzero & \gzero & 1 & 1 & \gzero & 1 & \gzero & \gzero & 1
    \end{array}  
    $
    }
};      


%
%
%
%
%
%
%
%

\end{tikzpicture}

\caption{\label{fig:itlmodels} Point-based ($\pi$) vs.\ interval-based ($\cV$) labelling over the same finite linear order.}

\end{figure}

In its full generality, ITL interval-based semantics does not impose any constraint on the relationships between the proposition letters that hold over a strict-interval and those that hold over the point-intervals that it includes, i.e., the set of proposition letters $\cV([x,y])$ that hold over the strict-interval $[x,y]$ may differ from the sets of proposition letters $\cV([x,x]), 
\ldots, \cV([y,y])$ that hold on the point-intervals \emph{in} $[x,y]$ (which, obviously, may differ from each other). Similarly, the set of proposition letters $\cV([x',y'])$ 
that hold on a  proper subinterval $[x',y']$ of $[x,y]$, that is, $x\leq x'< y'\leq y$ and $[x',y']\neq [x,y]$, may differ from those that hold on $[x,y]$. Consider the example of Figure~\ref{fig:itlmodels}, where $\pi$ and $\cV$ agree on the labelling of points $0,\ldots, 4$ (they are interpreted as the intervals $[0,0], \ldots, [4,4]$ in the ITL semantics). The evaluation of proposition letters $p$ and $q$ on strict-intervals does not depend on that on their sub-intervals.
As an example, the label of the interval $[1,4]$ in Figure~\ref{fig:itlmodels} is $\cV([1,4]) = \mathset{p,q}$ and it features all the possible subsets of $\{p,q\}$ as the labels of its point intervals $[1,1], \ldots [4,4]$. As for its  proper subintervals, it holds that
$\cV([1,2])= \cV([2,3])=\cV([3,4])= \emptyset$, $\cV([1,3]) = \{p\}$,
and $\cV([2,4]) = \mathset{p,q}$. 


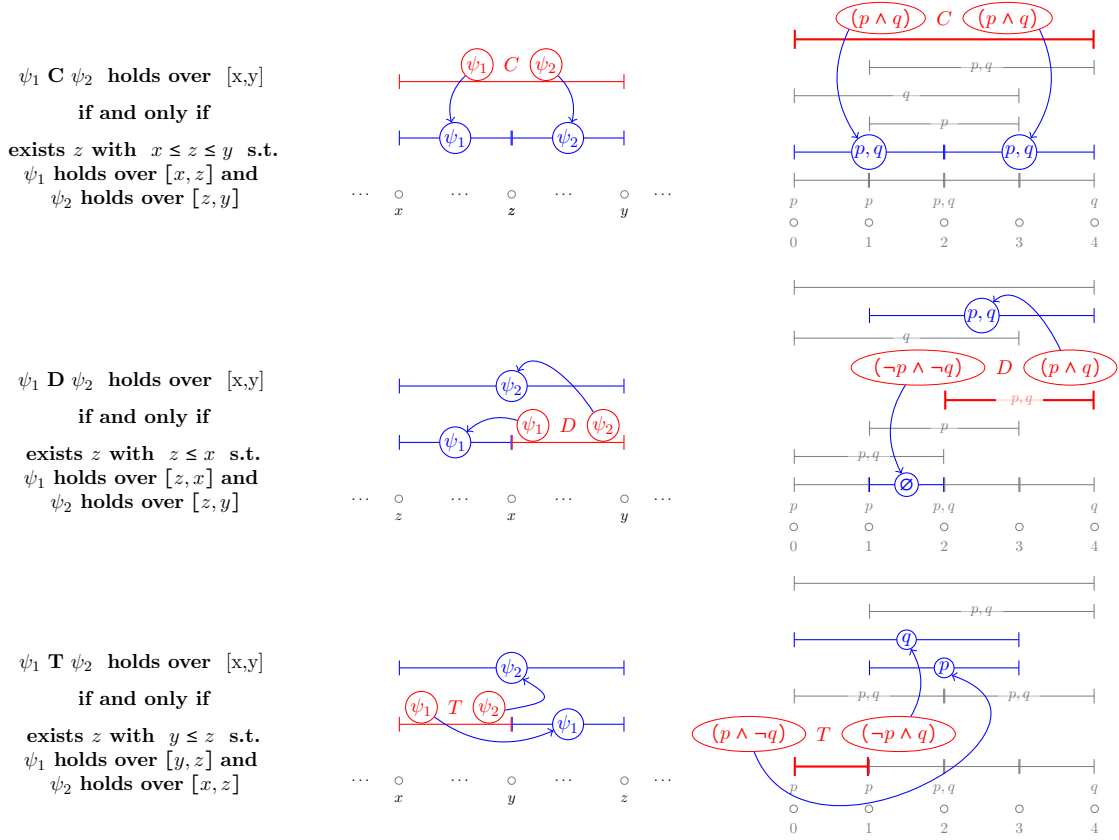
\begin{figure}
\begin{tikzpicture}[scale=1, node distance=2cm]

\begin{scope}[scale=.75, node distance=1cm]


\node[opacity=0.5,draw, circle, inner sep=1, label={[scale=1,opacity=0.5,]270:$\scalebox{0.525}{$ 0 $}$},
      label={[scale=1, opacity=0.5,]90:$\scalebox{0.525}{$ p $}$}](0-0) {};
      
\node[opacity=0.5,draw, circle, inner sep=1, label={[scale=1, opacity=0.5,]270:$\scalebox{0.525}{$ 1 $}$},
      label={[scale=1, opacity=0.5,]90:$\scalebox{0.525}{$ p $}$}](1-1) [right of=0-0] {};
      
\node[opacity=0.5,draw, circle, inner sep=1, label={[scale=1, opacity=0.5,]270:$\scalebox{0.525}{$ 2 $}$},
      label={[scale=1, opacity=0.5,]90:$\scalebox{0.525}{$ p,q $}$}](2-2) [right of=1-1] {};
      
\node[opacity=0.5,draw, circle, inner sep=1, label={[scale=1, opacity=0.5,]270:$\scalebox{0.525}{$ 3 $}$},
      label={[scale=1, opacity=0.5,]90:$\scalebox{0.525}{$  $}$}](3-3) [right of=2-2] {};
      
\node[opacity=0.5,draw, circle, inner sep=1, label={[scale=1, opacity=0.5,]270:$\scalebox{0.525}{$ 4 $}$},
      label={[scale=1, opacity=0.5,]90:$\scalebox{0.525}{$ q $}$}](4-4) [right of=3-3] {};
      
\draw[|-|, opacity=0.5] ($(0-0.center) + (0,0.75)$) -- ($(1-1.center)+ (0,0.75)$);


\draw[|-|, opacity=1, blue] ($(0-0.center) + (0,1.25)$) -- ($(2-2.center)+ (0,1.25)$)
node[pos=0.5, fill=white,  circle, draw, inner sep=0.5](PSI1) {$\scalebox{0.69}{$ p,q $}$};

\draw[|-|, opacity=0.5] ($(0-0.center) + (0,2.25)$) -- ($(3-3.center)+ (0,2.25)$)
node[pos=0.5, fill =white](47) {$\scalebox{0.525}{$ q $}$};

\draw[|-|, opacity=1, red, thick] ($(0-0.center) + (0,3.25)$) -- ($(4-4.center)+ (0,3.25)$)
node[pos=0.5, above=0.05cm](C) {\scalebox{0.69}{$C$}}
      node[anchor=east, ellipse, draw, line width=0.2pt, inner sep=1, draw, inner sep=0.5](C1) at (C.west) {\scalebox{0.69}{$(p\wedge q)$}}
      node[anchor=west, ellipse, draw, line width=0.2pt, inner sep=1, draw, inner sep=0.5](C2) at (C.east) {\scalebox{0.69}{$(p\wedge q)$}};

\draw[|-|, opacity=0.5] ($(1-1.center) + (0,0.75)$) -- ($(2-2.center)+ (0,0.75)$);

\draw[|-|, opacity=0.5] ($(1-1.center) + (0,1.75)$) -- ($(3-3.center)+ (0,1.75)$)
node[pos=0.5, fill =white](47) {$\scalebox{0.525}{$ p $}$};

\draw[|-|, opacity=0.5] ($(1-1.center) + (0,2.75)$) -- ($(4-4.center)+ (0,2.75)$)
node[pos=0.5, fill =white](47) {$\scalebox{0.525}{$ p,q $}$};

\draw[|-|, opacity=0.5] ($(2-2.center) + (0,0.75)$) -- ($(3-3.center)+ (0,0.75)$);


\draw[|-|, opacity=1, blue] ($(2-2.center) + (0,1.25)$) -- ($(4-4.center)+ (0,1.25)$)
node[pos=0.5, fill=white, circle, draw, inner sep=0.5](PSI2) {$\scalebox{0.69}{$ p,q $}$};

\draw[|-|, opacity=0.5] ($(3-3.center) + (0,0.75)$) -- ($(4-4.center)+ (0,0.75)$);

\draw (C1.south west) edge[->,bend right, blue] (PSI1);
\draw (C2.south east) edge[->,bend left, blue] (PSI2);

\pgftransformshift{\pgfpoint{-7cm}{0.5cm}}
 
\begin{scope}[node distance=1.5cm, opacity = 1] 

\node[opacity=0.5,draw, circle, inner sep=1, 
      label={[label distance=0.2cm]180:\scalebox{0.525}{$\ldots$}},
      label={[label distance=0.5cm]0:\scalebox{0.525}{$\ldots$}},
      label={[]270:\scalebox{0.525}{$x$}}](0-0) {};
      
\node[opacity=0.5,draw, circle, inner sep=1, 
      label={[]270:\scalebox{0.525}{$z$}},
      label={[]270:\scalebox{0.525}{$z$}}](1-1) [right of=0-0] {};
      
\node[opacity=0.5,draw, circle, inner sep=1, 
      label={[label distance=0.5cm]180:\scalebox{0.525}{$\ldots$}},
      label={[label distance=0.2cm]0:\scalebox{0.525}{$\ldots$}},
      label={[]270:\scalebox{0.525}{$y$}}](2-2) [right of=1-1] {};

\draw[|-|,  blue]
($(0-0.center) + (0,1)$) -- ($(1-1.center)+ (0,1)$)
node[pos=0.5, fill=white, scale=0.7, circle, draw, inner sep=0.5](PSI1)[]{$\psi_1$};

\draw[|-|,  blue]
($(1-1.center) + (0,1)$) -- ($(2-2.center)+ (0,1)$)
node[pos=0.5, fill=white, scale=0.7, circle, draw, inner sep=0.5](PSI2)[]{$\psi_2$};

\draw[|-|, red] 
      ($(0-0.center) + (0,2)$) -- ($(2-2.center) +(0,2)$)
      node[pos=0.5, above](C) {\scalebox{0.69}{$C$}}
      node[anchor=east, circle, draw, inner sep=0.5](C1) at (C.west) {\scalebox{0.69}{$\psi_1$}}
      node[anchor=west, circle, draw, inner sep=0.5](C2) at (C.east) {\scalebox{0.69}{$\psi_2$}};

\draw[->, blue] (C1) edge[bend right, blue] (PSI1);
\draw[->, blue] (C2) edge[bend left, blue] (PSI2);

\pgftransformshift{\pgfpoint{-4.5cm}{1cm}}

\node[scale=0.7]{
\begin{tabular}{c}
$\psi_1 \cdtC \psi_2 \mbox{ \textbf{ holds over } [x,y] }$ \\[0.25cm]
\textbf{ if and only if } \\[0.25cm]
$\textbf{ exists } z \textbf{ with }$
$x\leq z\leq y$ \textbf{ s.t. } \\
$\psi_1 \textbf{ holds over } [x,z] \textbf{ and }$ \\
$\psi_2 \textbf{ holds over } [z,y]$ 
\end{tabular}
};

\end{scope}

\pgftransformshift{\pgfpoint{7cm}{-5.9cm}}


\node[opacity=0.5,draw, circle, inner sep=1, label={[scale=1,opacity=0.5,]270:$\scalebox{0.525}{$ 0 $}$},
      label={[scale=1, opacity=0.5,]90:$\scalebox{0.525}{$ p $}$}](0-0) {};
      
\node[opacity=0.5,draw, circle, inner sep=1, label={[scale=1, opacity=0.5,]270:$\scalebox{0.525}{$ 1 $}$},
      label={[scale=1, opacity=0.5,]90:$\scalebox{0.525}{$ p $}$}](1-1) [right of=0-0] {};
      
\node[opacity=0.5,draw, circle, inner sep=1, label={[scale=1, opacity=0.5,]270:$\scalebox{0.525}{$ 2 $}$},
      label={[scale=1, opacity=0.5,]90:$\scalebox{0.525}{$ p,q $}$}](2-2) [right of=1-1] {};
      
\node[opacity=0.5,draw, circle, inner sep=1, label={[scale=1, opacity=0.5,]270:$\scalebox{0.525}{$ 3 $}$},
      label={[scale=1, opacity=0.5,]90:$\scalebox{0.525}{$  $}$}](3-3) [right of=2-2] {};
      
\node[opacity=0.5,draw, circle, inner sep=1, label={[scale=1, opacity=0.5,]270:$\scalebox{0.525}{$ 4 $}$},
      label={[scale=1, opacity=0.5,]90:$\scalebox{0.525}{$ q $}$}](4-4) [right of=3-3] {};
      
\draw[|-|, opacity=0.5] ($(0-0.center) + (0,0.75)$) -- ($(1-1.center)+ (0,0.75)$);


\draw[|-|, opacity=0.5] ($(0-0.center) + (0,1.25)$) -- 
($(2-2.center)+ (0,1.25)$)
node[pos=0.5, fill =white](47) {$\scalebox{0.525}{$ p, q $}$};

\draw[|-|, opacity=0.5] ($(0-0.center) + (0,3.35)$) -- 
($(3-3.center)+ (0,3.35)$)
node[pos=0.5, fill =white](47) {$\scalebox{0.525}{$ q $}$};

\draw[|-|, opacity=0.5] ($(0-0.center) + (0,4.25)$) -- 
($(4-4.center)+ (0,4.25)$);

\draw[|-|, opacity=1, blue] ($(1-1.center) + (0,0.75)$) -- 
($(2-2.center)+ (0,0.75)$)
node[pos=0.5, fill=white,  circle, draw, inner sep=0.5](PSI1) 
{$\scalebox{0.69}{$\emptyset$}$};


\draw[|-|, opacity=0.5] ($(1-1.center) + (0,1.75)$) -- ($(3-3.center)+ (0,1.75)$)
node[pos=0.5, fill =white](47) {$\scalebox{0.525}{$ p $}$};


\draw[|-|, opacity=1, blue] ($(1-1.center) + (0,3.75)$) -- ($(4-4.center)+ (0,3.75)$)
node[pos=0.5, fill=white, circle, draw, inner sep=0.5](PSI2) {$\scalebox{0.69}{$ p,q $}$};;

\draw[|-|, opacity=0.5] ($(2-2.center) + (0,0.75)$) -- ($(3-3.center)+ (0,0.75)$);


\draw[|-|, opacity=1, , opacity=1, red, thick] ($(2-2.center) + (0,2.25)$) -- ($(4-4.center)+ (0,2.25)$)
node[pos=0.4, above=0.2cm](D) {\scalebox{0.69}{$D$}}
      node[anchor=east, ellipse, draw, line width=0.2pt, inner sep=1](D1) at (D.west) {\scalebox{0.69}{$(\neg p\wedge \neg q)$}}
      node[anchor=west, ellipse, draw, inner sep=1, line width=0.2pt](D2) at (D.east) {\scalebox{0.69}{$(p\wedge q)$}}
node[pos=0.5, fill =white, opacity=0.9](47) {$\scalebox{0.525}{{\transparent{0.5} $ p, q $}}$}      ;

\draw[|-|, opacity=0.5] ($(3-3.center) + (0,0.75)$) -- ($(4-4.center)+ (0,0.75)$);

\draw[->, blue] (D1) edge[bend right, blue] (PSI1);
\draw[->, blue] (D2) edge[bend right, blue, in=-100] (PSI2);

\pgftransformshift{\pgfpoint{-7cm}{0.5cm}}
 
\begin{scope}[node distance=1.5cm, opacity = 1] 

\node[opacity=0.5,draw, circle, inner sep=1, 
      label={[label distance=0.2cm]180:\scalebox{0.525}{$\ldots$}},
      label={[label distance=0.5cm]0:\scalebox{0.525}{$\ldots$}},
      label={[]270:\scalebox{0.525}{$z$}}](0-0) {};
      
\node[opacity=0.5,draw, circle, inner sep=1,,
      label={[]270:\scalebox{0.525}{$x$}}](1-1) [right of=0-0] {};
      
\node[opacity=0.5,draw, circle, inner sep=1, 
      label={[label distance=0.5cm]180:\scalebox{0.525}{$\ldots$}},
      label={[label distance=0.2cm]0:\scalebox{0.525}{$\ldots$}},
      label={[]270:\scalebox{0.525}{$y$}}](2-2) [right of=1-1] {};

\draw[|-|,  blue]
($(0-0.center) + (0,1)$) -- ($(1-1.center)+ (0,1)$)
node[pos=0.5, fill=white, scale=0.7, circle, draw, inner sep=0.5](PSI1)[]{$\psi_1$};

\draw[|-|,  red]
($(1-1.center) + (0,1)$) -- ($(2-2.center)+ (0,1)$)
      node[pos=0.5, above](D) {\scalebox{0.69}{$D$}}
      node[anchor=east, circle, draw, inner sep=0.5](D1) at (D.west) {\scalebox{0.69}{$\psi_1$}}
      node[anchor=west, circle, draw, inner sep=0.5](D2) at (D.east) {\scalebox{0.69}{$\psi_2$}};

\draw[|-|, blue] ($(0-0.center) + (0,2)$) -- ($(2-2.center) +(0,2)$)
node[pos=0.5, fill=white, scale=0.7, circle, draw, inner sep=0.5](PSI2)[]{$\psi_2$};

\draw[->, blue] (D1) edge[bend right, blue] (PSI1);
\draw[->, blue] (D2) edge[bend right, blue, in=-90] (PSI2);

\pgftransformshift{\pgfpoint{-4.5cm}{1cm}}

\node[scale=0.7]{
\begin{tabular}{c}
$\psi_1 \cdtD \psi_2 \mbox{ \textbf{ holds over } [x,y] }$ \\[0.25cm]
\textbf{ if and only if } \\[0.25cm]
$\textbf{ exists } z \textbf{ with }$
$ z\leq x $ \textbf{ s.t. } \\
$\psi_1 \textbf{ holds over } [z,x] \textbf{ and }$ \\
$\psi_2 \textbf{ holds over } [z,y]$ 
\end{tabular}
};

\end{scope}

\pgftransformshift{\pgfpoint{7cm}{-5.5cm}}


\node[opacity=0.5,draw, circle, inner sep=1, label={[scale=1,opacity=0.5,]270:$\scalebox{0.525}{$ 0 $}$},
      label={[scale=1, opacity=0.5,]90:$\scalebox{0.525}{$ p $}$}](0-0) {};
      
\node[opacity=0.5,draw, circle, inner sep=1, label={[scale=1, opacity=0.5,]270:$\scalebox{0.525}{$ 1 $}$},
      label={[scale=1, opacity=0.5,]90:$\scalebox{0.525}{$ p $}$}](1-1) [right of=0-0] {};
      
\node[opacity=0.5,draw, circle, inner sep=1, label={[scale=1, opacity=0.5,]270:$\scalebox{0.525}{$ 2 $}$},
      label={[scale=1, opacity=0.5,]90:$\scalebox{0.525}{$ p,q $}$}](2-2) [right of=1-1] {};
      
\node[opacity=0.5,draw, circle, inner sep=1, label={[scale=1, opacity=0.5,]270:$\scalebox{0.525}{$ 3 $}$},
      label={[scale=1, opacity=0.5,]90:$\scalebox{0.525}{$  $}$}](3-3) [right of=2-2] {};
      
\node[opacity=0.5,draw, circle, inner sep=1, label={[scale=1, opacity=0.5,]270:$\scalebox{0.525}{$ 4 $}$},
      label={[scale=1, opacity=0.5,]90:$\scalebox{0.525}{$ q $}$}](4-4) [right of=3-3] {};
      
\draw[|-|, opacity=1, red, thick] ($(0-0.center) + (0,0.75)$) -- ($(1-1.center)+ (0,0.75)$)
node[pos=0.4, above=0.2](T) {\scalebox{0.69}{$T$}}
      node[anchor=east, ellipse, draw, line width=0.2pt, inner sep=1](T1) at (T.west) {\scalebox{0.69}{$(p\wedge \neg q)$}}
      node[anchor=west, ellipse, draw, line width=0.2pt, inner sep=1](T2) at (T.east) {\scalebox{0.69}{$(\neg p\wedge q)$}}
;


\draw[|-|, opacity=0.5] ($(0-0.center) + (0,2)$) -- ($(2-2.center)+ (0,2)$)
node[pos=0.5, fill =white](47) {$\scalebox{0.525}{$ p,q $}$};


\draw[|-|, opacity=1, blue] ($(0-0.center) + (0,3)$) -- ($(3-3.center)+ (0,3)$)
node[pos=0.5, fill =white, , circle, draw, inner sep=0.5](PSI2) {$\scalebox{0.69}{$ q $}$};

\draw[|-|, opacity=0.5] ($(0-0.center) + (0,4)$) -- ($(4-4.center)+ (0,4)$);

\draw[|-|, opacity=0.5] ($(1-1.center) + (0,0.75)$) -- ($(2-2.center)+ (0,0.75)$)
;


\draw[|-|, opacity=1, blue] ($(1-1.center) + (0,2.5)$) -- ($(3-3.center)+ (0,2.5)$)
node[pos=0.5, fill =white, circle, draw, inner sep=0.5](PSI1) {$\scalebox{0.69}{$ p $}$};


\draw[|-|, opacity=0.5] ($(1-1.center) + (0,3.5)$) -- ($(4-4.center)+ (0,3.5)$)
node[pos=0.5, fill=white] {$\scalebox{0.525}{$ p,q $}$};

\draw[|-|, opacity=0.5] ($(2-2.center) + (0,0.75)$) -- ($(3-3.center)+ (0,0.75)$);


\draw[|-|, opacity=0.5] ($(2-2.center) + (0,2)$) -- ($(4-4.center)+ (0,2)$)
node[pos=0.5, fill =white](47) {$\scalebox{0.525}{$ p, q $}$};

\draw[|-|, opacity=0.5] ($(3-3.center) + (0,0.75)$) -- ($(4-4.center)+ (0,0.75)$);

\draw (T1) edge[->, bend right, blue, looseness=1.8, out=270, in=-40] (PSI1.south east);
\draw (T2) edge[->, bend right, blue] (PSI2);

\pgftransformshift{\pgfpoint{-7cm}{0.5cm}}
 
\begin{scope}[node distance=1.5cm, opacity = 1] 

\node[opacity=0.5,draw, circle, inner sep=1, 
      label={[label distance=0.2cm]180:\scalebox{0.525}{$\ldots$}},
      label={[label distance=0.5cm]0:\scalebox{0.525}{$\ldots$}},
      label={[]270:\scalebox{0.525}{$x$}}](0-0) {};
      
\node[opacity=0.5,draw, circle, inner sep=1,,
      label={[]270:\scalebox{0.525}{$y$}}](1-1) [right of=0-0] {};
      
\node[opacity=0.5,draw, circle, inner sep=1, 
      label={[label distance=0.5cm]180:\scalebox{0.525}{$\ldots$}},
      label={[label distance=0.2cm]0:\scalebox{0.525}{$\ldots$}},
      label={[]270:\scalebox{0.525}{$z$}}](2-2) [right of=1-1] {};

\draw[|-|,  red]
($(0-0.center) + (0,1)$) -- ($(1-1.center)+ (0,1)$)
node[pos=0.5, above](T) {\scalebox{0.69}{$T$}}
      node[anchor=east, circle, draw, inner sep=0.5](T1) at (T.west) {\scalebox{0.69}{$\psi_1$}}
      node[anchor=west, circle, draw, inner sep=0.5](T2) at (T.east) {\scalebox{0.69}{$\psi_2$}}
;

\draw[|-|,  blue]
($(1-1.center) + (0,1)$) -- ($(2-2.center)+ (0,1)$)
 node[pos=0.5, fill=white, scale=0.7, circle, draw, inner sep=0.5](PSI1)[]{$\psi_1$}     ;

\draw[|-|, blue] ($(0-0.center) + (0,2)$) -- ($(2-2.center) +(0,2)$)
node[pos=0.5, fill=white, scale=0.7, circle, draw, inner sep=0.5](PSI2)[]{$\psi_2$};

\draw[->, blue] (T1) edge[bend right, blue] (PSI1);
\draw(T2) edge[->,bend right, blue, looseness=3, in=-90, out=-50] (PSI2.south east);

\pgftransformshift{\pgfpoint{-4.5cm}{1cm}}

\node[scale=0.7]{
\begin{tabular}{c}
$\psi_1 \cdtT \psi_2 \mbox{ \textbf{ holds over } [x,y] }$ \\[0.25cm]
\textbf{ if and only if } \\[0.25cm]
$\textbf{ exists } z \textbf{ with }$
$ y \leq z$ \textbf{ s.t. } \\
$\psi_1 \textbf{ holds over } [y,z] \textbf{ and }$ \\
$\psi_2 \textbf{ holds over } [x,z]$ 
\end{tabular}
};

\end{scope}

\end{scope}

\end{tikzpicture}

\caption{\label{fig:cdtsemantics} The semantics of the three $\mathsf{CDT}$ binary modalities $\mathsf{C}$, $\mathsf{D}$, and $\mathsf{T}$.}

\end{figure}

One of the first ITLs proposed in literature was $\mathsf{CDT}$ \cite{DBLP:journals/ndjfl/Venema90}, whose name comes from its three binary modalities $\cdtC$
(\emph{Chopping}), $\cdtD$ (\emph{Dawning}), and $\cdtT$ 
(\emph{Terminating}). Their semantics is graphically depicted in Figure~\ref{fig:cdtsemantics}. Intuitively speaking, if we take a point $z$ 
inside an interval $[x,y]$ and we consider the ternary relation 
$[x,y]$ \emph{ may be split into } $[x,z]$ \emph{ and }  $[z,y]$,
the three \LogicCDT modalities allow one to talk about 
the properties of such a relation starting from any of the three 
intervals. More precisely, a formula $\psi_1 \cdtC \psi_2$ (\emph{chopping  between $\psi_1$ and $\psi_2$}) holds over an
interval $[x,y]$ if $[x,y]$ can be split into  $[x,z]$ and $[z,y]$, $\psi_1$ holds over $[x,z]$, and $\psi_2$ holds over $[z,y]$ (topmost part of Figure~\ref{fig:cdtsemantics}). 
A formula $\psi_1 \cdtD \psi_2$ (\emph{$\psi_1$ dawning $\psi_2$}) holds over an interval $[x,y]$ if 
there exists an interval $[z,x]$ such that $\psi_1$ holds over $[z,x]$ 
and $\psi_2$ holds over the interval $[z,y]$ covering both $[z,x]$ and $[x,y]$ (middle part of
Figure~\ref{fig:cdtsemantics}). A formula $\psi_1 \cdtT 
\psi_2$ (\emph{$\psi_1$ terminating $\psi_2$}) holds 
over an interval $[x,y]$ if there exists an interval $[y,z]$ such that
$\psi_1$ holds over $[y,z]$ and $\psi_2$ holds over the interval $[x,z]$
covering both $[x,y]$ and $[y,z]$
(bottom part of Figure~\ref{fig:cdtsemantics}).

$\mathsf{CDT}$ turns out to be very expressive. It can be easily checked that it allows one to specify a number of meaningful properties in 
a rather straightforward way. As an example, it is easy to write a
$\mathsf{CDT}$ formula that forces one or more proposition
letters to behave like an equivalence relation over the points of 
the underlying linear order. However, such an expressivity 
is paid with an undecidable satisfiability problem on every 
interesting linear order, that is, any linear order but finite, bounded ones,
where the problem is trivially decidable. The same undecidability results hold for any of the $\mathsf{CDT}$ fragments that feature just one modality among $\cdtC\!$, $\cdtD\!$, and $\cdtT$ 
\cite{DBLP:journals/japll/GorankoMSS06,DBLP:conf/csl/HodkinsonMS08}.

A representative fragment of $\mathsf{CDT}$ is $\mathsf{HS}$ \cite{DBLP:journals/jacm/HalpernS91}, which includes a unary modality for each ordering relation between a pair of intervals (the so-called Allen's relations \cite{allen1981interval}). A graphical account of Allen's relations and the corresponding $\mathsf{HS}$ modalities is given in Figure~\ref{fig:allenRelations}. 
For the sake of simplicity, we deliberately omitted the modality for the inverse of each of the depicted relations, namely $\hsEAbar, \hsEBbar, \hsEDbar,\hsEEbar, \hsELbar$, and $\hsEObar$. 
The semantics of each $\mathsf{HS}$ modality can be expressed by means of a suitable combination of $\mathsf{CDT}$ modalities as shown in 
Appendix~\ref{appendix:cdt}.
The converse is not true. In Figure~\ref{fig:hsoperators}, we make an extensive use of the modal 
constant $\pi$, which holds over an interval $[x,y]$ if and only if 
$x = y$, that is, $[x,y]$ is a point-interval.
It immediately follows that $\neg \pi$ holds on all and only strict intervals.
It is worth pointing out that some $\mathsf{HS}$ modalities can be defined as suitable combinations of other ones (a complete account of definability equations for the most
significant classes of linear orders is given in \cite{DBLP:journals/tcs/BresolinMMSS14,DBLP:journals/iandc/BresolinMMSS19}). For what concerns the $\mathsf{HS}$ fragments we focus on in this paper, namely those featuring unary modalities $\hsEA, \hsEB$, and $\hsED$ (not to be confused with the binary modality $\cdtD$ of $\mathsf{CDT}$), it holds that modality $\hsEL$ can be defined in terms of modality $\hsEA$ and modality $\hsED$ can be expressed as a suitable combination of modalities $\hsEB$ and $\hsEE$. Notice that the opposite is not true, e.g., modality $\hsEA$ cannot be defined in terms of modalities $\hsEL, \hsED$, and $\hsEB$ and it is not possible to define modality $\hsEB$ (resp., $\hsED$) in terms of modalities $\hsEA$ and $\hsED$ (resp., $\hsEB$). 


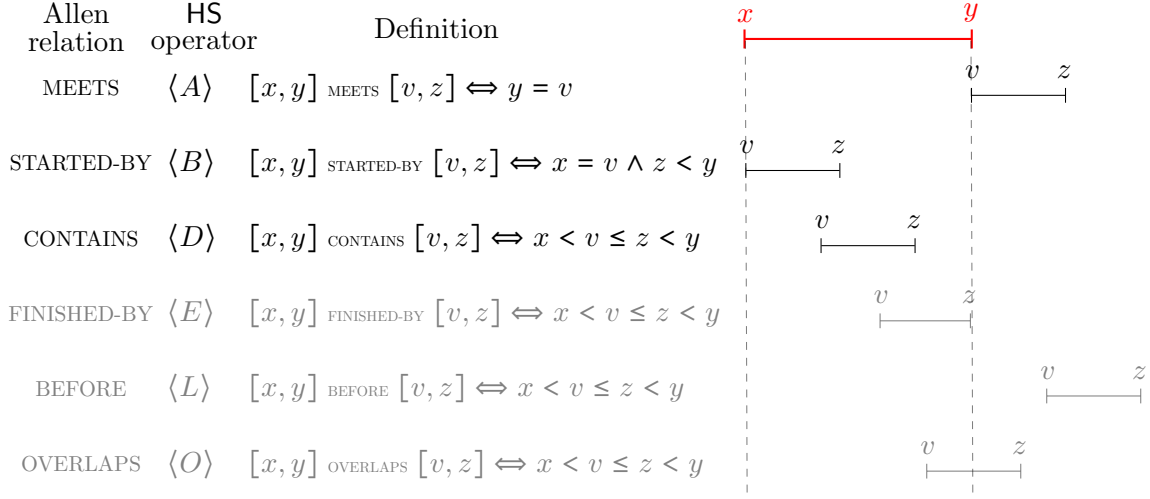
\begin{figure}
\centering

\begin{tikzpicture}[
	inner sep =0pt,
	anchor= west]

\node(Allen){Allen};
\node[anchor=north, yshift=-0.1cm]at (Allen.south){relation};

\node[xshift=1cm](HS) at (Allen.east){$\mathsf{HS}$};
\node[anchor=north, yshift=-0.1cm](HSop) at (HS.south) {operator};

\node[xshift=2cm, yshift=-0.2cm](Definition) at (HS.east){Definition};

\node[xshift=5cm](Example) at (Definition.east){\phantom{Example}};

\begin{scope}[node distance=1cm]

\node(A)[below of=Allen, node distance=1cm]{\allenTextA};
\node(B)[below of=A]{\allenTextB};
\node(D)[below of=B]{\allenTextD};

\begin{scope}[opacity=0.5]

\node(E)[below of=D]{\allenTextE};
\node(L)[below of=E]{\allenTextL};
\node(O)[below of=L]{\allenTextO};

\end{scope}

\path let \p1 = (A), \p2 = (HSop) in node[xshift=-0.5cm](HSA) at (\x2, \y1) {$\hsEA$};

\path let \p1 = (B), \p2 = (HSop) in node[xshift=-0.5cm](HSA) at (\x2, \y1) {$\hsEB$};

\path let \p1 = (D), \p2 = (HSop) in node[xshift=-0.5cm](HSA) at (\x2, \y1) {$\hsED$};

\begin{scope}[opacity=0.5]

\path let \p1 = (E), \p2 = (HSop) in node[xshift=-0.5cm](HSA) at (\x2, \y1) {$\hsEE$};

\path let \p1 = (L), \p2 = (HSop) in node[xshift=-0.5cm](HSA) at (\x2, \y1) {$\hsEL$};

\path let \p1 = (O), \p2 = (HSop) in node[xshift=-0.5cm](HSA) at (\x2, \y1) {$\hsEO$};

\end{scope}

\def\xshift{-2.5cm}

\path let \p1 = (A), \p2 = (Definition) in node[xshift=\xshift](DA) at (\x2, \y1) {$[x,y] \allenA  [v,z] \Leftrightarrow y = v$}; 

\path let \p1 = (B), \p2 = (Definition) in node[xshift=\xshift](DA) at (\x2, \y1) {$[x,y] \allenB  [v,z] \Leftrightarrow x = v \wedge z < y$}; 

\path let \p1 = (D), \p2 = (Definition) in node[xshift=\xshift](DA) at (\x2, \y1) {$[x,y] \allenD  [v,z] \Leftrightarrow x < v \leq z < y$}; 

\begin{scope}[opacity=0.5]

\path let \p1 = (E), \p2 = (Definition) in node[xshift=\xshift](DA) at (\x2, \y1) {$[x,y] \allenE  [v,z] \Leftrightarrow x < v \leq z < y$}; 

\path let \p1 = (L), \p2 = (Definition) in node[xshift=\xshift](DA) at (\x2, \y1) {$[x,y] \allenL  [v,z] \Leftrightarrow x < v \leq z < y$}; 

\path let \p1 = (O), \p2 = (Definition) in node[xshift=\xshift](DA) at (\x2, \y1) {$[x,y] \allenO  [v,z] \Leftrightarrow x < v \leq z < y$}; 

\end{scope}

\path let \p1 = (Example), \p2 = (Example) in 
	node[ yshift=-0.15cm, xshift=-2.5cm,
	      label={[label distance=0.2cm, red]90:$x$}
	    ](X) at (\x2, \y1) {};

\path let \p1 = (X), \p2 = (Example) in 
	node[ xshift=0.5cm,
	      label={[label distance=0.2cm, red]90:$y$}
		](Y) at (\x2, \y1) {};

\path let \p1 = (O), \p2 = (X) in 
	node[yshift=-0.5cm](Xproj) at (\x2, \y1) {};

\path let \p1 = (O), \p2 = (Y) in 
	node[yshift=-0.5cm](Yproj) at (\x2, \y1) {};

\draw[|-|, thick, red] (X.west) -- (Y.east);
\draw[dashed, opacity=0.5] (X) -- (Xproj);
\draw[dashed, opacity=0.5] (Y) -- (Yproj);


\path let \p1 = (A), \p2 = (Y) in 
	node[ 	xshift=0cm, yshift=-0.1cm,
			label={[label distance=0.2cm]90:$v$}
		](V) at (\x2, \y1) {};

\path let \p1 = (V), \p2 = (V) in 
	node[ xshift=1.19cm,
	      label={[label distance=0.2cm]90:$z$}
		](Z) at (\x2, \y1) {};

\draw[|-|] (V.west) -- (Z.east);


\path let \p1 = (B), \p2 = (X) in 
	node[ 	xshift=0cm, yshift=-0.1cm,
			label={[label distance=0.2cm]90:$v$}
		](V) at (\x2, \y1) {};

\path let \p1 = (V), \p2 = (V) in 
	node[ xshift=1.19cm,
	      label={[label distance=0.2cm]90:$z$}
		](Z) at (\x2, \y1) {};

\draw[|-|] (V.west) -- (Z.east);


\path let \p1 = (D), \p2 = (X) in 
	node[ 	xshift=1cm, yshift=-0.1cm,
			label={[label distance=0.2cm]90:$v$}
		](V) at (\x2, \y1) {};

\path let \p1 = (V), \p2 = (V) in 
	node[ xshift=1.19cm,
	      label={[label distance=0.2cm]90:$z$}
		](Z) at (\x2, \y1) {};

\draw[|-|] (V.west) -- (Z.east);

\begin{scope}[opacity=0.5]

\path let \p1 = (E), \p2 = (Y) in 
	node[ xshift=-0.05cm, yshift=-0.1cm,
	      label={[label distance=0.2cm]90:$z$}
		](Z) at (\x2, \y1) {};

\path let \p1 = (Z), \p2 = (Z) in 
	node[ 	xshift=-1.19cm,
			label={[label distance=0.2cm]90:$v$}
		](V) at (\x2, \y1) {};

\draw[|-|] (V.west) -- (Z.east);


\path let \p1 = (L), \p2 = (Y) in 
	node[ 	xshift=1cm, yshift=-0.1cm,
			label={[label distance=0.2cm]90:$v$}
		](V) at (\x2, \y1) {};

\path let \p1 = (V), \p2 = (V) in 
	node[ xshift=1.19cm,
	      label={[label distance=0.2cm]90:$z$}
		](Z) at (\x2, \y1) {};

\draw[|-|] (V.west) -- (Z.east);


\path let \p1 = (O), \p2 = (Y) in 
	node[ 	xshift=-0.595cm, yshift=-0.1cm,
			label={[label distance=0.2cm]90:$v$}
		](V) at (\x2, \y1) {};

\path let \p1 = (V), \p2 = (V) in 
	node[ xshift=1.19cm,
	      label={[label distance=0.2cm]90:$z$}
		](Z) at (\x2, \y1) {};

\draw[|-|] (V.west) -- (Z.east);

\end{scope}

\end{scope}

\end{tikzpicture}

\caption{\label{fig:allenRelations} 
Allen's relations and the corresponding $\mathsf{HS}$ modalities (the
 relations/modalities considered in this work are highlighted). }


\end{figure}

We conclude the section by showing that   $\mathsf{LTL}_f$ modalities \emph{Until} ($\psi_1 \ltlUntil \psi_2$)  and \emph{Next} ($\ltlNext \psi$) can be  easily encoded  by means of a combination of modalities $\hsEA$ and $\hsEB$ (no need to bring up modality $\hsED$).  
Formulas that define $\psi_1 \ltlUntil \psi_2$ and $\ltlNext \psi$ in $\mathsf{AB}$, together with a graphical account of how they ``operate'' on an interval model, are given in Figure~\ref{fig:abuntil}. 
The $\mathsf{LTL}_f$ formula into which $\psi_1 \ltlUntil \psi_2$ is mapped is the conjunction of formulas $\hsAB\hsEA(\pi \wedge \psi_1)$ and  $\hsEA(\pi \wedge \psi_2)$. By definition,  $\psi_1 \ltlUntil \psi_2$ holds at a point $x$ if there is a point $y$, with $x\leq y$, where $\psi_2$ holds and $\psi_1$ holds at $x_i$, for all $x \leq x_i < y$.
The proposed encoding forces the translation of $\psi_1 \ltlUntil \psi_2$ to hold over the interval $[x,y]$. The second conjunct of the resulting formula constrains $\psi_2$ to hold on  $[y,y]$ (the right endpoint of $[x,y]$). The formula $\hsEA(\pi \wedge \psi_2)$ indeed states that there is an interval $[y,y']$, that starts where $[x,y]$ ends (modality $\hsEA$), which is a point-interval (constant $\pi$) and satisfies $\psi_2$. It immediately follows that $[y,y']$ is equal to $[y,y]$. The first conjunct $\hsAB\hsEA(\pi \wedge \psi_1)$ forces the formula $\hsEA(\pi \wedge \psi_1)$ to hold on each proper prefix ($\hsAB= \neg \hsEB \neg$) of the interval $[x,y]$, that is, on each interval $[x,x_i]$, with $x \leq x_i < y$. Then, by the argument we already used for $\hsEA(\pi \wedge \psi_2)$, we can conclude that $\psi_1$ holds at each point-interval $[x_i,x_i]$, with $x \leq x_i <y$.  
As for modality $\ltlNext$, the $\mathsf{LTL}_f$ formula $\ltlNext \psi$ is mapped into the $\mathsf{AB}$ formula  $\hsEA(\neg \pi \wedge \hsAB \pi \wedge \hsEA(\pi \wedge \psi) )$. By definition, $\ltlNext \psi$ holds at a point $x$ if and only if $\psi$ holds at the point $x+1$. For the sake of generality and simplicity, the proposed encoding of $\ltlNext \psi$ holds on an interval $[x,y]$ if and only if $\psi$ holds at the point-interval $[y+1,y+1]$ regardless of the fact that $[x,y]$ is a strict- or a point-interval ($[x,y]$ can be forced to be a point-intervals by adding $\pi$ as a conjunct of the resulting formula, i.e., $\pi \wedge \hsEA(\neg \pi \wedge \hsAB \pi \wedge \hsEA(\pi \wedge \psi) )$). Whenever $\hsEA(\neg \pi \wedge \hsAB \pi \wedge \hsEA(\pi \wedge \psi) )$ holds over an interval $[x,y]$, the outermost modality $\hsEA$ imposes
the existence of an interval $[y,y']$, with $y\leq y'$, where $\neg \pi$, $\hsAB \pi$, and $\hsEA(\pi \wedge \psi)$ hold. The first two conjuncts  respectively force $y'>y$ ($\neg \pi$) and all proper prefixes $[y,y'']$ of $[y,y']$ to be point-intervals ($\hsAB \pi$). The only way to satisfy them is to constrain $y'$ to be equal to $y+1$.
The truth of $\hsEA(\pi \wedge \psi)$ on 
$[y, y+1]$ implies that there is an interval $[y+1, y']$ where both $\pi$ and $\psi$ hold, which amounts to say that $\psi$ holds over the point-interval $[y+1, y+1]$. 
 
It is worth pointing out that the truth values of proposition letters on strict-intervals do not come into play in the proposed encoding. It immediately follows that such an encoding still properly works under the homogeneity assumption.


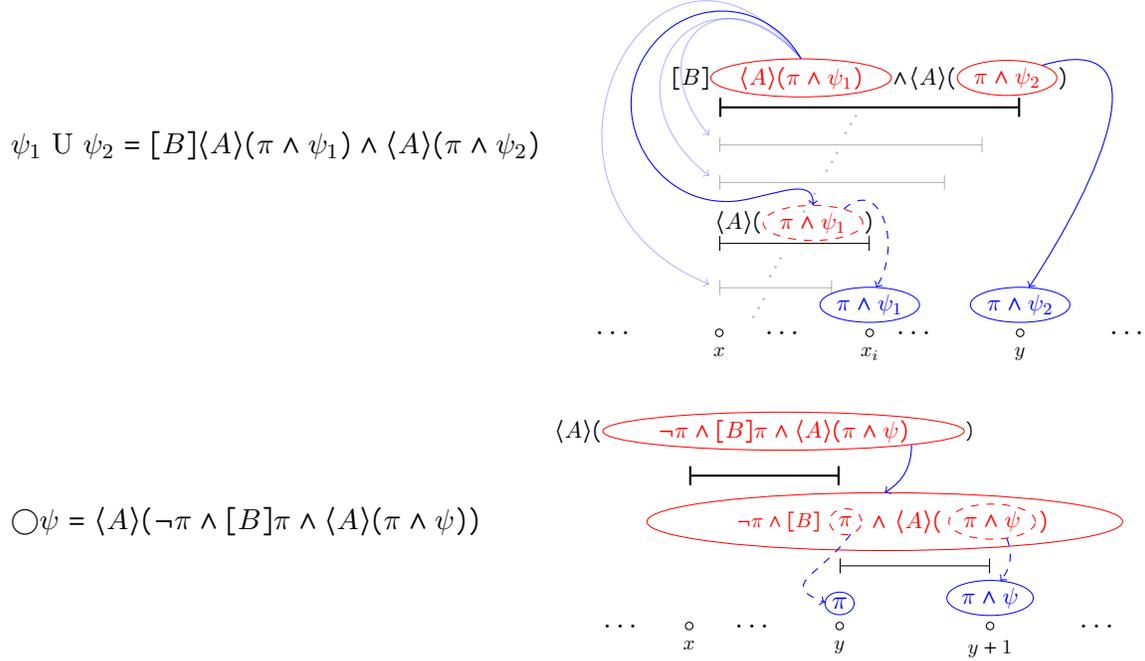
\begin{figure}
\vspace{-1.5cm}
\begin{tikzpicture}[scale=1, node distance=1cm, remember picture]
\node(UT){$\psi_1 \ltlUntil \psi_2 =  \hsAB\hsEA (\pi \wedge \psi_1) \wedge \hsEA(\pi \wedge \psi_2)$};

\pgftransformshift{\pgfpoint{5.9cm}{-2.5cm}}


\begin{scope}[node distance=2cm]

\node[draw, circle, inner sep=1, 
  label={[]270:$\scalebox{0.7}{$ x $}$},
  label={[label distance=1cm ]180:$\ldots$}, 
  label={[]90:$\scalebox{0.7}{$  $}$}](0-0) [] {$ $};
  
 \node[draw, circle, inner sep=1, 
  label={[]270:$\scalebox{0.7}{$ x_i $}$},
  label={[label distance=0.75cm ]180:$\ldots$}, 
  label={[]90:$\scalebox{0.7}{$  $}$}](1-1) [right of=0-0] {$ $};
  
\node[draw, circle, inner sep=1, 
  label={[]270:$\scalebox{0.7}{$ y $}$},
  label={[label distance=1cm ]180:$\ldots$}, 
  label={[label distance=1cm ]0:$\ldots$}, 
  label={[]90:$\scalebox{0.7}{$  $}$}](2-2) [right of=1-1] {$ $};

\draw[|-|, thick] ($(0-0.center) + (0,3)$) -- ($(2-2.center)+ (0,3)$)
        node[pos=0.5, above](47) {
        \begin{tikzpicture}[inner sep=0, anchor=west, node distance=0, remember picture]
               \node[scale=0.8](B) {$\hsAB$};
               \node[draw,ellipse,  red,  line width=0.2pt, inner sep=1, draw, scale=0.8](PSI1) at (B.east)  {$\hsEA (\pi \wedge \psi_1)$};
               \node[scale=0.8](AND) at (PSI1.east) {$\wedge$};
               \node[scale=0.8](A) at (AND.east) {$\hsEA($};
               \node[draw,ellipse,  line width=0.2pt, inner sep=1,  red, scale=0.8](PSI2) at (A.east) {$\pi \wedge \psi_2$};
               \node[scale=0.8] at(PSI2.east) {$)$};
        \end{tikzpicture}
        };

\node[anchor=south, draw, ellipse, blue,  line width=0.2pt, inner sep=1, scale=0.8, yshift=0.1cm](YL) at (2-2.north){$\pi \wedge \psi_2$};  
  
\draw (PSI2) edge[bend left, blue,->, out=110, looseness=1.5] (YL);

\draw[|-|, opacity=0.3] ($(0-0.center) + (0,2.5)$) -- ($(2-2.center)+ (-0.5,2.5)$)
    node[pos=0.5, above, scale=0.8] {\rotatebox{60}{$\ldots$}}
    node[pos=0](B1) {};
\draw[|-|, opacity=0.3] ($(0-0.center) + (0,2)$) -- ($(2-2.center)+ (-1,2)$)
    node[pos=0.5, above, scale=0.8] {\rotatebox{60}{$\ldots$}}
    node[pos=0.4, below, scale=0.8] {\rotatebox{60}{$\ldots$}}
    node[pos=0](B2) {};
\draw[|-|] ($(0-0.center) + (0,1.19)$) -- ($(2-2.center)+ (-2,1.19)$)
    node(BPSI1)[pos=0.5, above=-0.1cm] {
        \begin{tikzpicture}[inner sep=0, anchor=west, node distance=0]
               \node[scale=0.8](A) {$\hsEA($};
               \node[draw,ellipse,  red,scale=0.8 ,  line width=0.2pt, inner sep=1,dashed](PSI1A) at (A.east)  {$\pi \wedge \psi_1$};
              \node[scale=0.8] at (PSI1A.east) {$)$};
        \end{tikzpicture}
    };
\draw[|-|, opacity=0.3] ($(0-0.center) + (0,0.6)$) -- ($(2-2.center)+ (-2.5,0.6)$)
    node[pos=0.5, above, scale=0.8] {\rotatebox{60}{$\ldots$}}
    node[pos=0.35, below, scale=0.8] {\rotatebox{60}{$\ldots$}}
    node[pos=0](B3) {};;
  
\draw[opacity=0.3] (PSI1.north) edge[bend right, blue,->, looseness=2.5, out=-90, in=270] (B1);

\draw[blue] (PSI1.north) edge[bend right, looseness=1.19, in=-120, out=-60] ($(PSI1.north) + (-2.1,0)$);
\draw[blue] ($(PSI1.north) + (-2.1,0)$) edge[bend right, looseness=1.1, in=-120, out=-60] 
($(PSI1.north) + (-1,-1.9)$);
\draw[blue] ($(PSI1.north) + (-1,-1.9)$) edge[->, bend left, looseness=1, in=90, out=0] 
($(PSI1A.north)$);

\draw[opacity=0.3] (PSI1.north) edge[bend right, blue,->, looseness=2.5, out=-100, in=270] (B2);
\draw[opacity=0.3] (PSI1.north) edge[bend right, blue,->, looseness=2.5, out=-120, in=270] (B3);

\node[anchor=south, draw, ellipse, blue, ,  line width=0.2pt, inner sep=1, scale=0.8, yshift=0.1cm](XL) at (1-1.north){$\pi \wedge \psi_1$};  
  
\draw (PSI1A) edge[bend left, blue,->, out=110, looseness=1.5,dashed, in=125] (XL);

\end{scope}
  

\pgftransformshift{\pgfpoint{-6.3cm}{-2.5cm}}

\node(NT){$\ltlNext \psi = \hsEA ( \neg \pi \wedge \hsAB \pi \wedge \hsEA (\pi \wedge \psi))$};

\pgftransformshift{\pgfpoint{5.9cm}{-1.4cm}}


\begin{scope}[node distance=2cm]

\node[draw, circle, inner sep=1, 
  label={[]270:$\scalebox{0.7}{$ x $}$},
  label={[label distance=0.5cm ]180:$\ldots$}, 
  label={[]90:$\scalebox{0.7}{$  $}$}](0-0) [] {$ $};
  
 \node[draw, circle, inner sep=1, 
  label={[]270:$\scalebox{0.7}{$ y $}$},
  label={[label distance=0.75cm ]180:$\ldots$}, 
  label={[]90:$\scalebox{0.7}{$  $}$}](1-1) [right of=0-0] {$ $};
  
\node[draw, circle, inner sep=1, 
  label={[]270:$\scalebox{0.7}{$ y + 1 $}$},
  label={[label distance=1cm ]180:$ $}, 
  label={[label distance=1cm ]0:$\ldots$}, 
  label={[]90:$\scalebox{0.7}{$  $}$}](2-2) [right of=1-1] {$ $};

\draw[|-|, thick] ($(0-0.center) + (0,2)$) -- ($(1-1.center)+ (0,2)$)
        node[pos=0.5, above=0.2](47) {
        \begin{tikzpicture}[inner sep=0pt, anchor=west, node distance=0]
               \node[scale=0.8](B) {$\hsEA ( $};
               \node[draw,ellipse, ,  line width=0.2pt, inner sep=1,  red,scale=0.8](PSI1) at (B.east)  {$\neg \pi \wedge \hsAB \pi \wedge \hsEA (\pi \wedge \psi)$};
               \node[scale=0.7] at (PSI1.east) {$)$};
        \end{tikzpicture}
        };

\draw[|-|] ($(1-1.center) + (0,0.8)$) -- ($(2-2.center)+ (0,0.8)$)
node[ellipse, red, draw, ,  line width=0.2pt, inner sep=1, pos=0.3, above=0.2cm](PSI1A) {
\begin{tikzpicture}[inner sep=0, anchor=west, node distance=0]
                \node[scale=0.7](A) {$\neg \pi \wedge \hsAB$};
                \node[draw,ellipse, ,  line width=0.2pt, red,scale=0.8, dashed, inner sep=2pt, xshift=-0.15cm](PSI1L) at (A.east)  {$\pi$};
                \node[scale=0.8, xshift=-0.1cm](AA) at 
                (PSI1L.east)  {$\wedge\ \hsEA ($};
               \node[draw,ellipse,  red,scale=0.8,dashed, xshift=-0.15cm,  line width=0.2pt, inner sep=1](PSI1R) at (AA.east) {$\pi \wedge \psi$};
               \node[scale=0.8] at (PSI1R.east) {$)$};
         \end{tikzpicture}};

\node[draw, ellipse, blue,scale=0.9, ,  line width=0.2pt, inner sep=1, anchor=south, yshift=0.1cm] (PI) at (1-1.north) {$\pi$};         

\node[draw, ellipse, blue,scale=0.8,  line width=0.2pt, inner sep=1, anchor=south, yshift=0.1cm] (PSI) at (2-2.north) {$\pi \wedge \psi$};

\draw (PSI1.south east) edge[->, blue, bend left] (PSI1A.north);
\draw (PSI1L) edge[->, blue, dashed, bend right, looseness=1.8,  in=-90, out=-15] (PI);
\draw (PSI1R) edge[->, blue, dashed, bend left] (PSI);

\end{scope}
  

\end{tikzpicture}

\caption{\label{fig:abuntil} The encoding of $\mathrm{LTL}_f$ modalities $\ltlUntil$ and $\ltlNext$ in $\mathrm{AB}$.}

\end{figure}

In the appendix (see Figure~\ref{fig:abuntilAsExample}),
we show how to exploit such an encoding to translate formulas  $p \ltlUntil (\neg p \wedge \neg q)$ and $\ltlNext(\neg p \wedge \neg q)$)
into equivalent $\mathsf{AB}$ formulas and, by means of the example of Figure~\ref{fig:itlmodels},
we show how the interval model is constrained 
when the resulting formula holds over an interval.



\section{The logic \texorpdfstring{$\mathsf{BD}$}{BD} of prefixes and infixes}\label{sec:logic} 

In this section, we introduce the logic $\mathsf{BD}$ of prefixes and infixes,
we formally state the homogeneity assumption, and we define the relation of finite satisfiability
under such an assumption. We conclude the section with a short analysis
of the relationships between such a logic and a suitable restriction of 
generalized $*$-free regular expressions. 


$\mathsf{BD}$ formulas are built up from a countable  set $\Prop$ of proposition letters according to the grammar:
$
\varphi ::= p\ |\ \neg \psi\ |\  \psi \vee \psi \ |\ 
\hsEB \psi \ |\ \hsED \psi,
$
where $p \in \Prop$ and $\hsEB$ and $\hsED$ are the modalities for Allen's relations \emph{Begins} and \emph{During}, respectively.
In the following,  we denote by $|\varphi|$ the size of the parse tree 
for a formula $\varphi$ generated by 
the above grammar. It is easy to show that
$|\varphi|$ is less than or equal to the number of symbols used to encode $\varphi$. 

Let $N \in \bN$ be a natural number and let $\bI_N$ be the set of all intervals over the prefix $0\ldots N$ of $\bN$. A (finite) model for $\mathsf{BD}$ formulas is 
a pair $\bfM = (\bI_N, \cV)$, where  $\cV: \bI_N \rightarrow 
2^{\Prop}$ is a valuation that maps intervals in $\bI_N$ to sets of 
proposition letters. 
Given a model $\bfM$ and an interval $[x,y]$, the semantics of a $\mathsf{BD}$ formula is defined as follows:
\begin{compactitem}
\item $\bfM, [x,y] \models p$ iff $p \in \cV([x,y])$;
\item $\bfM, [x,y] \models \neg \psi$ iff 
$\bfM, [x,y] \not\models  \psi$;

\item $\bfM, [x,y] \models \psi_1 \wedge \psi_2$ iff 
$\bfM, [x,y] \models  \psi_1$ and $\bfM, [x,y] \models  \psi_2$;

\item $\bfM, [x,y] \models \hsEB \psi$ iff 
there is $y'$, with $x\leq y'<y$, such that 
$\bfM, [x,y'] \models  \psi$;

\item $\bfM, [x,y] \models \hsED \psi$ iff 
there exist $x', y'$, with $x<x'\leq y'<y$, such that $\bfM, [x',y'] \models  \psi$.
\end{compactitem}
The logical constants $\top$ (true) and $\bot$ (false), the Boolean connectives 
$\vee, \rightarrow$, and $\leftrightarrow$, and the (universal) dual modalities
$\hsAB$ and $\hsAD$ are defined in the usual way.
The modal constant $\pi$, that has been introduced in Section~\ref{sec:itlintro}, is defined as $\hsAB \bot$.


\begin{figure}
\centering

\begin{tikzpicture}[scale=1, node distance=0.9cm]

\node at (3, -6.9) {(a)};

\node[draw, circle, inner sep=1, 
label={[]270:$\scalebox{0.6}{$0$}$},
label={[]90:$\scalebox{0.6}{$\mathset{p,r}$}$}](0) {};

\node[draw, circle, inner sep=1, right of=0, 
label={[]270:$\scalebox{0.6}{$1$}$},
label={[]90:$\scalebox{0.6}{$\mathset{p}$}$}](1) {};

\node[draw, circle, inner sep=1, right of=1, 
label={[]270:$\scalebox{0.6}{$2$}$},
label={[]90:$\scalebox{0.6}{$\mathset{p,q,r}$}$}](2) {};

\node[draw, circle, inner sep=1, right of=2, label={[]270:$\scalebox{0.6}{$3$}$},
label={[]90:$\scalebox{0.6}{$\mathset{p,q,r}$}$}](3) {};

\node[draw, circle, inner sep=1, right of=3, label={[]270:$\scalebox{0.6}{$4$}$},
label={[]90:$\scalebox{0.6}{$\mathset{p,q,r}$}$}](4) {};

\node[draw, circle, inner sep=1, right of=4, label={[]270:$\scalebox{0.6}{$5$}$},
label={[]90:$\scalebox{0.6}{$\mathset{q,r}$}$}](5) {};

\node[draw, circle, inner sep=1, right of=5, label={[]270:$\scalebox{0.6}{$6$}$},
label={[]90:$\scalebox{0.6}{$\mathset{q}$}$}](6) {};

\node[draw, circle, inner sep=1, right of=6, label={[]270:$\scalebox{0.6}{$7$}$},
label={[]90:$\scalebox{0.6}{$\mathset{q}$}$}](7) {};


\draw[|-|] ($(1.center) + (0,2)$) -- ($(4.center)+ (0,2)$)
node[pos=0.5, above](03) {$\scalebox{0.6}{$\mathset{p}$}$};

\draw[|-|] ($(2.center) + (0,0.9)$) -- ($(4.center)+ (0,0.9)$)
node[pos=0.5, above](12) {$\scalebox{0.6}{$\mathset{p, q, r}$}$};

\draw[|-|] ($(2.center) + (0,1.5)$) -- ($(6.center)+ (0,1.5)$)
node[pos=0.6, above](47) {$\scalebox{0.6}{$\mathset{q}$}$};


\node(L1) at (0.6, -3.3){\scalebox{0.75}{$
\begin{array}{c|c}
[x,y] & \cV_a(x,y) \\
\hline
\hline \\

[0,0] & \mathset{p,r} \\

[0,1] & \mathset{p} \\

[0,2] & \mathset{p} \\

[0,3] & \mathset{p} \\

[0,4] & \mathset{p} \\

[0,5] & \emptyset \\

[0,6] & \emptyset \\

[0,7] & \emptyset \\

[1,1] & \mathset{p} \\

[1,2] & \mathset{p} \\

[1,3] & \mathset{p} \\

[1,4] & \mathset{p} \\ 
 
\end{array}$}};

\node[anchor=west](L2) at (L1.east){\scalebox{0.75}{$
\begin{array}{c|c}
[x,y] & \cV_a(x,y) \\
\hline
\hline \\

[1,5] & \emptyset \\

[1,6] & \emptyset \\

[1,7] & \emptyset \\

[2,2] & \mathset{p,q,r} \\

[2,3] & \mathset{p,q,r} \\

[2,4] & \mathset{p,q,r} \\

[2,5] & \mathset{q,r} \\

[2,6] & \mathset{q} \\

[2,7] & \mathset{q} \\

[3,3] & \mathset{p,q,r} \\

[3,4] & \mathset{p,q,r} \\

[3,5] & \mathset{q,r} 
\end{array}$}};

\node[anchor=west](L3) at (L2.east){\scalebox{0.75}{$
\begin{array}{c|c}
[x,y] & \cV_a(x,y) \\
\hline
\hline \\

[3,6] & \mathset{q} \\

[3,7] & \mathset{q} \\

[4,4] & \mathset{p,q,r} \\

[4,5] & \mathset{q,r} \\

[4,6] & \mathset{q} \\

[4,7] & \mathset{q} \\

[5,5] & \mathset{q,r} \\

[5,6] & \mathset{q} \\

[5,7] & \mathset{q} \\

[6,6] & \mathset{q} \\

[6,7] & \mathset{q} \\

[7,7] & \mathset{q} 
\end{array}$}};


\pgftransformshift{\pgfpoint{7.5cm}{0cm}}

\node at (3, -6.9) {(b)};

\node[draw, circle, inner sep=1, 
label={[]270:$\scalebox{0.6}{$0$}$},
label={[]90:$\scalebox{0.6}{$\mathset{p,r}$}$}](0) {};

\node[draw, circle, inner sep=1, right of=0, 
label={[]270:$\scalebox{0.6}{$1$}$},
label={[]90:$\scalebox{0.6}{$\mathset{p}$}$}](1) {};

\node[draw, circle, inner sep=1, right of=1, 
label={[]270:$\scalebox{0.6}{$2$}$},
label={[]90:$\scalebox{0.6}{$\mathset{p,q,r}$}$}](2) {};

\node[draw, circle, inner sep=1, right of=2, label={[]270:$\scalebox{0.6}{$3$}$},
label={[]90:$\scalebox{0.6}{$\mathset{p,q,r}$}$}](3) {};

\node[draw, circle, inner sep=1, right of=3, label={[]270:$\scalebox{0.6}{$4$}$},
label={[]90:$\scalebox{0.6}{$\mathset{p,q,r}$}$}](4) {};

\node[draw, circle, inner sep=1, right of=4, label={[]270:$\scalebox{0.6}{$5$}$},
label={[]90:$\scalebox{0.6}{$\mathset{q,r}$}$}](5) {};

\node[draw, circle, inner sep=1, right of=5, label={[]270:$\scalebox{0.6}{$6$}$},
label={[]90:$\scalebox{0.6}{$\mathset{q}$}$}](6) {};

\node[draw, circle, inner sep=1, right of=6, label={[]270:$\scalebox{0.6}{$7$}$},
label={[]90:$\scalebox{0.6}{$\mathset{q}$}$}](7) {};


\draw[|-|, dashed] ($(1.center) + (0,2)$) -- ($(4.center)+ (0,2)$)
node[pos=0.5, above](03) {$\scalebox{0.6}{$\mathset{r}$}$};

\draw[|-|, dashed] ($(2.center) + (0,0.9)$) -- ($(4.center)+ (0,0.9)$)
node[pos=0.5, above](12) {$\scalebox{0.6}{$\mathset{p}$}$};

\draw[|-|] ($(2.center) + (0,1.5)$) -- ($(6.center)+ (0,1.5)$)
node[pos=0.6, above](47) {$\scalebox{0.6}{$\mathset{q}$}$};

\node(L1) at (0.75, -3.385){\scalebox{0.75}{$
\begin{array}{c|c}
[x,y] & \cV_b(x,y) \\
\hline
\hline \\

[0,0] & \mathset{p,r} \\

[0,1] & \mathset{p} \\

[0,2] & \mathset{p} \\

[0,3] & \mathset{p} \\

[0,4] & \mathset{p} \\

[0,5] & \emptyset \\

[0,6] & \emptyset \\

[0,7] & \emptyset \\

[1,1] & \mathset{p} \\

[1,2] & \mathset{p} \\

[1,3] & \mathset{p} \\[0.25cm]

[1,4] & \mathset{\mathbf{r}} \\ 
 
\end{array}$}};

\node[draw, rectangle, dashed,anchor=north, thick,
     minimum width=2cm, minimum height=0.45cm, yshift=-4.76cm] at (L1.north)  {};

\node[anchor=west, yshift=-0.09cm](L2) at (L1.east){\scalebox{0.75}{$
\begin{array}{c|c}
[x,y] & \cV_b(x,y) \\
\hline
\hline \\

[1,5] & \emptyset \\

[1,6] & \emptyset \\

[1,7] & \emptyset \\

[2,2] & \mathset{p,q,r} \\

[2,3] & \mathset{p,q,r} \\[0.25cm]

[2,4] & \mathset{\mathbf{p}} \\[0.25cm]

[2,5] & \mathset{q,r} \\

[2,6] & \mathset{q} \\

[2,7] & \mathset{q} \\

[3,3] & \mathset{p,q,r} \\

[3,4] & \mathset{p,q,r} \\

[3,5] & \mathset{q,r} 
\end{array}$}};

\node[draw, rectangle, dashed,anchor=north, thick,
     minimum width=2cm, minimum height=0.45cm, yshift=-2.69cm] at (L2.north)  {};

\node[anchor=west, yshift=0.18cm](L3) at (L2.east){\scalebox{0.75}{$
\begin{array}{c|c}
[x,y] & \cV_b(x,y) \\
\hline
\hline \\

[3,6] & \mathset{q} \\

[3,7] & \mathset{q} \\

[4,4] & \mathset{p,q,r} \\

[4,5] & \mathset{q,r} \\

[4,6] & \mathset{q} \\

[4,7] & \mathset{q} \\

[5,5] & \mathset{q,r} \\

[5,6] & \mathset{q} \\

[5,7] & \mathset{q} \\

[6,6] & \mathset{q} \\

[6,7] & \mathset{q} \\

[7,7] & \mathset{q} 
\end{array}$}};

\end{tikzpicture}

\caption{\label{fig:homvsgen} 
A homogeneous model (a - left) vs.\ a general one (b - right).
}

\vspace{-0.5cm}

\end{figure}
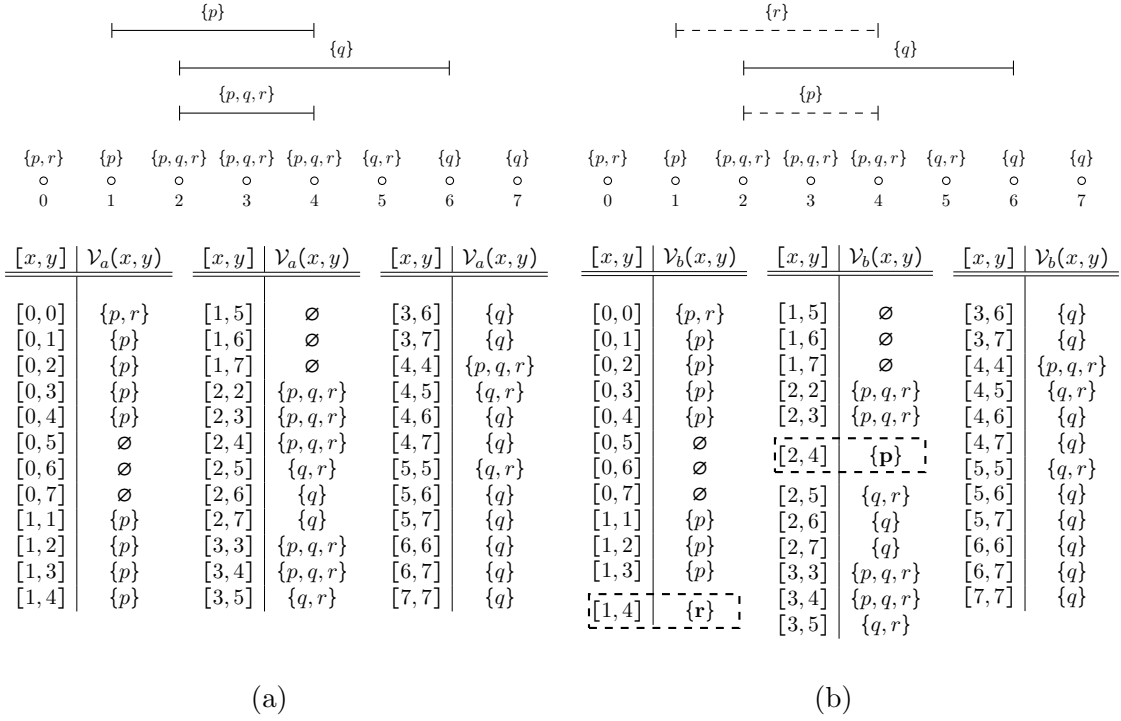

We say that a $\mathsf{BD}$ formula  $\varphi$ is \emph{finitely satisfiable} if and only if there exist
a (finite) model $\bfM$  and an interval $[x,y]$ 
such that $\bfM, [x,y] \models \varphi$ W.l.o.g., $[x,y]$ can be assumed to be the maximal interval $[0, N]$. 
Hereafter, whenever we use the term \emph{satisfiable}, we always mean finite satisfiability, that is,  satisfiability over the class of finite linear orders.

\begin{defi}[Homogeneity]\label{def:homogeneity}
	We say that a model $\bfM = (\bI_N, \cV)$ 
	satisfies the homogeneity property 
	($\bfM$ is \emph{homogeneous} for short) if and only if 
\vspace{-0.1cm}	
$$ \forall p\in\Prop ~~ \forall [x,y]\in\bI_N ~~
\Big( p\in\cV([x,y]) ~\Leftrightarrow~ \forall z\in[x,y] ~ p\in\cV([z,z]) \Big).$$
\end{defi}
\begin{exa}\label{ex:homogeneity}
In Figure~\ref{fig:homvsgen},\vspace{0.1cm} we give an example of a homogeneous model (a) and of an  arbitrary non-homogeneous  one (b). For the sake of readability, we will refer to them as
$\bfM_a = (\bI_7, \cV_a)$ and $\bfM_b= (\bI_7, \cV_b)$, respectively.
The complete definitions of $\cV_a$ and $\cV_b$
are given in Figure~\ref{fig:homvsgen} below the respective models. It is easy to check that 
the definition of $\cV_a$ satisfies  the homogeneity property as stated by Definition~\ref{def:homogeneity}.

To begin with, we observe that, in homogeneous models, 
the labelling $\cV$ of the intersection of two intervals contains the labellings of the two intervals. This is the case, e.g., with intervals $[1,4]$ and $[2,6]$ in Figure~\ref{fig:homvsgen} (a),  whose intersection is the interval $[2,4]$. This is not the case with arbitrary models. Consider, e.g., the very same intervals  in Fig.~\ref{fig:homvsgen} (b). Interval $[1,4]$ violates the homogeneity property because $r \in \cV_b(1,4)$
but $r \notin \cV_b(1,1)$, thus violating the left-to-right direction of 
Definition~\ref{def:homogeneity}. Interval $[2,4]$
violates the homogeneity property as well, because 
$q \in \cV_b(2,2) \cap \cV_b(3,3) \cap \cV_b(4,4) $,
but $q \notin \cV_b(2,4)$ (the same for $r$), thus violating the right-to-left direction of 
Definition~\ref{def:homogeneity}. 
All the other intervals in Figure~\ref{fig:homvsgen} (b), including $[2,6]$,  satisfy the homogeneity property, but this is obviously not sufficient to consider the model homogeneous, 
since \emph{each interval} of the model \emph{must} satisfy it.

It is worth pointing out that the homogeneity property \emph{does not} entail, in general, a similar containment property for formulas $\psi \notin \Prop$.
As an example, it is easy to check that in the homogeneus model of Figure~\ref{fig:homvsgen} (a) $\bfM_a, [1,4] \models \hsEB(p \wedge \neg q)$, but $\bfM_a, [2,4] \nmodels \hsEB(p \wedge \neg q)$, and $\bfM_a, [2,6] \models \hsED(q \wedge \neg r)$, but $\bfM_a, [2,4] \nmodels \hsED(q \wedge \neg r)$. 

To conclude, notice that, in homogeneous models, for proposition letters it holds that the labelling of point-intervals determines that of all intervals. 
This is not the case with arbitrary models. Counterexamples are intervals $[1,4]$ and $[2,4]$ in Figure~\ref{fig:homvsgen} (b). 


\end{exa}

Satisfiability can be relativized to homogeneous models as follows.
We say that a $\mathsf{BD}$ formula $\varphi$ is \emph{satisfiable under homogeneity} if there 
is a homogeneous model $\bfM$  
such that $\bfM, [0, N] \models \varphi$. Satisfiability under homogeneity is clearly more restrictive than plain satisfiability. 
We know from \cite{DBLP:journals/fuin/MarcinkowskiM14,DBLP:conf/icalp/MarcinkowskiMK10} that dropping the homogeneity 
assumption makes $\mathsf{D}$ undecidable. This is not the case with the fragment $\mathsf{B}$, whose expressive power is quite limited, which remains decidable \cite{DBLP:journals/jancl/GorankoMS04}. 
Hereafter, we will always refer to $\mathsf{BD}$ under the homogeneity assumption, denoted by $\mathsf{BD}_{hom}$.

\smallskip

We conclude the section by giving a short account of the relationships  between $\mathsf{BD}_{hom}$
and generalized $*$-free regular expressions. Let $\Sigma$ be a finite alphabet.
A \emph{generalized $*$-free regular expression} (hereafter, simply \emph{expression}) 
$e$ over $\Sigma$ is a term of the form: 
$$
e ::= \emptyset \ |\ a\ |\ \neg e\ |\ e + e \ |\ \ e \cdot e, \mbox{ for any $a \in \Sigma$.}
$$

We exclude the empty word $\epsilon$ from the syntax, as it makes the correspondence between finite words and finite models of $\mathsf{BD}_{hom}$ formulas easier (such a simplification is quite common in the literature). An expression $e$ 
defines a language $\Lang(e)\subseteq \Sigma^+$, which is inductively defined as follows:

\smallskip

\begin{compactitem}

\item $\Lang(\emptyset) = \emptyset$;
\smallskip

\item $\Lang(a) = \{ a \}$, for every $a \in \Sigma$;
\smallskip

\item $\Lang(\neg e) = \Sigma^+ \setminus \Lang(e)$;
\smallskip

\item $\Lang(e_1 + e_2) = \Lang(e_1) \cup \Lang(e_2)$;
\smallskip

\item $\Lang(e_1 \cdot e_2) =  \{ w_1w_2:   w_1 \in \Lang(e_1), w_2 \in \Lang(e_2)\}$.

\end{compactitem}

\smallskip

In \cite{stockmeyer1974complexity}, Stockmeyer shows that the problem of deciding the emptiness of  $\Lang(e)$, for a given expression $e$, is non-elementary hard. 

Let us  consider now the logic $\mathsf{C}$ of the chop operator (under the homogeneity assumption), that we introduced in Section~\ref{sec:itlintro}.
$\mathsf{C}$ features the chop modality $\cdtC$ and the modal constant $\pi$. 
%
$\mathsf{BD}_{hom}$ is a small fragment of $\mathsf{C}$, whose modalities $\hsEB$ and $\hsED$ can be defined in $\mathsf{C}$ as follows (see Appendix~\ref{appendix:cdt} for details): $\hsEB \psi = \psi \cdtC  \neg \pi$ and $\hsED \psi = \neg \pi \cdtC  (\psi \cdtC  \neg \pi)$. 

It can be shown that, for any expression $e$ over $\Sigma$, there exists a formula $\varphi_e$ of $\mathsf{C}$ whose set of models is the language $\Lang(e)$, that is, $\Lang(e) = \{\cV(0,0)\ldots\cV(N,N) : (\bI_N,\cV) \models \psi_e \}$. Such a formula is the conjunction of two sub-formulas $\psi_\Sigma$ and $\psi_e$. $\psi_\Sigma$ guarantees that 
each point interval $[x,x]$ of the model is labelled by exactly one proposition letter from $\Sigma$; $\psi_e$  
constrains the valuation on the basis of the inductive structure of (the translation of) $e$. As an example, if $e=e_1\cdot e_2$, then $\psi_e = \psi_{e_1} \cdtC  ((\neg \pi \wedge \neg (\neg \pi \cdtC  \neg \pi)) \cdtC \psi_{e_2})$.

Such a mapping of expressions in $\mathsf{C}$ formulas allows one to conclude that the satisfiability problem for $\mathsf{C}$ is non-elementary hard (its non-elementary decidability follows from the opposite mapping). A careful look at the expression-to-formula mapping reveals that  modality $\mathsf{C}$ only comes into play in the translation of expressions featuring the operator of concatenation. In view of that, it is worth looking for subclasses of generalized $*$-free regular expressions where the concatenation operation is used in a very restricted manner, so as to avoid the need of modality $\mathsf{C}$ in the translation. 

Let us focus our attention on the following class of \emph{restricted expressions}:
$$
e ::= \emptyset \ |\ a\ |\ \neg e\ |\ e + e \ |\ \ \Pre(e)\ |\ \Sub(e),  
\mbox{ for any $a \in \Sigma$},
$$

\noindent where $\Pre(e)$ and $\Sub(e)$ are respectively a shorthand for $e\cdot(\neg\emptyset)$ (thus defining the language $\Lang(\Pre(e)) = \{ w v:   w \in \Lang(e), v \in \Sigma^+\}$), and $(\neg\emptyset)\cdot e\cdot(\neg\emptyset)$ (thus defining the language $\Lang(\Sub(e)) = \{ u w v:  u,v \in \Sigma^+ , w \in \Lang(e)\}$).
Every restricted expression $e$ of the above form can be mapped into an equivalent formula $\varphi_e$ of $\mathsf{BD}_{hom}$ by applying the usual constructions for empty language, letters, negation, and union, plus the following two rules: (i) $\varphi_{\Pre(e)} = \hsEB \psi_e$, and (ii) $\varphi_{\Sub(e)} = \hsED \psi_e$.

\smallskip
In the following sections,
we will show that the satisfiability problem for $\mathsf{BD}_{hom}$ belongs to \expspace. From the above mapping, it immediately follows that the emptiness problem for the considered subclass of expressions, that only uses prefixes and infixes, can be decided in exponential space (rather than in non-elementary time). 



\section{Homogeneous compass structures}
\label{sec:compass}

In this section, we introduce a spatial representation of homogeneous models, 
called \emph{homogeneous compass structures}, that will considerably simplify 
the proofs of the next sections.
 
Let $\varphi$ be a $\mathsf{BD}_{hom}$ formula. We define the \emph{closure} of 
$\varphi$, denoted by $\closure$, as the set of all its sub-formulas and of their negations, 
plus formulas $\hsEB \top$ and $\hsAB \bot$. For every 
$\mathsf{BD}_{hom}$ formula $\varphi$, it holds that $\closure \leq 2|\varphi| + 2$. 

\begin{defi}\label{def:atoms}
A \emph{$\varphi$-atom} (\emph{atom} for short) is a maximal subset $F$ of $\closure$ that, 
for all $\psi \in \closure$, satisfies the following two conditions
(as usual, we identify every formula of the form $\neg\neg \psi$ as $\psi$):
\begin{compactitem}
\item[]  $\psi \in F$ if and only if $\neg \psi \notin F$ (consistency), and
\item[] if $\psi = \psi_1 \vee \psi_2$, then 
$\psi \in F$ if and only if $\{\psi_1 , \psi_2\}\cap F\neq \emptyset$ (completeness).
\end{compactitem}
\end{defi}

Let $\atoms$ be the set of all 
$\varphi$-atoms. We have that $|\atoms| 
\leq 2^{\frac{|\closure|}{2}}$, and, from
$\closure \leq 2|\varphi| + 2$, we have 
$|\atoms| \leq 2^{|\varphi| + 1}$.

	It is easy to see that, given a model $\bfM = (\bI_N, \cV)$, 
	we can uniquely associate an atom $F^{[x,y]}$ in $\atoms$ 
	with each interval $[x,y] \in \bI_N$ by simply letting 
	$F^{[x,y]} = \mathset{ \psi \in \closure : \bfM, [x,y] \models \psi }$.
	An example of such an extension of the labelling $\cV$ to atoms is given 
	in Figure~\ref{fig:reqExplanationBD} both in a graphical (top) 
	and in a tabular (bottom) form. 

A distinctive feature of atoms is that any atom $F$ is uniquely determined by its subset of formulas 
$(F \cap \Prop) \cup (F \cap \{ \hsEB \psi \in \closure  \}) \cup (F \cap \{ \hsED \psi \in \closure  \})$, that is, once we establish which proposition letters,  $\hsEB$ sub-formulas, and $\hsED$ sub-formulas belong to it, we can automatically  derive all the other formulas $\psi \in \closure$ belonging to it. This subset of its formulas can be viewed as the \emph{fingerprint} of the atom.
In Figure~\ref{fig:reqExplanationBD}, 
we exploit such a feature to identify the atoms associated with the various intervals (we simply specify the formulas in $(F \cap \Prop) \cup (F \cap \{ \hsER \psi \in \closure  \})$ belonging to them -- boxes in the middle of intervals). 


\begin{figure}
\centering

\begin{tikzpicture}

\begin{scope}[anchor=west, inner sep=0.1cm]
\node(F0){$\varphi=\hsEB($};
\node(F1) at (F0.east) {$p \wedge \neg r$};
\node(F2) at (F1.east){$) \wedge \hsED($};
\node(F3) at (F2.east){$\neg q \wedge \hsED q$};
\node(F4) at (F3.east){$)$};	

\draw[decorate, line width=1pt,
      decoration={brace,raise=0.3cm,
                  amplitude=5pt,mirror},
                  color=black,
                  ] 
(F1.east) -- (F1.west) node[above=0.53cm, midway]{$\psi_1$};

\draw[decorate, line width=1pt,
      decoration={brace,raise=0.3cm,
                  amplitude=5pt,mirror},
                  color=black,
                  ] 
(F3.east) -- (F3.west) node[above=0.53cm, midway]{$\psi_2$};

\end{scope}

\pgftransformshift{\pgfpoint{10cm}{-2.cm}}

\draw[step=1.0,black, opacity=0.25, very thin,xshift=-1cm,yshift=-1cm] (1,1)
grid (5,5);

\begin{scope}[opacity=0.5]

\node[yshift=-0.3cm](0) {$\scalebox{0.7}{$0$}$};
\node[ right of=0](0) {$\scalebox{0.7}{$1$}$};
\node[ right of=0](0) {$\scalebox{0.7}{$2$}$};
\node[ right of=0](0) {$\scalebox{0.7}{$3$}$};
\node[ right of=0](0) {$\scalebox{0.7}{$4$}$};
 
\node[xshift=4.3cm, yshift=-0.0cm](0) {$\scalebox{0.7}{$0$}$};
\node[ above of=0](0) {$\scalebox{0.7}{$1$}$};
\node[ above of=0](0) {$\scalebox{0.7}{$2$}$};
\node[ above of=0](0) {$\scalebox{0.7}{$3$}$};
\node[ above of=0](0) {$\scalebox{0.7}{$4$}$};

\end{scope}

\node[draw, circle, inner sep=1, 
label={[label distance=-0.5cm, yshift=0.3cm]0:\scalebox{0.7}{$F^{[0,4]}$}}](4) at (0, 4) {};

\node[draw, circle, inner sep=1, 
label={[label distance=-0.5cm, yshift=0.3cm]0:\scalebox{0.7}{$F^{[1,4]}$}}](4) at (1, 4) {};

\node[draw, circle, inner sep=1, 
label={[label distance=-0.5cm, yshift=0.3cm]0:\scalebox{0.7}{$F^{[2,4]}$}}](4) at (2, 4) {};

\node[draw, circle, inner sep=1, 
label={[label distance=-0.5cm, yshift=0.3cm]0:\scalebox{0.7}{$F^{[3,4]}$}}](4) at (3, 4) {};

\node[draw, circle, inner sep=1, 
label={[label distance=-0.5cm, yshift=0.3cm]0:\scalebox{0.7}{$F^{[0,3]}$}}](4) at (0, 3) {};

\node[draw, circle, inner sep=1, 
label={[label distance=-0.5cm, yshift=0.3cm]0:\scalebox{0.7}{$F^{[0,2]}$}}](4) at (0, 2) {};

\node[draw, circle, inner sep=1, 
label={[label distance=-0.5cm, yshift=0.3cm]0:\scalebox{0.7}{$F^{[0,1]}$}}](4) at (0, 1) {};

\node[draw, circle, inner sep=1, 
label={[label distance=-0.5cm, yshift=0.3cm]0:\scalebox{0.7}{$F^{[1,3]}$}}](3) at (1, 3) {};

\node[draw, circle, inner sep=1, 
label={[label distance=-0.5cm, yshift=0.3cm]0:\scalebox{0.7}{$F^{[1,2]}$}}](2) at (1, 2) {};

\node[draw, circle, inner sep=1, 
label={[label distance=-0.5cm, yshift=0.3cm]0:\scalebox{0.7}{$F^{[2,3]}$}}](1) at (2, 3) {};

\node[draw, circle, inner sep=1, 
label={[label distance=-0.5cm, yshift=0.3cm]0:\scalebox{0.7}{$F^{[0,0]}$}}](0) at (0, 0) {};

\node[draw, circle, inner sep=1, 
label={[label distance=-0.5cm, yshift=0.3cm]0:\scalebox{0.7}{$F^{[1,1]}$}}](11) at (1, 1) {};

\node[draw, circle, inner sep=1, 
label={[label distance=-0.5cm, yshift=0.3cm]0:\scalebox{0.7}{$F^{[2,2]}$}}](22) at (2, 2) {};

\node[draw, circle, inner sep=1, 
label={[label distance=-0.5cm, yshift=0.3cm]0:\scalebox{0.7}{$F^{[3,3]}$}}](33) at (3, 3) {};

\node[draw, circle, inner sep=1, 
label={[label distance=-0.5cm, yshift=0.3cm]0:\scalebox{0.7}{$F^{[4,4]}$}}](44) at (4, 4) {};

\draw[dashed, opacity=0.5] (0) -- (11) -- (22) -- (33) -- (44);

\pgftransformshift{\pgfpoint{-8.5cm}{-4.5cm}}

\begin{scope}[node distance=3cm]

\node[draw, circle, inner sep=1,  node 
distance=0,  
label={[]270:$\scalebox{0.7}{$0$}$},
label={[name=l0]90:$\scalebox{0.7}{$p,q,r$}$}](1) {};

\node[fit=(l0), draw, inner sep=-2]{};

\node[draw, circle, inner sep=1, right of=1, 
label={[]270:$\scalebox{0.7}{$1$}$},
label={[name=l1]90:$\scalebox{0.7}{$p$}$},
label={[xshift=0.4cm]90:$\scalebox{0.7}{$\psi_1$}$}
](2) {};

\node[fit=(l1), draw, inner sep=-2]{};

\node[draw, circle, inner sep=1, right of=2, label={[]270:$\scalebox{0.7}{$2$}$},
label={[name=l2]90:$\scalebox{0.7}{$p, q, r$}$}](3) {};

\node[fit=(l2), draw, inner sep=-2]{};

\node[draw, circle, inner sep=1, right of=3, label={[]270:$\scalebox{0.7}{$3$}$},
label={[name=l3]90:$\scalebox{0.7}{$p, r$}$}](4) {};

\node[fit=(l3), draw, inner sep=-2]{};

\node[draw, circle, inner sep=1, right of=4, label={[]270:$\scalebox{0.7}{$4$}$},
label={[name=l4]90:$\scalebox{0.7}{$ r$}$}](5) {};

\node[fit=(l4), draw, inner sep=-2]{};

\draw[|-|] ($(1.center) + (0,1)$) -- ($(2.center)+ (0,1)$)
node[pos=0.5, fill=white, draw, inner sep=2pt](12) {$\scalebox{0.7}{$p$}$}
node[pos=0.65, fill=white, inner sep=1pt](12) {$\scalebox{0.7}{$\psi_1$}$};
;

\draw[|-|] ($(1.center) + (0,1.5)$) -- ($(3.center)+ (0,1.5)$)
node[pos=0.5, fill=white, draw, inner sep=2pt](46) {$\scalebox{0.7}{$p,\hsEB \psi_1 $}$}
node[pos=0.65, fill=white, inner sep=1pt](46) {$\scalebox{0.7}{$ \psi_1 $}$}
;

\draw[|-|] ($(1.center) + (0,2.5)$) -- ($(4.center)+ (0,2.5)$)
node[pos=0.5, fill=white, draw, inner sep=2pt](16) {$\scalebox{0.7}{$p, \hsEB \psi_1,\hsED q$}$}
node[pos=0.65, xshift=0.1cm, fill=white, inner sep=1pt](16) {$\scalebox{0.7}{$ \psi_1, \psi_2$}$};

\draw[|-|] ($(1.center) + (0,3.5)$) -- ($(5.center)+ (0,3.5)$)
node[pos=0.5, fill=white, draw, inner sep=2pt](16) {$\scalebox{0.7}{$\hsEB\psi_1,\hsED q, \hsED \psi_2$}$}
node[pos=0.64, fill=white, inner sep=1pt](16) {$\scalebox{0.7}{$ \psi_2,  \varphi $}$};

\draw[-|] ($(2.center) + (0,1)$) -- ($(3.center)+ (0,1)$)
node[pos=0.5, fill=white,draw, inner sep=1.5pt, draw, inner sep=2pt](12) {$\scalebox{0.7}{$p, \hsEB \psi_1 $}$}
node[pos=0.8, fill=white, inner sep=1pt](12) {$\scalebox{0.7}{$\psi_1$}$};

\draw[|-|] ($(2.center) + (0,2)$) -- ($(4.center)+ (0,2)$)
node[pos=0.5, fill=white, draw, inner sep=2pt](47) {$\scalebox{0.7}{$p, \hsEB \psi_1, \hsED q$}$}
node[pos=0.74, fill=white, inner sep=1pt](47) {$\scalebox{0.7}{$\psi_1, \psi_2$}$};

\draw[|-|] ($(2.center) + (0,3)$) -- ($(5.center)+ (0,3)$)
node[pos=0.5, fill=white,draw, inner sep=1.5pt, draw, inner sep=2pt](16) {$\scalebox{0.7}{$\hsEB \psi_1,\hsED q  $}$}
node[pos=0.61, fill=white, inner sep=1pt](16) {$\scalebox{0.7}{$ \psi_2 $}$}
;

\draw[-|] ($(3.center) + (0,1)$) -- ($(4.center)+ (0,1)$)
node[pos=0.5, fill=white, draw, inner sep=2pt](12) {$\scalebox{0.7}{$p,r$}$};

\draw[|-|] ($(3.center) + (0,1.5)$) -- ($(5.center)+ (0,1.5)$)
node[pos=0.5, fill=white,draw, inner sep=1.5pt, draw, inner sep=2pt](46) {$\scalebox{0.7}{$r$}$};

\draw[-|] ($(4.center) + (0,1)$) -- ($(5.center)+ (0,1)$)
node[pos=0.5, fill=white, draw, inner sep=2pt](12) {$\scalebox{0.7}{$r$} $};

\end{scope}

\pgftransformshift{\pgfpoint{6cm}{-4cm}}

\node[inner sep=0]{\scalebox{0.7}{$\begin{array}{c||c|c|c|c|c|c||c|c|c||c|c|c|c} 
F^{[x, y]}  &  \vpairmath{ p \\  \gray{\neg p} }  &   \vpairmath{ q \\ \gray{\neg q} }  &   \vpairmath{ r \\ \gray{\neg r} }  &   \vpairmath{\hsEB \psi_1 \\ \gray{\hsAB\neg \psi_1}}  &   \vpairmath{ \hsED q \\ \gray{\hsAD \neg q} }  &   \vpairmath{\hsED \psi_2 \\ \gray{\hsAD\neg \psi_2}}  &  \vpairmath{ \psi_1 \\ \gray{\neg \psi_1} }  &   \vpairmath{ \psi_2 \\ \gray{\neg \psi_2} }  &  \vpairmath{\varphi \\ \gray{\neg \varphi}}  &   \scalebox{0.9}{$\reqB(F^{[x, y]} )$}  &  \scalebox{0.9}{$\obsB(F^{[x, y]} )$}  &  \scalebox{0.9}{$\reqD(F^{[x, y]} )$}  &  \scalebox{0.9}{$\obsD(F^{[x, y]} )$}  \\  
\hline \hline  
& & & & & & & & & & & & & \\ 
 
[0,0]   &  1       &  1        &  1        &  \gzero  &  \gzero  &  \gzero  &  \gzero   &  \gzero  &  \gzero  &  \emptyset  &  \emptyset         &  \emptyset  &  \mathset{q}   \\  
 
[0,1]   &  1       &  \gzero   &   \gzero  &  \gzero  &  \gzero  &  \gzero  &  1        &  \gzero	 &  \gzero  &  \emptyset  &  \mathset{\psi_1}  &  \emptyset  &  \emptyset  \\  
 
[0,2]   &  1       &  \gzero   &   \gzero  &  1       &  \gzero  &  \gzero  &  1        &  \gzero	 &  \gzero  &  \mathset{\psi_1}  &  \mathset{\psi_1}  &  \emptyset  &  \emptyset  \\  
 
[0,3]   &  1       &  \gzero   &  \gzero   &  1       &  1  &  \gzero  &  1             &  	1	 &  \gzero  &  \mathset{\psi_1}  &  \mathset{\psi_1}  &  \mathset{q}  &  \mathset{\psi_2}  \\  
 
[0,4]   &  \gzero  &  \gzero   &  \gzero   &  1       &  1  &  1	        &  \gzero        &  	1	 &  1  &  \mathset{\psi_1}  &  \emptyset         &  \mathset{q, \psi_2}  &  \mathset{\psi_2}  \\  
 
[1,1]   &  1       &  \gzero   &  \gzero   &  \gzero   &  \gzero  &  \gzero  &   1       &  \gzero	 &  \gzero  &  \emptyset  &  \mathset{\psi_1}  &  \emptyset   &  \emptyset  \\  
 
[1,2]   &  1       &  \gzero   &  \gzero   &  1        &  \gzero  &  \gzero  &   1       &  \gzero	 &  \gzero  &  \mathset{\psi_1}  &  \mathset{\psi_1}  &  \emptyset   &  \emptyset  \\  
 
[1,3]   &  1        &  \gzero   &  \gzero   &  1        &  1      &  \gzero  &   1       &  	1	 &  \gzero  &  \mathset{\psi_1}  &  \mathset{\psi_1}  &  \mathset{q}  &  \mathset{\psi_2}  \\  
 
[1,4]   &  \gzero   &  \gzero   &  \gzero   &  1        &  1      &  \gzero  &   \gzero  &  	1	 &  \gzero  &  \mathset{\psi_1}  &  \emptyset         &  \mathset{q}  &  \mathset{\psi_2}  \\  
 
[2,2]   &  1        &  1        &  1        &  \gzero  &  \gzero  &  \gzero  &  \gzero  &  \gzero	 &  \gzero  &  \emptyset  &  \emptyset               &  \emptyset  &  \mathset{q}  \\  
 
[2,3]   &  1       &  \gzero   &  1         &  \gzero  &  \gzero  &  \gzero  &  \gzero  &  \gzero	 &  \gzero  &  \emptyset  &  \emptyset               &  \emptyset  &  \emptyset  \\  
 
[2,4]   &  \gzero  &  \gzero   &  1         &  \gzero  & \gzero  &  \gzero  &  \gzero  &  	\gzero	 &  \gzero  &  \emptyset  &  \emptyset               & \emptyset   &  \emptyset  \\  
 
[3,3]   &  1       &  \gzero   &  1         &  \gzero  & \gzero  &  \gzero  &  \gzero  &  	\gzero	 &  \gzero  &  \emptyset  &  \emptyset               &  \emptyset  &  \emptyset  \\  
 
[3,4]   &  \gzero  &   \gzero  &  1        &  \gzero  &  \gzero  &  \gzero  &  \gzero  &  	\gzero	 &  \gzero  &  \emptyset  &  \emptyset              &  \emptyset  &  \emptyset  \\  
 
[4,4]   &  \gzero  &   \gzero  &  1        &  \gzero  & \gzero  &  \gzero  &  \gzero   &  \gzero	 &  \gzero  &  \emptyset  &  \emptyset              &  \emptyset  &  \emptyset  \\  
\end{array}$}};

\end{tikzpicture}

\caption{\label{fig:reqExplanationBD} 
A graphical (top) and tabular (bottom) account of  $\reqR(F)$ and  $\obsR(F)$, with $F \in \atoms$ and $R \in \mathset{B,D}$, for the formula $\varphi = \hsEB(p \wedge \neg r) \wedge \hsED(\neg q \wedge \hsED q)$.   
}

\end{figure}

The next definition introduces three 
functions that characterize the \emph{temporal behavior} of atoms.
    
\begin{defi}\label{def:atomsets}
Let $\varphi$ be a $\mathsf{BD}_{hom}$ formula and let $R \in \{B, D\}$.
We associate with each atom $F\in\atoms$ three subsets of $\closure$ defined as follows:
  \begin{compactitem}
\item $\reqR(F) =\{\psi \in \closure: \hsER \psi \in F\}$ (temporal requests); 
\item $\obsR(F) =\{\psi \in \closure: \hsER \psi \in \closure \wedge \psi \in F\}$ (observables);
\item $\boxR(F) =\{\psi \in \closure: \hsAR \psi \in F\}$ (universal conditions).
\end{compactitem}   
\end{defi}
Let us  now give a short account of the roles of the sets $\reqR(F)$, $\obsR(F)$, and $\boxR(F)$, and of their relationships. To start with, we observe that if an atom $F$ associated with an interval $[x,y]$ contains a formula $\hsER\psi$
(that is, $\hsER\psi$ is true at $[x,y]$), 
then there must be another interval $[x',y']$
such that $[x,y]$ and $[x',y']$ satisfy relation $R$ and $\psi$ belongs to the atom $F'$ associated with $[x',y']$ (i.e., $\psi$  is true at 
$[x',y']$). 
According to Definition~\ref{def:atomsets}, this amounts to say that $\psi \in \reqR(F)$ and $\psi \in \obsR(F')$.  
Notice that, for each atom $F$, if $\psi \in \obsR(F)$, then $\psi \in F$ as well, that is, $\obsR(F)\subseteq F$. In other words, $\obsR(F)$ allows us to ``mark'' the formulas in (the atom associated with) an interval which can possibly be used to satisfy some $\hsER$-requests of another interval. It is worth pointing out that 
$\psi \in \obsR(F)$ and $\psi \notin \reqR(F)$, i.e., $\hsER\psi \notin F$, is a perfectly legitimate configuration for an atom, where $\psi$ holds on the corresponding interval $[x,y]$, but not on any interval which is reachable from $[x,y]$ via relation $R$. In such a case, according to 
Definition~\ref{def:atoms} (consistency) and Definition~\ref{def:atomsets}, we have that $\neg \psi \in \boxR(F)$, i.e., $\neg\psi$ belongs to every atom $F'$ associated with an interval $[x',y']$ such that $[x,y]$ and $[x',y']$ satisfy relation $R$, and thus, again by consistency of atoms,
$\psi \notin \obsR(F')$. 
Last but not least, we observe that for each $F \in \atoms$ and $\psi \in \{\psi': \hsEB \psi' \in\closure\}$, 
by consistency and completeness of atoms, either $\psi\in\reqB(F)$ or $\neg\psi\in\boxB(F)$. Hence, $\boxB(\cdot)$ is not strictly necessary; we introduced it to simplify some proofs. The same holds with $\psi \in \{\psi': \hsED \psi' \in\closure\}$.

Let $[x,y]$ be an interval. We denote the atom associated with $[x,y]$ by $F^{[x,y]}$. 
In Figure~\ref{fig:reqExplanationBD} (top),  for each interval $[x,y]$, we report the values of $\obsB(F^{[x,y]})$ and
$\obsD(F^{[x,y]})$ to the right of the box containing the fingerprint of $F^{[x,y]}$; moreover, we add the formula $\varphi$ if (and only if) it holds on $[x,y]$.
In Figure~\ref{fig:reqExplanationBD} (bottom), we give a tabular account of the whole atom, e.g., the atom $F^{[1,2]}$ associated with the interval $[1,2]$ is $\mathset{p, \neg q, \neg r,\hsAD \neg q, \psi_1, \neg\psi_2, \hsEB \psi_1, \hsAD\neg \psi_2,\neg \varphi }$, where $\psi_1 = p \wedge \neg r$ and $\psi_2 = \neg q \wedge \hsED q$. 


Since sets $\reqR(F)$, $\obsR(F)$, and $\boxR(F)$ will be exploited to prove most of the results of the paper, we further illustrate their behavior by means of the example in Figure~\ref{fig:reqExplanationBD}.
\begin{exa}\label{ex:atomsets}
To start with, we observe that $\reqR(F)$, $\obsR(F)$, and $\boxR(F)$ are univocally determined by their argument $F$. However, while $\obsR(F)\subseteq F$, this is not the case with $\reqR(F)$ and $\boxR(F)$. With reference to the example in Figure~\ref{fig:reqExplanationBD}, it holds that $q \in \reqD(F^{[1,4]})$, but $q \notin F^{[1,4]}$. 

Let us focus on $\reqB(F)$, which extracts the arguments of the $\hsEB$-formulas in $F$. In the considered example, $\psi_1 (= p \wedge \neg r) \in \reqB(F^{[0,2]})$, as $\hsEB \psi_1 \in F^{[0,2]}$. Hence, there must be a proper prefix $[0,y]$ of $[0,2]$ such that $\psi_1 \in F^{[0,y]}$ (equivalently, $\psi_1 \in \obsB(F^{[0,y]})$), and this is actually the case with $y = 1$. In general, for each $\psi \in \reqB(F^{[x,y]})$ (resp., $\reqD(F^{[x,y]})$), there must be a prefix $[x,y']$ (resp., infix $[x',y']$ ) of $[x,y]$ such that $\psi \in \obsB(F^{[x,y']})$ (resp., $\obsD(F^{[x',y']})$). On the contrary, if $\hsER \psi \in \closure$ and $\psi \notin \reqR(F)$, then, necessarily, $\hsAR \neg \psi \in F$ and thus $\neg \psi \in \boxR(F)$, i.e., $\psi \not\in \boxR(F)$. 
Consider, for instance, the interval $[1,4]$ in Figure~\ref{fig:reqExplanationBD}. We have that $\reqD(F^{[1,4]})= \{q\}$ and, since $\{ \psi: \hsED\psi \in \closure  \} = \{q, \psi_2 (= \neg q \wedge \hsED q) \}$, it holds that $\boxD(F^{[1,4]})= \{\neg \psi_2\}$. 

Finally, unlike the case of $\reqR$, for every interval $[x,y]$, formula $\neg \psi \in \boxB(F^{[x,y]})$  (resp., $\neg\psi \in \boxD(F^{[x,y]})$),
and prefix $[x,y']$ (resp., infix $[x',y']$ ) of $[x,y]$, $\psi\!\notin\!\obsB(F^{[x,y']})$ (resp., $\obsD(F^{[x',y']})$).   
As an example, since $\neg\psi_2 \in \boxD(F^{[1,4]})$, we can conclude that $\psi_2 \notin \obsD(F^{[2,2]}) \cup \obsD(F^{[2,3]}) \cup 
\obsD(F^{[3,3]})$.
\end{exa}
In order to further explain the relation between $\reqR$ and $\obsR$, let us consider the example in Figure~\ref{fig:reqExplanationBD} from another angle. Suppose that, for a fixed $N \in \bN$ (in the example, $N=4$), one wants to find, for each $[x,y]$ in $\bI_N$, a ``labelling'' $F^{[x,y]}\in \atoms$ such that 

\vspace{0.15cm} 
$
\begin{array}{lc}
	(*_1) &  \bfM, [x,y] \models \psi  \mbox{ if and only if } \psi \in F^{[x,y]},
\end{array}
$ 
\vspace{0.15cm} 

where $\bfM = (\bI_4, \cV)$ and $\cV([x,y]) = F^{[x,y]} \cap \Prop$, and 

\vspace{0.15cm}
$
\begin{array}{lc}
	(*_2) &  \varphi \in F^{[0,N]},
\end{array}
$
\vspace{0.15cm}

Such a problem is the \emph{bounded satisfiability problem}, which is simpler than the \emph{finite satisfiability problem} we are addressing in this paper, where $N$ is not given as a parameter.

It can be shown that the labelings that, for all $[x,y] \in \bI_N$, satisfy the property: 

\vspace{0.15cm}
$
\begin{array}{ll}
       
   (*_3) &  
   \begin{array}{l}
   \reqB(F^{[x,y]})= 
   \hspace{-0.15cm}
   \bigcup\limits_{x \leq y'< y} 
   \hspace{-0.15cm}
   \obsB(F^{[x,y']})  \mbox{ and } \reqD(F^{[x,y]})=\hspace{0cm} \bigcup\limits_{x < x' \leq y'< y} \hspace{0cm} \obsD(F^{[x',y']})
   \end{array}	 
\end{array}
$

are \emph{all and  only} those labellings that satisfy property $(*_1)$. Notice that these labellings may only differ on interval $[0,N]$.

This means that all the requests that we associate with an interval $[x,y]$ by means of its labelling $F^{[x,y]}$ must be satisfied (fulfilled). Consider, for instance,  the prefix relation. It holds that $\reqB(F^{[x,y]}) \subseteq \bigcup_{x \leq y'< y} \obsB(F^{[x,y']})$. Moreover, there cannot be 
a formula $\psi$ such that $\psi \in \bigcup_{x \leq y'< y} \obsB(F^{[x,y']}) $ and $\psi \notin \reqB(F^{[x,y]})$, as, from the latter, it immediately follows
that $\neg \psi \in \boxB(F^{[x,y]})$ which implies $\psi \notin \bigcup_{x \leq y'< y} \obsB(F^{[x,y']})$ (contradiction).  This allows us to conclude that 
$\reqB(F^{[x,y]}) \supseteq \bigcup_{x \leq y'< y} \obsB(F^{[x,y']})$ as well (consistency), and thus $\reqB(F^{[x,y]}) = \bigcup_{x \leq y'< y} \obsB(F^{[x,y']})$. The same holds for modality $D$ as well as for all the other modalities of $\mathsf{HS}_{hom}$. In fact, this is a fairly general property whose validity does not depend on the homogeneity assumption. Thus, we can conclude that $(*_2)$ and $(*_3)$ are \emph{necessary and sufficient} conditions for $\bfM$ to satisfy $\varphi$.  

\medskip

Let us now introduce two meaningful binary relations $\thenB$ and $\thenD$ over $\atoms$.
\begin{defi}\label{def:intervalrelationsonatoms}
For all $F, G \in \atoms$ we let:
\begin{compactitem} 
\item $F \thenB G$ iff $\reqB(F) = \reqB(G) \cup \obsB(G)$;
\item $F \thenD G$ iff $\reqD(F) \supseteq \reqD(G) \cup \obsD(G)$.
\end{compactitem}
\end{defi}

Relations $\thenB$ and $\thenD$ are often referred to as \emph{view-to-type} dependencies since they constrain the labelling of an interval on the basis of the labellings of the intervals that it can access via certain (interval) relations. As already pointed out, for every $\psi \in \{ \psi':  \hsEB \psi' \in \closure\}$, either $\psi \in \reqB(F)$ or $\neg \psi \in \boxB(F)$ (and vice versa). Given two atoms $F$ and $G$, with $F \thenB G$, and a formula $\neg \psi \in \boxB(F)$ it immediately follows that $\psi \notin \reqB(F)$, and thus, from $\reqB(F) = \reqB(G) \cup \obsB(G)$, it follows $\psi \notin \obsB(G)$. 
Now, from $\neg \psi \in \boxB(F)$, it follows that $\hsEB \psi \in \closure$, and from $\hsEB \psi \in \closure$ and $\psi \notin \obsB(G)$, it follows that $\psi \notin G$. For the  maximality of atoms, it follows that $\neg \psi \in G$. This allows us to conclude that for every pair of atoms $F$ and $G$, with $F \thenB G$, $\boxB(F) \subseteq G$.
The same argument can be applied to the relation $\thenD$, and thus for every pair of atoms $F$ and $G$, with $F \thenD G$, it holds that $\boxD(F) \subseteq G$. In addition, relation $\thenD$ is transitive (by definition of atom, from $\reqR(F) \supseteq \reqR(G)$, it immediately follows that $\boxR(F) \subseteq \boxR(G)$), while $\thenB$ is not. Finally, we call an atom which is $\thenB$-related to itself a $B$-reflexive atom (the notion of $D$-reflexive atom is defined in an analogous way). Atoms which are not $B$-reflexive are called $B$-irreflexive ($D$-irreflexive atoms are defined analogously). Reflexive atoms will play a crucial role in the proof of the results of the next sections. 

An account of relations $\thenB$ and $\thenD$ is given by the examples in Figure~\ref{fig:brelation} and Figure~\ref{fig:drelation}, respectively. In Figure~\ref{fig:brelation} (resp., Figure~\ref{fig:drelation}), we denote $B$-reflexive (resp., $D$-reflexive) atoms by self-loops.


\begin{figure}
\centering

\begin{tikzpicture}

\begin{scope}[anchor=west, inner sep=0.1cm]
\node(F0){$\varphi=\hsEB($};
\node(F1) at (F0.east) {$r \wedge \neg p \wedge \neg q$};
\node(F2) at (F1.east){$) \wedge \hsEB($};
\node(F3) at (F2.east){$\neg p \wedge \neg q \wedge \neg r$};
\node(F4) at (F3.east){$) \wedge \hsEB\hsED($};
\node(F5) at (F4.east){$r \wedge \neg p \wedge \neg q$};	
\node(F6) at (F5.east){$)$};

\draw[decorate, line width=1pt,
      decoration={brace,raise=0.3cm,
                  amplitude=5pt,mirror},
                  color=black,
                  ] 
(F1.east) -- (F1.west) node[above=0.53cm, midway]{$\psi_1$};

\draw[decorate, line width=1pt,
      decoration={brace,raise=0.3cm,
                  amplitude=5pt,mirror},
                  color=black,
                  ] 
(F3.east) -- (F3.west) node[above=0.53cm, midway]{$\psi_2$};

\draw[decorate, line width=1pt,
      decoration={brace,raise=0.3cm,
                  amplitude=5pt,mirror},
                  color=black,
                  ] 
(F5.east) -- (F5.west) node[above=0.53cm, midway]{$\psi_1$};

\end{scope}

\pgftransformshift{\pgfpoint{-0.5cm}{-5cm}}

\begin{scope}[node distance=1.5cm]

\node[draw, circle, inner sep=1,  
label={[]270:$\scalebox{0.7}{$0$}$},
label={[name=l0]90:$\scalebox{0.7}{$p,q,r$}$}](0) {};

\node[fit=(l0), draw, inner sep=-2]{};

\node[draw, circle, inner sep=1, right of=0, 
label={[]270:$\scalebox{0.7}{$1$}$},
label={[name=l1]90:$\scalebox{0.7}{$p,r$}$}](1) {};

\node[fit=(l1), draw, inner sep=-2]{};

\node[draw, circle, inner sep=1, right of=1, label={[]270:$\scalebox{0.7}{$2$}$},
label={[name=l2]90:$\scalebox{0.7}{$q,r$}$}](2) {};

\node[fit=(l2), draw, inner sep=-2]{};

\node[draw, circle, inner sep=1, right of=2, label={[]270:$\scalebox{0.7}{$3$}$},
label={[]90:$\scalebox{0.7}{$\psi_2$}$}](3) {};

\node[draw, circle, inner sep=1, right of=3, label={[]270:$\scalebox{0.7}{$4$}$},
label={[]90:$\scalebox{0.7}{$\psi_2$}$}](4) {};


\draw[|-|] ($(0.center) + (0,1)$) -- ($(1.center)+ (0,1)$)
node[pos=0.5, fill=white,draw, inner sep=1.5pt](12) {$\scalebox{0.7}{$p,r$}$};

\draw[|-|] ($(0.center) + (0,2)$) -- ($(2.center)+ (0,2)$)
node[pos=0.5, fill=white,draw, inner sep=1.5pt](46) {$\scalebox{0.7}{$r$}$}
node[pos=0.625, fill=white, inner sep=1pt](46) {$\scalebox{0.7}{$\psi_1$}$};

\draw[|-|] ($(0.center) + (0,3)$) -- ($(3.center)+ (0,3)$)
node[pos=0.5, fill=white,draw, inner sep=1.5pt](12) {$\scalebox{0.7}{$\hsEB \psi_1,\hsED \psi_1$}$}
node[pos=0.75, fill=white, inner sep=1pt](12) {$\scalebox{0.7}{$\psi_2$}$};

\draw[|-|] ($(0.center) + (0,4)$) -- ($(4.center)+ (0,4)$)
node[pos=0.5, fill=white,draw, inner sep=1.5pt](46) {$\scalebox{0.7}{$\hsEB \psi_1,\hsEB \psi_2,\hsED \psi_1,\hsEB \hsED \psi_1$}$}
node[pos=0.9, fill=white, inner sep=1pt](46) {$\scalebox{0.7}{$\psi_2, \varphi$}$};

\draw[|-|] ($(1.center) + (0,1)$) -- ($(2.center)+ (0,1)$)
node[pos=0.5, fill=white,draw, inner sep=1.5pt](46) {$\scalebox{0.7}{$r$}$}
node[pos=0.75, fill=white, inner sep=1pt](46) {$\scalebox{0.7}{$\psi_1$}$};

\end{scope}

\pgftransformshift{\pgfpoint{9cm}{0cm}}

\draw[step=1.0,black, opacity=0.25, very thin,xshift=-1cm,yshift=-1cm] (1,1)
grid (5,5);

\begin{scope}[opacity=0.5]

\node[yshift=-0.3cm](0) {$\scalebox{0.7}{$0$}$};
\node[ right of=0](0) {$\scalebox{0.7}{$1$}$};
\node[ right of=0](0) {$\scalebox{0.7}{$2$}$};
\node[ right of=0](0) {$\scalebox{0.7}{$3$}$};
\node[ right of=0](0) {$\scalebox{0.7}{$4$}$};
 
\node[xshift=4.3cm, yshift=-0.0cm](0) {$\scalebox{0.7}{$0$}$};
\node[ above of=0](0) {$\scalebox{0.7}{$1$}$};
\node[ above of=0](0) {$\scalebox{0.7}{$2$}$};
\node[ above of=0](0) {$\scalebox{0.7}{$3$}$};
\node[ above of=0](0) {$\scalebox{0.7}{$4$}$};

\end{scope}

\node[draw, circle, inner sep=1, label={[xshift=0.1cm, yshift=0.09cm]-150:\scalebox{0.6}{$ $}}](4) at (0, 4) {};

\node[draw, circle, inner sep=1, label={[xshift=0.1cm]-150:\scalebox{0.6}{$ $}}](3) at (0, 3) {};

\node[draw, circle, inner sep=1, label={[xshift=0.1cm]-150:\scalebox{0.6}{$ $}}](2) at (0, 2) {};

\node[draw, circle, inner sep=1, label={[xshift=0.1cm]-150:\scalebox{0.6}{$ $}}](1) at (0, 1) {};

\node[draw, circle, inner sep=1, label={[xshift=0.1cm]-150:\scalebox{0.6}{$ $}}](0) at (0, 0) {};

\draw (4) edge[->, bend left, looseness=1.5, in=110] node[midway, fill=white, inner sep = 0] {\scalebox{0.6}{$B$}}(3) ; 
\draw (3) edge[->, bend left] node[midway, fill=white, inner sep = 0] {\scalebox{0.6}{$B$}}(2) ; 
\draw (2) edge[->, bend left] node[midway, fill=white, inner sep = 0] {\scalebox{0.6}{$B$}}(1) ; 
\draw (1) edge[->, bend left] node[midway, fill=white, inner sep = 0] {\scalebox{0.6}{$B$}}(0) ;

\draw (4) edge [->,out=90,in=10,looseness=25] node[midway, fill=white, inner sep = 0] {\scalebox{0.6}{$B$}} (4);
\draw (3) edge [dashed, ->,out=180,in=90,looseness=35] node[midway, fill=white, inner sep = 0] {\scalebox{0.6}{$\xcancel{B}$}} (3);
\draw (2) edge [dashed, ->,out=60,in=-10,looseness=35] node[midway, fill=white, inner sep = 0] {\scalebox{0.6}{$\xcancel{B}$}} (2);
\draw (1) edge [->,,out=60,in=-10,looseness=25] node[midway, fill=white, inner sep = 0] {\scalebox{0.6}{$B$}} (1);
\draw (0) edge [->,,out=60,in=-10,looseness=25] node[midway, fill=white, inner sep = 0] {\scalebox{0.6}{$B$}} (0);

\draw (4) edge[->, bend right, opacity=0.75, dashed, looseness=2, out=-70] node[midway, fill=white, inner sep = 0] {\scalebox{0.6}{$\xcancel{B}$}}(2) ;
\draw (4) edge[->, bend right, opacity=0.75, dashed, looseness=2, out=-70] node[midway, fill=white, inner sep = 0] {\scalebox{0.6}{$\xcancel{B}$}}(1) ;
\draw (4) edge[->, bend right, opacity=0.75, dashed, looseness=2, out=-70] node[midway, fill=white, inner sep = 0] {\scalebox{0.6}{$\xcancel{B}$}}(0) ;

\draw (3) edge[->, bend right, opacity=0.75, dashed, looseness=1.9, out=-70, in=-110] node[midway, fill=white, inner sep = 0, looseness=2] {\scalebox{0.6}{$\xcancel{B}$}}(1) ;
\draw (3) edge[->, bend right, opacity=0.75, dashed, looseness=2, out=-70, in=-90] node[midway, fill=white, inner sep = 0] {\scalebox{0.6}{$\xcancel{B}$}}(0) ;

\draw (2) edge[->, bend right, looseness=2, out=-70, in=-110] node[midway, fill=white, inner sep = 0] {\scalebox{0.6}{${B}$}}(0) ;

\node[draw, circle, inner sep=1, thick, label={[xshift=0.1cm]-90:\scalebox{0.6}{$  $}}](D) at (1, 2) {};

\draw (4) edge[->, bend left] node[midway, fill=white, inner sep = 0] {\scalebox{0.6}{$D$}}(D) ; 
 
 \draw (3) edge[->, bend left] node[midway, fill=white, inner sep = 0] {\scalebox{0.6}{$D$}}(D) ; 

\node[draw, circle, inner sep=1, label={[xshift=0.1cm]-90:\scalebox{0.6}{$  $}}](1) at (1, 1) {};

\node[draw, circle, inner sep=1, label={[xshift=0.1cm]-90:\scalebox{0.6}{$  $}}](2) at (2, 2) {};

\node[draw, circle, inner sep=1, label={[xshift=0.1cm]-90:\scalebox{0.6}{$  $}}](3) at (3, 3) {};

\node[draw, circle, inner sep=1, label={[xshift=0.1cm]-90:\scalebox{0.6}{$  $}}](4) at (4, 4) {};

\draw[dashed, opacity=0.5] (0) -- (1) -- (2) -- (3) -- (4);

\pgftransformshift{\pgfpoint{-2cm}{-3cm}}

\node(T) {\input{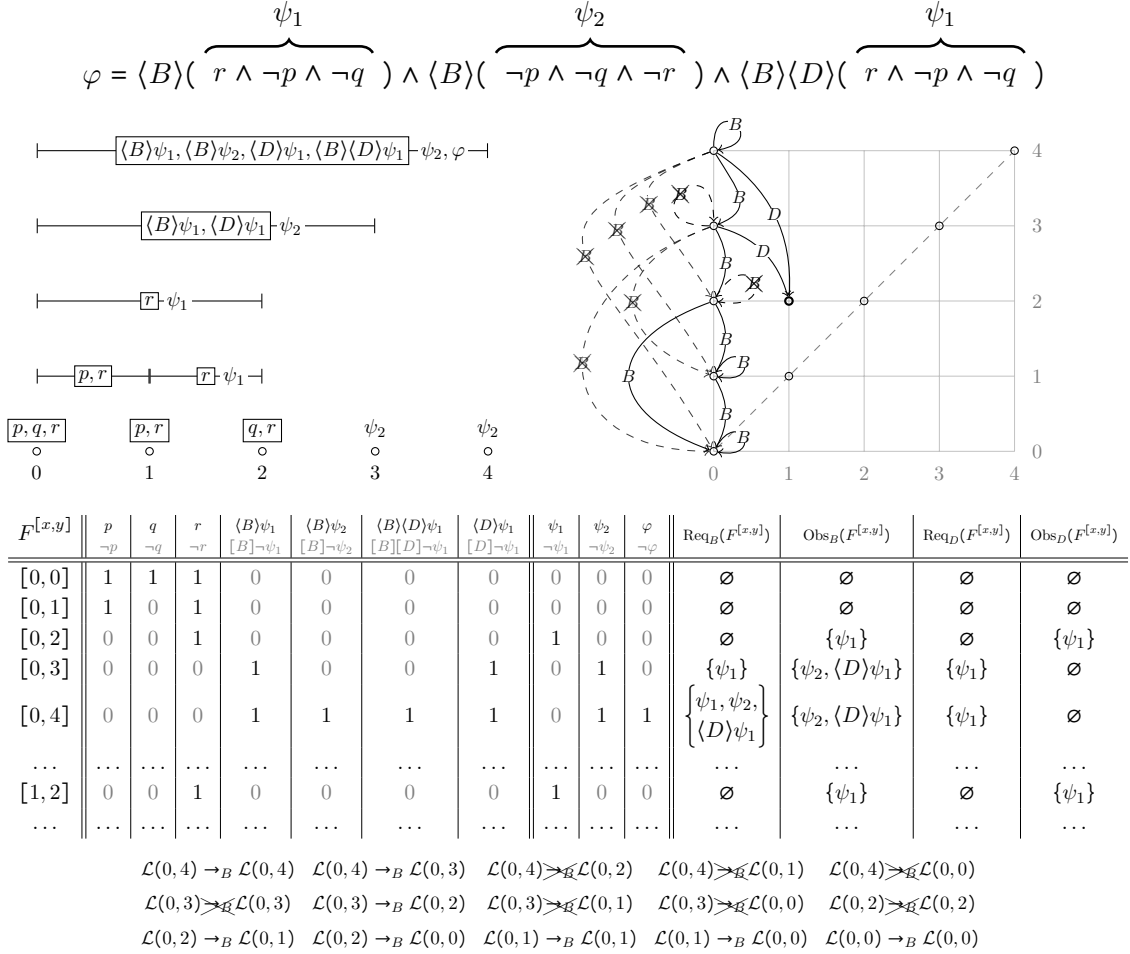}};
\node[anchor=north] at (T.south){
\scalebox{0.65}{$ \begin{array}{ccccccc}
 \cL(0,4)  \thenB \cL(0,4) & \cL(0,4)  \thenB \cL(0,3) &
 \cL(0,4) \thenNB \cL(0,2) &
 \cL(0,4) \thenNB \cL(0,1) &
 \cL(0,4) \thenNB \cL(0,0) \\[0.25cm]
  \cL(0,3) \thenNB \cL(0,3) &
 \cL(0,3) \thenB \cL(0,2) &
 \cL(0,3) \thenNB \cL(0,1) &
 \cL(0,3) \thenNB \cL(0,0) &
 \cL(0,2) \thenNB \cL(0,2) 	\\[0.25cm]
 \cL(0,2) \thenB \cL(0,1) &
 \cL(0,2) \thenB \cL(0,0) &
  \cL(0,1) \thenB \cL(0,1) &
 \cL(0,1) \thenB \cL(0,0) & 
  \cL(0,0) \thenB \cL(0,0) 

 \end{array}
$
}
};

\end{tikzpicture}

\caption{\label{fig:brelation} An account of the relation $\thenB$ from both the interval point of view (left) and the spatial one (right). }
\vspace{-0.5cm}

\end{figure}

\begin{exa}\label{ex:thenb}
We exemplify the behavior of the relation $\thenB$ by the example in Figure~\ref{fig:brelation}. First of all, we observe that it constrains the relationship between the $\reqB(F^{[x,y]})$ part of the labelling of an interval $[x,y]$ and that of its maximal proper prefix  $[x, y -1]$ (if any). In any ``consistent model'', for any strict-interval $[x, y]$, it indeed holds that $F^{[x,y]} \thenB F^{[x,y-1]}$ ($\thenB$ constrains the maximal prefix of an interval, not all its prefixes). 
By ``unravelling' the definition of $\thenB$, we can show that the following three conditions are satisfied:
\begin{compactenum}
\item from $\reqB(F^{[x,y-1]}) \subseteq \reqB(F^{[x,y]})$, it follows that $\boxB(F^{[x,y]}) \subseteq \boxB(F^{[x,y-1]})$, that is, $F^{[x,y-1]}$ features at least the universal requests in $F^{[x,y]}$;

\item from $\obsB(F^{[x,y-1]}) \subseteq \reqB(F^{[x,y]})$, it follows that $F^{[x,y-1]}$ possibly satisfies some of the requests in $\reqB(F^{[x,y]})$ and it
does not violate any of the universal requests in $\boxB(F^{[x,y]})$; otherwise, we would have $\obsB(F^{[x,y-1]}) \setminus \reqB(F^{[x,y]})\neq \emptyset$;

\item from $\reqB(F^{[x,y]}) = \reqB(F^{[x,y-1]}) \cup \obsB(F^{[x,y-1]})$, we have that, for each $\psi \in \reqB(F^{[x,y]})$, either $F^{[x,y-1]}$ satisfies $\psi$ ($\psi \in \obsB(F^{[x,y-1]})$) or it delegates its satisfaction to its prefixes ($\psi \in \reqB(F^{[x,y-1]})$). In the former case,  $F^{[x,y-1]}$ might potentially ask for $\psi$ again (whenever $\psi \in \reqB(F^{[x,y-1]})$ as well). This is actually the behaviour that one may expect from the labelling of the maximal prefix $[x,y-1]$ of an interval $[x,y]$: a request in $\reqB(F^{[x,y]})$ is either satisfied locally in $F^{[x,y-1]}$ (possibly once and for all) or delegated to its prefixes. 
\end{compactenum}

In Figure~\ref{fig:brelation} (both in the interval model to the left and in its graphical counterpart to the right),  we show the labelling of the pairs of atoms/intervals that satisfy the relation $\thenB$. The example formula features three $\hsEB$ requests, namely, $\psi_1 = r \wedge \neg p \wedge \neg q$, $\psi_2 = \neg p \wedge \neg q \wedge \neg r$, and $\psi_3 = \hsED\psi_1$. Let us focus our attention on the intervals starting at point $0$, namely, $[0,0], [0,1], [0,2], [0,3]$, and $[0,4]$. In Figure~\ref{fig:brelation} (bottom), we report the atoms associated with these intervals, plus the one for the interval $[1,2]$, which satisfies the request $\hsED \psi_1$ of intervals $[0,3]$ and $[0,4]$.
In addition, at the very bottom of Figure~\ref{fig:brelation}, we show when $F^{[0,x]}\thenB F^{[0,x']}$, with $0\leq x \leq x' \leq 4$, is true and when it is not (a graphical account of the same pieces of information is given in the top right part of Figure~\ref{fig:brelation}).

We now analyse the behavior of $\reqB(F^{[0,x]})$ and $\obsB(F^{[0,x]})$ for $x=0$ up to $x=4$. 

We first observe that both $F^{[0,0]}$ and $F^{[0,1]}$ satisfy neither $\psi_1$ nor $\psi_2$ (it suffices to observe that $p$ holds on both of them). Moreover, both $\reqB(F^{[0,0]}) = \reqB(F^{[0,1]}) = \emptyset$ and $\obsB(F^{[0,0]})= \obsB(F^{[0,1]})= \emptyset$, and
thus both $F^{[0,1]}\thenB F^{[0,0]}$ and $F^{[0,0]} \thenB F^{[0,1]}$ trivially hold, that is, the two labellings can be swapped without any consequence on the consistency of $B$ requests. For the very same reason, it holds that $F^{[0,1]}\thenB F^{[0,1]}$ as well as $F^{[0,0]} \thenB F^{[0,0]}$.

Atoms  $F^{[0,0]}$ and $F^{[0,1]}$ are trivially $B$-reflexive because both their $B$-requests and their $B$-observables are empty. However, there may exist $B$-reflexive atoms $F$ whose sets of $B$-requests and $B$-observables are both non-empty. A necessary and sufficient condition for an atom $F$ to be $B$-reflexive is indeed that $\obsB(F) \subseteq \reqB(F)$. In such a case, the $B$-requests that $F$ possibly satisfies by means of its observables are immediately reproduced by the $B$-requests themselves. This is the case with atom $F^{[0,4]}$ in Figure~\ref{fig:brelation} where both $\obsB(F^{[0,4]})$ and $\reqB(F^{[0,4]})$ are non-empty and
$\obsB(F^{[0,4]})\subseteq \reqB(F^{[0,4]})$.

Atom $F^{[0,2]}$ is a different story: it features $\psi_1$ and thus $\psi_1 \in \obsB(F^{[0,2]})$. However, $\psi_1 \notin \reqB(F^{[0,2]})$ and thus $F^{[0,2]}$ is not  $B$-reflexive. Clearly, $F^{[0,2]}\thenB F^{[0,1]}$, since both of them have no $\hsEB$-requests.

Atom $F^{[0,3]}$ is the first atom with at least one $\hsEB$ request, namely,  $\reqB(F^{[0,3]})= \{\psi_1\}$. It holds that $F^{[0,3]}\thenB F^{[0,2]}$, since $\reqB(F^{[0,3]}) = \reqB(F^{[0,2]}) \cup \obsB(F^{[0,2]})$ ($\{\psi_1\}= \emptyset \cup \{\psi_1\}$). On the contrary, $F^{[0,3]}\thenB F^{[0,1]}$ does not hold, since $F^{[0,1]}$ neither features $\hsEB \psi_1$ nor satisfies $\psi_1$ ($\{\psi_1\} \neq \emptyset \cup \emptyset$). 
As for the observables, compared to $F^{[0,2]}$, $F^{[0,3]}$ ``loses'' $\psi_1$, which is transferred to its $\hsEB\psi_1$ request, but both 
$\psi_2$ and $\hsED \psi_1$ are in $\obsB(F^{[0,3]})$, and they may satisfy $\hsEB$-requests coming from atoms labelling intervals that feature $[0,3]$ as a proper prefix.  

As for $F^{[0,4]}$, $\obsB(F^{[0,4]}) = \emptyset$ and $\reqB(F^{[0,4]}) = \{\psi_1, \psi_2, \hsED \psi_1\}$. It can be easily checked that
$F^{[0,4]} \thenB F^{[0,3]}$, i.e., $\reqB(F^{[0,4]}) = \reqB(F^{[0,3]}) \cup \obsB(F^{[0,3]})$, as $\{\psi_1, \psi_2, \hsED\psi_1\} = \{\psi_1\} \cup \{\psi_2, \hsED \psi_1 \}$. By $\reqB(F^{[0,3]})$, $\psi_1$ is transferred to the proper prefixes of $[0,3]$,  while both $\psi_2$ and  $\hsED \psi_1$ are in $\obsB(F^{[0,3]})$ and thus are locally satisfied.

We conclude by noticing that there may be atoms $F$ and $G$ such that $\reqB(F) = \reqB(G) \cup \obsB(G)$ (i.e., $F \thenB G$),
and $\reqB(G) \cap \obsB(G) \neq \emptyset$, that is, a $\hsEB$-request may be at the same time  locally satisfied by $G$ and featured as request for its proper prefixes.
\end{exa}

The next important proposition determines, for any atom $F^{[x,y]}$, the number of atoms $F^{[x,y+k]}$, with $k\geq 1$ and $(\reqB(F^{[x,y+k]}), \obsB(F^{[x,y+k]})) \neq (\reqB(F^{[x,y]}), \obsB(F^{[x,y]}))$, that 
may have $F^{[x,y]}$ as (the labeling of) a prefix. Its proof is given in Appendix~\ref{appendix:proofs}.

\begin{restatable}{prop}{lbbc}\label{lem:distinctsupprefixes}
Let $\varphi$ be a $\mathsf{BD}_{hom}$ formula. For any atom $F \in \atoms$ and any sequence of atoms $F_h \thenB \ldots \thenB  F_1 \thenB 
F_0 = F $, where, for each $0 \leq i \neq j \leq h$, $\reqB(F_i)\neq \reqB(F_j)$ or $\obsB(F_i) \setminus \reqB(F_i) \neq \obsB(F_j) \setminus \reqB(F_j)$, it holds that $h \leq 2|\{\psi: \hsEB \psi \in \closure \}| - (2|\reqB(F)| + |\obsB(F) \setminus \reqB(F)|)$. 
\end{restatable}

\noindent Intuitively, Proposition \ref{lem:distinctsupprefixes} provides a linear bound on the number of distinct atoms 
that may appear in a $\thenB$ chain of atoms.


\begin{figure}
\centering

\begin{tikzpicture}

\begin{scope}[anchor=west, inner sep=0.1cm]
\node(F0){$\varphi=\hsED($};
\node(F1) at (F0.east) {$p \wedge q$};
\node(F2) at (F1.east){$) \wedge \hsED($};
\node(F3) at (F2.east){$\neg p \wedge q$};
\node(F4) at (F3.east){$) \wedge \hsED($};
\node(F5) at (F4.east){$p \wedge \neg q$};    
\node(F6) at (F5.east){$)$};

\draw[decorate, line width=1pt,
      decoration={brace,raise=0.3cm,
                  amplitude=5pt,mirror},
                  color=black,
                  ] 
(F1.east) -- (F1.west) node[above=0.53cm, midway]{$\psi_1$};

\draw[decorate, line width=1pt,
      decoration={brace,raise=0.3cm,
                  amplitude=5pt,mirror},
                  color=black,
                  ] 
(F3.east) -- (F3.west) node[above=0.53cm, midway]{$\psi_2$};

\draw[decorate, line width=1pt,
      decoration={brace,raise=0.3cm,
                  amplitude=5pt,mirror},
                  color=black,
                  ] 
(F5.east) -- (F5.west) node[above=0.53cm, midway]{$\psi_3$};

\end{scope}

\pgftransformshift{\pgfpoint{-2.5cm}{-4.5cm}}

\begin{scope}[node distance=1.5cm]

\node[draw, circle, inner sep=1, 
label={[]270:$\scalebox{0.7}{$0$}$},
label={[]90:$\scalebox{0.7}{$ $}$}](0) {};

\node[draw, circle, inner sep=1, right of=0, 
label={[]270:$\scalebox{0.7}{$1$}$},
label={[name=l1]90:$\scalebox{0.7}{$p$}$},
label={[xshift=0.35cm]90:$\scalebox{0.7}{$\psi_3$}$}
](1) {};

\node[fit=(l1), draw, inner sep=-2]{};

\node[draw, circle, inner sep=1, right of=1, label={[]270:$\scalebox{0.7}{$2$}$},
label={[name=l2]90:$\scalebox{0.7}{$p,q$}$},
label={[xshift=0.45cm]90:$\scalebox{0.7}{$\psi_1$}$}
](2) {};

\node[fit=(l2), draw, inner sep=-2]{};

\node[draw, circle, inner sep=1, right of=2, label={[]270:$\scalebox{0.7}{$3$}$},
label={[name=l3]90:$\scalebox{0.7}{$q$}$},
label={[xshift=0.35cm]90:$\scalebox{0.7}{$\psi_2$}$}](3) {};

\node[fit=(l3), draw, inner sep=-2]{};

\node[draw, circle, inner sep=1, right of=3, label={[]270:$\scalebox{0.7}{$4$}$},
label={[]90:$\scalebox{0.7}{$ $}$}](4) {};


\draw[|-|] ($(0.center) + (0,2.8)$) -- ($(4.center)+ (0,2.8)$)
node[pos=0.5, fill=white,draw, inner sep=1.5pt](16) {$\scalebox{0.7}{$\hsED \psi_1,\hsED \psi_2,\hsED \psi_3$}$}
node[pos=0.75, fill=white, inner sep=1pt](16) {$\scalebox{0.7}{$\varphi$}$};

\draw[|-|] ($(1.center) + (0,1)$) -- ($(2.center)+ (0,1)$)
node[pos=0.5, fill=white,draw, inner sep=1.5pt](12) {$\scalebox{0.7}{$p$}$}
node[pos=0.75, fill=white, inner sep=1pt](12) {$\scalebox{0.7}{$\psi_3$}$};

\draw[|-|] ($(1.center) + (0,2.1)$) -- ($(3.center)+ (0,2.1)$)
node[pos=0.5, fill=white,draw, inner sep=1.5pt](47) {$\scalebox{0.7}{$\hsED \psi_1$}$};

\draw[|-|] ($(2.center) + (0,1.5)$) -- ($(3.center)+ (0,1.5)$)
node[pos=0.5, fill=white,draw, inner sep=1.5pt](46) {$\scalebox{0.7}{$q$}$}
node[pos=0.75, fill=white, inner sep=1pt](46) {$\scalebox{0.7}{$\psi_2$}$};

\end{scope}

\pgftransformshift{\pgfpoint{9cm}{-0.55cm}}

\draw[step=1.0,black, opacity=0.25, very thin,xshift=-1cm,yshift=-1cm] (1,1)
grid (5,5);

\begin{scope}[opacity=0.5]

\node[yshift=-0.3cm](0) {$\scalebox{0.7}{$0$}$};
\node[ right of=0](0) {$\scalebox{0.7}{$1$}$};
\node[ right of=0](0) {$\scalebox{0.7}{$2$}$};
\node[ right of=0](0) {$\scalebox{0.7}{$3$}$};
\node[ right of=0](0) {$\scalebox{0.7}{$4$}$};
\node[ right of=0](0) {$\scalebox{0.7}{$\ldots$}$};

\node[xshift=4.3cm, yshift=-0.0cm](0) {$\scalebox{0.7}{$0$}$};
\node[ above of=0](0) {$\scalebox{0.7}{$1$}$};
\node[ above of=0](0) {$\scalebox{0.7}{$2$}$};
\node[ above of=0](0) {$\scalebox{0.7}{$3$}$};
\node[ above of=0](0) {$\scalebox{0.7}{$4$}$};

\end{scope}

\node[draw, circle, inner sep=1, label={[xshift=0.1cm, yshift=0.09cm]90:\scalebox{0.6}{$ $}}](4) at (0, 4) {};
\node[draw, circle, inner sep=1, label={[xshift=-0.1cm]-90:\scalebox{0.6}{$ $}}](3) at (1, 3) {};
\node[draw, circle, inner sep=1, label={[xshift=0.1cm]-90:\scalebox{0.6}{$ $}}](2) at (1, 2) {};
\node[draw, circle, inner sep=1, label={[xshift=0.1cm]-90:\scalebox{0.6}{$ $}}](1) at (2, 3) {};
\node[draw, circle, inner sep=1, label={[xshift=0.1cm]-150:\scalebox{0.6}{$  $}}](0) at (0, 0) {};

\draw (4) edge[->] node[midway, fill=white, inner sep = 0] {\scalebox{0.6}{$D$}}(3) ; 
\draw (4) edge[->] node[midway, fill=white, inner sep = 0] {\scalebox{0.6}{$D$}}(1) ; 

\node[draw, circle, inner sep=1, label={[xshift=0.1cm]-90:\scalebox{0.6}{$ $}}](11) at (1, 1) {};

\node[draw, circle, inner sep=1, label={[xshift=0.1cm]-90:\scalebox{0.6}{$ $}}](22) at (2, 2) {};

\node[draw, circle, inner sep=1, label={[xshift=0.1cm]-90:\scalebox{0.6}{$ $}}](33) at (3, 3) {};

\node[draw, circle, inner sep=1, label={[xshift=0.1cm]-90:\scalebox{0.6}{$  $}}](44) at (4, 4) {};

\draw[dashed, opacity=0.5] (0) -- (11) -- (22) -- (33) -- (44);

 \draw[->, looseness=0.9] (4.west) to[out=230,in=120] node[pos =0.4,fill=white, inner sep = 0] {\scalebox{0.6}{$D$}}(2.west);

 \draw[->, looseness=0.9] (4.west) to[out=230,in=120] node[pos =0.4,fill=white, inner sep = 0] {\scalebox{0.6}{$D$}}(11.north);
 
\draw[->, looseness=0.7] (4) to[out=50,in=120] node[pos =0.4,fill=white, inner sep = 0] {\scalebox{0.6}{$D$}}(33);
\draw[->, looseness=0.7] (4) to[out=50,in=120] node[pos =0.4,fill=white, inner sep = 0] {\scalebox{0.6}{$D$}}(33);
\draw[->, looseness=1.5] (4) to[out=20,in=0] node[pos =0.4,fill=white, inner sep = 0] {\scalebox{0.6}{$D$}}(22.east);

\draw (4) edge [->,out=150,in=90,looseness=25] node[midway, fill=white, inner sep = 0] {\scalebox{0.6}{$D$}} (4);
\draw (3) edge [->,out=0,in=-60,looseness=25] node[midway, fill=white, inner sep = 0] {\scalebox{0.6}{$D$}} (3);
\draw (11) edge [dashed, ->,out=0,in=-60,looseness=35] node[midway, fill=white, inner sep = 0] {\scalebox{0.6}{$\xcancel{D}$}} (11);
\draw (22) edge [dashed, ->,out=0,in=-60,looseness=35] node[midway, fill=white, inner sep = 0] {\scalebox{0.6}{$\xcancel{D}$}} (22);
\draw (33) edge [dashed, ->,out=0,in=-60,looseness=35] node[midway, fill=white, inner sep = 0] {\scalebox{0.6}{$\xcancel{D}$}} (33);
\draw (2) edge [dashed, ->,out=0,in=-60,looseness=35] node[midway, fill=white, inner sep = 0] {\scalebox{0.6}{$\xcancel{D}$}} (2);
\draw (1) edge [dashed, ->,out=-60,in=-120,looseness=35] node[midway, fill=white, inner sep = 0] {\scalebox{0.6}{$\xcancel{D}$}} (1);

\pgftransformshift{\pgfpoint{-2cm}{-3.8cm}}

\node(T) {\input{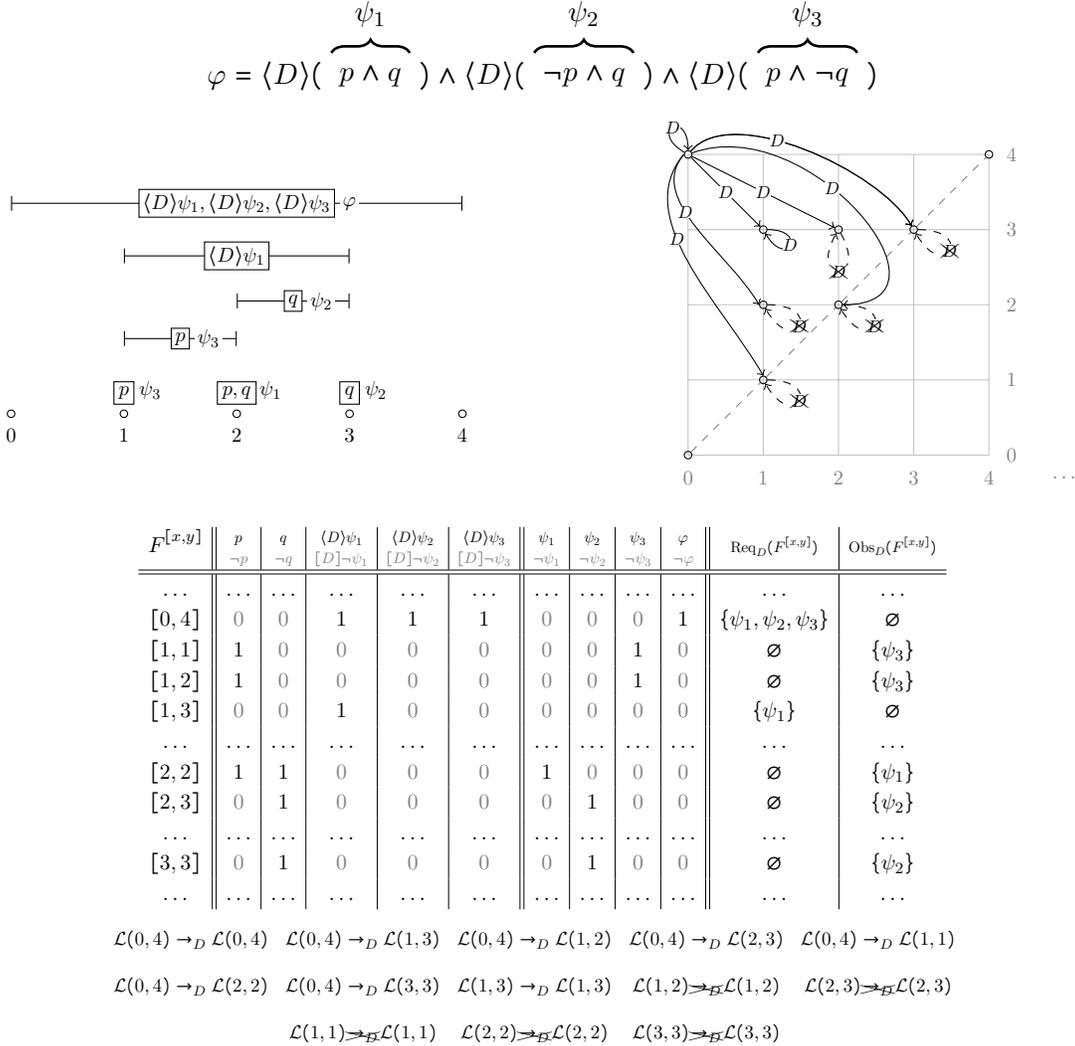}};
\node[anchor=north] at (T.south){
\scalebox{0.65}{$ \begin{array}{ccccccc}
 \cL(0,4)  \thenD \cL(0,4) & 
 \cL(0,4)  \thenD \cL(1,3) & 
 \cL(0,4)  \thenD \cL(1,2) &
 \cL(0,4)  \thenD \cL(2,3) &
 \cL(0,4)  \thenD \cL(1,1) \\[0.5cm]
 \cL(0,4)  \thenD \cL(2,2) &
 \cL(0,4)  \thenD \cL(3,3) &
 \cL(1,3)  \thenD \cL(1,3) &
 \cL(1,2)  \thenND \cL(1,2) & 
 \cL(2,3)  \thenND \cL(2,3) \\[0.5cm] 
 &
 \cL(1,1)  \thenND \cL(1,1) &
 \cL(2,2)  \thenND \cL(2,2) &
 \cL(3,3)  \thenND \cL(3,3) &

 \end{array}
$
}
};

\end{tikzpicture}

\caption{\label{fig:drelation} 
An account of relation $\thenD$  from both the interval point of
view (left) and the spatial one (right). For the sake of readability, we only highlight the sub-intervals of $[0,4]$.}

\end{figure}

Let us consider now relation $\thenD$. We first observe that, according to Definition~\ref{def:intervalrelationsonatoms}, given two atoms $F$ and $G$, the condition imposed by $F \thenD G$  is weaker than the one imposed by $\thenB$, that is, containement (superset) instead of full equality of the two sets. The reason is that by $F \thenD G$ we state that $G$ may label any sub-interval $[x,y]$ of an interval $[x',y']$ ($x'< x \leq y < y'$) and not just its maximal proper sub-interval $[x'+1, y'-1]$, while $G$ in $F \thenB G$ only refers to the maximal proper prefix $[x',y'-1]$ of $[x',y']$. We exemplify the behaviour of $\thenD$ by means of the following example.

\begin{exa}\label{ex:thend}
In Figure~\ref{fig:drelation}, we illustrate the labelling of intervals which are required to satisfy relation $\thenD$. We refer both to the interval model and to its graphical counterpart. Let us consider the following three $\hsED$-requests: $\psi_1 =  p \wedge q$, $\psi_2 = \neg p \wedge  q $, and $\psi_3 = p \wedge \neg q$, and all the proper sub-intervals of the largest interval in the model, namely, sub-intervals $[1,1],  [1,2], [1,3], [2,2], [2,3]$, and $[3,3]$ of interval $[0,4]$. In Figure~\ref{fig:drelation} (bottom), we report the cases where $F^{[x,y]}\thenD F^{[x',y']}$, for $0< x' \leq y' < 4$ and $[x,y] =[0,4]$, is true and those where it is not. A graphical account of these relations is given in Figure~\ref{fig:drelation} (top right).

Let us now describe the behavior of $\reqD(F^{[x,y]})$ and $\obsD(F^{[x,y]})$  moving from interval $[x,y]$ to its maximal sub-interval $[x+1, y-1]$ starting from the largest interval $[0,4]$. First, we observe that, since $\obsD(F^{[0,4]}) = \emptyset$, it trivially holds that $\reqD(F^{[0,4]}) \supseteq \reqD(F^{[0,4]}) \cup \obsD(F^{[0,4]})$, and thus $F^{[0,4]} \thenD F^{[0,4]}$, that is, $F^{[0,4]}$ is $D$-reflexive. Hence, its labeling $F^{[0,4]}$ may possibly be associated with a proper sub-interval of any interval labelled by $F^{[0,4]}$. The same holds with the $D$-reflexive atom $F^{[1,3]}$.
On the contrary, the relation $F^{[0,4]} \thenD F^{[1,3]} $ cannot be turned into the relation $F^{[1,3]} \thenD F^{[0,4]}$, as $\reqD(F^{[1,3]}) (=
\{\psi_1\}) \not\supseteq \obsD(F^{[0,4]}) \cup \reqD(F^{[0,4]}) (= \emptyset \cup \{\psi_1, \psi_2, \psi_3\})$.

It is worth pointing out that the following stronger consistency property, involving equality in place of containment, actually holds for $\hsED$-requests: for all $[x,y]$, it holds that 
$\reqD(F^{[x,y]}) = 
    \reqD(F^{[x + 1, y]}) \cup 
    \reqD(F^{[x, y - 1]}) \cup
    \obsD(F^{[x + 1, y]}) \cup 
    \obsD(F^{[x, y - 1]}) \cup 
    \obsD(F^{[x + 1, y - 1]})$. 
It states that the $\hsED$-requests that hold over an interval $[x,y]$ must be completely ``covered'' by those holding over its maximal proper prefix $[x + 1, y]$, its maximal proper suffix $[x, y - 1]$, and the union of all the observables of its maximal proper prefix, its maximal proper suffix, and its maximal proper sub-interval $[x + 1, y - 1]$.
As an example, in Figure~\ref{fig:drelation}, we have that $\reqD(F^{[0,4]}) = \reqD(F^{[1,4]}) \cup \reqD(F^{[0, 3]}) \cup \obsD(F^{[1, 4]}) \cup \obsD(F^{[0, 3]}) \cup \obsD(F^{[1, 3]}) = \{\psi_1, \psi_2\} \cup \{\psi_1, \psi_3\} \cup \emptyset \cup \emptyset \cup \emptyset = \{\psi_1, \psi_2, \psi_3\}$. Finally, notice that since  both the maximal prefix $[x,y-1]$ and the maximal suffix $[x+1, y]$) of an interval $[x,y]$ are not proper sub-intervals of it, it may be the case that $F^{[x,y]} \thenB F^{[x,y-1]}$ and/or $F^{[x,y]} \thenB F^{[x+1,y]}$ \emph{do not hold} in a consistent model. For instance, in Figure~\ref{fig:drelation}, neither $F^{[1,3]} \thenD F^{[1,2]}$ nor $F^{[1,3]} \thenD F^{[2,3]}$ holds. 
\end{exa}

The next proposition shows that to check the equality of atoms it suffices to check
the equality of their propositional components and  of their respective sets of $\hsEB$- and $\hsED$-requests.

\begin{prop}\label{lem:atomdeterminacy}
Let $F,G \in \atoms$. It holds that $F=G$ if and only if $\reqB(F) = \reqB(G)$, $\reqD(F) = \reqD(G)$, and $F \cap \Prop = G \cap \Prop$. 
\end{prop}

The statement of Proposition~\ref{lem:atomdeterminacy} immediately follows from the fact that, for each atom $F$ and each  $\psi \in F$, either  $\psi \in \Prop \cup \{\hsEB \psi': \psi' \in \reqB(F)  \} \cup \{\hsED \psi': \psi' \in \reqD(F)  \}$ or $\psi$ is a Boolean combination of  $\Prop \cup \{\hsEB \psi': \psi' \in \reqB(F)  \} \cup
\{\hsED \psi': \psi' \in \reqD(F)  \}$.

\smallskip

Given a formula $\varphi$, a \emph{$\varphi$-compass structure} 
(compass structure, when $\varphi$ is clear from the context) 
is a pair $\cG=(\bG_N, \cL)$, where $N \in \bN$, $\bG_N = \{(x,y) : 
0\le x\le y\le N\}$, and $\cL: \bG_N\rightarrow \atoms$ is a 
labelling function that satisfies the following conditions:

\smallskip

\begin{compactitem}
\item (\emph{initial formula}) $\varphi \in \cL(0,N)$;
\item ($B$-\emph{consistency}) for all $0 \leq x\leq y < N$, $\cL(x,y+1) \thenB \cL(x,y)$,
and for all $0 \leq x \leq N$, $\reqB(\cL(x,x)) = \emptyset$;
\item ($D$-\emph{consistency}) for all $0 \leq x < x' \leq y' < y \leq N$, $\cL(x,y) \thenD \cL(x',y')$;
\item ($D$-\emph{fulfilment}) for all $0 \leq x\leq y \leq N$ and all $\psi \in \reqD(\cL(x,y))$,
there exist $x < x' \leq y' < y$ such that  $\psi \in \cL(x',y')$.
\end{compactitem}

\smallskip

An example of compass structure is given in Figure~\ref{fig:reqExplanationBD} (top right), where the labelling with atoms of the second octant of a finite cartesian plane
is graphically depicted. Notice that the definition of $\thenB$ and $B$-consistency guarantee that all $B$-requests are fulfilled. 


Let $\cG=(\bG_N, \cL)$ be a compass structure and let $\cP: \bG_N\rightarrow 2^{\Prop}$ such that $\cP(x,y) = \{ p \in \Prop: p \in \cL(x',x') \mbox{ for all $x\leq x'\leq y$} \}$. We say that $\cG$ is \emph{homogeneous} if for all $(x,y)\in\bG_N$, $\cL(x,y) \cap \Prop = \cP(x,y)$. The proof of the next theorem is straightforward and thus omitted. 

\begin{thm}\label{thm:satthencompass}
A $\mathsf{BD}_{hom}$ formula $\varphi$ is satisfiable  
if and only if there is a homogeneous $\varphi$-compass structure.
\end{thm} 

Hereafter, we will often write compass structure for homogeneous 
$\varphi$-compass structure.

\noindent In the next sections, we will prove  a small model theorem about compass structures for an input $\mathsf{BD}_{hom}$ formula $\varphi$. In particular, we will prove that a model can be built by contracting a larger one in such a way that the resulting model is still a compass structure for $\varphi$. To this end, we need to preliminarily state some spatial properties of compass structures where the distinction between $B$-reflexive (resp., $D$-reflexive) and $B$-irreflexive (resp., $D$-irreflexive) atoms plays a major role. Intuitively, if a point is labelled with an atom which is both $B$-reflexive and $D$-reflexive, its only purpose is to ``fill the gaps'' in the model, as each $B$/$D$-request that it possibly solves for other points are transferred to its prefixes/sub-intervals. On the other hand, a $B$-irreflexive, $D$-irreflexive, or both $B$-irreflexive and $D$-irreflexive point must be dealt with carefully since its observables includes at least one $B$- or $D$-request that is solved once and for all, and it is not transferred to its prefixes/sub-intervals.


\section{Spatial properties of compass structures for 
\texorpdfstring{$\mathsf{BD}_{hom}$}{BD_hom} formulas}
\label{sec:properties}

In this section, we state a series of spatial properties of compass structures that turn out to be quite useful to prove the results of Sections \ref{sec:expspace} and \ref{sec:abdexpspace}.  
Each property is proved by making use of the 
previous one as follows.

In Section \ref{subsec:columns}, we show that for any compass structure and any of its $X$-axis coordinates $x$, the sequence $\cL(x,0)\ldots\cL(x,N)$ is monotonic, i.e., for any triplet $0\leq y_1 < y_2 < y_3 \leq N$, it cannot be the case that $\cL(x,y_1)= \cL(x,y_3)$ and $\cL(x, y_2) \neq \cL(x, y_1)$. Such a property allows us to represent relevant information associated with any column $x$ in space (polynomially) bounded in $|\varphi|$.
Next, in Section \ref{subsec:equivcolumns}, we define an equivalence relation over columns such that two columns are equivalent if they feature the same set of atoms. It is easy to check that such an equivalence relation is of finite index and its index is  exponentially  bounded in $|\varphi|$. By exploiting the representation of Section \ref{subsec:columns}, we first define a partial order over equivalent columns, and then we prove that, in a compass structure,  such  a relation totally orders them.
Finally, in Section \ref{subsec:covered}, by exploiting the total order of the elements of each equivalence class, we show a crucial property of the rows of a compass structure, which is the cornerstone of the proof. First, we associate with each point $(x,y)$ on row $y$, with $0\leq x \leq y$, a tuple consisting of (i) $\cL(x,y)$, (ii) the equivalence class $\sim_x$ of column $x$, and (iii) the set of pairs $(\cL(x',y), \sim_{x'})$, for all 
$x < x' \leq y$, and then we prove that, for every pair of points $(x,y), (x',y)$ that feature the same tuple, $\cL(x,y') = \cL(x', y')$ for all $y'>y$, that is, columns $x$ and $x'$ behave the same way (i.e., exhibit the same labelling) from $y$ to the upper end.


\subsection{A finite characterisation of columns and of their relationships}
\label{subsec:columns}

In this section, we first show that, in every compass structure, 
the atoms that appear in a column $x$ must respect a certain order, 
that is, they cannot be interleaved. Let $F, G,$ and $H$ be three 
pairwise distinct atoms. In Figure \ref{\detokenize{fig:step1picture}}(a), 
we give a graphical account of the property that we are going to prove, while, in Figure 
\ref{\detokenize{fig:step1picture}}(b), we show a violation of it (atom $H$ appears before and after atom $G$ moving upward along the column). To start with, we state a fundamental property of  $B$-irreflexive atoms (the proof is given in Appendix~\ref{appendix:proofs}). 

\begin{restatable}{lem}{lbstep}\label{lem:bstep}
Let $\cG=(\bG_N, \cL)$ be a compass structure. 
For all $x \leq y < N$, if $\reqB(\cL(x,y))\subset 
\reqB(\cL(x,y+1))$, then $\cL(x,y)$ is $B$-irreflexive.
\end{restatable}

Observe that, from Lemma~\ref{lem:bstep}, it follows that, given any sequence of points  $(x, x)(x,x +1)\ldots (x,x+k)$ in a compass structure, whenever we encounter a $B$-irreflexive atom, we have to drop at least one $B$-request, and thus the number of irreflexive atoms in the sequence is bounded by $|\{ \hsEB \psi \in \closure\}|$.

Pairing such an observation with the statement of Lemma~\ref{lem:distinctsupprefixes}, it is possible to provide a bound on the number of distinct atoms that can be placed above a given atom $F$ in a column, taking into account $B$-requests, $D$-requests, and negative literals in $F$.

Formally, we define a function $\Deltareq: \atoms \rightarrow \bN$ as follows:
\[
\begin{array}{rcl}
\Deltareq(F) &=& (2|\{\hsEB \psi \in \closure \}| - (2|\reqB(F)| + |\obsB(F) \setminus \reqB(F)|))+ \\
&& (|\{\hsED \psi \in \closure \}| - |\reqD(F)|) + \\
&& (|\{\neg p: p \in \closure \cap \Prop\}| - |\{ \neg p : p  \in \closure \cap \Prop \wedge  \neg p \in F  \}|)
\end{array}
\]
The proof of Proposition~\ref{lem:distinctsupprefixes} (see Section~\ref{sec:compass})
helps us to understand why the factor $2$ comes into play in the case of $B$-requestes.
Informally, from (the proof of) Proposition~\ref{lem:distinctsupprefixes}, it immediately follows that, in order to move down from an atom featuring $\hsEB \psi$ to an atom featuring $\neg \psi, \hsAB \neg \psi$, one must pass through an atom featuring $\psi, \hsAB \neg \psi$.
It can be easily checked that, for each $F \in \atoms$, it holds that $0 \leq \Deltareq(F) \leq 4 |\varphi| + 1$.

To illustrate how $\Deltareq$ works, we give a simple example.
\begin{exa}\label{ex:sequencebound}
Let $\{\psi: \hsEB \psi \in \closure \} = \{\psi_1\}$ and let $F_3 \thenB F_2 \thenB F_1$,
with $\reqB(F_3)=\{\psi_1\}$ and $\reqB(F_2) = \reqB(F_1) = \emptyset$ (by definition of $\thenB$, it immediately follows that $\obsB(F_2) = \{\psi_1\}$ and $\obsB(F_3) = \emptyset$). For simplicity, let us assume that $\{\psi: \hsED \psi \in \closure \} = \emptyset$, and thus $\reqD(F_3) = \reqD(F_2) = \reqD(F_1) = \emptyset$, and
$(F_3 \cap F_2 \cap F_1)\cap \Prop= \Prop = \{p\}$. 
It holds that $\Deltareq(F_1)= (2\cdot 1 - (2 \cdot 0 + 0)) + (0 - 0) + (1 - 0)= 3$, $\Deltareq(F_2)= (2\cdot 1 - (2 \cdot 0 + 1)) + (0 - 0) + (1 - 0)= 2$,
and $\Deltareq(F_3)= (2\cdot 1 - (2 \cdot 1 + 0)) + (0 - 0) + (1 - 0)= 1$.
\end{exa}

We say that an atom $F$ is \emph{initial} if and only 
if $\reqB(F)=\emptyset$. A \emph{$B$-sequence} is a sequence of atoms
$\shadingB= F_0\ldots F_n$ such that $F_0$ is initial
and for all $0< i \leq n$ we have $F_i\thenB F_{i-1}$,
$\reqD(F_i) \supseteq \reqD(F_{i-1})$, and
$F_i \cap \Prop \subseteq F_{i-1} \cap \Prop$. 
It is worth pointing out that atoms in a $B$-sequence 
are monotonically non-increasing 
in $\Deltareq$, that is, $\Deltareq(F_0)\geq\ldots\geq \Deltareq(F_n)$.

\begin{defi}\label{def:flatbsequence}
Let $\shadingB= F_0\ldots F_n$ be a $B$-sequence. We say that $\shadingB$ is \emph{flat}  if and only if it can be written as a sequence $\oF^{k_0}_0\ldots \oF^{k_m}_{m}$, where $k_i > 0$, for all $0\leq i \leq m$,  and $\oF_i \neq \oF_j$, for all $0\leq i<j\leq m$.
\end{defi}

\noindent For the sake of clarity, it is worth mentioning that here and in the following $\oF$ it is not used to denote the complement of $F$, but as a simple alias for atoms. 

We say that a flat $B$-sequence
$\oF^{k_0}_0 \ldots \oF^{k_m}_{m}$ is \emph{decreasing} 
if and only if $\Deltareq(\oF_0)>\ldots> \Deltareq(\oF_{m})$. Flat (decreasing) $B$-sequences are the key ingredient of the following important lemmas. They indeed provide a suitable abstraction of the labelling of a sequence of intervals $[x,y_1], \ldots, [x, y_n]$ with the same left endpoint. In particular, as we will show, if we ignore the $k_i$ exponents, the representation of a flat (decreasing) $B$-sequence is bounded by the size of the input formula $\varphi$. 

The following definition is at the basis of the abstraction of the sequence of intervals/points $[x,y_1], \ldots, [x, y_n]$
of a compass structure into a flat (decreasing) $B$-sequence.

\begin{defi}\label{def:shading}
Let $\cG=(\bG_N, \cL)$ be a compass structure for $\varphi$ and
$0 \leq x \leq N$. We define the \emph{shading of $x$} in $\cG$, 
written $\shadingG(x)$, as the sequence of atoms $\cL(x,x)\ldots\cL(x,N)$. 
\end{defi}

\noindent 
A shading $\shadingG(x)$ is nothing else but the word of atoms obtained ``by reading the atoms along the column'' 
$(x,x)\ldots(x,N)$ from bottom to top in a compass structure. As an example, in Figure \ref{fig:reqExplanationBD},
$\shadingG(1) = F^{[1,1]}F^{[1,2]}F^{[1,3]}F^{[1,4]}$. The next lemma easily follows from Definition 
\ref{def:flatbsequence} and Definition \ref{def:shading}. It allows us to abstract the shadings in a compass 
structure into flat $B$-sequences. The easy proof is omitted.

\begin{figure}
    \centering
    \begin{tikzpicture}[node distance=0.5cm]

\node[inner sep=0](B) {$\vdots$};

\node[draw, circle, inner sep=1, above of=B, fill=white, 
label={[]180:$\scalebox{0.7}{$F$}$}](1) {};

\node[draw, circle, inner sep=1, above of=1, fill=white, 
label={[]180:$\scalebox{0.7}{$F$}$}](2) {};

\node[draw, circle, inner sep=1, above of=2, fill=white, 
label={[]180:$\scalebox{0.7}{$G$}$}](3) {};

\node[draw, circle, inner sep=1, above of=3, fill=white, 
label={[]180:$\scalebox{0.7}{$H$}$}](4) {};

\node[draw, circle, inner sep=1, above of=4, fill=white, 
label={[]180:$\scalebox{0.7}{$H$}$}](5) {};

\node[draw, circle, inner sep=1, above of=5, fill=white, 
label={[]180:$\scalebox{0.7}{$H$}$}](6) {};

\node[inner sep=0, above of=6](T) {$\vdots$};

\draw($(B)+ (-0.75,0)$) edge[<-] node[sloped, above] {$\scalebox{0.6}{$F\cap Prop \supseteq G \cap Prop 
\supseteq H \cap Prop $}$}  ($(T)+ (-0.75,0)$) ; 

\draw($(B)+ (-1.5,0)$) edge[->] node[sloped, above] {$\scalebox{0.6}{$ \reqB(F) \subseteq \reqB(G) \subseteq \reqB(H) $}$}  ($(T)+ (-1.5,0)$) ; 

\draw($(B)+ (-2.25,0)$) edge[->] node[sloped, above] {$\scalebox{0.6}{$\reqD(F) \subseteq \reqD(G) \subseteq \reqD(H)$}$}  ($(T)+ (-2.25,0)$) ;

\draw[|-|] ($(1)+(0.5,0)$) -- ($(1)+(1.2,0)$) node[pos =0.5, above,label={[]270:$\scalebox{0.7}{$\vdots$}$}] {\scalebox{0.75}{$F$}};  

\draw[|-|] ($(1)+(0.5,0.5)$) -- ($(1)+(1.5,0.5)$) node[pos =0.5, above] {\scalebox{0.75}{$F$}}; 

\draw[|-|] ($(1)+(0.5,1)$) -- ($(1)+(1.8,1)$) node[pos =0.5, above] {\scalebox{0.75}{$G$}};

\draw[|-|] ($(1)+(0.5,1.5)$) -- ($(1)+(2.1,1.5)$) node[pos =0.5, above] {\scalebox{0.75}{$H$}};

\draw[|-|] ($(1)+(0.5,2)$) -- ($(1)+(2.4,2)$) node[pos =0.5, above] {\scalebox{0.75}{$H$}};

\draw[|-|] ($(1)+(0.5,2.5)$) -- ($(1)+(2.7,2.5)$) node[pos =0.5, above,label={[]90:$\scalebox{0.7}{$\vdots$}$}] {\scalebox{0.75}{$H$}};

\pgftransformshift{\pgfpoint{0cm}{-0.7cm}}

\node[inner sep=0](C1) {$(a)$};

\pgftransformshift{\pgfpoint{5cm}{0cm}}

\node[inner sep=0](C2) {$(b)$};

\pgftransformshift{\pgfpoint{0cm}{0.7cm}}


\node[inner sep=0](B) {$\vdots$};

\node[draw, circle, inner sep=1, above of=B, fill=white, 
label={[]180:$\scalebox{0.7}{$F$}$}](1) {};

\node[draw, circle, inner sep=1, above of=1, fill=white, 
label={[]180:$\scalebox{0.7}{$F$}$}](2) {};

\node[draw, circle, inner sep=1, above of=2, fill=white, 
label={[]180:$\scalebox{0.7}{$H$}$}](3) {};

\node[draw, circle, inner sep=1, above of=3, fill=white, 
label={[]180:$\scalebox{0.7}{$H$}$}](4) {};

\node[draw, circle, inner sep=1, above of=4, fill=white, 
label={[]180:$\scalebox{0.7}{$G$}$}](5) {};

\node[draw, circle, inner sep=1, above of=5, fill=white, 
label={[]180:$\scalebox{0.7}{$H$}$}](6) {};

\node[inner sep=0, above of=6](T) {$\vdots$};

\end{tikzpicture}
    \caption{\label{fig:step1picture}
    $(a)$ Monotonicity of atoms along a column in a compass structure, 
    together with a graphical account of the corresponding intervals and 
    of how proposition letters and $B$- and $D$-requests 
    must behave. $(b)$ An example of a violation of monotonicity.
    }
\end{figure}
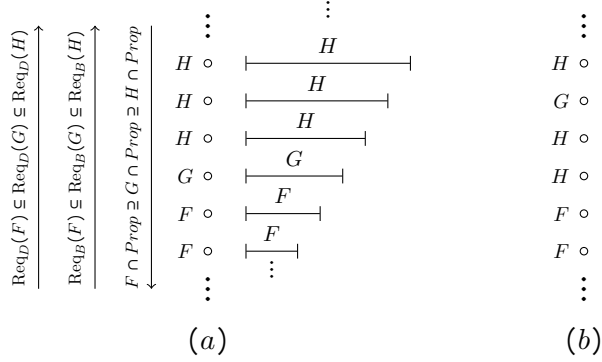 

\begin{lem}\label{lem:shadingthensequence}
Let $\cG=(\bG_N, \cL)$ be a compass structure
and $0 \leq x \leq N$. It holds that $\shadingG(x)$
is a $B$-sequence.
\end{lem}
    
\noindent The next lemma is the missing piece that allows us to restrict our attention to decreasing flat $B$-sequences when abstracting shadings in a compass-structure. 

\begin{restatable}{lem}{lbdet}\label{lem:bdeterminization}
Let $\cG=(\bG_N, \cL)$ be a compass structure for a formula $\varphi$. For all $x \leq y < N$, it holds that $\cL(x,y) = \cL(x,y + 1)$ if and only if $\cL(x,y)$ is $B$-reflexive, $\Prop \cap \cL(x,y) = \Prop \cap \cL(x, y + 1)$, and $\reqD(\cL(x, y)) = \reqD(\cL(x,y+1))$.
\end{restatable}

The proof of Lemma~\ref{lem:bdeterminization} is given in Appendix~\ref{appendix:proofs}. From Lemma \ref{lem:bdeterminization},
it follows that, given a shading $\shadingG(x)= \cL(x,x)\ldots\cL(x,N)$, to change the atom moving from $(x,y)$ to $(x, y + 1)$
 we need to ``sacrifice'' at least one element in $\Prop \cap \cL(x,y)$ or to ``acquire'' one element in
$\reqD(x, y)$ or in $\reqB(x,y)$.
 
The next corollary immediately follows from Lemma~\ref{lem:distinctsupprefixes} and Lemma~\ref{lem:bdeterminization}. It allows us to give a bound on the distinct atoms that may appear in a shading. More precisely, it states that the shading of each column $x$ in $\cG$ is a  flat decreasing $B$-sequence, and it gives a polynomial bound on the number of distinct atoms occurring in it.

\begin{cor}\label{cor:compassflatshadings}
Let $\cG=(\bG_N, \cL)$ be a compass structure for a formula $\varphi$. Then, for all $0\leq x \leq N$,  $\shadingG(x)$ is a flat decreasing $B$-sequence $\oF^{k_0}_0\ldots \oF^{k_m}_{m}$, with $0 \leq m \leq 4 |\varphi| + 1$.
\end{cor}

%


\subsection{A suitable equivalence relation over columns of a compass structure}\label{subsec:equivcolumns}

By exploiting the above (finite) characterisation of columns,
we can define a natural equivalence relation of finite index over columns.

First, we observe that, thanks to Corollary \ref{cor:compassflatshadings},
if multiple copies of the same atom are present in a column, their occurrences are consecutive, and thus can be represented as blocks.
Moreover, if two columns feature the same set of atoms, the (blocks of) atoms appear in the same order in the two columns because of the monotonicity of $\reqB$, $\reqD$, and $Prop$, the latter being forced by the homogeneity assumption 
(see Figure \ref{\detokenize{fig:step1picture}} (a)).
 
We will consider two columns $x,x'$ equivalent if and only if they feature the same set of atoms. Moreover, we will define a partial order relation over the shadings of equivalent columns that will allow us to totally order them. More precisely, for any two equivalent columns $x$ and $x'$, we will say that $\shadingG(x) < \shadingG(x')$ if and only if for every row $y$, with $y \geq x'$, the atom $\cL(x',y)$ is equal to the atom $\cL(x, y')$ for some row $y'$, with $x \leq y' \leq y$. 
Intuitively, this means that moving up along column $x'$ an atom cannot appear until it has appeared on column $x$. In Fig. \ref{\detokenize{fig:step2picture}} (a), we depict two equivalent columns that satisfy such a condition.
In general, when moving upward, atoms on column $x'$ are often ‘‘delayed'' with respect to atoms in column $x$, the limit case being when atoms on the same row are equal.
In Fig. \ref{\detokenize{fig:step2picture}} (b), a violation of the condition
(boxed atoms) is shown. 
In the following, we will prove that the latter situation never occurs in a compass structure.

Let us now define an equivalence relation $\sim$ over flat decreasing $B$-sequences as follows. 

\begin{defi}\label{def:equivalence}
 Two flat decreasing $B$-sequences $\shadingB= \oF^{k_0}_0\ldots \oF^{k_m}_{m}$ and $\shadingB'= \oG^{h_0}_0\ldots \oG^{h_{m'}}_{m'}$ are \emph{equivalent}, written $\shadingB\sim \shadingB'$, if and only if $m = m'$ and, for all $0\leq i \leq m$, $\oF_i = \oG_i$.  
\end{defi}
 Definition \ref{def:equivalence} says that two flat decreasing $B$-sequences are equivalent if and only if they feature exactly the same sequence of atoms regardless of their exponents. We choose as the representative of an equivalence class a flat decreasing $B$-sequence where each exponent is equal to one, e.g., the $B$-sequence $\oF^{k_0}_0\ldots \oF^{k_m}_{m}$ belongs to the equivalence class $[\oF_0\ldots \oF_{m}]_\sim$.  
Given an equivalence class $[\oF_0\ldots \oF_{m}]_\sim$ and $0\leq i\leq m$, we denote by $[\oF_0\ldots \oF_{m}]^i_\sim$ the $i^{th}$ atom in its sequence, that is, $[\oF_0\ldots \oF_{m}]^i_\sim = \oF_i$ for all $0 \leq i \leq m$. 
In addition, we define a function $next$ that, given an equivalence class $[\oF_0\ldots \oF_{m}]_\sim$ and one of its atom $\oF_i$,
returns the successor of $\oF_i$ in the sequence $[\oF_0\ldots \oF_{m}]_\sim$ (for $i=n$, it is undefined). 
It can be easily checked that $\sim$ is of finite index. From Corollary~\ref{cor:compassflatshadings}, it follows that its index 
is (roughly) bounded by $|\atoms|^{4 |\varphi| + 2} = 2^{(|\varphi| 
+ 1)(4 |\varphi| + 2)} = 2^{4|\varphi|^2 + 6 |\varphi| + 2}$.

\begin{figure}
    \centering
    \vspace{-0.1cm}
    \begin{tikzpicture}[node distance=0.5cm]

\draw[opacity=0.5] (-0.5,-0.5) -- (1.5,1.5);

\node[draw, circle, inner sep=1,  fill=white,
label={[]180:$\scalebox{0.7}{$F_1$}$},
label={[]270:$\scalebox{0.7}{$x$}$}](B) {};

\node[draw, circle, inner sep=1, above of=B, fill=white,
label={[]180:$\scalebox{0.7}{$F_2$}$}](1) {};

\node[draw, circle, inner sep=1, above of=1, fill=white, 
label={[]180:$\scalebox{0.7}{$F_3$}$}](2) {};

\node[draw, circle, inner sep=1, above of=2, fill=white, 
label={[]180:$\scalebox{0.7}{$F_3$}$}](3) {};

\node[draw, circle, inner sep=1, above of=3, fill=white, 
label={[]180:$\scalebox{0.7}{$F_3$}$}](4) {};

\node[draw, circle, inner sep=1, above of=4, fill=white, 
label={[]180:$\scalebox{0.7}{$F_4$}$}](5) {};

\node[draw, circle, inner sep=1, above of=5, fill=white, 
label={[]180:$\scalebox{0.7}{$F_4$}$}](6) {};

\node[inner sep=0, above of=6](T) {$\vdots$};

\pgftransformshift{\pgfpoint{1cm}{1cm}}

\node[draw, circle, inner sep=1,  fill=white,
label={[]0:$\scalebox{0.7}{$F_1$}$},
label={[]270:$\scalebox{0.7}{$x'$}$}](RB) {};

\node[draw, circle, inner sep=1, above of=RB, fill=white,
label={[]0:$\scalebox{0.7}{$F_2$}$}](R1) {};

\node[draw, circle, inner sep=1, above of=R1, fill=white, 
label={[]0:$\scalebox{0.7}{$F_3$}$}](R2) {};

\node[draw, circle, inner sep=1, above of=R2, fill=white, 
label={[]0:$\scalebox{0.7}{$F_3$}$}](R3) {};

\node[draw, circle, inner sep=1, above of=R3, fill=white, 
label={[]0:$\scalebox{0.7}{$F_4$}$}](R4) {};

\node[inner sep=0, above of=R4](T) {$\vdots$};

\draw[opacity=0.2] ($(2) + (-0.25,0)$) -- (2) -- (RB);
\draw[opacity=0.2] ($(3) + (-0.25,0)$) -- (3) -- (R1) -- ($(R1) + (0.25,0)$);
\draw[opacity=0.2] ($(4) + (-0.25,0)$) -- (4) -- (R2) -- ($(R2) + (0.25,0)$);
\draw[opacity=0.2] ($(5) + (-0.25,0)$) -- (5) -- (R3) -- ($(R3) + (0.25,0)$);
\draw[opacity=0.2] ($(6) + (-0.25,0)$) -- (6) -- (R4) -- ($(R4) + (0.25,0)$);

\pgftransformshift{\pgfpoint{-0.3cm}{-2cm}}

\node[inner sep=0](C1) {$(a)$};

\pgftransformshift{\pgfpoint{4.5cm}{0cm}}

\node[inner sep=0](C2) {$(b)$};

\pgftransformshift{\pgfpoint{-0.5cm}{1cm}}

\draw[opacity=0.5] (-0.5,-0.5) -- (1.5,1.5);

\node[draw, circle, inner sep=1,  fill=white,
label={[]180:$\scalebox{0.7}{$F_1$}$},
label={[]270:$\scalebox{0.7}{$x$}$}](B) {};

\node[draw, circle, inner sep=1, above of=B, fill=white,
label={[]180:$\scalebox{0.7}{$F_2$}$}](1) {};

\node[draw, circle, inner sep=1, above of=1, fill=white, 
label={[]180:$\scalebox{0.7}{$F_3$}$}](2) {};

\node[draw, circle, inner sep=1, above of=2, fill=white, 
label={[]180:$\scalebox{0.7}{$F_3$}$}](3) {};

\node[draw, circle, inner sep=1, above of=3, fill=white, 
label={[]180:$\scalebox{0.7}{$F_3$}$}](4) {};

\node[draw, circle, inner sep=1, above of=4, fill=white, 
label={[]180:$\scalebox{0.7}{$F_3$}$}](5) {};

\node[draw, circle, inner sep=1, above of=5, fill=white, 
label={[]180:$\scalebox{0.7}{$F_4$}$}](6) {};

\node[inner sep=0, above of=6](T) {$\vdots$};

\pgftransformshift{\pgfpoint{1cm}{1cm}}

\node[draw, circle, inner sep=1,  fill=white,
label={[]0:$\scalebox{0.7}{$F_1$}$},
label={[]270:$\scalebox{0.7}{$x'$}$}](RB) {};

\node[draw, circle, inner sep=1, above of=RB, fill=white,
label={[]0:$\scalebox{0.7}{$F_2$}$}](R1) {};

\node[draw, circle, inner sep=1, above of=R1, fill=white, 
label={[]0:$\scalebox{0.7}{$F_3$}$}](R2) {};

\node[draw, circle, inner sep=1, above of=R2, fill=white, 
label={[]0:$\scalebox{0.7}{$F_4$}$}](R3) {};

\node[draw, circle, inner sep=1, above of=R3, fill=white, 
label={[]0:$\scalebox{0.7}{$F_4$}$}](R4) {};

\node[inner sep=0, above of=R4](T) {$\vdots$};

\draw[opacity=0.2] ($(2) + (-0.25,0)$) -- (2) -- (RB);
\draw[opacity=0.2] ($(3) + (-0.25,0)$) -- (3) -- (R1) -- ($(R1) + (0.25,0)$);
\draw[opacity=0.2] ($(4) + (-0.25,0)$) -- (4) -- (R2) -- ($(R2) + (0.25,0)$);
\draw[opacity=0.2] ($(5) + (-0.25,0)$) -- (5) -- (R3) -- ($(R3) + (0.25,0)$);
\draw[opacity=0.2] ($(6) + (-0.25,0)$) -- (6) -- (R4) -- ($(R4) + (0.25,0)$);

\node[fit=(5)(R3), draw] {};

\end{tikzpicture}
    \vspace{-0.1cm}
    \caption{\label{fig:step2picture}Two equivalent columns that respect the order $(a)$ and  two equivalent columns that violate it $(b)$.
    \vspace{-0.1cm}}
\end{figure}
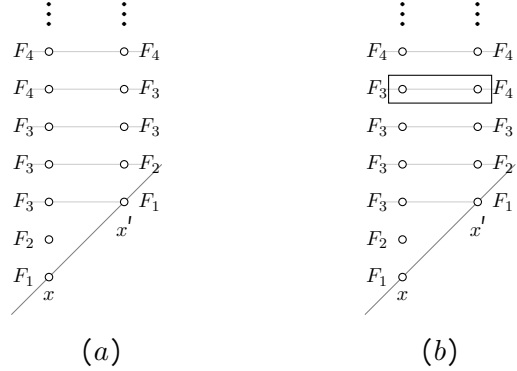

Let $\shadingB = \oF^{k_0}_0\ldots \oF^{k_m}_{m}$ be a flat decreasing $B$-sequence. We define the length of $\shadingB$, written $|\shadingB|$, as $\sum_{0\leq i\leq m} k_i$. A partial order $<$ over the elements of each equivalence class $[\shadingB]_\sim$ can be defined as follows. 

\begin{defi}\label{def:domination}
Let $\shadingB= \oF^{k_0}_0\ldots \oF^{k_m}_{m}$ and $\shadingB'= \oF^{h_0}_0\ldots \oF^{h_m}_{m}$ be two equivalent flat decreasing $B$-sequences. We say that $\shadingB$ is dominated by $\shadingB'$, written $\shadingB < \shadingB'$, if and only if (i) $|\shadingB| > |\shadingB'|$ and, (ii) for all $0 \leq i \leq m$, $\Sigma_{0 \leq j \leq i} k_j \leq (|\shadingB| - |\shadingB'|) + \Sigma_{0 \leq j \leq i} h_j$. 
\end{defi}

\begin{exa}\label{ex:columnorder}
Let us consider the equivalent flat decreasing $B$-sequences in Figure~\ref{fig:step2picture}. From left to right, they are
$\shadingB^0=F_1F_2F^3_3F^2_4$ , $\shadingB^1=F_1F_2F^2_3F_4$, $\shadingB^2=F_1F_2F^4_3F_4$, and $\shadingB^3=F_1F_2F_3F^2_4$ (for the sake of clarity, the exponent $1$ is omitted).
Let us consider first $\shadingB^0$ and $\shadingB^1 $. By condition $(i)$ of Definition~\ref{def:domination}, the only possible  domination relation is $\shadingB^0 < \shadingB^1$. Let us now check if condition $(ii)$ of Definition~\ref{def:domination} is satisfied.
To this end, let us consider the following representation of flat shadings. In general, whenever $\shadingB^0 $ is equivalent to $\shadingB^1$, in order to prove that $\shadingB^0 < \shadingB^1$, it  suffices to take the alignment of $\shadingB^1$ (the shorter sequence) as a \emph{suffix} of $\shadingB^0$ (the longer sequence). Such an alignment is obtained  by prefixing $\shadingB^1$  with a word of length $|\shadingB^0|- |\shadingB^1|$ only featuring the blank symbol `\_\!\_'. In the example, the required alignment of $\shadingB^1$ is $\hshadingB^1=\_\!\_^2F_1F_2F^2_3F_4$. It holds that $\shadingB^0 <\shadingB^1$ if and only if the position of the first occurrence of each atom $F_i$ in $\shadingB^0$ is not strictly smaller than that of the first occurrence of $F_i$ in $\hshadingB^1$. In the considered case, we have that:
\begin{compactitem} 
\item $F_1$ occurs for the first time at position $0$ in $\shadingB^0$ 
and at position $2$ in $\hshadingB^1$;
\item $F_2$ occurs for the first time at position $1$ in $\shadingB^0$ 
and at position $3$ in $\hshadingB^1$;
\item $F_3$ occurs for the first time at position $2$ in $\shadingB^0$ 
and at position $4$ in $\hshadingB^1$;
\item $F_4$ occurs for the first time at position $5$ in $\shadingB^0$ 
and at position $6$ in $\hshadingB^1$.
\end{compactitem}
Hence, we can conclude that $\shadingB^0 <\shadingB^1$. On the contrary, we have that  $\shadingB^2$ is equivalent  to $\shadingB^3$, but $\shadingB^2\not<\shadingB^3$. Indeed, if we consider the alignment $\hshadingB^3 = \_\!\_^2F_1F_2F_3F^2_4$,it holds that atom $F_4$ occurs for the first time at position $5$ in $\hshadingB^3$ and at position $6$ in $\shadingB^2$. Lemma~\ref{lem:shadingorder} below proves that such a scenario cannot occur in the case of compass structures.
\end{exa}

Let $\shadingB= \oF^{k_0}_0\ldots \oF^{k_m}_{m}$ be a flat decreasing $B$-sequence and let $0 \leq i \leq |\shadingB|$.
We introduce a notation for atom retrieval by letting $\shadingB[i]= \oF_{j}$, where $j$ is such that $\sum_{0\leq {j'} < j} k_{j'} < i \leq \sum_{0\leq {j'} \leq j} k_{j'}$.  
The next lemma constrains the relationships between pairs of equivalent shadings (flat decreasing $B$-sequences) that appear in a compass structure.

\begin{lem}\label{lem:shadingorder}
Let $\cG=(\bG_N, \cL)$ be a compass structure.  
For every pair of columns $0\leq x<x'\leq N$ such that $\shadingG(x) \sim \shadingG(x')$, it holds that $\shadingG(x) < \shadingG(x')$.
\end{lem}

\begin{proof}
Let $\Delta = x'- x$, $\shadingG(x)= \oF^{k_0}_0\ldots \oF^{k_m}_{m}$, and $\shadingG(x')=\oF^{h_0}_0\ldots \oF^{h_m}_{m}$. By contradiction, let us assume  that $\shadingG(x) \not{<} \shadingG(x')$. 
From $\shadingG(x) \sim \shadingG(x')$, it follows that both $B$-sequences feature the same atoms in the same order; they may only differ in their numerousness, i.e., in the exponents of some atoms. Since $|\shadingG(x)| > |\shadingG(x')|$ ($x'$ is closer to $N$ than $x$ and thus it is a shorter column), from $\shadingG(x) \sim \shadingG(x')$ and $\shadingG(x) \not{<} \shadingG(x')$, 
it follows that there exists  $0 < i \leq  N - x'$ such that one of the following conditions holds:
\begin{compactenum}
\item $\shadingG(x)[\Delta + i] \cap \Prop \supset \shadingG(x')[i] \cap \Prop$;
\item $\reqD(\shadingG(x)[\Delta + i]) \subset
\reqD(\shadingG(x')[i])$;
\item 
$\shadingG(x)[\Delta + i]$ is $B$-irreflexive; 
\item $\reqB(\shadingG(x)[ \Delta + i]) \subset 
\reqB(\shadingG(x')[i])$.
\end{compactenum}

The above cases stem from the fact that we are assuming, by contradiction, that
for a certain index $i$ there exists an index $j$ such that 
$\shadingG(x')[i] = \oF_j$ and $\shadingG(x)[\Delta + i] = \oF_{j-1}$
(and thus $\shadingG(x')[i]  \neq \shadingG(x)[\Delta + i]$).
This is the case, for instance, with $x$ and $x'$
in Figure~\ref{fig:step2picture} (b), where $\Delta=2$,
$\shadingG(x')[3] = F_4$, and $\shadingG(x)[5] = F_3$.

Let $i$ is the minimum index which satisfies one of the above conditions. In the following, we will assume 
that $\shadingG(x')[i] = \oF_j$ and $\shadingG(x)[\Delta + i] = \oF_{j-1}$ for some $0 < j \leq m$.

Before proving that all the above cases lead to a contradiction, we would like to explain how 
they have been identified. Since $\shadingG(x) \sim \shadingG(x')$, but $\shadingG(x) \not{<} \shadingG(x')$, and $|\shadingG(x)| < |\shadingG(x')|$, the situation is analogous to the one depicted in Figure \ref{fig:step2picture} (b). By construction, $\shadingG(x)$ is the first starting the unraveling of the sequence of atoms $\oF_0\ldots \oF_{m}$ ($F_1\ldots F_4$ in Figure \ref{fig:step2picture} (b)), later followed by $\shadingG(x')$, that unravels the same sequence of atoms (since $\shadingG(x) \sim \shadingG(x')$). Then, 
either (i) for every $i$ there exists $k'\leq k$ such that $\shadingG(x')[i] = \oF_{k'}$ and $\shadingG(x)[i + \Delta] = \oF_k$, i.e., $\shadingG(x')$ ``waits'' for $\shadingG(x)$ before displaying any new atom in the sequence, 
or (ii) there exist $i$ and $k'> k$ such that $\shadingG(x')[i] = \oF_{k'}$ and $\shadingG(x)[i + \Delta] = \oF_k$.
As an example, this is the case with $i=3$, $k=3$, and $k'=4$ in Figure~\ref{fig:step2picture} (b).
Condition (i) suffices to conclude that $\shadingG(x) < \shadingG(x')$, while $\shadingG(x) \not{<} \shadingG(x')$ when condition (ii) holds.
One can easily check that if $i$ is the  minimal index that satisfies condition (ii), then $k'= k + 1$. Hence, it holds that $F_{k'} \thenB F_{k}$, but $F_{k'} \neq F_{k}$.

Cases (1)-(4) above are all the possible ways in which we may have $F_{k'} \thenB F_{k}$ and $F_{k'} \neq F_{k}$, with the additional constraint  that $F_{k'} = \cL(x', x' + i)$ and 
$F_{k'} = \cL(x, x + \Delta + i)$. 

Let $\shadingG(x)[i+\Delta]  (= \oF_{j-1}) = \cL(x, x + \Delta + i)$ and $\shadingG(x')[i] (= \oF_j) = \cL(x', x' + i)$.


As for case (1), since the considered compass structures satisfy the homogeneity assumption and $[x', x'+ i]$ finishes $[x, x + \Delta + i]$, it holds that $\Prop\cap \cL(x', x' + i) \supseteq \Prop\cap \cL(x, x + \Delta + i)$.
On the other hand, since  $k' > k$, there is $i'>i$ such that $\Prop\cap \cL(x,x + \Delta + i') = \Prop\cap \cL(x', x' + i)$,
and, since $[x, x + \Delta + i]$ begins $[x, x + \Delta + i']$, $\Prop\cap \cL(x + \Delta + i) \supseteq \Prop\cap \cL(x, x + \Delta + i') = \Prop\cap \cL(x', x' + i)$. It follows that $\Prop\cap \cL(x + \Delta + i) = \Prop\cap \cL(x', x' + i)$ (contradiction).

As for case (2), since $[x', x'+ i]$  is a proper suffix of $[x,x+ \Delta + i]$, all the proper sub-intervals of $[x', x'+ i]$ are proper sub-intervals of $[x,x+ \Delta + i]$ as well, and, 
since $\cG$ is a homogeneous compass structure, $\reqD(\cL(x', x' + i)) \subseteq \reqD(\cL(x, x + \Delta + i))$ (contradiction).



As for case (3), let us assume that 
$\shadingG(x)[\Delta + i] = \cL(x, x + \Delta + i)$ is $B$-irreflexive.
%
From $\shadingG(x) \sim \shadingG(x')$ and minimality of $i$, it follows that $\shadingG(x)[ \Delta + i - 1] (= \shadingG(x')[i - 1]) = \cL(x, x + \Delta + i) = \oF_{j - 1}$.
%
From Lemma~\ref{lem:bdeterminization} (left-to-right direction), it immediately follows that $\cL(x, x + \Delta + i)$ is B-reflexive (contradiction).

Finally, let us assume $\reqB(\shadingG(x)[ \Delta + i]) (= \oF_{j-1}) \subset \reqB(\shadingG(x')[i]) (= \oF_j)$ (case 4). We prove that such an assumption contradicts minimality of $i$. We distinguish two cases: either (i) $\shadingG(x')[i - 1] = \oF_{j-1}$ or (ii) $\shadingG(x')[i - 1] = \oF_j$. In case (i), from Lemma~\ref{lem:bstep}, it follows that $\oF_{j-1}$ is $B$-irreflexive. Hence, $\shadingG(x)[ \Delta + i - 1] \neq \shadingG(x)[ \Delta + i] (= \oF_{j-1})$, thus violating minimality of $i$.
In case (ii), it immediately follows that the minimum index must be less than or equal to $j-1$.
\end{proof} 

Even if it is susceptible to a  quite straightforward spatial interpretation, the statement of 
Lemma~\ref{lem:shadingorder} is rather technical, and its role in the general picture may be a little obscure. As a matter of fact, it is an intermediate result that contributes a specific tile to the whole puzzle. The next example summarizes the properties of compass structures we proved so far, and gives a preview of the property we are going to demonstrate in the
next subsection.

\begin{exa}\label{ex:columnlemma}
In Figure~\ref{fig:columnlemma}, we represent the shading of 
some columns belonging to the same equivalence class from a 
certain row $y$ up to the top row $N$ of a compass structure. 
More precisely, we consider 5 columns $x_1, \ldots, x_5$, 
with $\shadingG(x_1) \sim \ldots \sim \shadingG(x_5)$ and 
$x_1 < \ldots < x_5$, and a row $y \leq N$, with $x_i\leq y$ 
for all $i \in \{1\ldots 5\}$. Then, for $i \in \{1,\ldots, 5\}$, 
we restrict our attention to the suffix of length $N - y$ of 
column $x_i$, that is, $Suf_y(x_i) = \shadingG(x_i)[y]\ldots
\shadingG(x_i)[N]$, for $i \in \{1,\ldots, 5\}$. 
In Figure~\ref{fig:columnlemma}, we have 
$Suf_y(x_1) = F_3^4 F_4^3$, 
$Suf_y(x_2) = F_2F_3^4 F_4^2$, 
$Suf_y(x_3) = Suf_y(x_4) = F_2^3F_3^2 F_4^2$, and 
$Suf_y(x_5) = F_2^5F_3 F_4$.

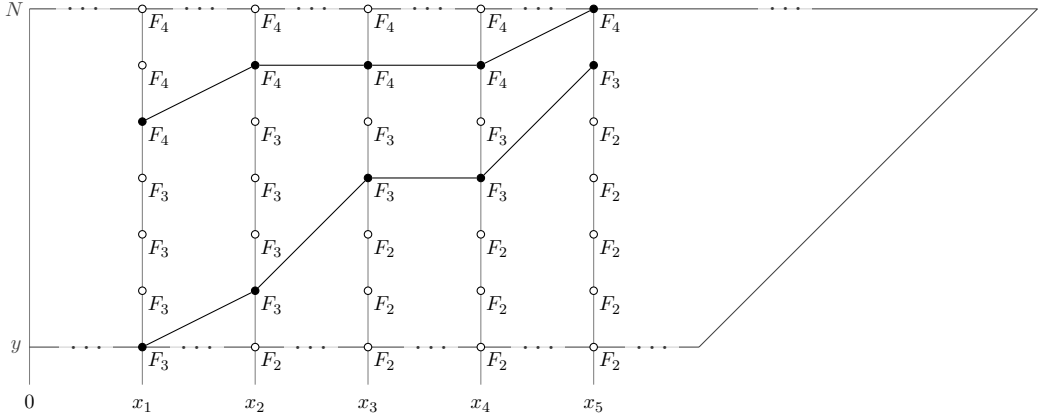
\begin{figure}
    \centering
    \vspace{-0.1cm}
    \begin{tikzpicture}[node distance=0.75cm]

\draw[opacity=0.5] ($(0,0) + (0,-0.5)$) -- ($(0,4.5)$);

\draw[opacity=0.5] ($(-1.5,-0.5)$) -- ($(-1.5,4.5)$)
node[pos=-0.045, opacity=1] {\scalebox{0.7}{$0$}};

\draw[opacity=0.75] ($(0,0) + (-1.5,0)$) -- ($(7.4
,0)$) 
node[fill=white, pos=0.085] {$\ldots$}
node[fill=white, pos=0.25] {$\ldots$}
node[fill=white, pos=0.42] {$\ldots$}
node[fill=white, pos=0.6] {$\ldots$}
node[fill=white, pos=0.76] {$\ldots$}
node[fill=white, pos=0.93] {$\ldots$}
node[pos=-0.02] {\scalebox{0.7}{$y$}}
;

\draw[opacity=0.75] ($(0,0) + (-1.5,4.5)$) -- ($(11.9
,4.5)$)
node[fill=white, pos=0.053] {$\ldots$}
node[fill=white, pos=0.169] {$\ldots$}
node[fill=white, pos=0.28] {$\ldots$}
node[fill=white, pos=0.395] {$\ldots$}
node[fill=white, pos=0.505] {$\ldots$}
node[fill=white, pos=0.75] {$\ldots$}
node[pos=-0.015] {\scalebox{0.7}{$N$}}

;

\draw[opacity=0.75] ($(7.4,0)$) -- ($(11.9
,4.5)$);

\node[draw, circle, fill=black, inner sep=1,
label={[label distance=-0.15cm]315:$\scalebox{0.7}{$F_3$}$},
label={[label distance=0.5cm]270:$\scalebox{0.7}{$x_1$}$}](F31) {};

\node[draw, circle, inner sep=1, above of=F31, fill=white,
label={[label distance=-0.15cm]315:$\scalebox{0.7}{$F_3$}$}](1) {};

\node[draw, circle, inner sep=1, above of=1, fill=white, 
label={[label distance=-0.15cm]315:$\scalebox{0.7}{$F_3$}$}](2) {};

\node[draw, circle, inner sep=1, above of=2, fill=white, 
label={[label distance=-0.15cm]315:$\scalebox{0.7}{$F_3$}$}](3) {};

\node[draw, circle, inner sep=1, above of=3, fill=black, 
label={[label distance=-0.15cm]315:$\scalebox{0.7}{$F_4$}$}](F41) {};

\node[draw, circle, inner sep=1, above of=F41, fill=white, 
label={[label distance=-0.15cm]315:$\scalebox{0.7}{$F_4$}$}](5) {};

\node[draw, circle, inner sep=1, above of=5, fill=white, 
label={[label distance=-0.15cm]315:$\scalebox{0.7}{$F_4$}$}](6) {};

\pgftransformshift{\pgfpoint{1.5cm}{0cm}}

\draw[opacity=0.5] ($(0,0) + (0,-0.5)$) -- ($(0,4.5)$);

\node[draw, circle, inner sep=1,  fill=white,
label={[label distance=-0.15cm]315:$\scalebox{0.7}{$F_2$}$},
label={[label distance=0.5cm]270:$\scalebox{0.7}{$x_2$}$}](B) {};

\node[draw, circle, inner sep=1, above of=B, fill=black,
label={[label distance=-0.15cm]315:$\scalebox{0.7}{$F_3$}$}](F32) {};

\draw (F31) -- (F32);

\node[draw, circle, inner sep=1, above of=F32, fill=white, 
label={[label distance=-0.15cm]315:$\scalebox{0.7}{$F_3$}$}](2) {};

\node[draw, circle, inner sep=1, above of=2, fill=white, 
label={[label distance=-0.15cm]315:$\scalebox{0.7}{$F_3$}$}](3) {};

\node[draw, circle, inner sep=1, above of=3, fill=white, 
label={[label distance=-0.15cm]315:$\scalebox{0.7}{$F_3$}$}](4) {};

\node[draw, circle, inner sep=1, above of=4, fill=black, 
label={[label distance=-0.15cm]315:$\scalebox{0.7}{$F_4$}$}](F42) {};

\draw (F41) -- (F42);

\node[draw, circle, inner sep=1, above of=F42, fill=white, 
label={[label distance=-0.15cm]315:$\scalebox{0.7}{$F_4$}$}](6) {};

\pgftransformshift{\pgfpoint{1.5cm}{0cm}}

\draw[opacity=0.5] ($(0,0) + (0,-0.5)$) -- ($(0,4.5)$);

\node[draw, circle, inner sep=1,  fill=white,
label={[label distance=-0.15cm]315:$\scalebox{0.7}{$F_2$}$},
label={[label distance=0.5cm]270:$\scalebox{0.7}{$x_3$}$}](B) {};

\node[draw, circle, inner sep=1, above of=B, fill=white,
label={[label distance=-0.15cm]315:$\scalebox{0.7}{$F_2$}$}](1) {};

\node[draw, circle, inner sep=1, above of=1, fill=white, 
label={[label distance=-0.15cm]315:$\scalebox{0.7}{$F_2$}$}](2) {};

\node[draw, circle, inner sep=1, above of=2, fill=black, 
label={[label distance=-0.15cm]315:$\scalebox{0.7}{$F_3$}$}](F33) {};

\draw (F32) -- (F33);

\node[draw, circle, inner sep=1, above of=F33, fill=white, 
label={[label distance=-0.15cm]315:$\scalebox{0.7}{$F_3$}$}](4) {};

\node[draw, circle, inner sep=1, above of=4, fill=black, 
label={[label distance=-0.15cm]315:$\scalebox{0.7}{$F_4$}$}](F43) {};

\draw (F42) -- (F43);

\node[draw, circle, inner sep=1, above of=F43, fill=white, 
label={[label distance=-0.15cm]315:$\scalebox{0.7}{$F_4$}$}](6) {};

\pgftransformshift{\pgfpoint{1.5cm}{0cm}}

\draw[opacity=0.5] ($(0,0) + (0,-0.5)$) -- ($(0,4.5)$);

\node[draw, circle, inner sep=1,  fill=white,
label={[label distance=-0.15cm]315:$\scalebox{0.7}{$F_2$}$},
label={[label distance=0.5cm]270:$\scalebox{0.7}{$x_4$}$}](B) {};

\node[draw, circle, inner sep=1, above of=B, fill=white,
label={[label distance=-0.15cm]315:$\scalebox{0.7}{$F_2$}$}](1) {};

\node[draw, circle, inner sep=1, above of=1, fill=white, 
label={[label distance=-0.15cm]315:$\scalebox{0.7}{$F_2$}$}](2) {};

\node[draw, circle, inner sep=1, above of=2, fill=black, 
label={[label distance=-0.15cm]315:$\scalebox{0.7}{$F_3$}$}](F34) {};

\draw (F33) -- (F34);

\node[draw, circle, inner sep=1, above of=F34, fill=white, 
label={[label distance=-0.15cm]315:$\scalebox{0.7}{$F_3$}$}](4) {};

\node[draw, circle, inner sep=1, above of=4, fill=black, 
label={[label distance=-0.15cm]315:$\scalebox{0.7}{$F_4$}$}](F44) {};

\draw (F43) -- (F44);

\node[draw, circle, inner sep=1, above of=F44, fill=white, 
label={[label distance=-0.15cm]315:$\scalebox{0.7}{$F_4$}$}](6) {};

\pgftransformshift{\pgfpoint{1.5cm}{0cm}}

\draw[opacity=0.5] ($(0,0) + (0,-0.5)$) -- ($(0,4.5)$);

\node[draw, circle, inner sep=1,  fill=white,
label={[label distance=-0.15cm]315:$\scalebox{0.7}{$F_2$}$},
label={[label distance=0.5cm]270:$\scalebox{0.7}{$x_5$}$}](B) {};

\node[draw, circle, inner sep=1, above of=B, fill=white,
label={[label distance=-0.15cm]315:$\scalebox{0.7}{$F_2$}$}](1) {};

\node[draw, circle, inner sep=1, above of=1, fill=white, 
label={[label distance=-0.15cm]315:$\scalebox{0.7}{$F_2$}$}](2) {};

\node[draw, circle, inner sep=1, above of=2, fill=white, 
label={[label distance=-0.15cm]315:$\scalebox{0.7}{$F_2$}$}](3) {};

\node[draw, circle, inner sep=1, above of=3, fill=white, 
label={[label distance=-0.15cm]315:$\scalebox{0.7}{$F_2$}$}](4) {};

\node[draw, circle, inner sep=1, above of=4, fill=black, 
label={[label distance=-0.15cm]315:$\scalebox{0.7}{$F_3$}$}](F35) {};

\draw (F34) -- (F35);

\node[draw, circle, inner sep=1, above of=F35, fill=black, 
label={[label distance=-0.15cm]315:$\scalebox{0.7}{$F_4$}$}](F45) {};

\draw (F44) -- (F45);

\end{tikzpicture}
    \vspace{-0.1cm}
    \caption{\label{fig:columnlemma} A graphical account of the property stated
    by Lemma~\ref{lem:shadingorder} for suffixes of equal length belonging to columns in the same equivalence class.
    \vspace{-0.1cm}}
\end{figure}

First, we observe that a natural weakening of relation $<$ of Definition~\ref{def:domination}, where condition $(i)$ is dropped,
can be exploited to compare suffixes of equal length. The resulting relation is no more asymmetric, and to distinguish it from the original one, we replace the symbol $<$ by $\leq$. From Lemma~\ref{lem:shadingorder}, it immediately follows that, in a compass structure, $Suf_y(x_i) \leq Suf_y(x_j)$ whenever $\shadingG(x_i) \sim \shadingG(x_j)$ and $x_i < x_j$. 
%

In Figure~\ref{fig:columnlemma}, the lowest 
occurrences of atoms $F_3$ and $F_4$ on each suffix have been connected by a broken line. We observe that, by moving from $x_1$ to $x_5$, the two broken lines do not intersect, that is, the broken line for $F_3$ is always below the broken line for $F_4$, and both are non-decreasing. This does not happen by chance, as we will prove in the next section (see Lemma \ref{lem:stepconstraint}). 
Moreover, Lemma~\ref{lem:shadingorder} guarantees that identical 
suffixes are contiguous in the equivalence class, and can thus be grouped together as one unique block. 
This is the case with $Suf_y(x_3)$ and $Suf_y(x_4)$ in Figure~\ref{fig:columnlemma}.  
These features of column suffixes will be exploited to obtain a finite characterization of rows bounded by some function of $|\varphi|$. 

%

\end{exa}

In general, the number of distinct suffixes depends on the distance $N - y$. Such a dependency may be problematic for a finite characterization of rows, as $N$ can be arbitrary large. Think of a scenario where $y$ is far away from $N$ and all suffixes belong to the same equivalence class and differ from each other, that is, $Suf_y(0) <  \ldots < Suf_y(y)$.  Luckily, in the next section, we will prove that for every row $y$ and every set of points  $x_1< \ldots < x_n \leq y$, whose columns belong to the same equivalence class, the length of the maximum strictly ascending chain $Suf_y(x_{j_1}) < \ldots < Suf_y(x_{j_m})$, with $\{j_1, \ldots j_m\} \subseteq \{ 1, \ldots, n\}$ (that is, the maximum number of distinct suffixes), is bounded by some function of $|\varphi|$. 


\subsection{A spatial property of rows in homogeneous compass structures}\label{subsec:covered}

We conclude the section by stating a covering property of points in a row of a compass structure that allows us provide a characterization of  rows bounded in the size of the formula $\varphi$ under consideration. A graphical account of such a property is given in Figure~\ref{mfcsstep3picture}.

To begin with, we introduce the fundamental notions of intersection and fingerprint.

\begin{defi}\label{def:intersection}
Let $\cG=(\bG_N, \cL)$ be a compass structure and let $0\leq x \leq y$. The intersection of row $y$ and column $x$ is the pair $([\shadingG(x)]_\sim, \cL(x,y))$ consisting of the equivalence class of $x$ and the labelling of $(x,y)$.
\end{defi}

The intersection $(x,y)$ is basically the equivalence class of $x$, that is, $[\shadingG(x)]_\sim$, where we have pinpointed the atom in $\shadingG(x)$ that appears at height $y$. As an example, in the example of Figure~\ref{fig:reqExplanationBD}, the intersection of $(1,3)$ is the
pair $([F^{[1,1]}F^{[1,2]}F^{[1,3]}F^{[1,4]}]_\sim, F^{[1,3]})$.

Let us now define the notion of fingerprint for a given a point $(x,y)$, which basically pairs the intersection of row $y$ and column $x$ with the set of intersections of row $y$ and column $x'$, for all $x < x' \leq y$, that is, the intersections of row $y$ with the columns
occurring to the ``right'' of column $x$. Formally, let $\future(x,y)$ be the set $\{ ([\shadingG(x')]_\sim, \cL(x',y)): x'>x \}$. As an example, in Figure~\ref{fig:reqExplanationBD}, we have that $\future(1,3) = \{([F^{[2,2]}F^{2,3}F^{[2,4]}]_\sim, F^{[2,3]}), ([F^{[3,3]}F^{[3,4]}]_\sim, F^{[3,3]})\}$. As a general rule, $\future(x,y)$ collects the equivalence classes of $\sim$ which are witnessed to the right of $x$ on row $y$ plus a ``pointer'' to the ``current atom'', that is, the atom they are exposing on $y$. 
If $\cG=(\bG_N, \cL)$ is homogeneous (as it is in our setting), for all $0 \leq x \leq y\leq N$, the number of possible sets $\future(x,y)$ is bounded by $2^{2^{4|\varphi|^2 + 6 |\varphi| + 2} \cdot 2^{|\varphi| + 1}} = 2^{2^{4|\varphi|^2 + 7 |\varphi| + 3}}$, that 
is, it is doubly exponential in the size of $|\varphi|$. Such a bound is obtained as follows. The number of possible equivalence classes for columns, that is, the number of possible flat decreasing $B$-sequences, is $2^{4|\varphi|^2 + 6 |\varphi| + 2}$ and the number of possible atoms is $2^{|\varphi| + 1}$. Their product gives a bound on the number of pairs $([\shadingG(x')]_\sim, \cL(x',y))$. Then, the  number of all  possible sets on such pairs is bounded by $2^{2^{4|\varphi|^2 + 6 |\varphi| + 2} \cdot 2^{|\varphi| + 1}}$.

\begin{defi}\label{def:fingerprint}
Let $\cG=(\bG_N, \cL)$ be a compass structure and let $0\leq x \leq y$. The fingerprint of $(x,y)$, denoted by $fp(x,y)$, is the triplet 
consisting of the intersection of row $y$ and column $x$ plus the set $\future(x,y)$, that is, $fp(x,y)= ([\shadingG(x)]_\sim, \cL(x,y), \future(x,y))$.
\end{defi}

\noindent As an example, the fingerprint of row $3$ and column $1$ in Figure~\ref{fig:reqExplanationBD} is: 
$$fp(1,3) = \left(\begin{array}{c}[F^{[1,1]}F^{[1,2]}F^{[1,3]}F^{[1,4]}]_\sim, F^{[1,3]},\\
\{([F^{[2,2]}F^{[2,3]}F^{[2,4]}]_\sim, 
F^{[2,3]}), ([F^{[3,3]}F^{[3,4]}]_\sim,\allowbreak 
F^{[3,3]})\}\end{array}\right).$$

The next lemma constrains the way in which two columns $x, x'$, with $x<x'$ and $\shadingG(x) \sim \shadingG(x')$, 
evolve from a given row $y$ on when  $\future(x,y) = \future(x',y)$.

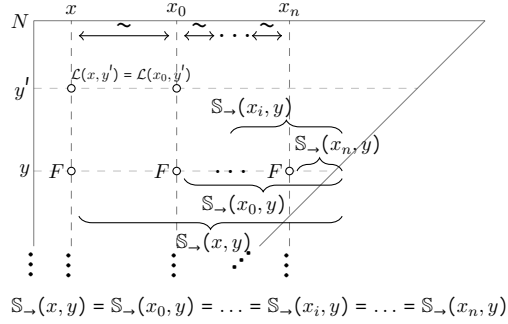
\begin{figure}
    \centering
    \begin{tikzpicture}[node distance=0.5cm]

\draw[opacity=0.5] (3,3) -- (6,6) node[sloped, pos=-0.07, opacity=1] {$\cdots$};
\draw[opacity=0.25, dashed] (0,4) -- (4,4) node[opacity=1, pos = -0.03] {\scalebox{0.7}{$y$}};
\draw[opacity=0.25, dashed] (0,5.1) -- (5.1,5.1) node[opacity=1, pos = -0.03] {\scalebox{0.7}{$y'$}};
\draw[opacity=0.5] (0,3) -- (0,6) node[sloped, pos=-0.08, opacity=1] {$\cdots$};
\draw[opacity=0.5] (0,6) -- (6,6) node[opacity=1, pos = -0.03] {\scalebox{0.7}{$N$}};

\draw[opacity=0.5, dashed] (0.5,3) -- (0.5,6) node[opacity=1, pos = 1.05] {\scalebox{0.7}{$x$}} node[sloped, pos=-0.085, opacity=1] {$\cdots$};

\draw[opacity=0.5, dashed] (1.9,3) -- (1.9,6) node[opacity=1, pos = 1.05] {\scalebox{0.7}{$x_0$}} node[sloped, pos=-0.085, opacity=1] {$\cdots$};

\draw[opacity=0.5, dashed] (3.4,3) -- (3.4,6) node[opacity=1, pos = 1.05] {\scalebox{0.7}{$x_n$}} node[sloped, pos=-0.085, opacity=1] {$\cdots$};

\draw[<->] (0.6,5.8) -- (1.8, 5.8) node[pos=0.5, above=-0.1cm] {$\sim$};
\draw[<->] (2,5.8) -- (2.4, 5.8) node[pos=0.5, above=-0.1cm,
label={[yshift=-0.1cm, xshift=-0.15cm]0:$\scalebox{1}{$\ldots$}$}] {$\sim$};
\draw[<->] (2.9,5.8) -- (3.3, 5.8) node[pos=0.5, above=-0.1cm] {$\sim$};

\pgftransformshift{\pgfpoint{0.5cm}{4cm}}


\node[draw, circle, inner sep=1,  fill=white,
label={[label distance=-0.1cm]180:$\scalebox{0.7}{$F$}$}](X) {};

\node[draw, circle, inner sep=1,  fill=white,
label={[label distance=0.3cm, name= XI]0:$\scalebox{1}{$\ldots$}$},
label={[label distance=-0.1cm]180:$\scalebox{0.7}{$F$}$}, right of=X, node distance=1.4cm](X0) {};

\node[draw, circle, inner sep=1,  fill=white,
label={[label distance=-0.1cm, xshift=-0.1cm, yshift=0.2cm]0:$\scalebox{0.5}{$\cL(x, y')= \cL(x_0, y')$}$}, above of=X, node distance=1.1cm](AX) {};

\node[draw, circle, inner sep=1,  fill=white, above of=X0, node distance=1.1cm](AX) {};

\node[draw, circle, inner sep=1,  fill=white,
label={[label distance=-0.1cm]180:$\scalebox{0.7}{$F$}$}, right of=X0, node distance=1.5cm](XN) {};

\draw [decorate,decoration={brace,amplitude=5pt}]
($(XN)+ (0.1,0)$) -- (3.6,0) 
node [black,midway,yshift=0.35cm, xshift =0.2cm](HN) {\scalebox{0.7}{
$\future(x_n, y)$}};

\draw [decorate,decoration={brace,amplitude=5pt}]
($(XI)+ (0,0.5)$) -- (3.6,0.5) 
node [black,midway,yshift=0.35cm, xshift=-0.5cm](HN) {\scalebox{0.7}{
$\future(x_i, y)$}};

\draw [decorate,decoration={brace,amplitude=5pt, mirror}]
($(X0)+ (0.1,-0.1)$) -- (3.6,-0.1) 
node [black,midway,yshift=-0.35cm, xshift=-0.3cm](HN) {\scalebox{0.7}{
$\future(x_0, y)$}};

\draw [decorate,decoration={brace,amplitude=5pt, mirror}]
($(X)+ (0.1,-0.6)$) -- (3.6,-0.6) 
node [black,midway,yshift=-0.35cm, xshift=0cm](HN) {\scalebox{0.7}{
$\future(x, y)$}};
\pgftransformshift{\pgfpoint{2.5cm}{-1.8cm}}

\node {\scalebox{0.7}{$\future(x, y) = \future(x_0, y)= \ldots = \future(x_i, y)= \ldots = \future(x_n, y)$}};

\end{tikzpicture}
    \caption{\label{mfcsstep3picture}A graphical account of the behaviour of covered points: $x$ is covered by $x_0 < \ldots < x_n$ on row $y$, and thus the labelling of points on column $x$ above $(x,y)$ is exactly the same as that of the corresponding points on column $x_0$ above $(x_0,y)$, that is, $\cL(x,y')=\cL(x_0,y')$, for all $y\leq y'\leq N$.}
\end{figure}

\begin{restatable}{lem}{lemstepconstraint}\label{lem:stepconstraint}
Let $\cG=(\bG_N, \cL)$ be a compass structure and let
$0\leq x < x'\leq y \leq N $. If $fp(x,y) = fp(x',y)$
and $y'$ is the smallest point greater than $y$ such that $\cL(x,y')\neq \cL(x,y)$, if any, and $N$ otherwise, 
then, for all $y \leq y''\leq y'$, $\cL(x,y'') = \cL(x',y'')$.
\end{restatable}

\noindent From Lemma~\ref{lem:stepconstraint}, the next corollary easily follows.

\begin{cor}\label{cor:inbetweeners}
Let $\cG=(\bG_N, \cL)$ be a compass structure and let
$0\leq x < x'\leq y \leq N $.
If $fp(x,y) = fp(x',y)$ and $y'$ is the smallest point greater than $y$ such that $\cL(x,y')\neq \cL(x,y)$, if any, and $N$ otherwise, 
then, for every pair of points $\ox, \ox'$, with
$x < \ox < x' < \ox'$, with $\cL(\ox,y)= \cL(\ox', y)$ and
$\shadingG(\ox)\sim \shadingG(\ox') \not\sim 
\shadingG(x)$, it holds that $\cL(\ox, y'')= \cL(\ox', y'')$, for all $y \leq y'' \leq y'$.
\end{cor}

The above results lead us to the identification of those points $(x,y)$ whose behaviour 
perfectly reproduces that of a number of points $(x',y)$
on their right with $fp(x,y) = fp(x',y)$.
These points $(x,y)$, like all points ``above'' them, are useless with respect to fulfilment in a compass structure. We call them \emph{covered points}.

 \begin{defi}\label{def:coveredpoint} Let $\cG=(\bG_N, \cL)$ be a compass structure and let $0\leq x\leq y\leq N$. We say that $(x,y)$
 is a \emph{covered} point  if and only if there exist $n+1 = \Deltareq(\cL(x,y))$ distinct points $x_0 < \ldots < x_n \leq y$, with  $x < x_0$, such that for all $0\leq i \leq n$, $fp(x,y) = fp(x_i,y)$.
 In such a case, we say that $x$ is covered by $x_0 < \ldots < x_n$ on $y$.
\end{defi}

The following lemma holds.

\begin{lem}\label{lem:coveredstability}
  Let $\cG=(\bG_N, \cL)$ be a compass structure  and let $0 \leq x \leq y \leq N$ be such that $x$ is covered by 
  $x_0 < \ldots < x_n$ on $y$. Then, for all $y \leq y'  \leq N$, it holds that  $\shadingG(x)[y'] = 
  \shadingG(x_0)[y']$. 
\end{lem}

\begin{proof}

Let $\shadingG(x,y)= F^{k_0}_0\ldots F^{k_m}_m$. The proof is by induction on $n = \Deltareq(\cL(x,y))$. 

If $n=0$, then $\cL(x,y)=F_{m}$, and since $\cL(x,y) = \cL(x_0, y)$, $F_{m}= \cL(x_0,y)$. Since we are on the last atom of the sequence $\shadingG(x,y)$ and $\shadingG(x,y)\sim \shadingG(x_0,y)$, it holds that $\cL(x, y')= \cL(x_0,y')$ for all $y < y' \leq N$. 

If $n>0$, let $\cL(x,y)=F_{i}$, with $0\leq i < m$ (if $i=m$, we can follow the same reasoning path as of the inductive basis). By Lemma~\ref{lem:stepconstraint}, there exists a minimum point  $y'>y$ such that $\cL(x,y')= \cL(x_0, y')=\ldots= \cL(x_n, y') =F_{i+1}$, 
and thus, for all $y\leq y''\leq y'$, it holds that $\cL(x,y'')= \cL(x_0, y'')$.
Moreover, by Corollary~\ref{cor:inbetweeners}, for every $\ox> x_n$ such that $\shadingG(\ox)\not\sim\shadingG(x)$ if
there exists $x<\ox'<x_n$ such that $\shadingG(\ox')\sim\shadingG(\ox)$ and $\cL(\ox,y)=\cL(\ox',y)$, it holds that $\cL(\ox,y')= \cL(\ox',y')$.
Then, it follows that $\future(x,y')= \future(x_i,y')$ for every $0 \leq i<n$ (every one but $x_n$). Since $\Deltareq(F_{i}) < \Deltareq(F_{i+1})$,
and $x$ is covered by $x_0< \ldots < x_{n - 1}$ on $y'$, we can apply the inductive hypothesis and conclude that, for every $y'\leq y''\leq N$, $\cL(x,y'')= \cL(x_0, y'')$.
\end{proof}

In Figure~\ref{fig:coveredpoints}, we give an intuitive account of the notion of covered point and of the statement of Lemma~\ref{lem:coveredstability}.
First of all, we observe that, since $\future(x,y) = \future(x_0,y) = \ldots = \future(x_n,y)$ and, for all $0\leq j,j'\leq n$, it holds that $(\shadingG(x_j), \cL(x_j,y))=(\shadingG(x_{j'}), \cL(x_{j'},y))$, there exists $x_n <\hx \leq y$ such that  $\hx$ is the smallest point greater than $x_n$ that satisfies the condition $(\shadingG(x_n), \cL(x_n,y)) = (\shadingG(\hx),$ $\cL(\hx,y))$. 
Now, it can be the case that $\future(x_n,y) \supset \future(\hx,y)$, and all points $\ox'> x_n$ such that $(\shadingG(\ox'), \cL(\ox',y)) = (\shadingG(\ox), \cL(\ox,y))$, for some $x < \ox < x_n$, are in between $x_n$ and $\hx$, that is,  $x_n < \ox' < \hx$. Then, it may happen that, for all $0\leq i \leq n$, $\cL(x_i, y')= F_{i+1}$, as all points $(x_i, y')$ satisfy some $D$-request $\psi$ that only belongs to $\cL(\ox', y' - 1)$.
In such a case, as shown in Figure~\ref{fig:coveredpoints}, $\cL(\hx, y') = F_i$, because for all points
$(\hx', \hy')$, with $\hx <\hx' \leq \hy' < y'$, $\psi \notin \cL(\hx', \hy')$. 
Hence, $(\shadingG(x_n), F_{i+1}) \in \future(x_j, y')$ for all $0 \leq j <n$, but $(\shadingG(x_n), F_{i+1}) \notin \future(x_n, y')$. Then, by applying Corollary~\ref{cor:inbetweeners}, we have that $\future(x_0, y')= \future(x_{n-1}, y')$. Since
 $\Deltareq(F_{i+1}) < \Deltareq(F_{i}) (=n)$, it holds that $ \Deltareq(F_{i+1}) \leq  n-1$ 
The same argument can then be applied to  $x, x_0, \ldots, x_{n-1}$ on $y'$, and so on.


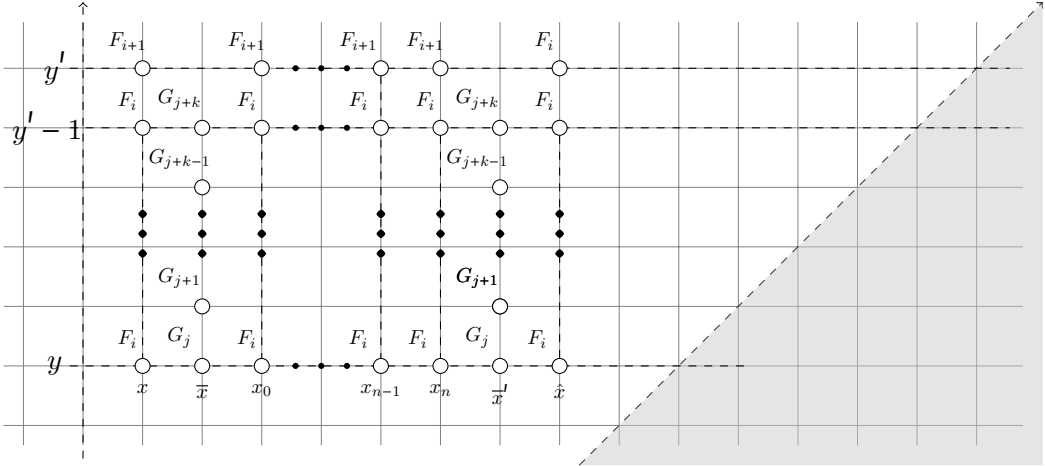
\begin{figure}
\begin{tikzpicture}[scale=0.88]


\draw[step=0.9,black, opacity=0.5, very thin] (-1.2,-1.2)
grid (14.2,5.2);

\draw[->,dashed] (0,-1.4) -- (0,5.5);

\draw[->,dashed] (7.5,-1.5) -- (14.5,5.5);

\fill[ black!20, opacity=0.5] (7.5,-1.5) -- (14.5,5.5) -- (14.5,-1.5);

\draw[dashed]  (-0.2, 0) -- (10,0) node[pos=0, xshift=-0.2cm] {$y$};
\draw[dashed]  (-0.2, 4.5) -- (14,4.5) node[pos=0, xshift=-0.2cm] {$y'$};
\draw[dashed]  (-0.2, 3.6) -- (14,3.6) node[pos=0, xshift=-0.3cm] {$y' - 1$};


\node[draw, circle, inner sep=2, 
label={[]270:$\scalebox{0.7}{$x$}$},
label={[xshift=-0.2cm]90:$\scalebox{0.7}{$F_i$}$}, fill=white](X) at (0.9,0) {};

\node[draw, circle, inner sep=2, 
label={[]270:$\scalebox{0.7}{$\ox$}$},
label={[xshift=-0.3cm]90:$\scalebox{0.7}{$G_j$}$}, fill=white](OX) at (1.8,0)
{};

\node[draw, circle, inner sep=2, 
label={[]270:$\scalebox{0.7}{$x_0$}$},
label={[xshift=-0.2cm]90:$\scalebox{0.7}{$F_i$}$}, fill=white](X1) at (2.7,0)
{};

\node[](D) at (3.6,0)
{$\scalebox{2}{$\ldots$}$};

\node[draw, circle, inner sep=2, 
label={[]270:$\scalebox{0.7}{$x_{n-1}$}$},
label={[xshift=-0.3cm]90:$\scalebox{0.7}{$F_i$}$}, fill=white](XNN) at (4.5,0)
{};

\node[inner sep=0](XNND) at (4.5,2) {$\scalebox{2}{$\vdots$}$};

\node[draw, circle, inner sep=2, 
label={[]270:$\scalebox{0.7}{$ $}$},
label={[xshift=-0.3cm]90:$\scalebox{0.7}{$F_{i+1}$}$}, fill=white](XNN1) at (4.5,4.5)
{};

\draw[dashed] (XNN1) -- ($(XNND)+(0,0.1)$);
\draw[dashed] (XNND) -- (XNN);

\node[draw, circle, inner sep=2, 
label={[]270:$\scalebox{0.7}{$ $}$},
label={[xshift=-0.3cm]90:$\scalebox{0.7}{$F_i$}$}, fill=white](XNN2) at (4.5,3.6)
{};

\node[draw, circle, inner sep=2, 
label={[]270:$\scalebox{0.7}{$x_n$}$},
label={[xshift=-0.3cm]90:$\scalebox{0.7}{$F_i$}$}, fill=white](XN) at (5.4,0)
{};

\node[draw, circle, inner sep=2, 
label={[]270:$\scalebox{0.7}{$\ox'$}$},
label={[xshift=-0.3cm]90:$\scalebox{0.7}{$G_j$}$}, fill=white](OXP) at (6.3,0)
{};

\node[draw, circle, inner sep=2, 
label={[]270:$\scalebox{0.7}{$\hx$}$},
label={[xshift=-0.3cm]90:$\scalebox{0.7}{$F_i$}$}, fill=white](HX) at (7.2,0)
{};


\node[draw, circle, inner sep=2, 
label={[]270:$\scalebox{0.7}{}$},
label={[xshift=-0.2cm]90:$\scalebox{0.7}{$F_i$}$}, fill=white](XYP1) at (0.9,3.6)
{};

\node[inner sep=0](DX) at (0.9,2) {$\scalebox{2}{$\vdots$}$};

\node[draw, circle, inner sep=2, 
label={[]270:$\scalebox{0.7}{$ $}$},
label={[xshift=-0.3cm]90:$\scalebox{0.7}{$G_{j+k}$}$}, fill=white](OXYP1) at (1.8,3.6)
{};

\draw[dashed] (XYP1) -- ($(DX)+(0,0.1)$);
\draw[dashed] (DX) -- (X);

\node[draw, circle, inner sep=2, 
label={[]270:$\scalebox{0.7}{$ $}$},
label={[xshift=-0.3cm]90:$\scalebox{0.7}{$G_{j+k-1}$}$}, fill=white](OXYPM1) at
(1.8,2.7)
{};

\node[draw, circle, inner sep=2, 
label={[]270:$\scalebox{0.7}{$ $}$},
label={[xshift=-0.3cm]90:$\scalebox{0.7}{$G_{j+1}$}$}, fill=white](OXYP1)
at
(1.8,0.9)
{};

\node[inner sep=0](DX) at (1.8,2) {$\scalebox{2}{$\vdots$}$};

\node[draw, circle, inner sep=2, 
label={[]270:$\scalebox{0.7}{}$},
label={[xshift=-0.2cm]90:$\scalebox{0.7}{$F_i$}$}, fill=white](X1YP1) at (2.7,3.6)
{};

\node[inner sep=0](DX1) at (2.7,2) {$\scalebox{2}{$\vdots$}$};

\draw[dashed] (X1YP1) -- ($(DX1)+(0,0.1)$);
\draw[dashed] (DX1) -- (X1);

\node[](D) at (3.6,3.6)
{$\scalebox{2}{$\ldots$}$};

\node[](D) at (3.6,4.5)
{$\scalebox{2}{$\ldots$}$};

\node[draw, circle, inner sep=2, 
label={[]270:$\scalebox{0.7}{}$},
label={[xshift=-0.2cm]90:$\scalebox{0.7}{$F_i$}$}, fill=white](XNYP1) at
(5.4,3.6)
{};

\node[inner sep=0](DXN) at (5.4,2) {$\scalebox{2}{$\vdots$}$};

\draw[dashed] (XNYP1) -- ($(DXN)+(0,0.1)$);
\draw[dashed] (DXN) -- (XN);

\node[draw, circle, inner sep=2, 
label={[]270:$\scalebox{0.7}{$ $}$},
label={[xshift=-0.3cm]90:$\scalebox{0.7}{$G_{j+1}$}$}, fill=white](OXP1)
at (6.3,0.9)
{};

\node[draw, circle, inner sep=2, 
label={[]270:$\scalebox{0.7}{$ $}$},
label={[xshift=-0.3cm]90:$\scalebox{0.7}{$G_{j+1}$}$}, fill=white](OXP1)
at (6.3,0.9)
{};

\node[inner sep=0](DOXP) at (6.3,2) {$\scalebox{2}{$\vdots$}$};

\node[draw, circle, inner sep=2, 
label={[]270:$\scalebox{0.7}{$ $}$},
label={[xshift=-0.3cm]90:$\scalebox{0.7}{$G_{j+k-1}$}$}, fill=white](OXYPM1)
at (6.3,2.7)
{};

\node[draw, circle, inner sep=2, 
label={[]270:$\scalebox{0.7}{$ $}$},
label={[xshift=-0.3cm]90:$\scalebox{0.7}{$G_{j+k}$}$}, fill=white](OXYP1)
at (6.3,3.6)
{};

\node[draw, circle, inner sep=2, 
label={[]270:$\scalebox{0.7}{}$},
label={[xshift=-0.2cm]90:$\scalebox{0.7}{$F_i$}$}, fill=white](HXP1) at
(7.2,3.6)
{};

\node[inner sep=0](HXPD) at (7.2,2) {$\scalebox{2}{$\vdots$}$};

\draw[dashed] (HXP1) -- ($(HXPD)+(0,0.1)$);
\draw[dashed] (HX) -- (HXPD);


\node[draw, circle, inner sep=2, 
label={[]270:$\scalebox{0.7}{}$},
label={[xshift=-0.2cm]90:$\scalebox{0.7}{$F_{i+1}$}$}, fill=white](XYP) at (0.9,4.5)
{};

\node[draw, circle, inner sep=2, 
label={[]270:$\scalebox{0.7}{}$},
label={[xshift=-0.2cm]90:$\scalebox{0.7}{$F_{i+1}$}$}, fill=white](XYP) at
(2.7,4.5)
{};

\node[draw, circle, inner sep=2, 
label={[]270:$\scalebox{0.7}{}$},
label={[xshift=-0.2cm]90:$\scalebox{0.7}{$F_{i+1}$}$}, fill=white](XYP) at
(5.4,4.5)
{};

\node[draw, circle, inner sep=2, 
label={[]270:$\scalebox{0.7}{}$},
label={[xshift=-0.2cm]90:$\scalebox{0.7}{$F_{i}$}$}, fill=white](XYP) at
(7.2,4.5)
{};

\end{tikzpicture}
\caption{\label{fig:coveredpoints} An intuitive account of the statement of Lemma~\ref{lem:coveredstability}. }
\vspace{-0.2cm}

\end{figure}


\section{The satisfiability problem for
\texorpdfstring{$\mathsf{BD}_{hom}$}{BD_hom} belongs to \expspace}
\label{sec:expspace}

In this section, by exploiting the properties proved in Section~\ref{sec:properties}, we show that the problem of checking whether a $\mathsf{BD}_{hom}$ formula $\varphi$ is satisfied by some homogeneous model can be decided in exponential space. 
First, by means of a suitable small model theorem, we prove
that either $\varphi$ is unsatisfiable or it is satisfied 
by a model (a compass structure) of at most doubly-exponential  size in $|\varphi|$; then, we show that this model of doubly-exponential size can be guessed in single exponential space.

\begin{thm}\label{thm:bound2}
The problem of deciding whether or not a $\mathsf{BD}_{hom}$ formula $\varphi$ is satisfiable, over finite linear orders, belongs to \expspace.
\end{thm}
The proof of Theorem \ref{thm:bound2} follows from
Corollary \ref{cor:coveredismonotone}, Lemma \ref{thm:smallcompass}, and Lemma \ref{thm:bound1} below.

First of all, thanks to the property proved in Section \ref{subsec:covered}, we know that, for every row $y$, there is a 
bounded number of columns $C_y = \{x_0, x_1, \ldots, x_n\}$ that may behave pairwise differently for the portion of the compass structure above $y$, where $n$ only depends on $|\varphi|$. This means that each column $0\leq x\leq y$, with $x \notin C_y$, behaves exactly as some $x_i \in C_y$ 
    above $y$, that is, for all $y'> y$, $\cL(x,y')=\cL(x_i,y')$.
Moroever, by exploiting Lemma \ref{lem:coveredstability}, we can show that, 
for each row $y$, the cardinality of the set of columns $x_0, x_1, \ldots, 
x_m$ which are not covered on $y$ is at most exponential in $|\varphi|$. 
Then, the bound on the length of the sequence of triplets for non-covered points that appear on $y$ is exponential in $|\varphi|$.
The latter claim stems from the fact that the third component $\future(x,y)$ of each triplet
$fp(x,y)= ([\shadingG(x)]_\sim, \cL(x,y), \future(x,y))$ satisfies the condition $\future(x,y) \subseteq \future(x',y)$
for all $x' \leq x$.
Thanks to such a monotonic behavior of the third component of fingerprints on any row $y$, there are at most 
an exponential number of distinct $\future(x,y)$. Finally the number of identical non-covered points associated with the same triplet $([\shadingG(x)]_\sim, \cL(x,y), \future(x,y))$ is (polynomially) bounded by $\Deltareq(\cL(x,y))$.

The set of possible characterizations of rows in a compass structure may thus vary in the set of possible sequences of triplets, which are exponentially long in $|\varphi|$. Since the latter can be viewed as symbols of a finite alphabet, whose cardinality is exponential in $|\varphi|$, we can conclude that the number of distinct characterizations/sequences is bounded by a double exponential in $|\varphi|$.
It immediately follows that, in a compass structure whose size is more than doubly exponential in $|\varphi|$, there exist two rows  $y, y'$, with $y < y'$, such that the sequences of the triplets for non-covered points that appear on $y$ and $y'$ are exactly the same. This allows us to apply a ‘‘contraction'' between $y$  and $y'$ on the compass structure.

An example of how contraction works is given in Figure \ref{\detokenize{fig:contractionpicture}}.
First of all, notice that rows $7$ and $11$ feature the same sequences for triplets of non-covered points.
Notice that the two sequences are the same even though the second occurrence (from the right) of $(Sh',F_2)$
precedes the second occurrence (from the right) of $(Sh,F_3)$ on row $7$ and follows it on row 11.
Moreover, on any row, each covered point is connected by an edge to the non-covered point that ‘‘behaves'' in the same way.
More precisely, we have that column $2$ behaves as column $4$ between $y=7$ and $y'=15$, columns $3, 5$, and $7$ behave as column $8$ between $y=11$ and $y'=15$,
and column $4$ behaves as column $6$ between $y=11$ and $y'=15$.
The compass structure in Figure \ref{\detokenize{fig:contractionpicture}}.(a) can thus be shrinked into the compass structure in Figure \ref{\detokenize{fig:contractionpicture}}.(b), where
each column of non-covered points $x$ on $y'$ is copied 
above the corresponding non-covered point $x'$ on $y$. Moreover, 
the column of a non-covered point $x$ on $y'$ is copied
over all the points which are covered by the non-covered point $x'$ corresponding to $x$ on $y$. This is the case with point $2$ in  Figure \ref{\detokenize{fig:contractionpicture}}.(b)
which takes the new column of its ‘‘covering'' point $4$.
The resulting compass structure is $y'-y$ shorter than the original one, and we can repeatedly apply such a contraction  until we achieve the desired bound.

The next corollary, which easily follows from Lemma~\ref{lem:coveredstability},
is crucial for the proof of Theorem \ref{thm:bound2}. It basically states  that the property of ``being covered'' propagates upward.

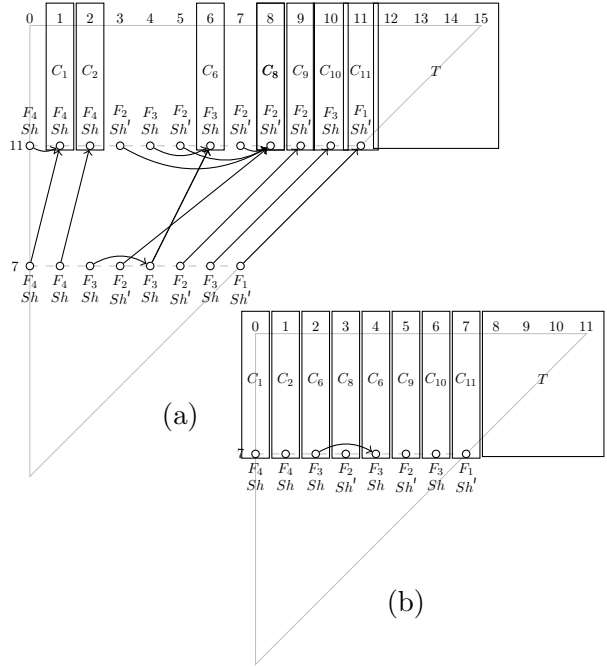
\begin{figure}
    \centering

\scalebox{1.1}{\begin{tikzpicture}[node distance=0.4cm]

\node[draw, circle, inner sep=1,  fill=white,
label={[label distance=-0.1cm]270:$\scalebox{0.5}{\vpair{
F_4 \\ Sh}}$}](0) {};

\node[draw, circle, inner sep=1,  fill=white,
label={[label distance=-0.1cm]270:$\scalebox{0.5}{\vpair{
F_4 \\ Sh}}$}, right of=0](1) {};

\node[draw, circle, inner sep=1,  fill=white,
label={[label distance=-0.1cm]270:$\scalebox{0.5}{\vpair{
F_3 \\ Sh}}$}, right of=1](2) {};

\node[draw, circle, inner sep=1,  fill=white,
label={[label distance=-0.1cm]270:$\scalebox{0.5}{\vpair{
F_2 \\ Sh'}}$}, right of=2](3) {};

\node[draw, circle, inner sep=1,  fill=white,
label={[label distance=-0.1cm]270:$\scalebox{0.5}{\vpair{
F_3 \\ Sh}}$}, right of=3](4) {};

\node[draw, circle, inner sep=1,  fill=white,
label={[label distance=-0.1cm]270:$\scalebox{0.5}{\vpair{
F_2 \\ Sh'}}$}, right of=4](5) {};

\node[draw, circle, inner sep=1,  fill=white,
label={[label distance=-0.1cm]270:$\scalebox{0.5}{\vpair{
F_3 \\ Sh}}$}, right of=5](6) {};

\node[draw, circle, inner sep=1,  fill=white,
label={[label distance=-0.1cm]270:$\scalebox{0.5}{\vpair{
F_1 \\ Sh'}}$}, right of=6](7) {};

\node[below of=5, node distance=2cm] {\small(a)};

\pgftransformshift{\pgfpoint{0cm}{1.6cm}}

\node[draw, circle, inner sep=1,  fill=white,
label={[label distance=-0.1cm]90:$\scalebox{0.5}{\vpair{
F_4 \\ Sh}}$}](A0) {};

\node[draw, circle, inner sep=1,  fill=white,
label={[label distance=-0.1cm]90:$\scalebox{0.5}{\vpair{
F_4 \\ Sh}}$}, right of=A0](A1) {};

\node[draw, circle, inner sep=1,  fill=white,
label={[label distance=-0.1cm]90:$\scalebox{0.5}{\vpair{
F_4 \\ Sh}}$}, right of=A1](A2) {};

\node[draw, circle, inner sep=1,  fill=white,
label={[label distance=-0.1cm]90:$\scalebox{0.5}{\vpair{
F_2 \\ Sh'}}$}, right of=A2](A3) {};

\node[draw, circle, inner sep=1,  fill=white,
label={[label distance=-0.1cm]90:$\scalebox{0.5}{\vpair{
F_3 \\ Sh}}$}, right of=A3](A4) {};

\node[draw, circle, inner sep=1,  fill=white,
label={[label distance=-0.1cm]90:$\scalebox{0.5}{\vpair{
F_2 \\ Sh'}}$}, right of=A4](A5) {};

\node[draw, circle, inner sep=1,  fill=white,
label={[label distance=-0.1cm]90:$\scalebox{0.5}{\vpair{
F_3 \\ Sh}}$}, right of=A5](A6) {};

\node[draw, circle, inner sep=1,  fill=white,
label={[label distance=-0.1cm]90:$\scalebox{0.5}{\vpair{
F_2\\ Sh'}}$}, right of=A6](A7) {};

\node[draw, circle, inner sep=1,  fill=white,
label={[label distance=-0.1cm]90:$\scalebox{0.5}{\vpair{
F_2 \\ Sh'}}$}, right of=A7](A8) {};

\node[draw, circle, inner sep=1,  fill=white,
label={[label distance=-0.1cm]90:$\scalebox{0.5}{\vpair{
F_2 \\ Sh'}}$}, right of=A8](A9) {};

\node[draw, circle, inner sep=1,  fill=white,
label={[label distance=-0.1cm]90:$\scalebox{0.5}{\vpair{
F_3 \\ Sh}}$}, right of=A9](A10) {};

\node[draw, circle, inner sep=1,  fill=white,
label={[label distance=-0.1cm]90:$\scalebox{0.5}{\vpair{
F_1 \\ Sh'}}$}, right of=A10](A11) {};

\draw[->] (4) -- (A6);
\draw[->] (4) -- (A6);
\draw[->] (5) -- (A9);
\draw[->] (6) -- (A10);
\draw[->] (7) -- (A11);
\draw[->] (1) -- (A2);
\draw[->] (0) -- (A1);
\draw[->] (3) -- (A8);

\draw (A7) edge[bend right, ->] (A8);
\draw (A5) edge[bend right, ->] (A8);
\draw (A4) edge[bend right, ->] (A6);
\draw (A3) edge[bend right, ->] (A8);
\draw (A0) edge[bend right, ->] (A1);

\draw (2) edge[bend left, ->] (4);

\pgftransformshift{\pgfpoint{0cm}{-4.4cm}}

\begin{pgfonlayer}{background}

\draw[opacity=0.25] (0,0) -- (6,6) -- (0,6) -- cycle;
\draw[opacity=0.25, dashed] (0,4.4) -- (4.4,4.4) node[opacity=1, pos = -0.04] {\scalebox{0.7}{$\scalebox{0.7}{11}$}};
\draw[opacity=0.25, dashed] (0,2.8) -- (2.8,2.8) node[opacity=1, pos = -0.07] {\scalebox{0.7}{$\scalebox{0.7}{7}$}};

\node[yshift=0.1cm](L0) at (0,6) {$\scalebox{0.5}{0}$};
\node[right of=L0](L1) {$\scalebox{0.5}{1}$};
\node[right of=L1](L2) {$\scalebox{0.5}{2}$};
\node[right of=L2](L3) {$\scalebox{0.5}{3}$};
\node[right of=L3](L4) {$\scalebox{0.5}{4}$};
\node[right of=L4](L5) {$\scalebox{0.5}{5}$};
\node[right of=L5](L6) {$\scalebox{0.5}{6}$};
\node[right of=L6](L7) {$\scalebox{0.5}{7}$};
\node[right of=L7](L8) {$\scalebox{0.5}{8}$};
\node[right of=L8](L9) {$\scalebox{0.5}{9}$};
\node[right of=L9](L10) {$\scalebox{0.5}{10}$};
\node[right of=L10](L11) {$\scalebox{0.5}{11}$};
\node[right of=L11](L12) {$\scalebox{0.5}{12}$};
\node[right of=L12](L13) {$\scalebox{0.5}{13}$};
\node[right of=L13](L14) {$\scalebox{0.5}{14}$};
\node[right of=L14](L15) {$\scalebox{0.5}{15}$};

\end{pgfonlayer}

\node(T) [below of=L12, node distance =1.6cm] {} ; 

\node[draw, fit=(A1)(L1), inner sep=0] {$\scalebox{0.5}{$C_1$}$};
\node[draw, fit=(A2)(L2), inner sep=0] {$\scalebox{0.5}{$C_2$}$};
\node[draw, fit=(A6)(L6), inner sep=0] {$\scalebox{0.5}{$C_6$}$};
\node[draw, fit=(A8)(L8), inner sep=0] {$\scalebox{0.5}{$C_8$}$};
\node[draw, fit=(A8)(L8), inner sep=0] {$\scalebox{0.5}{$C_8$}$};
\node[draw, fit=(A9)(L9), inner sep=0] {$\scalebox{0.5}{$C_9$}$};
\node[draw, fit=(A10)(L10), inner sep=0] {$\scalebox{0.5}{$C_{10}$}$};
\node[draw, fit=(A11)(L11), inner sep=0] {$\scalebox{0.5}{$C_{11}$}$};
\node[draw, fit=(L12)(L15)(T), inner sep=0] {$\scalebox{0.5}{$T$}$};

\pgftransformshift{\pgfpoint{3cm}{-2.5cm}}

\draw[opacity=0.25] (0,0) -- (4.4,4.4) -- (0,4.4) -- cycle;
\draw[opacity=0.25, dashed] (0,2.8) -- (2.8,2.8) node[opacity=1, pos = -0.07] {\scalebox{0.7}{$\scalebox{0.7}{7}$}};

\node[yshift=0.1cm](L0) at (0,4.4) {$\scalebox{0.5}{0}$};
\node[right of=L0](L1) {$\scalebox{0.5}{1}$};
\node[right of=L1](L2) {$\scalebox{0.5}{2}$};
\node[right of=L2](L3) {$\scalebox{0.5}{3}$};
\node[right of=L3](L4) {$\scalebox{0.5}{4}$};
\node[right of=L4](L5) {$\scalebox{0.5}{5}$};
\node[right of=L5](L6) {$\scalebox{0.5}{6}$};
\node[right of=L6](L7) {$\scalebox{0.5}{7}$};
\node[right of=L7](L8) {$\scalebox{0.5}{8}$};
\node[right of=L8](L9) {$\scalebox{0.5}{9}$};
\node[right of=L9](L10) {$\scalebox{0.5}{10}$};
\node[right of=L10](L11) {$\scalebox{0.5}{11}$};

\pgftransformshift{\pgfpoint{0cm}{2.8cm}}

\node[draw, circle, inner sep=1,  fill=white,
label={[label distance=-0.1cm]270:$\scalebox{0.5}{\vpair{
F_4 \\ Sh}}$}](0) {};

\node[draw, circle, inner sep=1,  fill=white,
label={[label distance=-0.1cm]270:$\scalebox{0.5}{\vpair{
F_4 \\ Sh}}$}, right of=0](1) {};

\node[draw, circle, inner sep=1,  fill=white,
label={[label distance=-0.1cm]270:$\scalebox{0.5}{\vpair{
F_3 \\ Sh}}$}, right of=1](2) {};

\node[draw, circle, inner sep=1,  fill=white,
label={[label distance=-0.1cm]270:$\scalebox{0.5}{\vpair{
F_2 \\ Sh'}}$}, right of=2](3) {};

\node[draw, circle, inner sep=1,  fill=white,
label={[label distance=-0.1cm]270:$\scalebox{0.5}{\vpair{
F_3 \\ Sh}}$}, right of=3](4) {};

\node[draw, circle, inner sep=1,  fill=white,
label={[label distance=-0.1cm]270:$\scalebox{0.5}{\vpair{
F_2 \\ Sh'}}$}, right of=4](5) {};

\node[draw, circle, inner sep=1,  fill=white,
label={[label distance=-0.1cm]270:$\scalebox{0.5}{\vpair{
F_3 \\ Sh}}$}, right of=5](6) {};

\node[draw, circle, inner sep=1,  fill=white,
label={[label distance=-0.1cm]270:$\scalebox{0.5}{\vpair{
F_1 \\ Sh'}}$}, right of=6](7) {};

\draw (2) edge[bend left, ->] (4);

\node[below of=5, node distance=2cm] {\small(b)};

\node(T) [below of=L8, node distance =1.6cm] {} ; 

\node[draw, fit=(0)(L0), inner sep=0] {$\scalebox{0.5}{$C_1$}$};
\node[draw, fit=(1)(L1), inner sep=0] {$\scalebox{0.5}{$C_2$}$};
\node[draw, fit=(2)(L2), inner sep=0] {$\scalebox{0.5}{$C_6$}$};
\node[draw, fit=(3)(L3), inner sep=0] {$\scalebox{0.5}{$C_8$}$};
\node[draw, fit=(4)(L4), inner sep=0] {$\scalebox{0.5}{$C_6$}$};
\node[draw, fit=(5)(L5), inner sep=0] {$\scalebox{0.5}{$C_9$}$};
\node[draw, fit=(6)(L6), inner sep=0] {$\scalebox{0.5}{$C_{10}$}$};
\node[draw, fit=(7)(L7), inner sep=0] {$\scalebox{0.5}{$C_{11}$}$};
\node[draw, fit=(L8)(L11)(T), inner sep=0] {$\scalebox{0.5}{$T$}$};

\end{tikzpicture}}
    \caption{\label{fig:contractionpicture} An example of contraction, where compass structure (a) is contracted into compass structure (b).}
\end{figure}

\begin{cor}\label{cor:coveredismonotone}
Let $\cG=(\bG_N, \cL)$ be a compass structure. Then, for every covered point $(x,y)$, it holds that, for all $y \leq y' \leq N$, point $(x,y')$ is covered as well. 
\end{cor}
From Corollary~\ref{cor:coveredismonotone}, it immediately follows that, for every covered point $(x,y)$ and every $y \leq y' \leq N$, there exists $x'> x$ such that $\cL(x',y')=\cL(x,y')$. 
Hence, for all $\ox, \oy$, with $\ox < x \leq y' < \oy$,
and any $D$-request $\psi\in \reqD(\cL(\ox, \oy)) \cap \obsD(\cL(x,y))$, we have that $\psi \in \cL(x',y)$, with $x'>x$. This allows us to conclude that if $(x,y)$ is covered, then all points $(x,y')$, with $y' \geq y$,
are ``useless'' from the point of view of $D$-requests.

Let $\cG=(\bG_N, \cL)$ be a compass structure and $0\leq y \leq N$.  We define the set of \emph{witnesses} of $y$ as the set 
$\witnesses(y)=\{x: (x,y) \mbox{ is not covered}\}$.
Corollary~\ref{cor:coveredismonotone} guarantees that, for any row $y$, the shading $\shadingG(x)$ and the labelling $\cL(x,y)$ of witnesses $x \in \witnesses(y)$ are sufficient, bounded, and unambiguous pieces of information that one needs to maintain about $y$. 

Given a compass structure $\cG=(\bG_N, \cL)$  and $0\leq y \leq N$, we define the \emph{row blueprint} of $y$ in $\cG$, 
written $\rowG(y)$, as the sequence $\rowG(y)= ([\shadingB^0]_\sim, F_0)\ldots ([\shadingB^m]_\sim, F_m)$ such that $m + 1 = |\witnesses(y)|$ and there exists a bijection $b: \witnesses(y)\rightarrow \{0,\ldots, m\} $ such that, for every $x \in \witnesses(y)$, it holds that $\shadingG(x) \in [\shadingB^{b(x)}]_\sim$ and $F_{b(x)} = \cL(x,y)$, and for every $x,x'$ in $\witnesses(y)$, $b(x)<b(x') \leftrightarrow x <x'$.

Let $\cG=(\bG_N, \cL)$ be  a compass structure. The next lemma proves that  if $\cG$ features two distinct rows $y <y'$ which share the same blueprint, then there exists a compass structure $\cG=(\bG_{N'}, \cL')$ with $N' < N$.

\begin{lem}\label{thm:smallcompass}
Let $\cG=(\bG_N, \cL)$ be a compass structure.  If there exist two points $y, y'$, with $0 \leq y < y'\leq N$, such that
$\rowG(y) = \rowG(y')$, then there exists a compass structure $\cG'=(\bG_{N'}, \cL')$ with $N'= N - (y'-y)$.
\end{lem}

\begin{proof}
From $\rowG(y) = \rowG(y')$, by composing bijections,it follows that
there exists a bijection  $\ob: \witnesses(y) \rightarrow \witnesses(y')$
such that, for every $x \in \witnesses(y)$, $\cL(x,y) = \cL(\ob(x)
,y')$, $\shadingG(x) \sim \shadingG(\ob(x))$, and
$\future(x,y)= \future(\ob(x), y')$. Moreover, for every $x,x'\in \witnesses(y)$, 
$x \leq x' \leftrightarrow \ob(x) \leq \ob(x')$. 
For every point $0 \leq x\leq y$, we define the function 
$\Closest: \{0,\ldots, y\}\rightarrow \{0,\ldots, y\}$
as follows:  $$\Closest(x)= 
\left\{
\begin{array}{cc} 
x & \mbox{if $x \in \witnesses(y)$} \\
\min\left\{\begin{array}{c}x' : x' > x, x' \in \witnesses(y), \cL(x',y) = \cL(x,y),\\
\shadingG(x') \sim \shadingG(x), 
 \future(x') = \future(x) \end{array}\right\}  & \mbox{ otherwise.}
\end{array}\right.$$  
Let $\delta = y' - y$.
We define $\cL'$ as follows: \begin{compactenum} 
\item 
$\cL'(\ox,\oy) = \cL(\ox,\oy)$,
for all $0 \leq \ox \leq \oy \leq y$;
\item $\cL'(\ox,\oy) = \cL(\ox + \delta ,\oy + \delta)$,
for all $y < \ox \leq \oy \leq N$;
\item $\cL'(\ox,\oy) = \cL(\Closest(\ox),\oy + \delta)$,
for all points $(x,y)$ with  
$0 \leq  \ox  \leq y  $ and  $  y < \oy \leq N$.
To complete the proof it suffices to show that the resulting structure
$\cG'=(\bG_{N'}, \cL')$ is a homogeneous compass structure. This part is 
omitted, since it is pretty simple, but extremely long (it can be proved by exploiting 
Corollary~\ref{cor:coveredismonotone} and the definition of witnesses for a row $y$). 
\end{compactenum}
\vspace{-0.5cm}
\end{proof}

To conclude the proof of Theorem \ref{thm:bound2}, we only need to show that if a $\mathsf{BD}_{hom}$ formula is satisfiable, then it is satisfied by a doubly exponential compass structure, whose existence can be checked in exponential space. The following lemma provides both  the \emph{small model} theorem and the \expspace\ membership.

\begin{restatable}{lem}{thmbound}\label{thm:bound1}
Let $\varphi$ be a $\mathsf{BD}_{hom}$ formula. It holds that $\varphi$ is satisfiable if and only if there is a compass structure $\cG=(\bG_N, \cL)$ for it
such that $N \leq 2^{2(|\varphi| + 1)(4|\varphi|^2 + 7 |\varphi| + 3)2^{8|\varphi|^2 + 14 |\varphi| + 6}}$, whose existence can be checked in \expspace.
\end{restatable}

\noindent The proof of Lemma~\ref{thm:bound1} can be found in Appendix~\ref{appendix:proofs}. To compute the bound, it heavily relies on the small model property stated by Lemma~\ref{thm:smallcompass}, while the argument of the proof of the \expspace\ membership is a suitable adaptation of the classical on-the-fly reachability check used, for instance, to prove that $\mathsf{LTL}$ satisfiability belongs to \pspace\  \cite{DBLP:journals/fmsd/CourcoubetisVWY92}.

\section{The satisfiability problem for 
\texorpdfstring{$\mathsf{BDA}_{hom}$}{BDA_hom} is decidable in \expspace}\label{sec:abdexpspace}

In this section, we focus on the logic $\mathsf{BDA}_{hom}$, that extends $\mathsf{BD}_{hom}$ with modality $\hsEA$. We first specify its syntax and semantics; then, we go through the definitions and proofs of Sections \ref{subsec:columns}, \ref{subsec:equivcolumns},
and \ref{subsec:covered}, and show what changes must be done to transfer them from $\mathsf{BD}_{hom}$ to $\mathsf{BDA}_{hom}$. 


Formally, syntax and semantics of $\mathsf{BDA}_{hom}$ are obtained from those of $\mathsf{BD}_{hom}$ by simply adding the syntactic rule and the semantic clause for modality $\hsEA$, respectively.
$\mathsf{BDA}_{hom}$ formulas are built up from a countable set $\Prop$ of proposition letters according to the grammar:
$
\varphi ::= p\ |\ \neg \psi\ |\  \psi \vee \psi \ |\ 
\hsEB \psi \ |\ \hsED \psi \ |\ \hsEA \psi.
$
The semantics of a $\mathsf{BDA}_{hom}$ formula is specified by the semantic clauses for $\mathsf{BD}_{hom}$, given in Section~\ref{sec:logic}, plus the following one:
\\
\begin{compactitem}
\item $\bfM, [x,y] \models \hsEA \psi$ iff 
there is $y'$, with $y'\geq y$, such that
$\bfM, [y,y']\models \psi$.
\end{compactitem}
\vspace{0.45cm}
Hereafter, we will denote by $\TFA$ the set $\{\psi: \hsEA \psi \in \closure\}$.

Notice that modality $\hsEA$ allows one to refer to the formulas that hold at the right endpoint of an interval. As an example, to force formula $\psi$ to hold at the right endpoint of the current interval $[x,y]$ (point-interval $[y,y]$), it suffices to 
state that $\hsEA(\pi \wedge \psi)$ holds at $[x,y]$ (it is useful to remark that $\mathsf{BD}_{hom}$ already allowed us to constrain a formula $\psi$ to hold at the left endpoint $[x,x]$ of an interval $[x,y]$ by means of the formula $\hsEB(\pi \wedge \psi)$). 

Such an ability to force formulas to hold at the left or the right endpoint of an interval makes a significant difference between $\mathsf{BD}_{hom}$ and $\mathsf{BDA}_{hom}$. It will play a crucial role in proving that the \expspace\ complexity bound for the 
satisfiability problem of $\mathsf{BDA}_{hom}$ is tight. In fact, it is possible to show that it is the only capability that we need 
to add to $\mathsf{BD}_{hom}$ to prove \expspace\ hardness (to make it clear, we would need to suitably rewrite the encoding of Section~\ref{sec:hardness} into a lengthier one).\footnote{In \cite{BMPS21}, the satisfiability problem for $\mathsf{BD}_{hom}$ has been shown to be \pspacecomplete.}

In Appendix \ref{appendix:bda}, we give an example of a model of a $\mathsf{BDA}_{hom}$ formula that make use of all the three operators of \LogicBDAhom. Moreover, in Appendix \ref{appendix:aregexp}, we show that $\textsf{BDA}_{hom}$ may encode a very expressive fragment of generalized $*$-free regular expressions that features prefix, infix, and lookahead. In contrast to its non-elementarily-hard complexity in the case of full generalized $*$-free regular expressions \cite{stockmeyer1974complexity}, the emptiness problem  turns out to be \expspacecomplete\ for such a fragment.

Let us now prove decidability of $\mathsf{BDA}_{hom}$ in \expspace. To begin with, we state a lemma that establishes a basic property of modality $\hsEA$, which will be extensively used in the following definitions and proofs.
\begin{lem}\label{lem:aprop}
For every interval structure $\bfM = (\bI_N, \cV)$,
every triplet of points $x\leq y\leq z$ in 
$\{0, \ldots, N\}$, and every $\mathsf{HS}$ formula $\psi$
$\bfM, [x,z] \models \hsEA \psi$ if and only if 
$\bfM, [y,z] \models \hsEA \psi$.
\end{lem}

A proof of Lemma \ref{lem:aprop} can be found in \cite{DBLP:conf/stacs/BresolinMS07}. Here we give a graphical account of it in Figure~\ref{fig:alemma}, which shows that intervals (resp., points) sharing their right endpoint (resp., laying on the same row) must feature the same $\mathsf{A}$-requests. 


\begin{figure}
\centering
\scalebox{1.1}{ 
\begin{tikzpicture}
[scale=1,
node distance=0.8cm]


\node at (-0.73,3.2) {\small(a)};

\node[draw, circle, inner sep=1, 
label={[]270:$\scalebox{0.7}{$0$}$}](0) {};

\node[draw, circle, inner sep=1, right of=0, 
label={[]270:$\scalebox{0.7}{$1$}$}](1) {};

\node[draw, circle, inner sep=1, right of=1, 
label={[]270:$\scalebox{0.7}{$2$}$}](2) {};

\node[draw, circle, inner sep=1, label={[]270:$\scalebox{0.7}{$3$}$}, right of=2](3) {};

\node[draw, circle, inner sep=1, right of=3, label={[]270:$\scalebox{0.7}{$4$}$}, 
label={[xshift=-0.2cm]90:$\scalebox{0.7}{$\hsAA\neg \psi,\neg \psi$}$}](4) {};

\node[draw, circle, inner sep=1, right of=4, label={[]270:$\scalebox{0.7}{$5$}$}](5) {};

\node[draw, circle, inner sep=1, right of=5, label={[]270:$\scalebox{0.7}{$6$}$}](6) {};

\node[draw, circle, inner sep=1, right of=6, label={[]270:$\scalebox{0.7}{$7$}$}](7) {};


\draw[|-|,dashed] ($(4.center) + (0,1.1)$) -- ($(5.center)+ (0,1.1)$)
node[pos=0.5, above](02) {$\scalebox{0.7}{$\neg \psi$}$};

\draw[|-|,dashed] ($(4.center) + (0,1.9)$) -- ($(6.center)+ (0,1.9)$)
node[pos=0.5, above](12) {$\scalebox{0.7}{$\neg \psi$}$};

\draw[|-|, dashed] ($(4.center) + (0,2.7)$) -- ($(7.center)+ (0,2.7)$)
node[pos=0.5, above](46) {$\scalebox{0.7}{$\neg \psi$}$};

\draw[|-|] ($(3.center) + (0,0.7)$) -- ($(4.center)+ (0,0.7)$)
node[pos=0.3, above](03) {$\scalebox{0.7}{$\hsAA \neg \psi$}$};

\draw[|-|] ($(2.center) + (0,1.5)$) -- ($(4.center)+ (0,1.5)$)
node[pos=0.5, above](47) {$\scalebox{0.7}{$\hsAA \neg \psi$}$};

\draw[|-|] ($(1.center) + (0,2.3)$) -- ($(4.center)+ (0,2.3)$)
node[pos=0.5, above](16) {$\scalebox{0.7}{$\hsAA \neg \psi$}$};

\draw[|-|] ($(0.center) + (0,3)$) -- ($(4.center)+ (0,3)$)
node[pos=0.5, above](16) {$\scalebox{0.7}{$\hsAA \neg \psi$}$};



\pgftransformshift{\pgfpoint{8cm}{0.5cm}}

\draw[step=0.5,black, opacity=0.5, very thin,xshift=-0.5cm,yshift=-1cm] (0.4,0.4)
grid (4,4);
\pgftransformshift{\pgfpoint{0cm}{-0.5cm}}

\draw[dashed] (-0.1,-0.1) -- (3.6,3.6);

\begin{scope}[opacity=0.5, node distance=0.5cm]

\node[yshift=-0.3cm](0) {$\scalebox{0.7}{$0$}$};
\node[ right of=0](0) {$\scalebox{0.7}{$1$}$};
\node[ right of=0](0) {$\scalebox{0.7}{$2$}$};
\node[ right of=0](0) {$\scalebox{0.7}{$3$}$};
\node[ right of=0](0) {$\scalebox{0.7}{$4$}$};
\node[ right of=0](0) {$\scalebox{0.7}{$5$}$};
\node[ right of=0](0) {$\scalebox{0.7}{$6$}$};
\node[ right of=0](0) {$\scalebox{0.7}{$7$}$};

\node[xshift=-0.3cm, yshift=-0.0cm](0) {$\scalebox{0.7}{$0$}$};
\node[ above of=0](0) {$\scalebox{0.7}{$1$}$};
\node[ above of=0](0) {$\scalebox{0.7}{$2$}$};
\node[ above of=0](0) {$\scalebox{0.7}{$3$}$};
\node[ above of=0](0) {$\scalebox{0.7}{$4$}$};
\node[ above of=0](0) {$\scalebox{0.7}{$5$}$};
\node[ above of=0](0) {$\scalebox{0.7}{$6$}$};
\node[ above of=0](0) {$\scalebox{0.7}{$7$}$};
\end{scope}

\node[draw, circle, fill=black,  inner sep=0.7, 
      label={[xshift=-0.1cm,yshift=-0.2cm]0:
      	\scalebox{0.6}{$\begin{array}{l}
      		\hsAA \neg \psi, \\
      		\neg \psi
      		\end{array}$}
      	}
     ](5) at (2, 2) {};

\node[draw, circle,inner sep=1.5] at (5){};

\node[draw, circle, fill=black,  inner sep=1](3) at (1.5, 2) {};    
\node[draw, circle, fill=black,  inner sep=1](2) at (1, 2) {};  

\node[draw, circle, fill=black,  inner sep=1](1) at (0.5, 2) {};    
\node[draw, circle, fill=black,  inner sep=1](0) at (0, 2) {};

\draw [decorate,decoration={brace,amplitude=5pt, mirror}]
($(0)+ (-0.1,-0.1)$) -- ($(3)+ (0.1,-0.1)$) 
node [black,midway,yshift=-0.45cm, xshift=-0cm](HN) {\scalebox{0.7}{$\hsAA \neg \psi$}};

\node[draw, circle, fill=black,  inner sep=1,label={[xshift=0.15cm,yshift=-0.0cm]180:
      	\scalebox{0.6}{
      		$\begin{array}{l}
      			\neg \psi
      		\end{array}$}
      	}
     ](5) at (2, 2.5) {}; 
\node[draw, circle, fill=black,  inner sep=1,label={[xshift=0.15cm,yshift=-0.0cm]180:
      	\scalebox{0.6}{
      		$\begin{array}{l}
      			\neg \psi
      		\end{array}$}
      	}
     ](6) at (2, 3) {};
\node[draw, circle, fill=black,  inner sep=1,label={[xshift=0.15cm,yshift=-0.0cm]180:
      	\scalebox{0.6}{
      		$\begin{array}{l}
      			\neg \psi
      		\end{array}$}
      	}
     ](7) at (2, 3.5) {};      

\pgftransformshift{\pgfpoint{-8cm}{-4.5cm}}
\node at (-0.73,3.2) {\small(b)};

\node[draw, circle, inner sep=1, 
label={[]270:$\scalebox{0.7}{$0$}$}](0) {};

\node[draw, circle, inner sep=1, right of=0, 
label={[]270:$\scalebox{0.7}{$1$}$}](1) {};

\node[draw, circle, inner sep=1, right of=1, 
label={[]270:$\scalebox{0.7}{$2$}$}](2) {};

\node[draw, circle, inner sep=1, label={[]270:$\scalebox{0.7}{$3$}$}, right of=2](3) {};

\node[draw, circle, inner sep=1, right of=3, label={[]270:$\scalebox{0.7}{$4$}$}, 
label={[xshift=-0.2cm]90:$\scalebox{0.7}{$\hsAA\neg \psi,\neg \psi$}$}](4) {};

\node[draw, circle, inner sep=1, right of=4, label={[]270:$\scalebox{0.7}{$5$}$}](5) {};

\node[draw, circle, inner sep=1, right of=5, label={[]270:$\scalebox{0.7}{$6$}$}](6) {};

\node[draw, circle, inner sep=1, right of=6, label={[]270:$\scalebox{0.7}{$7$}$}](7) {};


\draw[|-|,dashed] ($(4.center) + (0,1.1)$) -- ($(5.center)+ (0,1.1)$)
node[pos=0.5, above](02) {$\scalebox{0.7}{$\neg \psi$}$};

\draw[|-|,dashed] ($(4.center) + (0,1.9)$) -- ($(6.center)+ (0,1.9)$)
node[pos=0.5, above=0.09cm, rectangle, draw, solid](12) {$\scalebox{0.7}{$\neg \psi,\ \  \red{\psi}$}$};

\node[right of=12, node distance=2cm, inner sep=0.5](C){\scalebox{0.7}{contradiction}};
\draw[->](C) -- (12);

\draw[|-|, dashed] ($(4.center) + (0,2.7)$) -- ($(7.center)+ (0,2.7)$)
node[pos=0.5, above](46) {$\scalebox{0.7}{$\neg \psi$}$};

\draw[|-|] ($(3.center) + (0,0.7)$) -- ($(4.center)+ (0,0.7)$)
node[pos=0.3, above](03) {$\scalebox{0.7}{$\hsAA \neg \psi$}$};

\draw[|-|] ($(2.center) + (0,1.5)$) -- ($(4.center)+ (0,1.5)$)
node[pos=0.5, above](47) {$\scalebox{0.7}{$\hsAA \neg \psi$}$};

\draw[|-|] ($(1.center) + (0,2.3)$) -- ($(4.center)+ (0,2.3)$)
node[pos=0.5, above, red](16) {$\scalebox{0.7}{$\hsEA \psi$}$};

\draw[|-|] ($(0.center) + (0,3)$) -- ($(4.center)+ (0,3)$)
node[pos=0.5, above](16) {$\scalebox{0.7}{$\hsAA \neg \psi$}$};



\pgftransformshift{\pgfpoint{8cm}{0.5cm}}

\draw[step=0.5,black, opacity=0.5, very thin,xshift=-0.5cm,yshift=-1cm] (0.4,0.4)
grid (4,4);
\pgftransformshift{\pgfpoint{0cm}{-0.5cm}}

\draw[dashed] (-0.1,-0.1) -- (3.6,3.6);

\begin{scope}[opacity=0.5, node distance=0.5cm]

\node[yshift=-0.3cm](0) {$\scalebox{0.7}{$0$}$};
\node[ right of=0](0) {$\scalebox{0.7}{$1$}$};
\node[ right of=0](0) {$\scalebox{0.7}{$2$}$};
\node[ right of=0](0) {$\scalebox{0.7}{$3$}$};
\node[ right of=0](0) {$\scalebox{0.7}{$4$}$};
\node[ right of=0](0) {$\scalebox{0.7}{$5$}$};
\node[ right of=0](0) {$\scalebox{0.7}{$6$}$};
\node[ right of=0](0) {$\scalebox{0.7}{$7$}$};

\node[xshift=-0.3cm, yshift=-0.0cm](0) {$\scalebox{0.7}{$0$}$};
\node[ above of=0](0) {$\scalebox{0.7}{$1$}$};
\node[ above of=0](0) {$\scalebox{0.7}{$2$}$};
\node[ above of=0](0) {$\scalebox{0.7}{$3$}$};
\node[ above of=0](0) {$\scalebox{0.7}{$4$}$};
\node[ above of=0](0) {$\scalebox{0.7}{$5$}$};
\node[ above of=0](0) {$\scalebox{0.7}{$6$}$};
\node[ above of=0](0) {$\scalebox{0.7}{$7$}$};
\end{scope}

\node[draw, circle, fill=black,  inner sep=0.7, 
      label={[xshift=-0.1cm,yshift=-0.2cm]0:
      	\scalebox{0.6}{$\begin{array}{l}
      		\hsAA \neg \psi, \\
      		\neg \psi
      		\end{array}$}
      	}
     ](5) at (2, 2) {};

\node[draw, circle,inner sep=1.5] at (5){};

\node[draw, circle, fill=black,  inner sep=1](3) at (1.5, 2) {};    
\node[draw, circle, fill=black,  inner sep=1](2) at (1, 2) {};  

\node[draw, circle, fill=black,  inner sep=1,
label={[]90:\scalebox{0.7}{$\red{\hsEA \psi}$}}](1) at (0.5, 2) {};

\node[draw, circle, fill=black,  inner sep=1,
label={[]270:\scalebox{0.5}{$\hsAA \neg \psi$}}](0) at (0, 2) {};

\draw [decorate,decoration={brace,amplitude=5pt, mirror}]
($(2)+ (-0.1,-0.1)$) -- ($(3)+ (0.1,-0.1)$) 
node [black,midway,yshift=-0.3cm, xshift=-0.1cm](HN) {\scalebox{0.5}{$\hsAA \neg \psi$}};

\node[draw, circle, fill=black,  inner sep=1,label={[xshift=-0.25cm,yshift=-0.0cm]0:
      	\scalebox{0.6}{
      		$\begin{array}{l}
      			\neg \psi
      		\end{array}$}
      	}
     ](5) at (2, 2.5) {}; 
\node[draw, circle, fill=black,  inner sep=1
     ](6) at (2, 3) {};

\node[rectangle, draw, solid, left of=6, inner sep=0.05, xshift=0.1cm](L6) {
      	\scalebox{0.6}{
      		$\begin{array}{l}
      			\neg \psi, \ \ \red{\psi}
      		\end{array}$}
      	};
\draw[->](C) -- (L6);

\node[draw, circle, fill=black,  inner sep=1,label={[xshift=-0.25cm,yshift=-0.0cm]0:
      	\scalebox{0.6}{
      		$\begin{array}{l}
      			\neg \psi
      		\end{array}$}
      	}
     ](7) at (2, 3.5) {};  

\end{tikzpicture}}

\caption{\label{fig:alemma} A graphical account of the statement of Lemma~\ref{lem:aprop} from both an interval point of view and a spatial one.}
\vspace{-0.5cm}

\end{figure}
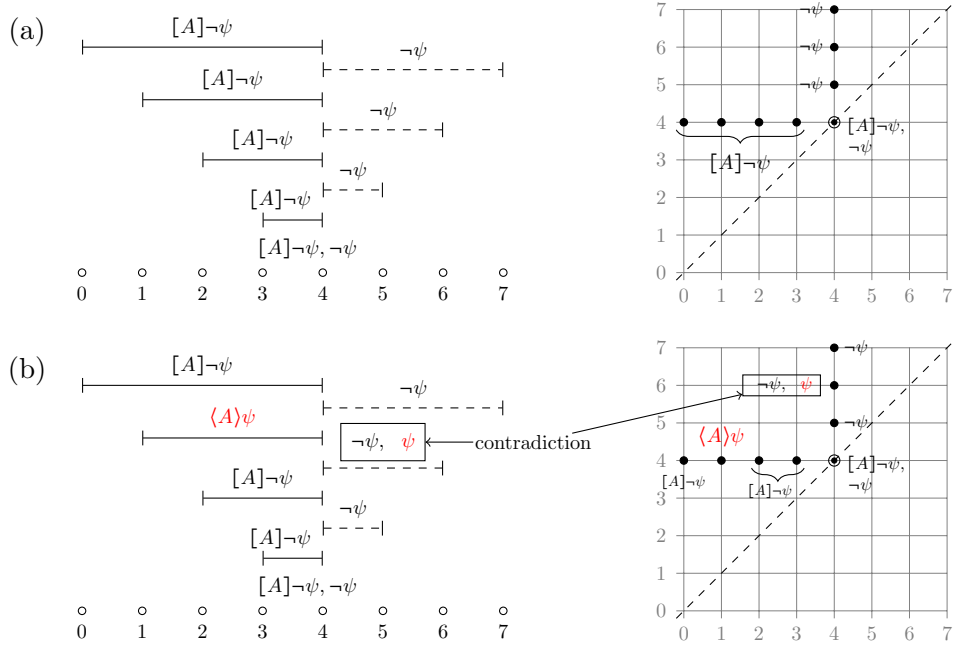

Before moving to formal definitions, let us dig deeper in the consequences of Lemma \ref{lem:aprop}, the constraints that can be imposed by means of modality $\hsEA$ in a compass structure, and the way in which we can deal with them by extending the labelling of a compass structure.

First, for every row $y$ in a compass structure, Lemma \ref{lem:aprop} states that all the points $(0,y)\ldots (y,y)$ share the very  same set of $\mathsf{A}$-requests, which must be satisfied on the column starting in $y$, that is, by points $(y,y)\ldots (y,N)$. 
Intuitively, if we consider the $90$ degree angle with vertex in $(y,y)$ we have a set of $\mathsf{A}$-requests associated with its 
horizontal edge which must be satisfied on its vertical edge. Notice that modality $\hsEA$ alone is not able to constrain the 
order according to which requests have to be satisfied on column $x$. To impose it, a suitable combination of modalities $\hsEA$ and $\hsEB$ is necessary.

Let us focus our attention on the sequence of points $(y,y)\ldots (y,N)$ on column $y$. For  each $y \leq y'\leq N$ and  each 
$\psi \in \TFA$, the following cases are the only admissible ones in a consistent compass structure: 

\begin{compactitem}
    \item  $\psi \in \reqA(\cL(y,y))$ and $\psi$ belongs to the atom $\cL(y,y'')$, for some $y\leq y''\leq y'$. In such a case, there are not pending requests for the suffix 
    $(y,y')\ldots(y,N)$ (as far as $\psi$ is concerned).
    We add the symbol  $\psi^{\checkmark}$ to the labelling of point $(y,y')$ to record that $\psi$ has been fulfilled on column $y$.

    \item   $\psi \in \reqA(\cL(y,y))$ and $\psi$ \emph{does not} belong to the atom $\cL(y,y'')$, for all $y\leq y''\leq y'$. In such a case, $\psi$ is not satisfied in the prefix $(y,y)\ldots(y,y')$ and, in order to guarantee the fulfillment of the $\mathsf{A}$-requests of $(y,y)$ we have to impose that $\psi$ appears in the labelling of some point in the suffix
    $(y,y'+1)\ldots(y,N)$. We add the symbol $\psi^{\areq}$ to the labelling of point $(y,y')$  to record that 
    the fulfillment of $\psi$ on the column $y$ has been 
    delegated to the suffix $(y,y'+1)\ldots(y,N)$, i.e., the ``upward'' part of column $y$.

    \item  $\psi \notin \reqA(\cL(y,y))$. In such a case, from the consistency of atoms, it follows that $\hsAA \neg \psi \in \cL(y,y)$, and thus $\psi \notin \cL(y,y'')$, for all $y\leq y''\leq y'$. We add the symbol $\psi^{\abox}$ to the labelling of point $(y,y')$  to record that $\psi$ does not belong to $\cL(y,y')$ because of the constraint $\hsAA \neg \psi \in 
    \cL(y,y)$ (in fact, such a constraint must be associated with all points in the sequence $(y,y)\ldots(y,N)$). 
    
\end{compactitem}

It is worth noticing that the above three conditions are mutually exclusive. The next example gives a graphical account of the 
behaviour of the labellings $\psi^{\areq}$, $\psi^{\asat}$, and
$\psi^{\abox}$.


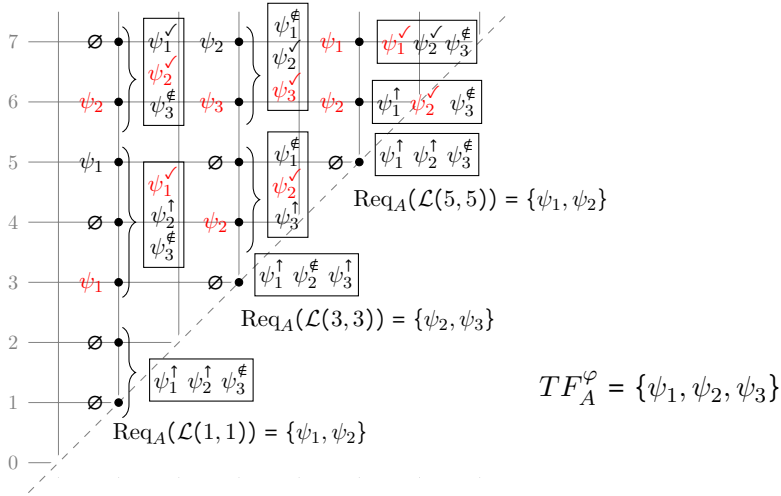
\begin{figure}
\centering
\begin{tikzpicture}
[scale=1, node distance=0.8cm]

\draw[step=0.8,black, opacity=0.5, very thin,xshift=-0.5cm,yshift=-1cm] (0.4,0.4)
grid (6.8,6.8);

\fill[white] (0,-0.4) -- (7,-0.4) -- (7,6.6) -- cycle;

\fill[white] (0,-0.4) -- (0,-0.7) -- (7,-0.7) --
(7,-0.4) -- cycle;

\pgftransformshift{\pgfpoint{0cm}{-0.5cm}}

\draw[dashed, opacity=0.5] (-0.1,-0.1) -- (6.3,6.3);

\begin{scope}[opacity=0.5, node distance=0.8cm]

\node[xshift=-0.3cm, yshift=0.3cm](0) {$\scalebox{0.7}{$0$}$};
\node[ above of=0](0) {$\scalebox{0.7}{$1$}$};
\node[ above of=0](0) {$\scalebox{0.7}{$2$}$};
\node[ above of=0](0) {$\scalebox{0.7}{$3$}$};
\node[ above of=0](0) {$\scalebox{0.7}{$4$}$};
\node[ above of=0](0) {$\scalebox{0.7}{$5$}$};
\node[ above of=0](0) {$\scalebox{0.7}{$6$}$};
\node[ above of=0](0) {$\scalebox{0.7}{$7$}$};

\end{scope}

\pgftransformshift{\pgfpoint{0.3cm}{0.3cm}}

 \node[draw, circle, fill=black,  inner sep=1,
 label={[]180:\scalebox{0.8}{$\emptyset$}},
 label={[xshift=3.5cm, yshift=-0.4cm]180:\scalebox{0.8}{$\reqA(\cL(1,1))= \{\psi_1, \psi_2\}$}},
 ](0) at (0.8,0.8) {};

 \node[draw, circle, fill=black,  inner sep=1,
 label={[]180:\scalebox{0.8}{$\emptyset$}},
 label={[xshift=-1.4cm]180:\scalebox{0.8}{$ $}},
 ](0) at (0.8,1.6) {};

\draw [decorate,decoration={brace,amplitude=5pt, mirror, raise=-0.05cm}]
($(0.9,0.6)$) -- ($(0.9, 1.8)$) 
node [black,midway, xshift=1cm,
inner sep=2pt, yshift=-0.1cm, draw](HN) {
\scalebox{0.8}{$
 \vpairmath{ \psi_1^{\areq}\  \psi_2^{\areq}\  \psi_3^{\abox}}$}
};

 \node[draw, circle, fill=black,  inner sep=1,
 label={[]180:\scalebox{0.8}{${\color{red}\psi_1}$}},
 label={[xshift=5.2cm, yshift=-0.5cm]180:\scalebox{0.8}{$
 \reqA(\cL(3,3))= \{\psi_2, \psi_3\}
 $}},
 ](0) at (0.8,2.4) {};

  \node[draw, circle, fill=black,  inner sep=1,
 label={[]180:\scalebox{0.8}{$\emptyset$}},
 label={[xshift=-1.4cm]180:\scalebox{0.8}{$ $}},
 ](0) at (0.8,3.2) {};
 
\draw [decorate,decoration={brace,amplitude=5pt, mirror,
raise=-0.05cm}]
($(0.9,2.2)$) -- ($(0.9, 4.2)$) 
node [black,midway, xshift=0.5cm,
inner sep=2pt, yshift=0.1cm, draw](HN) {
\scalebox{0.8}{$
 \vpairmath{ {\color{red}\psi_1^{\asat}}\\  \psi_2^{\areq} \\ \psi_3^{\abox}}$}
};

  \node[draw, circle, fill=black,  inner sep=1,
 label={[]180:\scalebox{0.8}{$\psi_1 $}},
 label={[xshift=6.7cm, yshift=-0.5cm]180:\scalebox{0.8}{$\reqA(\cL(5,5))=\{ \psi_1, \psi_2\} $}},
 ](0) at (0.8,4) {};

   \node[draw, circle, fill=black,  inner sep=1,
 label={[]180:\scalebox{0.8}{$ {\color{red}\psi_2} $}},
 label={[xshift=-1.4cm]180:\scalebox{0.8}{$ $}},
 ](0) at (0.8,4.8) {};

   \node[draw, circle, fill=black,  inner sep=1,
 label={[]180:\scalebox{0.8}{$\emptyset$}},
 label={[xshift=-1.4cm]180:\scalebox{0.8}{$ $}},
 ](0) at (0.8,5.6) {};

\draw [decorate,decoration={brace,amplitude=5pt, mirror, raise=-0.05cm}]
($(0.9,4.4)$) -- ($(0.9, 5.8)$) 
node [black,midway, xshift=0.5cm,
inner sep=2pt, yshift=0.1cm, draw](HN) {
\scalebox{0.8}{$
 \vpairmath{ {\psi_1^{\asat}}\\  {\color{red}
 \psi_2^{\asat}} \\ \psi_3^{\abox}}$}
};


\node [black, xshift=0.9cm,
inner sep=2pt, yshift=0.1cm, draw](HN) at (2.4,2.4) {
\scalebox{0.8}{$
 \vpairmath{ {\psi_1^{\areq}}\  {\color{black}
 \psi_2^{\abox}}\  \psi_3^{\areq}}$}
};

   \node[draw, circle, fill=black,  inner sep=1,
 label={[]180:\scalebox{0.8}{$\emptyset$}},
 label={[xshift=-1.4cm]180:\scalebox{0.8}{$ $}},
 ](0) at (2.4,2.4) {};

    \node[draw, circle, fill=black,  inner sep=1,
 label={[xshift=0.05cm]180:\scalebox{0.8}{${\color{red}\psi_2}$}},
 label={[xshift=-1.4cm]180:\scalebox{0.8}{$ $}},
 ](0) at (2.4,3.2) {};

    \node[draw, circle, fill=black,  inner sep=1,
 label={[]180:\scalebox{0.8}{$\emptyset$}},
 label={[xshift=-1.4cm]180:\scalebox{0.8}{$ $}},
 ](0) at (2.4,4) {};

   \node[draw, circle, fill=black,  inner sep=1,
 label={[]180:\scalebox{0.8}{${\color{red}\psi_3}$}},
 label={[xshift=-1.4cm]180:\scalebox{0.8}{$ $}},
 ](0) at (2.4,4.8) {};

   \node[draw, circle, fill=black,  inner sep=1,
 label={[]180:\scalebox{0.8}{$\psi_2$}},
 label={[xshift=-1.4cm]180:\scalebox{0.8}{$ $}},
 ](0) at (2.4,5.6) {};

\draw [decorate,decoration={brace,amplitude=5pt, mirror, raise=0.1cm}]
($(2.4,2.8)$) -- ($(2.4, 4.2)$) 
node [black,midway, xshift=0.65cm,
inner sep=2pt, yshift=0.2cm, draw](HN) {
\scalebox{0.8}{$
 \vpairmath{ {\psi_1^{\abox}}\\  {\color{red}
 \psi_2^{\asat}} \\ \psi_3^{\areq}}$}
};

\draw [decorate,decoration={brace,amplitude=5pt, mirror, raise=0.1cm}]
($(2.4,4.5)$) -- ($(2.4, 5.8)$) 
node [black,midway, xshift=0.65cm,
inner sep=2pt, yshift=0.25cm, draw](HN) {
\scalebox{0.8}{$
 \vpairmath{ {\psi_1^{\abox}}\\  {
 \psi_2^{\asat}} \\ {\color{red}\psi_3^{\asat}}}$}
};


\node [black, xshift=0.9cm,
inner sep=2pt, yshift=0.1cm, draw](HN) at (4,4) {
\scalebox{0.8}{$
 \vpairmath{ {\psi_1^{\areq}}\  {\color{black}
 \psi_2^{\areq}}\  \psi_3^{\abox}}$}
};

\node [black, xshift=0.9cm,
inner sep=2pt, yshift=0.8cm, draw](HN) at (4,4) {
\scalebox{0.8}{$
 \vpairmath{ {\psi_1^{\areq}}\  {\color{red}
 \psi_2^{\asat}}\  \psi_3^{\abox}}$}
};

\node [black, xshift=0.9cm,
inner sep=2pt, yshift=1.6cm, draw](HN) at (4,4) {
\scalebox{0.8}{$
 \vpairmath{ {\color{red} \psi_1^{\asat}}  {\color{black}
 \psi_2^{\asat}}  \psi_3^{\abox}}$}
};

\node[draw, circle, fill=black,  inner sep=1,
 label={[]180:\scalebox{0.8}{$\emptyset$}},
 label={[xshift=-1.4cm]180:\scalebox{0.8}{$ $}},
 ](0) at (4,4) {};

\node[draw, circle, fill=black,  inner sep=1,
 label={[]180:\scalebox{0.8}{${\color{red}\psi_2}$}},
 label={[xshift=-1.4cm]180:\scalebox{0.8}{$ $}},
 ](0) at (4,4.8) {};

\node[draw, circle, fill=black,  inner sep=1,
 label={[]180:\scalebox{0.8}{${\color{red}\psi_1}$}},
 label={[xshift=-1.4cm]180:\scalebox{0.8}{$ $}},
 ](0) at (4,5.6) {};

\node[xshift=8cm, yshift=1cm]{$\TFA=\{\psi_1, \psi_2, \psi_3\}$};

\end{tikzpicture}

\caption{\label{fig:alabelling} An
 example of how the additional labels $\areq$, $\asat$, and $\abox$ behave in a compass structure. To the left of the highlighted points $(x,y)$ we put the set $\cL(x,y)\cap\{\psi_1, \psi_2, \psi_3\}$, while to their right we put their additional marking.}
\vspace{-0.5cm}

\end{figure}

\begin{exa}\label{ex:alabelling}

In Figure~\ref{fig:alabelling}, we show how the additional marking of atoms behave in a compass structure. Let $\TFA =\{\psi_1, \psi_2, \psi_3\}$. For the sake of brevity, we restrict our attention to the $\mathsf{A}$-requests transferred to points $(1,1), (3,3),$ and $(5,5)$.
Let $\reqA(\cL(1,1)) = \{\psi_1, \psi_2\}$, $\reqA(\cL(3,3))
= \{\psi_2, \psi_3\}$, and $\reqA(\cL(5,5))
= \{\psi_1, \psi_2\}$. By Lemma~\ref{lem:aprop}, it holds that $\reqA(\cL(0,1))
= \reqA(\cL(1,1))$,
$\reqA(\cL(0,3))
= \reqA(\cL(1,3))
= \reqA(\cL(2,3))
= \reqA(\cL(3,3))
$, and
$\reqA(\cL(0,5))
= \reqA(\cL(1,5))
= \reqA(\cL(2,5))
= \reqA(\cL(3,5))
= \reqA(\cL(4,5))
= \reqA(\cL(5,5))
$. The above requests must be satisfied on columns
$1$, $3$, and $5$, respectively.

Let us consider the case of column $1$. To start with, we label point  
$(1,1)$ with the set $\{\psi_1^{\areq}, \psi_2^{\areq}, 
\psi_3^{\abox}\}$. The presence of $\psi_3^{\abox}$ in the labelling
stems from $\psi_3 \notin \reqA(\cL(1,1))$. By consistency of atoms, it follows that $\hsAA \neg \psi_3 \in \cL(1,1)$, and thus 
$\neg \psi_3 \in \cL(1,y)$, for all $1\leq y\leq 7$. By associating 
$\psi_3^{\abox}$ with all points on column $1$, we make it explicit that $\neg \psi_3$, and not $\psi_3$, must belong to all atoms.
Notice that, in general, for every column $x$ and $\psi \in \TFA$,  either $\psi_{\abox}$ labels all points on the column or it labels none of them. Besides $\psi_3^{\abox}$, point $(1,1)$ is labeled by 
$\psi_1^{\areq}$ and $\psi_2^{\areq}$, to keep track of the fact that 
$\psi_1, \psi_2 \in \reqA(\cL(1,1))$. Both $\psi_1$ and $\psi_2$ must indeed be witnessed by at least one point on column $1$. Climbing the column, we mark all points $(1,y)$ above $(1,1)$ with $\psi_1^{\areq}$ (resp., $\psi_2^{\areq}$) until we encounter the first point that satisfies $\psi_1$ (resp., $\psi_2$). In Figure~\ref{fig:alabelling},
$\psi_1$ (resp., $\psi_2$) is satisfied ``for the first time'' at point
$(1,3)$ (resp., point $(1,6)$). From point $(1,3)$ (resp., $(1,6)$) on,
the label $\psi_1^{\asat}$ (resp., $\psi_2^{\asat}$), instead of the label $\psi_1^{\areq}$ (resp., $\psi_2^{\areq}$), is used to make it explicit that formula $\psi_1$ (resp., $\psi_2$) has already been satisfied on column $1$. Additional occurrences of $\psi_1$ in the labelling above $(1,3)$, like, for instance, the one at point $(1,5)$, does not affect the additional labelling. Similar considerations apply to columns $3$ and $5$.

We conclude the illustration of the example by highlighting some 
points of interests.
\begin{compactitem}

\item While the presence of $\psi^{\abox}$ in the additional labelling of a point $(x,y)$ \emph{forces} $\neg \psi$ to belong to atom $\cL(x,y)$, the presence of either $\psi^{\areq}$ or $\psi^{\asat}$ implies neither $\psi \in \cL(x,y)$ nor $\neg \psi \in \cL(x,y)$ 
(see, e.g.,  the cases of points $(1,4)$, $(1,5)$, and $(1,7)$ in Figure~\ref{fig:alabelling}).

\item When $\psi^{\areq}$ is associated with on the bottom of a column,
sooner or later we make a transition from $\psi^{\areq}$ to $\psi^{\asat}$, that is, the labelling is monotone (such a behavior can be observed on columns $1$, $3$, and $5$ in Figure~\ref{fig:alabelling}.

\item For every column  $x$, the set of all the requests  $\psi$ in $\TFA$ that must be satisfied on the column (as recorded by the additional labels $\psi^{\areq}$/$\psi^{\asat}$) 
is given  by the set  $\reqA(\cL(x,x))$ 
at the bottom of the column. 
Different columns may obviously feature different sets of requests
(this is the case with columns $1$ and $3$ in Figure~\ref{fig:alabelling}), and columns with the same set of requests
may fulfil them in a different order (this is the case with columns  $1$ and $5$ in Figure~\ref{fig:alabelling}).

\item For each column $x$ and request $\psi \in \TFA$, either $\psi^{\abox}$ or
$\psi^{\asat}$ belongs to the label of the top point $(x,N)$. Indeed, to guarantee that all the requests transferred to $\reqA(x,x)$ are sooner or later fulfilled, no $\mathsf{A}$-request can be still pending on the top row $(x,N)$. Such a condition is exemplified by points $(1,7),(3,7)$, and $(5,7)$ in Figure~\ref{fig:alabelling}.

\end{compactitem}

\end{exa}





\noindent We now refine the notion of {$\varphi$-atom} (see Section~\ref{sec:compass}) into that of \emph{marked-{$\varphi$-atom}}.
\begin{defi}\label{def:markedatom}
A
 \emph{marked $\varphi$-atom} (hereafter simply \emph{atom}) is a pair $F_{\alpha}=(F,\alpha)$, where:
\begin{compactenum} 

\item $F$  is a maximal subset of $\closure$ that, for all $\psi \in \closure$, satisfies the following 3 conditions:
(i) $\psi \in F$ if and only if $\neg \psi \notin F$, 
(ii) if $\psi = \psi_1 \vee \psi_2$, then 
$\psi \in F$ if and only if $\{\psi_1 , \psi_2\}\cap F\neq \emptyset$, and
(iii) if $\pi \in F$, then, for all $\hsAA \psi \in F$, $\psi \in F$;

\item $\alpha$ is a function $\alpha: \TFA \rightarrow \{ \areq, \asat, \abox\}$ that, for all $\psi\in \TFA$,
satisfies the following 4 conditions: (i)  if $\alpha(\psi) = \abox$, then $\neg \psi \in F$;
(ii) if $\psi \in F$, then $\alpha(\psi)= \asat$; (iii) if $\pi \in F$ 
and  $\alpha(\psi) = \areq$, then  $\hsEA \psi \in F$ and $\psi \notin F$;
(iv) if $\pi \in F$ 
and  $\alpha(\psi) = \asat$, then  $\psi \in F$.

\end{compactenum}

\end{defi}

It is worth pointing out that we can safely characterize the additional labelling as a function, because labels are mutually exclusive.

The second component of an atom keeps track of formulas $\psi$ that are forced to appear negated in every interval starting at $x$ due to the presence of $\hsAA \psi$ in the labelling of $[x,x]$ that is, formulas marked by $\areq$. Since we cannot consider a model fulfilled until all $\mathsf{A}$-requests are satisfied for all points $x$ in the model, we introduce the notion of \emph{final atom}. An atom $F_\alpha$ is \emph{final} if and only if for every $\psi \in \TFA$, it holds that  $\alpha(\psi) \in\{\asat, \abox\}$.

For the sake of simplicity, from now on, when we refer to $F_\alpha$ as a set, we refer to its first component $F$, e.g., when we write 
$\psi \in F_\alpha$,  we mean $\psi \in F$.
 
Let $\atoms$ be the set of all
$\varphi$-atoms.  We have that $|\atoms| \leq 2^{|\varphi| + 1}\cdot 2^{|\varphi| - 1} =  2^{2|\varphi|}$, where $|\varphi| =|\closure|/2$.
When we restrict our attention to the  first component of $F_\alpha$, functions $\reqR$, $\obsR$, and $\boxR$, with $R \in \{A,B, D\}$, 
can be defined exactly as in Section~\ref{sec:compass}. In addition, we specialize the relations $\thenB$ and $\thenD$: as follows:

\begin{compactitem} 
\item $F_\alpha \thenB G_\beta$ if and only if $\reqB(F_\alpha) = \reqB(G_\beta) \cup 
\obsB(G_\beta)$ and
for all $\psi \in \TFA$, $\alpha(\psi) = \beta(\psi)$ if
$\beta(\psi) \in \{\asat, \abox\}$ or $\psi \notin F$;
\item $F_\alpha \thenD G_\beta$ if and only if $\reqD(F_\alpha) \supseteq \reqD(G_\beta) \cup \obsD(G_\beta)$.
\end{compactitem}

The statement about labelings made in Section~\ref{sec:compass} can be extended to 
$\mathsf{BDA}_{hom}$ formulas by claiming that  the labelings that satisfy condition $(*_1)$ are all and only those claims that satisfy the following property: 
\[\begin{array}{ll}
       
   (*_3\mbox{-}b) &    \reqB(F^{[x,y]})= \bigcup_{x \leq y'< y} \obsB(F^{[x,y']}) \mbox{, }    
                       \reqD(F^{[x,y]})= \bigcup_{x \leq x' \leq y'< y} \obsD(F^{[x',y']}), \\
                  &    \mbox{ and } { \reqA(F^{[x,y]}) = \bigcup_{y \leq y'} \obsA(F^{[y,y']}) }  \mbox{, for all $[x,y] \in \bI_N$.}
	 
\end{array}
\]

Compass structures for  $\mathsf{BDA}_{hom}$ formulas can be defined by extending those for 
$\mathsf{BD}_{hom}$ ones with the following requirements:
\begin{compactitem}
\item (\emph{initial formula}) $\varphi \in \cL(0,N)$;

\item ($A$-\emph{consistency}) for all $0 \leq  x \leq  y \leq N$,
$\reqA(\cL(x,y)) = \reqA(\cL(y,y))$;

\item ($A$-\emph{fulfilment}) for every 
$0 \leq x \leq N$,  atom $\cL(x,N)$ is final.
\end{compactitem}

The result about compass structures for  $\mathsf{BD}_{hom}$ formulas can be lifted to
 $\mathsf{BDA}_{hom}$ ones.

\begin{thm}\label{thm:asatthencompass}
A $\mathsf{BDA}_{hom}$ formula $\varphi$ is satisfiable if and only if there is a homogeneous $\varphi$-compass structure.
\end{thm} 

\noindent In Appendix~\ref{appendix:bda}, we 
give a couple of examples of a consistent atom labelling in the case of $\mathsf{BDA}_{hom}$ formulas.

We are now ready to illustrate the (minor) changes that must be done in order to adapt the small model theorem proved in Section~\ref{sec:properties} to the logic $\mathsf{BDA}_{hom}$. 

First of all, by exploiting Lemma \ref{lem:aprop}, it can be easily proved that Lemma \ref{lem:bstep} holds for $\mathsf{BDA}_{hom}$ homogeneous compass
structures as well.

Second, to take into account the second component of atoms, we redefine the function $\Deltareq: \atoms \rightarrow 
\bN$ as follows:
\[
\begin{array}{rcl}
\Deltareq(F_\alpha) &=& (2|\{\hsEB \psi \in \closure \}| - 2|\reqB(F_\alpha)| - \\
&&  |\obsB(F_\alpha) \setminus \reqB(F_\alpha)|)+ \\
&& (|\{\hsED \psi \in \closure \}| - |\reqD(F_\alpha)|) + \\
&& (|\{\neg p: p \in \closure \cap \Prop\}|  - \\
&& |\{ \neg p : p  \in \closure \cap \Prop \wedge  \neg p \in F_\alpha  \}|) + \\
&& |\{\ \psi \in \TFA: \alpha(\psi)=\areq \}|
\end{array}
\]

The main complication caused by the introduction of modality $\hsEA$ is that a $B$-sequence sequence 
in a compass structure may be not flat (while it is still forced to be  decreasing). To cope with such a complication, we introduce the concept of minimal $B$-sequence.
A \emph{$B$-sequence} $\shadingB= F^0_{\alpha_0}\ldots F^n_{\alpha_n}$ is \emph{minimal} if and only if for every $0 \leq i < n$, $\Deltareq(F^i_{\alpha_i}) > \Deltareq(F^{i+1}_{\alpha_{i+1}})$. Let us observe that 
for every minimal $B$-sequence $\shadingB= F^0_{\alpha_0}\ldots F^n_{\alpha_n}$, it holds that
$n \leq 5|\varphi|$, that is, the length of a minimal $B$-sequence is at most $5|\varphi|+1$. 
A minimal $B$-sequence does not represent the whole sequence of atoms on a ``column'' $x$ of a given compass structure, as it happened for flat decreasing $B$-sequences in Section \ref{sec:properties}. Here, a minimal $B$-sequence represents the labellings of the sequence of points sharing the same ``column'' $x$ where the 
function $\Deltareq(F^i_{\alpha_i})$
decreases as long as we move up on $y$.
To capture such a behaviour, we define the following notion of \emph{shading}.

Let $\cG=(N, \cL)$ be a compass structure for $\varphi$ and $0 \leq x \leq N$. The \emph{shading of $x$} in $\cG$, written $\shadingG(x)$, is the sequence of pairs 
$(\cL(x,y_0), y_0)\ldots(\cL(x,y_m),y_m)$ such that:
\begin{compactenum}
\item $y_i < y_{i+1}$ for all $0\leq i < m$;
\item $\{ \Deltareq(\cL(x,y)): 0 \leq y\leq N\}= \{ \Deltareq(\cL(x,y_i)): 0 \leq i\leq m\}$;
\item for all $0\leq i \leq m$, $y_i = \min\left\{ 0 \leq y\leq N: 
  \Deltareq(\cL(x,y_i)) = \Deltareq(\cL(x,y))\right\}$, that is,  $y_i$ is the minimum
  height on the column $x$ that exhibits its value for $\Deltareq$.
\end{compactenum} 

For all  $0 \leq x \leq N$,  let $\shadingG(x) = \cL(x,y_0)\ldots\cL(x,y_m)$.
We denote by $\shadingGB(x)$
the sequence of atoms  $\cL(x,y_0)\ldots\cL(x,y_m)$,
and by  $\shadingGN(x)$  the sequence of natural  numbers $y_0\ldots y_m$, that is, 
the projections of $\shadingG(x)$ on its first and second component, respectively.

\smallskip

 The next lemma is the $\mathsf{BDA}_{hom}$ counterpart of Lemma~\ref{lem:shadingthensequence}.

\begin{lem}\label{lem:abdshadingthensequence}
Let $\cG=(N, \cL)$ be a compass structure and 
$0 \leq x \leq N$. Then, $\shadingGB(x)$
is a minimal $B$-sequence.
\end{lem}

The above (finite) characterisation works just as well as the one provided in Section~\ref{sec:properties} to define a
 natural equivalence relation of finite index over columns: 
we say that two columns $x, x'$ are equivalent, written $x \sim x'$, if and only if $\shadingGB(x)=\shadingGB(x')$.
Then, by using Lemma~\ref{lem:aprop}, we can prove that Lemma~\ref{lem:shadingorder} also holds for
$\mathsf{BDA}_{hom}$ compass structures.
The definitions of $\future(x,y)$ and 
of fingerprint  $fp(x,y)$, for all $0\leq x\leq y\leq N$, are the same as the ones given in Section~\ref{sec:properties}. 
Given the specialization  of atoms,  the number of possible sets $\future(x,y)$  turns out to be bounded by $2^{6^{5|\varphi|^2 + 2 |\varphi|} \cdot \frac{2}{3}^{5|\varphi| + 2}}$. Given two atoms $F_\alpha$ and $G_\beta$, we say that 
they are \emph{equivalent modulo 
$\textsf{A}$}, written $F_\alpha \equivmodA G_\beta$ 
if and only if $F \setminus \reqA(F_\alpha) = G \setminus \reqA(G_\beta)$
and $\alpha= \beta$, that is, $F_\alpha$ and $G_\beta$
may only differ in their $\hsEA$ requests.
We can prove the analogous of Lemma~\ref{lem:stepconstraint} and  
Corollary~\ref{cor:inbetweeners} for $\mathsf{BDA}_{hom}$ compass structures.

\begin{lem}\label{lem:astepconstraint}
Let $\cG=(N, \cL)$ be a compass structure and
$0\leq x < x'\leq y \leq N $. If $fp(x,y) = fp(x',y)$
and $y'$ is the smallest point greater than $y$ such that
$\cL(x,y')\nequivmodA \cL(x,y)$, if any, and $N$ otherwise, 
then, for all $y \leq y''\leq y'$, $\cL(x,y'') = \cL(x',y'')$.
\end{lem}

\begin{cor}\label{cor:ainbetweeners}
Let $\cG=(N, \cL)$ be a compass structure and
$0\leq x < x'\leq y \leq N $. If $fp(x,y) = fp(x',y)$ and
$y'$ is the smallest point greater than $y$ such that
$\cL(x,y')\nequivmodA \cL(x,y)$, if any, and $N$ otherwise, 
then, for all pairs of points $\ox, \ox'$, with
$x < \ox < x' < \ox'$, with $\cL(\ox,y)= \cL(\ox', y)$ and
$\ox\sim \ox' \not\sim x$, 
it holds that $\cL(\ox, y'')= \cL(\ox', y'')$, for all $y \leq y'' \leq y'$.
\end{cor}

The definitions of covered point, set of witnesses 
($\witnesses(y)$), and row blueprint ($\rowG(y)$)
for $\mathsf{BDA}_{hom}$ compass structures 
are exactly the same as the ones for $\mathsf{BD}_{hom}$ ones (see Definition~\ref{def:coveredpoint}
and Section~\ref{sec:expspace}).
As a consequence, Lemma~\ref{lem:coveredstability}, 
Corollary~\ref{cor:inbetweeners}, 
and Theorem~\ref{thm:smallcompass}
hold for $\mathsf{BDA}_{hom}$ compass structures as well. 
Thanks to them, we can devise an algorithm very similar to the one exploited in the proof of Theorem~\ref{thm:bound1}, that yields the following  result.

\begin{thm}\label{thm:abound1}
Let $\varphi$ be a $\mathsf{BDA}_{hom}$ formula. It holds that $\varphi$ is satisfiable if and only if there is a compass structure $\cG=(N, \cL)$ for it,
with $N \leq 2^{5|\varphi|\cdot(6^{10 |\varphi|^2+4 |\varphi|}\cdot \frac{2}{3}^{10 |\varphi| + 4})}$, whose existence can be checked in $EXPSPACE$.
\end{thm}

It is not difficult to show that the doubly-exponential upper bound on the size of models of satisfiable formulas is tight. One can indeed build a satisfiable formula of size $\cO(|n|^2)$, featuring $n + 1$ propositional letters, whose models are all of size at least $2^{2^n}$. Such a formula $\varphi$ can be obtained by making use of the notions introduced in Section~\ref{sec:expspace}, which were instrumental in proving the \expspacehardness\ of the satisfiability problem for $\mathsf{BDA}_{hom}$.

}


\section{ \expspace-hardness of
\texorpdfstring{$\mathsf{BDA}_{hom}$}{BDA_hom}}\label{sec:hardness}

In this section, we prove that the satisfiability problem for 
$\mathsf{BDA}_{hom}$ is \expspacehard. 
The proof consists of a reduction from the acceptance problem for  (non-deterministic) Turing Machines working in exponential space.\footnote{An alternative reduction from a well-known tiling problem can be found in \cite{DBLP:journals/corr/abs-2109-08320}.}

Let $Tm = (\Sigma, Q, 
\Delta, q_0, q_f)$ be a Turing machine,
where the alphabet $\Sigma$ contains the special blank symbol $\#$, $q_0, q_f \in Q$, and $\Delta\subseteq 
((Q \times \Sigma) \times (Q, \times \Sigma\setminus\{\#\}) \times \{\leftarrow,\rightarrow\})$. As an example,
the transition $((q, \sigma),(q', \sigma'), 
\rightarrow)$ is enabled if $Tm$ is 
at state $q$ and its head points to a cell containing $\sigma$, and if it is fired, then symbol $\sigma$ is replaced by symbol $\sigma'$,
state $q$ is replaced by state $q'$, and  the head  is moved to the next cell.

Notice that, according to the above definition, $Tm$ never writes the symbol $\#$, that can only be rewritten/consumed when encountered. Then, if we assume that the initial configuration of $Tm$ consists of the input word $w_{in} \in (\Sigma\setminus\{\#\})^*$, written in 
the first $0, \ldots,|w_{in}| - 1$ cells of the (infinite) tape, and all the other cells contain the symbol $\#$, each 
configuration of $Tm$ can be represented as a tuple $(w, q, i)$, where $w$ is the word consisting of all and only the symbols different from $\#$ in the tape, $q$ is the current state, and $0\leq i \leq |w|$ is the current position of the head (if $i = |w|$, then the symbol in the underlying cell is $\#$).

Let $\rightarrow_\Delta$ be the standard transition relation  over configurations and $\rightarrow^*_\Delta$ be its transitive and reflexive closure.
Given $w_{in} \in (\Sigma\setminus\{\#\})^*$, $Tm$ accepts $w_{in}$ if and only if $(w_{in},q_0, 0) \rightarrow^*_\Delta (w, q_f, i)$ for some $w\in (\Sigma\setminus\{\#\})^*$ and some $i \in \bbN$. Notice that $|w|$ at the end of the computation is exactly the number of used cells, that is, the space consumption.  

The language of $Tm$ is the set $L(Tm) =\{ w \in (\Sigma\setminus\{\#\})^*: \mbox{$Tm$ accepts 
$w$}\}$. We say that $Tm$ works in exponential space if and only if for every $w_{in} \in L(Tm)$ there exists a
computation $(w_{in},q_0, 0) \rightarrow^*_\Delta (w, q_f, i)$
such that $|w| \leq 2^{|w_{in}|}$.\footnote{
We do not care of words $w \notin L(Tm)$. In those cases, $Tm$ either would exceed the exponential space bound without reaching $q_f$ or would get stuck in a loop without passing through $q_f$.} 
From now on, we assume $|w_{in}| = n$, and thus the computation must take at most $2^n$ cells. Moreover, we assume that $Tm$ never writes on the left of the cell $0$ when it starts from the configuration $(w, q_0, 0)$.  It is trivial to check that we can make all these 
assumptions without any loss of generality. 

We are now ready to provide the reduction. More precisely, given a word $w_{in}\in (\Sigma\setminus\{\#\})^*$ and a Turing Machine which works in exponential space, we write a \LogicBDAhom formula $\varphi$ which is satisfiable if and only if $w_{in} \in L(Tm)$. 

To start with, we introduce some useful shorthands and some basic formulas that will be used as the building blocks of the proposed encoding. First, let 
$len_1$ be the formula $\hsAB \pi$, which holds at $[x, y]$ if and only if $y= x + 1$,  that is,
the length of $[x, y]$ is exactly 1 (the constant $\pi$ 
for $0$-length intervals has been introduced in
Section~\ref{sec:logic}).
Then, we introduce the global modality $[G] \psi$, whose semantics is as follows: given a model $\bfM=(\bI_N, \cV)$, $\bfM, [0,y] \models [G] \psi$ if and only if $\psi \in \cV([x',y'])$ for every $0\leq x' 
\leq y' \leq N$, i.e., $[G] \psi$ holds at an initial interval of the model (an interval whose left endpoint is $0$) if and only if $\psi$ holds at every interval of the model. In $\LogicABDhom$, $[G]\psi$ can be expressed by means of  the formula $\psi \wedge \hsAA \psi \wedge \hsAA\hsAA \psi \wedge \hsAB \psi \wedge \hsAB\hsAA \psi$. 
Finally, we write $\hsEB_\pi \psi$ and $\hsEA_\pi \psi$
for $\hsEB(\pi \wedge \psi)$ and $\hsEA(\pi \wedge \psi)$
to directly access the first and the last point of an interval, respectively.  

Let us consider now the problem of identifying (indexing) the cells of every configuration. Let $N = 2^n$. We make use of $n$ bits represented as proposition letters $b_0, \ldots, b_{n - 1}$. For each (point-)interval $[x,x]$ in the model, the index of the cell it represents in a configuration is binary-encoded as $id_x = \sum_{b_i \in \cV([x,x])} 2^i$. The first and the last cell of a computation are the cells indexed by $0$ and with $2^n - 1$, respectively. We will refer to them by means of the 
shorthands $cell_0 = \pi  \wedge \bigwedge_0^{n-1} \neg b_i$ and $cell_N = \pi \wedge\bigwedge_0^{n-1} b_i$, respectively.

To encode both the correct order of the cells, mimicking the tape on a linear temporal model, and the transitions of $TM$, we make use of the following formulas that check basic arithmetic properties of the bits associated 
with the endpoints of an interval:

\begin{compactitem}

\item $\psi^i_= = (\hsEB_\pi b_i \leftrightarrow \hsEA_\pi b_i) 
\wedge \psi^{i + 1}_=$ for $0\leq i < n -1$  and $\psi^{n-1}_= 
= (\hsEB_\pi b_{n-1} \leftrightarrow \hsEA_\pi b_{n-1})$. It is easy to check that $\psi^i_=$ holds over an interval $[x,y]$ if the bits from 
$i$ to $N$ for the cell associated with $[x,x]$ are equal to the bits
for the cell associated with $[y,y]$. Hence, $\psi^0_=$ holds
at $[x,y]$ if and only if the cells $[x,x]$ and $[y,y]$ have the same 
index, that is, $id_x = id_y$;

\item $\psi^i_+ = (\hsEB_\pi \neg b_i \wedge  \hsEA_\pi b_i \wedge
\psi^{i + 1}_= ) \vee (\hsEB_\pi  b_i \wedge  \hsEA_\pi \neg b_i 
\wedge \psi^{i + 1}_+)$ for $0\leq i < n - 1 $  and $\psi^{{n-1}}_+ 
(\hsEB_\pi \neg b_{n-1} \wedge \hsEA_\pi b_{n-1})$. It can be easily checked that $\psi^0_+$ 
holds over an interval $[x,y]$  if and only if $id_y = id_x + 1$;

\item $\psi^i_- = (\hsEB_\pi  b_i \wedge  \hsEA_\pi \neg b_i 
\wedge \psi^{i + 1}_= ) \vee (\hsEB_\pi \neg  b_i \wedge  
\hsEA_\pi b_i \wedge \psi^{i + 1}_- )$ for $0\leq i < n - 1$  and 
$\psi^{n-1}_- = (\hsEB_\pi  b_{n-1} \wedge \hsEA_\pi \neg b_{n-1})$. 
We have that $\psi^0_+$ holds over an interval $[x,y]$ 
if and only if $id_y = id_x - 1$.

\end{compactitem}

\noindent By means of the above formulas, we 
use point $x$ to encode the value of the cell of index  $x\ mod\
N$, where 
$mod$ denotes the remainder of the integer division, in the configuration whose number is given by the integer division $x / N$. 

We now write a formula $\varphi$ whose models encodes all and only the successful computations of $Tm$ on $w$, thus guaranteeing that $\varphi$ is satisfiable if and only if $w \in L(Tm)$

First, for each computation we encode the correct indexing of the tape by means of the formula $\psi_{tape} = [G]( len_1  \rightarrow \psi^0_+ \vee (\hsEB cell_N \wedge \hsEA 
cell_0 ) \vee \hsAA (\pi \wedge cell_N)  )$.

We constrain each cell/point to be labelled by \emph{exactly one element} of $\Sigma$ by means of the formula $\psi_{\Sigma} = [G]( \pi \rightarrow ( \bigvee_{\sigma \in \Sigma} \sigma ) \wedge  (\bigwedge_{\sigma \in \Sigma} (\sigma \rightarrow \bigwedge_{\sigma' \in \Sigma \setminus \{\sigma\} } \neg \sigma'
)  )$.

To encode the transitions in $\Delta$, we introduce some convenient notation.
Let $\delta = ((\sigma, q), (\sigma', q'), *)$ be any transition in $\Delta$, with  $* \in \{\leftarrow, \rightarrow\}$. We
put $\Sigma_{in}(\delta) = \sigma$, $\Sigma_{out}(\delta)= 
\sigma'$, $Q_{in}(\delta) = q$, $Q_{out}(\delta) = q'$. 
Moreover, we add some symbols for some meaningful subsets of $\Delta$: (i) $\overleftarrow{\Delta} = \{((\sigma, q), (\sigma', q'), \leftarrow) \in \Delta \}$; (ii) $\overrightarrow{\Delta} = \{((\sigma, q), (\sigma', q'), \rightarrow) \in \Delta \}$; (iii) $\Delta_0 =  \{ 
(\sigma, q_0), (\sigma', q'), *) \in \Delta, * \in \{ \leftarrow, \rightarrow\} 
\}$, and (iv) $\Delta_f =  \{ 
(\sigma, q), (\sigma', q_f), *) \in \Delta, * \in \{ \leftarrow, \rightarrow\} 
\}$. 
For all $\overline{\Delta} \in\{ \Delta, \overrightarrow{\Delta}, \overleftarrow{\Delta}, 
{\Delta_0}, \Delta_f \}$, we denote by $\overline{\Delta}$ the formula $\pi \wedge \bigvee_{\delta \in \overline{\Delta}} \delta$.
Finally, we define a relation $Next_\Delta \subseteq \Delta \times \Delta$ that 
includes all and only those pairs of transitions that can be fired one after the other, that is, $Next_\Delta = \{ (\delta, \delta') \in \Delta \times \Delta: Q_{out}(\delta) = Q_{in}(\delta') \}$.

We will also make use of $|\Delta|$ additional proposition letters $\delta \in\Delta$ to identify the transition which is fired in the 
current configuration. More precisely,  we will impose the following condition: if $\delta \in \cV([x,x])$, then transition $\delta$ is fired 
in configuration $x / N$, where the head is at position $x\ mod\ N$.

Next, we constrain every cell to be labelled by \emph{at most one element} of $\Delta$
by means of the formula $\psi_{\Delta_1} = [G]( \pi \rightarrow \bigwedge_{\delta \in \Delta} 
(\delta \rightarrow \bigwedge_{\delta' \in \Delta \setminus \{\delta\}} 
\neg \delta'))$. Moreover, we constrain every 
configuration to contain exactly one point 
labelled by an element in $\Delta$, which identifies the position of the head of $Tm$ in the current configuration. Such a condition is forced by the formula 
$\psi_{\Delta_2} =
[G]( \hsEB cell_0  \wedge \hsEA cell_N  \rightarrow (\hsEB 
\hsEA \Delta) \wedge \hsAB( \hsEA \Delta \rightarrow \hsAB\hsAA 
\neg \Delta))$.\footnote{To simplify the encoding, w.l.o.g, we assume that the last cell, that is, the cell of index $N$, of each configuration is never reached in a successful computation.}

Then, we encode the correct transition among consecutive configurations by means of the formula
\begin{displaymath}
\psi_{\Delta_3} = [G]\left( \begin{array}{l}\hsEB \Delta \wedge \hsEA \Delta \wedge 
\hsAB( \neg \pi \rightarrow \hsAA \neg \Delta) \rightarrow \\
\left(\bigvee_{(\delta, \delta') \in Next_{\Delta}} \hsEB \delta \wedge \hsEA \delta') \wedge (\hsEB\overleftarrow{\Delta} \rightarrow \psi^0_-)
\wedge (\hsEB\overrightarrow{\Delta} \rightarrow \psi^0_+)
\right)\end{array}\right).\end{displaymath} 

Finally, we correctly reproduce the symbols between any two 
consecutive configurations and we guarantee that each $\delta \in \Delta$ labelling a point is consistent with its $\sigma_{in} (\delta)$ (resp., $\sigma_{out}(\delta)$) symbol in the current (resp., next) configuration. This is encoded by means of the formula 
\begin{displaymath}\psi_{\Sigma\Delta} =
[G]\left( \begin{array}{l}
\psi^0_= \wedge \hsAB \neg \psi^0_= \rightarrow \\
\left(\begin{array}{l} (\neg \hsEB \Delta \wedge \bigwedge_{\sigma \in \Sigma} 
( \hsEB \sigma \leftrightarrow \hsEA \sigma ) ) \vee\\
( \hsEB \Delta \wedge \bigwedge_{\delta \in \Delta} 
( \hsEB \delta \rightarrow \hsEB \sigma_{in}(\delta) \wedge \hsEA \sigma_{out}(\delta) ) \end{array}\right)\end{array}\right).\end{displaymath}

To complete the encoding, it suffices to specify the initial configuration and the accepting condition. Let $w_{in} = w_0\ldots w_{n-1}$. The former is captured by the formula \begin{displaymath}\psi_{init} = \begin{array}{l}
\hsEB_\pi (w_0\wedge \Delta_0) \wedge \bigwedge_{i=1}^{i = n - 1 } \hsEB( \hsEA w_i \wedge \hsAB^{i + 1} \bot  \wedge \hsEB^{i} \top) 
\wedge \\ \hsAB(\hsAB\hsAA \neg cell_N \wedge \hsEB^{n} \top 
\rightarrow \hsEA \# )\end{array},\end{displaymath} where, for any $k \in \bbN$, $\hsEB^k \psi$ 
(resp., $\hsAB^k \psi$) is recursively defined as follows: $\psi$ if $k=0$, and $\hsEB\hsEB^{k-1} \psi$ (resp.,  $\hsAB\hsAB^{k-1} \psi$), otherwise. As for the latter,
it suffices the formula $\psi_{accept} = \hsEB\hsEA \Delta_f$. It is easy to prove that $w \in L(Tm)$ if and only if 
the \LogicABhom formula \begin{displaymath}\varphi = \psi_{tape} \wedge 
\psi_{\Sigma} \wedge \psi_{\Delta_1} \wedge \psi_{\Delta_2} 
\wedge \psi_{\Delta_3} \wedge \psi_{\Sigma\Delta} \wedge 
\psi_{init} \wedge \psi_{accept}\end{displaymath} is satisfiable. Moreover, 
it is easy to prove that each conjunct can be generated in \logspace.

 \begin{restatable}{thm}{thmabhomhard}\label{thm:abhomhard}
The satisfiability problem for the logic $\textsf{AB}_{hom}$  over finite linear orders is \expspacehard.
\end{restatable}

\tempcut{


\section{Conclusions}\label{sec:conclusions}

In this paper, we proved that the addition of modality $\hsEA$ to the logic $\mathsf{BD}_{hom}$ of prefixes and sub-intervals increases the complexity of the  problem of satisfiability checking over finite linear orders, under the homogeneity assumption. Indeed, while the addition of either modality $\hsEB$ or modality $\hsEE$ to the logic $\mathsf{D}_{hom}$ of sub-intervals ~\cite{DBLP:journals/lmcs/BozzelliMMPS22} does not affect the complexity of the problem, that remains \pspacecomplete\ \cite{DBLP:journals/iandc/BozzelliMPS23}, we showed that the addition of modality $\hsEA$ to $\mathsf{BD}_{hom}$ makes the problem for the resulting logic $\mathsf{BDA}_{hom}$ \expspacecomplete.
The same complexity has been recently proved, following a rather different path, for the logic  $\mathsf{BE}_{hom}$ \cite{DBLP:conf/lics/MonicaMPS23}.

The emerging picture is as follows. Both the addition of modality $\hsEA$ to the logic $\mathsf{BD}_{hom}$ and
the replacement of modality $\hsED$ with modality $\hsEE$ make the resulting logics $\mathsf{BDA}_{hom}$ and  $\mathsf{BE}_{hom}$
\expspacecomplete. However, as for expressiveness, the two logics turn out to be not comparable \cite{DBLP:journals/tcs/BresolinMMSS14}.
As a matter of fact, $\mathsf{BDA}_{hom}$ captures a fragment of $\mathsf{BE}_{hom}$, that is, $\mathsf{BD}_{hom}$ extended with a weaker variant $\hsEE_{\pi}$ of modality $\hsEE$, defined as $\hsEE_{\pi} \psi = \hsEA(\pi \wedge \psi)$, that allows one to constrain the truth of formulas on the right endpoint of an interval. As shown in Section~\ref{sec:hardness}, such an ability is the key property that causes the increase in complexity from \pspacecomplete\ to \expspacecomplete. 

%
%
%
%



As for future work, we aim at addressing the problem of establishing the exact complexity of the  problem of satisfiability checking for full $\mathsf{HS}_{hom}$ over finite linear orders, under the homogeneity assumption. At the moment, we only know that it is non-elementarily decidable \cite{DBLP:journals/acta/MolinariMMPP16}. We would also like to determine whether the problem remains \expspacecomplete\ for the logic $\mathsf{BEA}_{hom}$, that merges $\mathsf{BDA}_{hom}$ and  $\mathsf{BE}_{hom}$. Finally, we would like to study the model checking problem for the logics $\mathsf{BDA}_{hom}$ and  $\mathsf{BE}_{hom}$ as well as $\mathsf{BDA}_{hom}$ and $\mathsf{HS}_{hom}$.


\section*{Acknowledgements}\label{sec:acknowledgements}
We would like to thank the  reviewers for their comments that helped us a lot to improve the paper. Angelo Montanari and Adriano Peron would like to acknowledge the support from the  
2023 Italian INdAM-GNCS project \lq\lq Symbolic and numerical analysis of cyber-physical systems\rq\rq, ref. no. CUP E53C22001930001. Angelo Montanari would also like to acknowledge the support from the MUR PNRR project FAIR - Future AI Research (PE00000013) funded by the European Union NextGenerationEU.

\bibliographystyle{alphaurl}
\bibliography{biblio}

\appendix

\section{About the encoding of \LogicHS\ in 
\LogicCDT\ and of $\mathsf{LTL}_f$ in
$\mathsf{AB}$}
\label{appendix:cdt}


\begin{figure}
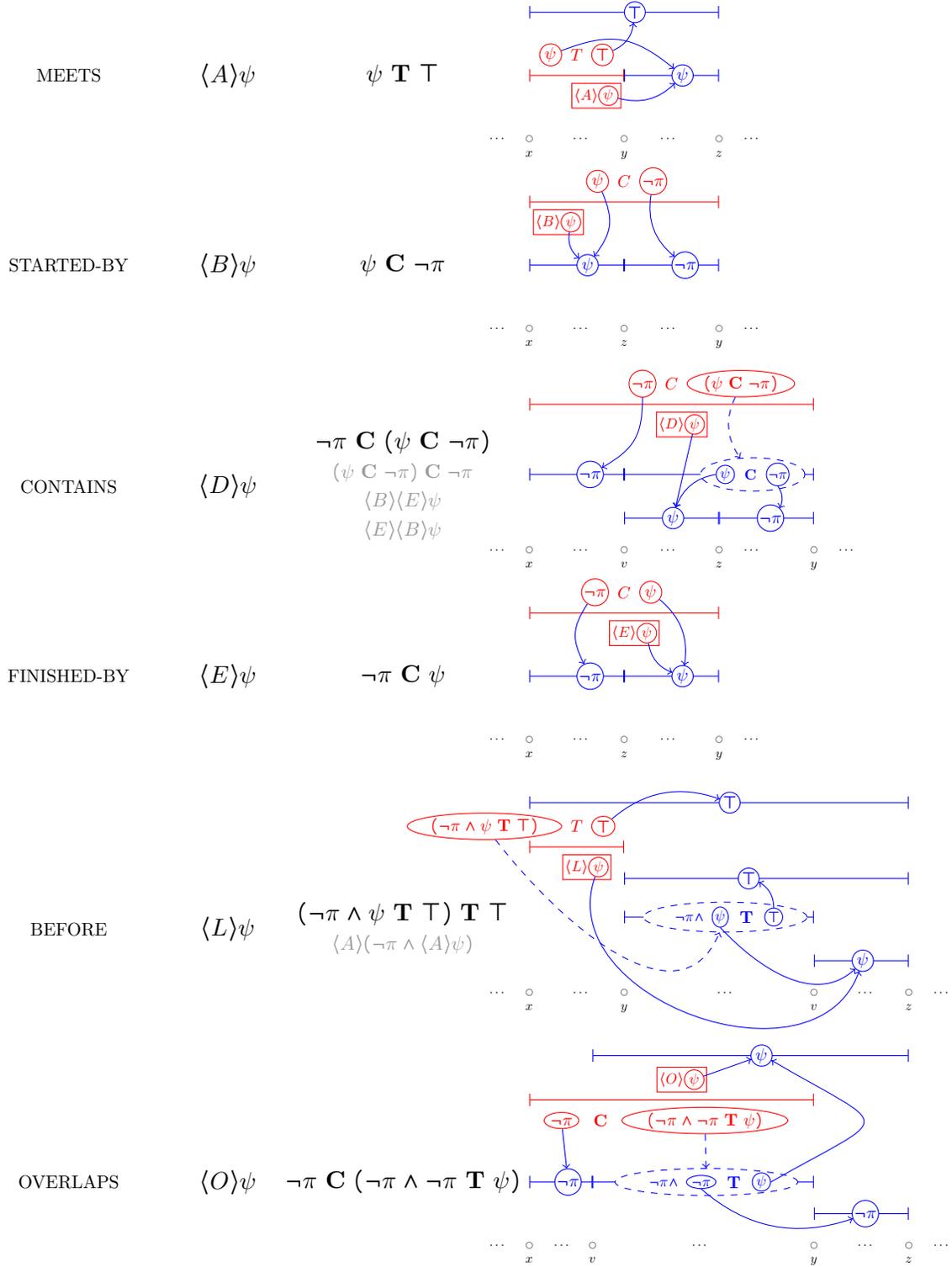


\begin{tikzpicture}[anchor=west, remember picture]

\node[minimum width=2.5cm](Allen){};
\node[minimum width=2.5cm](HS) at (Allen.east){};
\node[minimum width=3cm](CDT) at (HS.east) {};

\node(AT)[below of=Allen, node distance=1cm]{$\allenTextA$};
\node(AOP)[below of=HS] {$\hsEA \psi$};
\node(ACDT)[below of=CDT] {$\psi \cdtT \top$};

\begin{scope}[node distance=1.5cm, opacity = 1]

\node[opacity=0.5,draw, circle, inner sep=1, right of= ACDT, node distance=2cm, yshift=-1cm, 
      label={[label distance=0.2cm]180:\scalebox{0.525}{$\ldots$}},
      label={[label distance=0.5cm]0:\scalebox{0.525}{$\ldots$}},
      label={[]270:\scalebox{0.525}{$x$}}](0-0) {};
      
\node[opacity=0.5,draw, circle, inner sep=1,,
      label={[]270:\scalebox{0.525}{$y$}}](1-1) [right of=0-0] {};
      
\node[opacity=0.5,draw, circle, inner sep=1, 
      label={[label distance=0.5cm]180:\scalebox{0.525}{$\ldots$}},
      label={[label distance=0.2cm]0:\scalebox{0.525}{$\ldots$}},
      label={[]270:\scalebox{0.525}{$z$}}](2-2) [right of=1-1] {};

\draw[|-|,  red]
($(0-0.center) + (0,1)$) -- ($(1-1.center)+ (0,1)$)
	node[pos=0.5, above=0.1cm](T) {\scalebox{0.69}{$T$}}
      node[anchor=east, circle, draw, inner sep=0.5](T1) 
      	at (T.west) {\scalebox{0.69}{$\psi$}}
      node[anchor=west, circle, draw, inner sep=0.5](T2) 
      	at (T.east) {\scalebox{0.69}{$\top$}}
      node[pos=0.6, below, yshift=-0.2cm, inner sep=0pt](A) {\scalebox{0.6}{$\hsEA$}}
      node[anchor=west, circle, draw, inner sep=0.5, xshift=-0.00cm](A1) 
      	at (A.east) {\scalebox{0.6}{$\psi$}}	  	
;

\node[fit=(A)(A1), draw, red, rectangle, inner sep=1pt]  {};

\draw[|-|,  blue]
($(1-1.center) + (0,1)$) -- ($(2-2.center)+ (0,1)$)
 node[pos=0.5, fill=white, scale=0.7, circle, draw, inner sep=0.5](PSI1)[]{$\psi$};

\draw[|-|, blue] ($(0-0.center) + (0,2)$) -- ($(2-2.center) +(0,2)$)
node[pos=0.5, fill=white, scale=0.7, circle, draw, inner sep=0.5](PSI2)[]{$\top$};

\draw[->, blue] (T1) edge[bend left, blue, looseness=1.19] (PSI1);
\draw[->, blue] (A1) edge[bend right, blue] (PSI1);

\draw[->, blue] (T2) edge[bend right, blue] (PSI2);

\end{scope}

\begin{scope}[node distance = 3cm]

\node(BT)[below of=AT]{$\allenTextB$};
\node(BOP)[below of=AOP] {$\hsEB \psi$};
\node(BCDT)[below of=ACDT] {$\psi \cdtC \neg \pi$};

\end{scope}

\begin{scope}[node distance=1.5cm, opacity = 1]

\node[opacity=0.5,draw, circle, inner sep=1, right of= BCDT, node distance=2cm, yshift=-1cm, 
      label={[label distance=0.2cm]180:\scalebox{0.525}{$\ldots$}},
      label={[label distance=0.5cm]0:\scalebox{0.525}{$\ldots$}},
      label={[]270:\scalebox{0.525}{$x$}}](0-0) {};
      
\node[opacity=0.5,draw, circle, inner sep=1,,
      label={[]270:\scalebox{0.525}{$z$}}](1-1) [right of=0-0] {};
      
\node[opacity=0.5,draw, circle, inner sep=1, 
      label={[label distance=0.5cm]180:\scalebox{0.525}{$\ldots$}},
      label={[label distance=0.2cm]0:\scalebox{0.525}{$\ldots$}},
      label={[]270:\scalebox{0.525}{$y$}}](2-2) [right of=1-1] {};

\draw[|-|,  red]
($(0-0.center) + (0,2)$) -- ($(2-2.center)+ (0,2)$)
      node[pos=0.5, above=0.1cm](T) {\scalebox{0.69}{$C$}}
      node[anchor=east, circle, draw, inner sep=0.5](T1) 
            at (T.west) {\scalebox{0.69}{$\psi$}}
      node[anchor=west, circle, draw, inner sep=0.5](T2) 
            at (T.east) {\scalebox{0.69}{$\neg \pi$}}
         node[pos=0.1, below, yshift=-0.2cm, inner sep=0pt](A) {\scalebox{0.6}{$\hsEB$}}
      node[anchor=west, circle, draw, inner sep=0.5, xshift=-0.00cm](A1) 
            at (A.east) {\scalebox{0.6}{$\psi$}}            
;

\node[fit=(A)(A1), draw, red, rectangle, inner sep=1pt]  {};

\draw[|-|,  blue]
($(0-0.center) + (0,1)$) -- ($(1-1.center)+ (0,1)$)
 node[pos=0.5, fill=white, scale=0.7, circle, draw, inner sep=0.5](PSI1)[]{$\psi$};

\draw[|-|, blue] ($(1-1.center) + (0,1)$) -- ($(2-2.center) +(0,1)$)
node[pos=0.5, fill=white, scale=0.7, circle, draw, inner sep=0.5](PSI2)[]{$\neg \pi$};

\draw[->, blue] (T1) edge[bend left, blue, looseness=1.19] (PSI1);
\draw[->, blue] (A1) edge[bend right, blue] (PSI1);

\draw[->, blue] (T2) edge[bend right, blue] (PSI2);

\end{scope}

\begin{scope}[node distance = 3.5cm]

\node(DT)[below of=BT]{$\allenTextD$};
\node(DOP)[below of=BOP] {$\hsED \psi$};
\node(DCDT)[below of=BCDT] {$\begin{array}{c}
      \neg \pi \cdtC (\psi \cdtC \neg \pi) \\ 
      \scalebox{0.8}{$\transparent{0.4} (\psi \cdtC \neg \pi) \cdtC \neg \pi$} \\
      \scalebox{0.8}{$\transparent{0.4} \hsEB \hsEE \psi$} \\
      \scalebox{0.8}{$\transparent{0.4} \hsEE \hsEB \psi$} \\
\end{array}$};

\end{scope}

\begin{scope}[node distance=1.5cm, opacity = 1]

\node[opacity=0.5,draw, circle, inner sep=1, right of= DCDT, node distance=2cm, yshift=-1cm, 
      label={[label distance=0.2cm]180:\scalebox{0.525}{$\ldots$}},
      label={[label distance=0.5cm]0:\scalebox{0.525}{$\ldots$}},
      label={[]270:\scalebox{0.525}{$x$}}](0-0) {};
      
\node[opacity=0.5,draw, circle, inner sep=1,,
      label={[]270:\scalebox{0.525}{$v$}}](1-1) [right of=0-0] {};
      
\node[opacity=0.5,draw, circle, inner sep=1, 
      label={[label distance=0.5cm]180:\scalebox{0.525}{$\ldots$}},
      label={[label distance=0.2cm]0:\scalebox{0.525}{$\ldots$}},
      label={[]270:\scalebox{0.525}{$z$}}](2-2) [right of=1-1] {};

\node[opacity=0.5,draw, circle, inner sep=1, 
      label={[label distance=0.2cm]0:\scalebox{0.525}{$\ldots$}},
      label={[]270:\scalebox{0.525}{$y$}}](3-3) [right of=2-2] {};

\draw[|-|,  red]
($(0-0.center) + (0,2.3)$) -- ($(3-3.center)+ (0,2.3)$)
      node[pos=0.5, above=0.1cm](T) {\scalebox{0.69}{$C$}}
      node[anchor=east, circle, draw, inner sep=0.5](T1) 
            at (T.west) {\scalebox{0.69}{$\neg \pi$}}
      node[anchor=west, ellipse, draw, inner sep=0.5](T2) 
            at (T.east) {\scalebox{0.69}{$(\psi \cdtC \neg \pi)$}}
         node[pos=0.5, below, yshift=-0.2cm, inner sep=0pt](A) {\scalebox{0.6}{$\hsED$}}
      node[anchor=west, circle, draw, inner sep=0.5, xshift=-0.00cm](A1) 
            at (A.east) {\scalebox{0.6}{$\psi$}}            
;

\node[fit=(A)(A1), draw, red, rectangle, inner sep=1pt]  {};

\draw[|-|,  blue]
($(0-0.center) + (0,1.19)$) -- ($(1-1.center)+ (0,1.19)$)
 node[pos=0.5, fill=white, scale=0.7, circle, draw, inner sep=0.5](PSI1)[]{$\neg \pi$};


\draw[|-|, blue] ($(1-1.center) + (0,1.19)$) -- ($(3-3.center) +(0,1.19)$)
node[pos=0.4, fill=white, inner sep=0, ellipse, draw, dashed] (PSI2)
{ \begin{tikzpicture}[remember picture]
      \node(PSI22)[scale=0.6]{$ \cdtC $};
\node[anchor=east,fill=white, text=blue, circle, draw, solid, inner sep=0.5, scale=0.6](PSI2L) at (PSI22.west)  {$\psi$};
\node[anchor=west,fill=white, text=blue, circle, draw, inner sep=0.5, solid, scale=0.6](PSI2R) at (PSI22.east)  {$\neg \pi$};
\end{tikzpicture}
};

\draw[|-|, blue] ($(1-1.center) + (0,0.5)$) -- ($(2-2.center) +(0,0.5)$)
node[pos=0.4, fill=white, scale=0.7, circle, draw, inner sep=0.5](PSI)[]{$\psi$}; 
\draw[|-|, blue] ($(2-2.center) + (0,0.5)$) -- ($(3-3.center) +(0,0.5)$)
node[pos=0.4, fill=white, scale=0.7, circle, draw, inner sep=0.5](NPI)[]{$\neg \pi$};     

\draw[->, blue] (T1) edge[bend left, blue, looseness=1.19] (PSI1);

\draw[->, blue, dashed] (T2) edge[bend right, blue] (PSI2);

\draw (PSI2L.west) edge[bend right, blue,->] (PSI);

\draw (PSI2R.south) edge[bend left, blue,->] (NPI);

\draw (A1) edge[->, blue] (PSI);

\end{scope}
\input{sections/appendixCDT/figHSOperatorsE}
\input{sections/appendixCDT/figHSOperatorsL}
\input{sections/appendixCDT/figHSOperatorsO}

\end{tikzpicture}

\caption{\label{fig:hsoperators}
A graphical account of the encoding of $\mathsf{HS}$ modalities in $\mathsf{CDT}$. 
}	
\end{figure}


In this appendix, we focus on  the encoding of \LogicHS\ in \LogicCDT\ and of $\mathsf{LTL}_f$ in $\mathsf{AB}$. As for \LogicHS, in Figure~\ref{fig:hsoperators}, we show how to encode \LogicHS\ modalities in $\mathsf{CDT}$. 
The encoding of $\mathsf{LTL}_f$ modalities  in $\mathsf{AB}$ has been already illustrated in Section \ref{sec:itlintro}. Here, we show how to exploit it in order to deal with the example model in Figure \ref{fig:itlmodels}. 
Let us first consider the $\mathsf{LTL}_f$ formula $p \ltlUntil (\neg p \wedge \neg q)$ (Figure \ref{fig:abuntilAsExample} - top), which is true at time point $0$ (according to the point-based semantics $\pi$),
which is mapped into the $\mathsf{AB}$ formula $\psi = (\hsAB\hsEA (\pi \wedge p) \wedge \hsEA(\pi \wedge \neg p \wedge \neg q))$. We show that the latter holds over the interval $[0, 3]$ (according to the interval-based semantics $\cV$).
Let us evaluate the formula $\psi$ over the interval $[0, 3]$. The second conjunct $\hsEA(\pi \wedge \neg p \wedge \neg q)$ forces the existence of an interval $[3,y]$, with $y \geq 3$, where $\pi$, $\neg p$, and $\neg q$ hold. The truth of $\pi$ on $[3,y]$ restricts the number of  candidate intervals to $[3,3]$ only. Since $\cV([3,3])= \emptyset$, it immediately follows that both $\neg p$ and $\neg q$ hold over $[3,3]$ as well. 
The outermost modality of the first conjunct $\hsAB\hsEA (\pi \wedge p)$ forces the formula $\hsEA (\pi \wedge p)$ to be true over each proper prefix of $[0,3]$, namely, $[0,2], [0,1]$, and $[0,0]$. 
This amounts to say that, for each interval $[0, x']$, with $x' \in \{0,1,2\}$, $\hsEA (\pi \wedge p)$ holds 
on $[0,x']$ if and only if there is an interval $[x',y]$, with $y \geq x'$, which makes both
$\pi$ and  $p$ true. Since $\pi$ is true over $[x',y]$ if and only if $y = x'$, it immediately follows that $p$ belongs to $\cV([x',x'])$ for all point-intervals $[x',x']$, with $x' \in \{0,1,2\}$, that is, $p$ belongs to $\cV([0,0]), \cV([1,1])$, and $\cV([2,2])$.   



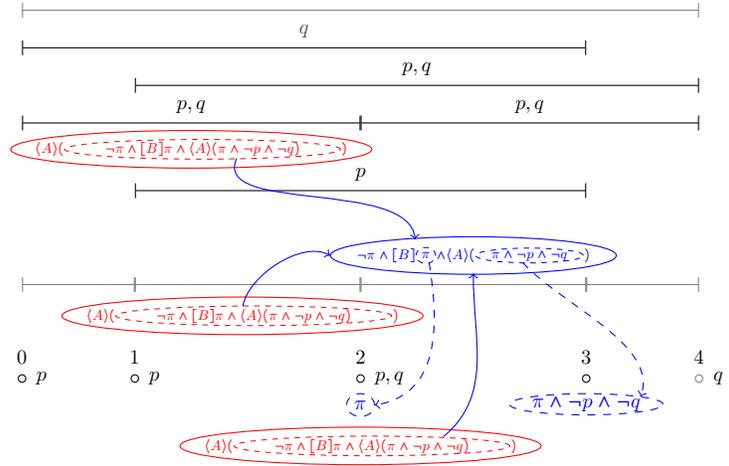
\begin{figure}
\begin{tikzpicture}[scale=1, node distance=1cm, remember picture]
\node(UT){$p \ltlUntil (\neg p \wedge \neg q) =  \hsAB\hsEA (\pi \wedge p) \wedge \hsEA(\pi \wedge \neg p \wedge \neg q)$};

\pgftransformshift{\pgfpoint{2.7cm}{-5.5cm}}

\scalebox{1.1}{

\tikzset{scaled/.style={scale=0.7, inner sep=0}}

\begin{scope}[node distance = 1.5cm]

\node[draw, circle, inner sep=1, label={[]90:$\scalebox{0.7}{$ 0 $}$},
      label={[]0:$\scalebox{0.7}{$ p $}$},  
      label={[blue]270:\scalebox{0.6}{$ $}}](0-0) {};
\node[scaled,blue, draw, ellipse, yshift=-0.5cm, pos=0.5, opacity=1](0-0-APSI1) at (0-0.south)  
 {\begin{tikzpicture}[inner sep=0, anchor=west]
      \node[scaled](A) {$\hsEA($};
      \node[scaled, draw, ellipse, dashed](PSI10) at (A.east) {$\pi \wedge p$};
      \node[scaled] at (PSI10.east) {$)$};
\end{tikzpicture}}; 

\node[scaled, draw, ellipse, dashed, blue,yshift=-0.2cm, anchor=north](0-0-PSI1) at (0-0-APSI1.south){$\pi \wedge p$};     
      
\node[draw, circle, inner sep=1, label={[]90:$\scalebox{0.7}{$ 1 $}$},
      label={[]0:$\scalebox{0.7}{$ p $}$}](1-1) [right of=0-0] {};

\node[scaled, draw, ellipse, dashed, blue,yshift=-0.2cm, anchor=north](1-1-PSI1) at (1-1.south){$\pi \wedge p$};

\node[draw, circle, inner sep=1, label={[]90:$\scalebox{0.7}{$ 2 $}$},
      label={[]0:$\scalebox{0.7}{$ p,q $}$}](2-2) [right of=1-1] {};

\node[scaled, draw, ellipse, dashed, blue,yshift=-0.2cm, anchor=north](2-2-PSI1) at (2-2.south){$\pi \wedge p$};     

\node[draw, circle, inner sep=1, label={[]90:$\scalebox{0.7}{$ 3 $}$},
      label={[]0:$\scalebox{0.7}{$  $}$}](3-3) [right of=2-2] {};

\node[scaled, draw, ellipse, dashed, red,yshift=-0.2cm, anchor=north](3-3-PSI2) at (3-3.south){$\pi \wedge \neg p \wedge \neg q $};     
      
\node[opacity=0.5,draw, circle, inner sep=1, label={[]90:$\scalebox{0.7}{$ 4 $}$},
      label={[]0:$\scalebox{0.7}{$ q $}$}](4-4) [right of=3-3] {};
      
\def\yshift{0.5}

\def\yuntil{0.5}

\draw[|-|, opacity=0.5] ($(0-0.center) + (0,0.75 + \yshift)+ (0, \yuntil)$) -- ($(1-1.center)+ (0,0.75 + \yshift)+ (0, \yuntil)$)
node[pos=0.5, above](47) {$\scalebox{0.7}{$  $}$}
node[scaled,blue, draw, ellipse, below=0.1cm, pos=0.5, opacity=1] (A01) {
\begin{tikzpicture}[inner sep=0, anchor=west]
\node[scaled](A) {$\hsEA($};
\node[scaled, draw, ellipse, dashed](PSI11) at (A.east) {$\pi \wedge p$};
\node[scaled] at (PSI11.east) {$)$};
\end{tikzpicture}
};

\def\yuntil{1.1}

\draw[|-|, opacity=1] ($(0-0.center) + (0,1.25+ \yshift)+ (0, \yuntil)$) -- ($(2-2.center)+ (0,1.25+ \yshift)+ (0, \yuntil)$)
node[pos=0.5, above](47) {$\scalebox{0.7}{$ p,q $}$}
node[scaled,blue, draw, ellipse, below=0.1cm, pos=0.5] (A02) {
\begin{tikzpicture}[inner sep=0, anchor=west, remember picture]
\node[scaled](A) {$\hsEA($};
\node[scaled, draw, ellipse, dashed](PSI12) at (A.east) {$\pi \wedge p$};
\node[scaled] at (PSI12.east) {$)$};
\end{tikzpicture}
};

\def\yuntil{1}

\draw[|-|, opacity=1] ($(0-0.center) + (0,2.75+ \yshift) + (0, \yuntil)$) -- ($(3-3.center)+ (0,2.75+ \yshift) + (0, \yuntil)$)
node[pos=0.5, above, opacity=0.7](47) {$\scalebox{0.7}{$ q $}$}
node[pos=0.5, below](47) {
      \begin{tikzpicture}[inner sep=0, anchor=west]
            \node[scaled](B) {$\hsAB$};
            \node[scaled,blue, draw, ellipse](A) at (B.east) {$\hsEA(\pi \wedge p)$};
            \node[scaled](W) at (A.east) {$\wedge$};
            \node[scaled](AA) at (W.east) {$\hsEA($}; 
            \node[scaled,red, draw, ellipse](PSI2) at (AA.east) {$\pi \wedge \neg p \wedge \neg q$};
            \node[scaled](C) at (PSI2.east) {$)$};    
      \end{tikzpicture}
};

\draw[|-|, opacity=0.5] ($(0-0.center) + (0,3.4+ \yshift)+ (0, \yuntil)$) -- ($(4-4.center)+ (0,3.4+ \yshift)+ (0, \yuntil)$)
node[pos=0.5, above](47) {$\scalebox{0.7}{$  $}$};

\def\yuntil{2.9}

\draw[|-|, opacity=0.5] ($(1-1.center) + (0,0.75+ \yshift)$) -- ($(2-2.center)+ (0,0.75+ \yshift)$)
node[pos=0.5, above](47) {$\scalebox{0.7}{$  $}$};

\draw[|-|, opacity=1] ($(1-1.center) + (0,1.75+ \yshift)$) -- ($(3-3.center)+ (0,1.75+ \yshift)$)
node[pos=0.5, above](47) {$\scalebox{0.7}{$ p $}$};

\draw[|-|, opacity=1] ($(1-1.center) + (0,\yuntil+ \yshift)$) -- ($(4-4.center)+ (0,\yuntil+ \yshift)$)
node[pos=0.5, above](47) {$\scalebox{0.7}{$ p,q $}$};

\draw[|-|, opacity=0.5] ($(2-2.center) + (0,0.75+ \yshift)$) -- ($(3-3.center)+ (0,0.75+ \yshift)$)
node[pos=0.5, above](47) {$\scalebox{0.7}{$  $}$};

\draw[|-|, opacity=1] ($(2-2.center) + (0,1.25+ \yshift)$) -- ($(4-4.center)+ (0,1.25+ \yshift)$)
node[pos=0.5, above](47) {$\scalebox{0.7}{$ p,q $}$};

\draw[|-|, opacity=0.5] ($(3-3.center) + (0,0.75+ \yshift)$) -- ($(4-4.center)+ (0,0.75+ \yshift)$)
node[pos=0.5, above](47) {$\scalebox{0.7}{$  $}$};

\draw (A.south west) edge[->, bend right, blue, in=-90, looseness=3, out=-40] (A02.west);
\draw (A.south west) edge[->, bend right, blue, in=-90, looseness=1, out=-40] (A01.west);
\draw (A.south west) edge[->, bend right, blue, in=-120, looseness=1, out=-40] (0-0-APSI1);

\draw ($(PSI12) + (-0.1,-0.1) $) edge[->,  blue,  dashed] (2-2-PSI1);
\draw ($(PSI11) + (-0.2,-0.15) $) edge[->,  blue,  dashed] (1-1-PSI1.north west);
\draw ($(PSI10) + (-0.2,-0.15) $) edge[->,  blue,  dashed, bend left, in=90, looseness=2] (0-0-PSI1.east);
\draw ($(PSI2) + (-0.2,-0.15) $) edge[->,  blue,  dashed, bend left, in=90, looseness=1,red] (3-3-PSI2);

\end{scope}}

\pgftransformshift{\pgfpoint{-2.7cm}{-3cm}}

\node(NT){$\ltlNext (\neg p \wedge \neg q) = \hsEA ( \neg \pi \wedge \hsAB \pi \wedge \hsEA (\pi \wedge \neg p \wedge \neg q))$};

\pgftransformshift{\pgfpoint{-0.3cm}{-5cm}}

\scalebox{1.1}{

\tikzset{scaled/.style={scale=0.7, inner sep=0}}

\begin{scope}[node distance = 1.5cm]

\node[] at (0,4.5) {};

\node[draw, circle, inner sep=1, label={[]90:$\scalebox{0.7}{$ 0 $}$},
      label={[]0:$\scalebox{0.7}{$ p $}$},  
      label={[blue]270:\scalebox{0.6}{$ $}}](0-0) {};

\node[draw, circle, inner sep=1, label={[]90:$\scalebox{0.7}{$ 1 $}$},
      label={[]0:$\scalebox{0.7}{$ p $}$}](1-1) [right of=0-0] {};


\node[draw, circle, inner sep=1, label={[]90:$\scalebox{0.7}{$ 2 $}$},
      label={[]0:$\scalebox{0.7}{$ p,q $}$}](2-2) [right of=1-1, node distance=3cm] {};

\node[scaled, draw, ellipse, dashed, blue,yshift=-0.2cm, anchor=north, inner sep=2pt](2-2-PI) at (2-2.south){$\pi$};     

\node[scaled,  red,yshift=-01cm, anchor=north](2-2-PSI1) at (2-2.south){ 
      \begin{tikzpicture}[inner sep=0, anchor=west]
            \node[scaled](A) {$\hsEA ( $};
            \node[scaled, draw, ellipse, dashed](PSI22) at (A.east) {$\neg \pi \wedge \hsAB \pi \wedge \hsEA (\pi \wedge \neg p \wedge \neg q)$};
            \node[scaled] at (PSI22.east) {$)$};
      \end{tikzpicture}
};     
\draw[red] (2-2-PSI1) ellipse (2.4cm and 0.25cm);

\node[draw, circle, inner sep=1, label={[]90:$\scalebox{0.7}{$ 3 $}$},
      label={[]0:$\scalebox{0.7}{$  $}$}](3-3) [right of=2-2, node distance=3cm] {};

\node[scaled, draw, ellipse, dashed, blue,yshift=-0.2cm, anchor=north](3-3-PSI2) at (3-3.south){$\pi \wedge \neg p \wedge \neg q $};     
      
\node[opacity=0.5,draw, circle, inner sep=1, label={[]90:$\scalebox{0.7}{$ 4 $}$},
      label={[]0:$\scalebox{0.7}{$ q $}$}](4-4) [right of=3-3] {};
      
\def\yshift{0.5}

\def\yuntil{0.5}

\draw[|-|, opacity=0.5] ($(0-0.center) + (0,0.25 + \yshift)+ (0, \yuntil)$) -- ($(1-1.center)+ (0,0.25 + \yshift)+ (0, \yuntil)$)
node[pos=0.5, above](47) {$\scalebox{0.7}{$  $}$}
;

\def\yuntil{1.1}

\draw[|-|, opacity=1] ($(0-0.center) + (0,1.8+ \yshift)+ (0, \yuntil)$) -- ($(2-2.center)+ (0,1.8+ \yshift)+ (0, \yuntil)$)
node[pos=0.5, above](0-2) {$\scalebox{0.7}{$ p,q $}$};

\node[scaled,  red,yshift=-0.3cm, anchor=north](0-2-PSI1) at (0-2.south){ 
      \begin{tikzpicture}[inner sep=0, anchor=west]
            \node[scaled](A) {$\hsEA ( $};
            \node[scaled, draw, ellipse, dashed](PSI02) at (A.east) {$\neg \pi \wedge \hsAB \pi \wedge \hsEA (\pi \wedge \neg p \wedge \neg q)$};
            \node[scaled] at (PSI02.east) {$)$};
      \end{tikzpicture}
};     
\draw[red] (0-2-PSI1) ellipse (2.4cm and 0.25cm);

\def\yuntil{1}

\draw[|-|, opacity=1] ($(0-0.center) + (0,2.9+ \yshift) + (0, \yuntil)$) -- ($(3-3.center)+ (0,2.9+ \yshift) + (0, \yuntil)$)
node[pos=0.5, above, opacity=0.7](47) {$\scalebox{0.7}{$ q $}$};

\draw[|-|, opacity=0.5] ($(0-0.center) + (0,3.4+ \yshift)+ (0, \yuntil)$) -- ($(4-4.center)+ (0,3.4+ \yshift)+ (0, \yuntil)$)
node[pos=0.5, above](47) {$\scalebox{0.7}{$  $}$};

\def\yuntil{2.9}

\draw[|-|, opacity=0.5] ($(1-1.center) + (0,0.75+ \yshift)$) -- ($(2-2.center)+ (0,0.75+ \yshift)$)
node[pos=0.5](1-2) {};

\node[scaled,  red,yshift=-0.2cm, xshift=-0.1cm, anchor=north](1-2-PSI1) at (1-2.south){ 
      \begin{tikzpicture}[inner sep=0, anchor=west]
            \node[scaled](A) {$\hsEA ( $};
            \node[scaled, draw, ellipse, dashed](PSI12) at (A.east) {$\neg \pi \wedge \hsAB \pi \wedge \hsEA (\pi \wedge \neg p \wedge \neg q)$};
            \node[scaled] at (PSI12.east) {$)$};
      \end{tikzpicture}
};     
\draw[red] (1-2-PSI1) ellipse (2.4cm and 0.25cm);

\draw[|-|, opacity=1] ($(1-1.center) + (0,2+ \yshift)$) -- ($(3-3.center)+ (0,2+ \yshift)$)
node[pos=0.5, above](47) {$\scalebox{0.7}{$ p $}$};

\draw[|-|, opacity=1] ($(1-1.center) + (0,0.5+\yuntil+ \yshift)$) -- ($(4-4.center)+ (0,0.5+\yuntil+ \yshift)$)
node[pos=0.5, above](47) {$\scalebox{0.7}{$ p,q $}$};


\draw[|-|, opacity=0.5] ($(2-2.center) + (0,0.75+ \yshift)$) -- ($(3-3.center)+ (0,0.75+ \yshift)$)
node[pos=0.5, above](2-3) {$\scalebox{0.7}{$  $}$};

\node[scaled,  blue,yshift=-0cm, anchor=south](2-3-PSI1) at (2-3.north){ 
      \begin{tikzpicture}[inner sep=0, anchor=west]
            \node[scaled](A) {$\neg \pi \wedge \hsAB $};
            \node[scaled, draw, ellipse, dashed, inner sep=2pt](PI) at (A.east) {$ \pi $};
            \node[scaled](AA) at (PI.east) {$ \wedge \hsEA ( $};
            \node[scaled, draw, ellipse, dashed](PSI23) at (AA.east) {$ \pi \wedge \neg p \wedge \neg q$};
            \node[scaled] at (PSI23.east) {$)$};
      \end{tikzpicture}
};     
\draw[blue] (2-3-PSI1) ellipse   (1.9cm and 0.25cm);
\node[inner sep=0] (2-3-PSI1-A22) at ($(2-3-PSI1) + (-90:1.5cm and 0.25cm)$) {};
\node[inner sep=0] (2-3-PSI1-A12) at ($(2-3-PSI1) + (180:1.5cm and 0.25cm)$) {};
\node[inner sep=0] (2-3-PSI1-A02) at ($(2-3-PSI1) + (120:1.5cm and 0.25cm)$) {};

\draw[|-|, opacity=1] ($(2-2.center) + (0,2.9+ \yshift)$) -- ($(4-4.center)+ (0,2.9 + \yshift)$)
node[pos=0.5, above](47) {$\scalebox{0.7}{$ p,q $}$};

\draw[|-|, opacity=0.5] ($(3-3.center) + (0,0.75+ \yshift)$) -- ($(4-4.center)+ (0,0.75+ \yshift)$)
node[pos=0.5, above](47) {$\scalebox{0.7}{$  $}$};

\draw ($(PSI22.north) + (0, -3.5pt)$) edge[->, bend right, blue, in=190, looseness=1, out=-40] ($(2-3-PSI1-A22.south)+ (0, 1pt)$);

\draw ($(PSI12.north) + (-1.1cm, -3.5pt)$) edge[->, bend left, blue, in=120, looseness=0.8, out=50] ($(2-3-PSI1-A12.west)+ (-0.38cm, 0pt)$);

\draw ($(PSI02.south) + (-0.5cm, 0.4pt)$) edge[->, bend left, blue, in=120, looseness=0.8, out=-90] ($(2-3-PSI1-A02.west)+ (0, 0pt)$);

\draw ($(PI.south) + (-0.35cm, 0.45pt)$) edge[->, bend left, blue, in=120, looseness=1.1, out=30, dashed] ($(2-2-PI.east)+ (0, 0pt)$);
\draw ($(PSI23.south west) + (-0.35cm, -1pt)$) edge[->, bend left, blue, in=120, looseness=0.8, out=0, dashed] ($(3-3-PSI2.north east)+ (0, 0pt)$);



\end{scope}}

\end{tikzpicture}

\vspace*{1.5cm}

\caption{\label{fig:abuntilAsExample} The proposed translation at work on the model of Figure~\ref{fig:itlmodels}.}


\end{figure}

Let us consider now the $\mathsf{LTL}_f$ formula $\ltlNext (\neg p \wedge \neg q)$ (Figure \ref{fig:abuntilAsExample} - bottom), which is true at time point $2$, which is mapped into the $\mathsf{AB}_f$ formula $\psi = \hsEA(\neg \pi \wedge \hsAB \pi \wedge \hsEA(\pi \wedge \neg p \wedge \neg q) )$. We prove that the latter holds over the interval $[0, 2]$. 
The outermost modality $\hsEA$ constrains
the three conjuncts $\neg \pi$, $\hsAB \pi$, and 
$\hsEA(\pi \wedge \neg p \wedge \neg q)$ to simultaneously hold over an interval $[2, y]$.
The truth of $\neg \pi$ imposes $y > 2$, and that of $\hsAB \pi$ allows us to conclude that $y=3$.
From the truth of $\hsEA(\pi \wedge \neg p \wedge \neg q)$ over $[2,3]$, it follows that there is $3\leq y$ such that the conjuncts $\pi$, $\neg p$, and $\neg q$ simultaneously hold over $[3,y]$. Once more, $\pi$ is true on $[3,y]$ if and only if $y=3$ $[3,3]$, and thus both $\neg p$ and $\neg q$ hold over $[3,3]$.
\section{Proofs of Auxiliary Results}\label{appendix:proofs}

In this section, we provide the proofs of some technical lemmas and propositions which are instrumental to the demonstration of the main results 
in the paper.

\lbbc*

\begin{proof}
Let us consider the sequence of pairs\  
$(\reqB(F_h),\ \obsB(F_h)\ \setminus\  \reqB(F_h))\ \ldots\
(\reqB(\allowbreak F_0),\allowbreak \obsB( F_0\allowbreak) \setminus \reqB(F_0))$
induced by $F_h \thenB \ldots \thenB  F_1 \thenB F_0 = F$. 
By Definition~\ref{def:intervalrelationsonatoms}, it holds 
that $\reqB(F_i) = \reqB(F_{i-1}) \cup \obsB(F_{i-1})$, 
for every $0 < i \leq h$. Moreover, by recursively unravelling 
the right part of the equation $\reqB(F_i) = \reqB(F_{i-1}) 
\cup \obsB(F_{i-1})$ by replacing $\reqB(F_{i-j})$ by 
$\reqB(F_{i-j-1}) \cup \obsB(F_{i-j-1})$, for $1\leq j < i$, we
obtain an alternative formulation of $\reqB(F_i)$ as $\reqB(F_0)
\cup \bigcup_{0\leq j < i} \obsB(F_{j})$. 

Now, for each $\psi \in \reqB(F_h)$, let us define the 
request index $reqIdx: \reqB(F_h) \rightarrow 
\{0, \ldots, h\}$ as the function: $$reqIdx(\psi) = 
\begin{cases}
0  & \mbox{if } \psi \in  \reqB(F_{0}) \mbox{;} \\
i  & \mbox{if there exists $i>0$ s.t. } 
      \psi \in  \reqB(F_{i}) 
      \setminus \reqB(F_{i -1}).
\end{cases}$$

The fact that $reqIdx$ is well defined immediately follows from  $\reqB(F_i) \supseteq \reqB(F_{i - 1})$, for all $0 < i \leq h$.

Similarly, for each $\psi \in \reqB(F_h) \cup \obsB(F_h)$, let us define the observable index $obsIdx: \reqB(F_h) \cup \obsB(F_h) \rightarrow \{0, \ldots, h\}$ as the function:
$$obsIdx(\psi) = 
\begin{cases}
0  & \mbox{if }\psi \in  \obsB(F_0) \cup \reqB(F_0) \\
i  & \mbox{if there exists $i>0$ s.t. }\psi \in  \obsB(F_{i}) \setminus \reqB(F_{i}) \mbox{.} 
\end{cases}$$

The fact that $reqObs$ is well defined follows from 
$\reqB(F_i) \supseteq \reqB(F_{i - 1})$, for all 
$0 < i \leq h$, and $\reqB(F_i) = \reqB(F_0) 
\cup \bigcup_{0\leq j < i} \obsB(F_{j})$,
for all $0 \leq i \leq h$.

For both functions $reqIdx$ and $obsIdx$ we define their 
images the standard way as follows: 
(i) $Img(reqIdx)=\{ i: \exists \psi \in \reqB(F_h) \mbox{ s.t. } 
reqIdx(\psi) = i \}$; (ii) $Img(obsIdx)=\{ i: \exists \psi \in \reqB(F_h) \cup \obsB(F_h) \mbox{ s.t. } 
obsIdx(\psi) = i \}$.

We now prove that there does not exist an index $i > 0$ such that $i \notin Img(reqIdx) \cup Img(obsIdx)$. By contradiction, let us assume that such an index 
exists (let us assume $i >0$; the case $i=0$ is symmetric). It follows that:
 
\begin{compactitem}

\item  from $i \notin Img(reqIdx)$, it follows that, for each $\psi \in \reqB(F_h)$,  either $reqIdx(\psi) > i $, and thus $\psi \notin \reqB(F_i) \cup \reqB(F_{i-1})$,
or $i > reqIdx(\psi)$, and thus $\psi \in \reqB(F_i) \cap \reqB(F_{i-1})$, and then $\reqB(F_i) = \reqB(F_{i-1})$;

\item  from $i \notin Img(obsIdx)$, it follows that, for each $\psi \in \reqB(F_h) \cup \obsB(F_h)$, either
$obsIdx(\psi) > i $, and thus $\psi \notin \obsB(F_i) \cup \obsB(F_{i-1}) \cup \reqB(F_{i}) \cup \reqB(F_{i-1})$, or $i > obsIdx(\psi)$, and then $i - 1 > obsIdx(\psi)$, because if $\psi \in \obsB(F_{i-1}) \setminus \reqB(F_{i-1})$, then $reqIdx(\psi) = i$ (contradiction). Hence,  $\obsB(F_{i}) \setminus \reqB(F_{i}) = \obsB(F_{i-1}) \setminus \reqB(F_{i-1}) = \emptyset$.

\end{compactitem}

From the above two cases, we can conclude that $(\reqB(F_{i}),\obsB(F_{i}) \setminus \reqB(F_{i}))= (\reqB(\allowbreak F_{i-1}),\obsB(F_{i-1}) 
\setminus \reqB(\allowbreak F_{i-1} \allowbreak ))$, and thus we obtain a contradiction.
Finally, we have that $h \leq |Img(reqIdx)| + |Img(obsIdx)|$, with $|Img(reqIdx)| \leq |\{\psi: \hsEB \psi \in \closure \}| - |\reqB(F_0)|$ and $|Img(obsIdx)| \leq |\{\psi: \hsEB \psi \in \closure \}| - |\reqB(F_0)| - |\obsB(F_0)\allowbreak \setminus \reqB(F_0)|$, and thus $h \leq 2|\{ \psi: \hsEB \psi \in \closure \}| - 2|\reqB(F_0)| - |\obsB(F_0)  \setminus \reqB(F_0)|$. 
\end{proof}

\lbstep*

\begin{proof}
Let us assume by contradiction that $\cL(x,y))$ 
is $B$-reflexive. This means that $\boxB(\allowbreak\cL(x,y))
\subseteq \cL(x,y)$. 
Since $\reqB(\cL(x,y))\subset \reqB(\allowbreak \cL(x,y +1 ))$, there exists a formula $\psi \in  \reqB(\cL(x,y +1 )) \setminus \reqB(\cL(x,y))$ and thus we have $\neg \psi \in \boxB(\cL(x,y))$ and, by $B$-reflexivity of  $\cL(x,y)$, $\neg \psi \in \cL(x,y)$. Since $\cG$ is a compass structure, it holds that $\cL(x,y)\thenB \cL(x,y - 1) \thenB \ldots \thenB \cL(x,x)$, and thus $\neg \psi \in \boxB(\cL(x,y'))$ and $\neg \psi \in \cL(x,y')$, for all $x\leq y'\leq y$. Since, by definition of $\thenB$, all $B$-requests are fulfilled in a compass structure, we can conclude that $\psi \notin\reqB(\cL(x,y +1 ))$ (contradiction). 
\end{proof}

\lbdet*

\begin{proof}
The left-to-right direction is proved via a case analysis. 
If $\cP(x,y) \neq  \cP(x, y + 1)$ or $\reqD(x,y) \neq  \reqD(x, y+1)$, 
then $\cL(x,y) \neq \cL(x,y+1)$ immediately follows.
If $\cL(x,y)$ is $B$-irreflexive, then one gets a contradiction by observing that having two occurrences of the same $B$-irreflexive atom stacked one above the other violates the consistency of the compass structure
(with respect to the $\thenB$ relation).

Let us prove now the right-to-left direction.
Suppose, by way of contradiction, that 
$\cL(x,y) \neq \cL(x,y + 1)$. Then, there exists a formula $\psi \in \closure$ such that $\psi \in \cL(x,y + 1)$ and $\neg \psi \in \cL(x,y)$. 
By Lemma \ref{lem:atomdeterminacy}, for all $0\leq x\leq y \leq N$, the truth of $\psi\in \cL(x,y)$ is uniquely determined by the truth values of $\cP(x,y)$, $\reqB(x,y)$, and $\reqD(x,y)$. By the assumption, we get $\reqB(x,y+1) \supset \reqB(x,y)$. To reach the contradiction, we then proceed as in the proof of Lemma \ref{lem:bstep}.
\end{proof}

\lemstepconstraint*

\begin{proof}
Let $\oy$ be the minimum point $\oy > y$ such that
$\cL(x',y)\neq \cL(x',\oy)$.
Let us assume by contradiction that $\oy \neq y'$.
 By Lemma~\ref{lem:shadingorder}
we have that $\oy > y'$. Let $\shadingG(x)= 
\oF^{k_0}_0\ldots\oF^{k_m}_{m}$, and 
 let $0 \leq i < m$ be the index such that $\shadingG(x)[y - x] 
 = \shadingG(x)[y - x'] = \oF_i$. Then we have 
 $\cL(x, y)= \oF_i = \cL(x',y)$, 
 $\cL(x, y')= \oF_{i+1}$, and 
 $\cL(x',y')=\oF_i$. Moreover 
 for every $y\leq y''< y'$ we have $\cL(x,y'')=\cL(x',y'')= \oF_i$, then
 $\oF_i$ is $B$-reflexive.
 Let us notice that  $\cP(x,y'-1) = \cP(x', y')= \oF_i \cap \Prop$ 
 then  we have that $\cP(x,y')= \cP(x',y')$.
 Since $\cL(x,y'-1)$ is $B$-reflexive we have that 
 $\reqD(\cL(x,y')) \supset  \reqD(\cL(x,y'-1))=\reqD(\cL(x',y'))=\reqD(\cL(x',y' - 1))$,
 otherwise conditions for Lemma~\ref{lem:bdeterminization} apply
 and $\cL(x,y')=  \cL(x,y'-1)$ (contradiction).
 This means that there exists $x < \ox < x'$
 such that $\psi \in (\cL(\ox,y'-1) 
 \cap \reqD(\cL(x, y')) \setminus \reqD(\cL(x', y')))$ and
for every $x' \leq x'' \leq y'-1$
we have $\psi \not \in \cL(x'',y'-1)$.
The simpler case is when $y' = y + 1$. 
In such a case from $\future(x,y)= \future(x',y)$
we have that there exists $\ox'> x'$ 
such that $\cL(\ox', y)= \cL(\ox, y)$ (contradiction).
Let us consider now the case in which  $y' > y+1$.
Since $\neg \psi \in\boxD(\cL(x,y' -1))$
we have that 
$\psi \notin \cL(x'', y'')$ for every $x<x''\leq y''<y'-1$.
Two cases arise:
\begin{compactitem}
\item there exists $y \leq y'' < y'-1$ such that 
$\cL(\ox, y'')$ is $B$-reflexive. If it is the case since
$\oF_i \cap \Prop =
 \cP(x, \oy') \subseteq \cP(\ox,\oy') \subseteq \cP(x', \oy') 
 = \oF_i \cap \Prop$ and 
 $\reqD(\oF_i)  =
 \reqD(\cL(x, \oy') \supseteq \reqD(\cL(\ox,\oy')) \supseteq 
 \reqD(\cL(x', \oy')) 
 = \reqD(\oF_i)$ for every 
$y \leq \oy \leq y'- 1$
we have
$\cP(\ox,\oy') = \cP(x, \oy')=\cP(x', \oy')$ and 
$\reqD(\cL(\ox,\oy')) = \reqD(\cL(x, \oy'))=\reqD(\cL(x', \oy'))$
for every 
$y'' \leq \oy \leq y'- 1$.
Then for Lemma~\ref{lem:bdeterminization}
we have that $\cL(\ox, y' -1) = \cL(\ox, y' -2)= \ldots= \cL(\ox, y'')$
this means that $\cL(\ox, y' -1)$ is not the first 
atom featuring $\psi$ on the 
column $\ox$ (contradiction);
\item for every $y \leq y'' < y'-1$ we have that 
$\cL( \ox, y'')$ is $B$-irreflexive. Then, from
$\future(x,y)= \future(x',y)$ there exists $\ox'> x'$
such that $\shadingG(\ox') \sim \shadingG(\ox)$ and
$\cL(\ox', y)= \cL(\ox,y)$. Let us observe that, by definition of 
$B$-sequence, for every $B$-sequence
$F_0^{h_0}\ldots F_n^{h_n}$ and  for every $1\leq i \leq n$ 
if $F_i$ is $B$-irreflexive then 
$h_i= 1$ (i.e., $B$-irreflexive atoms are unique in every $B$-sequence).
Then, we have that for every $y \leq y'' \leq y'-1$ we have 
$\cL(\ox, y'')= \cL(\ox', y'')$ and thus 
$\psi \in \cL(\ox', y'-1)$ this implies $\psi \in \reqD(\cL(x',y'))$
(contradiction).
\qedhere
\end{compactitem}
\end{proof}

\thmbound*

\begin{proof}
To start with, let us consider the problem of determining 
how many possible different $\rowG(y)$ we can have
in a compass structure $\cG=(\bG_N, \cL)$. Let us first observe that
 for the monotonicity of the function $\future$ we have, for every 
$0 \leq y \leq N$, 
$\future(0,y) \supseteq \ldots \supseteq \future(y,y)$.
Then, since we cannot have two incomparable, w.r.t. $\subseteq$ relation, 
$\future(x,y)$ and $\future(x',y)$  
we have at most $2^{4|\varphi|^2 + 6 |\varphi| + 2}\cdot 2^{|\varphi|+ 1} =
2^{4|\varphi|^2 + 7 |\varphi| + 3}$ possible 
distinct $\future(x,y)$, that is an upper bound of
the length of the longest possible 
$\subseteq$-ascending sequence in the set of pairs $([\shadingG]_{\sim}, F)$
(i.e., equivalence class and atom).

 Moreover,  each one of the possible witnesses 
is a pair $([\shadingG]_{\sim}, F)$
and, since $\witnesses(y)$ does not contain covered points, each fingerprint
$fp(x,y) = ([\shadingG(x)]_{\sim}, \cL(x,y),\allowbreak \future(x,y))$ can be 
associated to at most  
$4|\varphi| + 2$ (i.e.,   the maximum 
value for  $\Deltareq$ plus one) distinct points in $\witnesses(y)$. 
Summing up, we
have that the maximum length for $\rowG(y)$ is 
bounded by 
$2^{4|\varphi|^2 + 7 |\varphi| + 3} \cdot 2^{4|\varphi|^2 + 7 |\varphi| + 3} \cdot (4|\varphi| + 2)=
(4|\varphi| + 2) 2^{8|\varphi|^2 + 14 |\varphi| + 6}$.
In each of such position we can put a  pair $([\shadingG]_{\sim}, F)$
and thus the cardinality of the set of all possible $\rowG(y)$
is bounded by $(2^{4|\varphi|^2 + 7 |\varphi| + 3})^{(4|\varphi| + 2) 2^{8|\varphi|^2 + 14 |\varphi| + 6}}$ that is 
$2^{2(|\varphi| + 1)(4|\varphi|^2 + 7 |\varphi| + 3)2^{8|\varphi|^2 
+ 14 |\varphi| + 6}}$ which is doubly exponential in $|\varphi|$. 
Finally,
given a $\varphi$-compass structure $\cG=(\bG_N, \cL)$,
by repeatedly applying Theorem~\ref{thm:smallcompass},
we can obtain a $\varphi$-compass structure  $\cG=(\bG_{N'}, \cL')$
such that for every $0\leq y < y'\leq N$ we have 
$\rowG(y) \neq  \rowG(y')$, then, by means of the above considerations on the maximum 
 cardinality for the set of all possible $\rowG(y)$,  we 
may conclude that $\varphi$ is satisfiable iff there is a compass structure $\cG=(\bG_N, \cL)$ for it
such that $N \leq 2^{2(|\varphi| + 1)(4|\varphi|^2 + 7 |\varphi| + 3)2^{8|\varphi|^2 + 14 |\varphi| + 6}}$.

To complete the proof, it suffices to show that checking the existence of such a doubly exponential compass structure can be done in exponential space.

Let $M = 2^{2(|\varphi| + 1)
(4|\varphi|^2 + 7 |\varphi| + 3)2^{8|\varphi|^2 
+ 14 |\varphi| + 6}} + 1$ be the 
bound (plus  $1$)
on the size of a candidate compass structure for the input 
$\mathsf{BD}_{hom}$ formula $\varphi$, according to the small model theorem just proved.
In the following, we briefly describe a decision procedure that decides, for some $N \leq M$, whether or not there exists a compass structure 
$\cG= (\bG_N, \cL)$   
for the input $\mathsf{BD}_{hom}$
formula $\varphi$. 
If such a procedure works 
in exponential space with respect to $|\varphi|$, 
we can immediately conclude that
the satisfiability problem for $\mathsf{BD}_{hom}$ belongs to the
EXPSPACE complexity class.
The decision procedure begins at step $y=0$  by guessing 
$\rowG(y) = ([\shadingB^0]_\sim, F^0)$ where $F^0 = 
[\shadingB^0]^0_\sim$ and updates $y$ to $y+1$. For every $y>0$, the procedure 
 proceeds inductively as follows (let $\rowG(y) = 
 ([\shadingB^0]_\sim,\allowbreak F^0)\ldots ([\shadingB^k]_\sim, F^k)$):

\begin{compactenum}
\item if there exists $i$ for which $\varphi \in F^i$, then return $true$;
\item if $y = M$, then return $false$;  
\item non-deterministically guess a pair $([\shadingB^{k+1}]_\sim,\oF^{k+1})$
such that $\oF^{k+1} = [\shadingB^{k+1}]^0_\sim$;
\item for every $0 \leq i\leq k$, let 
$$\oF^i = \left\{\begin{array}{cc}
F^i  &  \mbox{if
$\left(\begin{array}{c}   
\reqD(F^i) = \bigcup\limits_{i < j \leq k } \obsD(F^j) \cup \reqD(F^j)\\ \wedge\\ 
\reqB(F^i) = \obsB(F^i) \cup \reqB(F^i) \\ \wedge \\  
F^i \cap \Prop = \oF^{k+1} \cap F^i \cap \Prop
\end{array}\right)$
} \\
\\
\hF  &\mbox{\begin{tabular}{l} 
where $\hF = next([\shadingB^{i}]_\sim, F^i)$,  
$\reqD(\hF) = \bigcup\limits_{i < j \leq k } \obsD(F^j) \cup \reqD(F^j)$, \\
$\reqB(\hF) = \reqB(F^i) \cup \obsB(F^i)$, and 
$\hF \cap \Prop = \oF^{k+1} \cap \hF \cap \Prop$ 
\end{tabular}
}\\ \\
\bot & \mbox{otherwise}.
\end{array}\right.$$

By Lemma~\ref{lem:bdeterminization}, $\oF^i$ is well defined.
\item if there exists $i$ for which $\oF^i = \bot$, then return $false$;
\item let $i_0< \ldots <i_h$ be the maximal sub-sequence of indexes 
in $0\ldots k+1$
such that,
for every $0 \leq j \leq h$, 
$([\shadingB^{i_j}]_\sim, \oF^{i_j})$ is not covered 
in $([\shadingB^{0}]_\sim, \oF^{0})\ldots 
([\shadingB^{k+1}]_\sim, \oF^{k+1})$,
then we define $\rowG(y+1)=\allowbreak ([\shadingB^{i_0}]_\sim, 
\oF^{i_0})\ldots ([\shadingB^{i_h}]_\sim, \oF^{i_h})$;
\item update $y$ to $y +1$ and restart from step $1$.
\end{compactenum}

\noindent Soundness and completeness of the above procedure can be proved using the result given  in this section. In particular, Corollary~\ref{cor:coveredismonotone} comes into play in the completeness proof (item $6$ keeps track of  all and only the not covered points on row $y+1$). Moreover, notice that, for each step $0 \leq y \leq M$, we have to keep track of:
\begin{compactenum}
\item  the current value of $y$, which cannot exceed $2^{2(|\varphi| + 1) (4|\varphi|^2 + 7 |\varphi| + 3)2^{8|\varphi|^2 + 14 |\varphi| + 6}} +1$ and can be logarithmically encoded using an exponential number of bits;
\item two rows, namely $\rowG(y)$ and $\rowG(y+1)$, whose maximum length is bounded by 
$2^{4|\varphi|^2 + 7 |\varphi| + 3} \cdot 2^{4|\varphi|^2 + 7 |\varphi| + 3} \cdot (4|\varphi| + 2)=
(4|\varphi| + 2) 2^{8|\varphi|^2 + 14 |\varphi| + 6}$ (exponential in $|\varphi|$). 
Moreover, each 
position in such sequences holds a pair $([\oF_{0}\ldots\oF_{m}], \oF_i)$.
Since $m \leq 4|\varphi| +1$, we have that each position holds at most 
$4|\varphi| + 3$ atoms.  Each atom can be represented using exactly $|\varphi| + 1$ bits.
Summing up, we have that the total space needed for keeping the two rows $y$ and $y+1$
(step $0 \leq y \leq M$) consists of  $2\cdot (|\varphi| + 1)(4|\varphi| + 3)\cdot2^{4|\varphi|^2 + 7 |\varphi| + 3} \cdot 2^{4|\varphi|^2 + 7 |\varphi| + 3} \cdot (4|\varphi| + 2)$ bits, which, simplified, turns out to be 
$4(8|\varphi|^3 + 18 |\varphi|^2
+ 13 |\varphi| +3) 2^{8|\varphi|^2 + 14 |\varphi| + 6}$ bits
that is still exponential in $|\varphi|$.   
\end{compactenum}
This shows that we can decide the satisfiability of $\varphi$  in exponential space regardless of the fact that the above procedure is non-deterministic thanks to Savitch's Theorem \cite{DBLP:journals/jcss/Savitch70}.
\end{proof}
\section{An example of a model for a \texorpdfstring{\LogicBDAhom}{BDA_hom} formula
}\label{appendix:bda}


\begin{figure}
\centering

\begin{tikzpicture}

\begin{scope}[anchor=west, inner sep=0.1cm]
\node (F0)               {$\varphi=\hsAA($};
\node (F1) at  (F0.east) {$\hsEB$};          
\node[xshift=-0.2cm]  (F2) at  (F1.east) {$\hsEB q$};
\node (F3) at  (F2.east) {$\longrightarrow $};
\node (F4) at  (F3.east) {$\hsED p$};
\node (F5) at  (F4.east) {$)$};

\draw[decorate, line width=1pt,
      decoration={brace,raise=0.3cm,
                  amplitude=5pt,mirror},
                  color=black,
                  ] 
(F5.west) -- (F1.west) node[above=0.53cm, midway]{$\psi_1$};

\node[xshift=2cm] (F0) at  (F5.east)             {$\neg\varphi=\hsEA($};
\node (F1) at  (F0.east) {$\hsEB$};          
\node[xshift=-0.2cm]  (F2) at  (F1.east) {$\hsEB q$};
\node (F3) at  (F2.east) {$\wedge $};
\node (F4) at  (F3.east) {$\hsAD  \neg p$};
\node (F5) at  (F4.east) {$)$};

\draw[decorate, line width=1pt,
      decoration={brace,raise=0.3cm,
                  amplitude=5pt,mirror},
                  color=black,
                  ] 
(F5.west) -- (F1.west) node[above=0.53cm, midway]{$\neg \psi_1$};

\end{scope}

\pgftransformshift{\pgfpoint{0.5cm}{-4.5cm}}

\begin{scope}[node distance=3.3cm]

\node[draw, circle, inner sep=1,  node 
distance=0,  
label={[]270:$\scalebox{0.7}{$0$}$},
label={[]90:$\scalebox{0.7}{$\psi_1$}$}](0) {};

\node[draw, circle, inner sep=1, right of=0, 
label={[]270:$\scalebox{0.7}{$1$}$},
label={[]90:$\scalebox{0.7}{$q,\psi_1
$}$}](1) {};

\node[draw, circle, inner sep=1, right of=1, label={[]270:$\scalebox{0.7}{$2$}$},
label={[]90:$\scalebox{0.7}{$p,q,\psi_1,\hsEA \neg \psi_1
$}$}](2) {};

\node[draw, circle, inner sep=1, right of=2, label={[]270:$\scalebox{0.7}{$3$}$},
label={[]90:$\scalebox{0.7}{$q,\psi_1
$}$}](3) {};

\node[draw, circle, inner sep=1, right of=3, label={[]270:$\scalebox{0.7}{$4$}$},
label={[]90:$\scalebox{0.7}{$p,\psi_1
$}$}](4) {};

\draw[|-|] ($(0.center) + (0,1)$) -- ($(1.center)+ (0,1)$)
node[pos=0.5, fill=white](12) {$\scalebox{0.7}{$\psi_1$}$};

\draw[|-|] ($(0.center) + (0,1.5)$) -- ($(2.center)+ (0,1.5)$)
node[pos=0.5, fill=white](46) {$\scalebox{0.7}{$\psi_1,\hsEA \neg \psi_1$}$};

\draw[|-|] ($(0.center) + (0,2.5)$) -- ($(3.center)+ (0,2.5)$)
node[pos=0.5, fill=white](16) {$\scalebox{0.7}{$\hsED p,\psi_1
$}$};

\draw[|-|] ($(0.center) + (0,3.5)$) -- ($(4.center)+ (0,3.5)$)
node[pos=0.5, fill=white](16) {$\scalebox{0.7}{$\hsED p,\psi_1
$}$};

\draw[-|] ($(1.center) + (0,1)$) -- ($(2.center)+ (0,1)$)
node[pos=0.5, fill=white](12) {$\scalebox{0.7}{$q,\hsEB q,\psi_1,\hsEA \neg \psi_1$}$};

\draw[|-|] ($(1.center) + (0,2)$) -- ($(3.center)+ (0,2)$)
node[pos=0.5, fill=white](47) {$\scalebox{0.7}{$q,\hsEB q,\hsED p,\hsEB \hsEB q,\psi_1
$}$};

\draw[|-|] ($(1.center) + (0,3)$) -- ($(4.center)+ (0,3)$)
node[pos=0.5, fill=white](16) {$\scalebox{0.7}{$\hsEB q,\hsED p,\hsEB \hsEB q,\psi_1
$}$};

\draw[-|] ($(2.center) + (0,1)$) -- ($(3.center)+ (0,1)$)
node[pos=0.5, fill=white](12) {$\scalebox{0.7}{$q,\hsEB q,\psi_1
$}$};

\draw[|-|] ($(2.center) + (0,1.5)$) -- ($(4.center)+ (0,1.5)$)
node[pos=0.5, fill=white](46) {$\scalebox{0.7}{$\hsEB q,\hsEB \hsEB q
$}$};

\draw[-|] ($(3.center) + (0,1)$) -- ($(4.center)+ (0,1)$)
node[pos=0.5, fill=white](12) {$\scalebox{0.7}{$\hsEB q,\psi_1
$} $};

\end{scope}

\pgftransformshift{\pgfpoint{6.4cm}{-4cm}}

\node(T){ \input{sections/aexpspace/figReqExplanationBDAatoms}};

\node[anchor=north] at (T.south) {
{\renewcommand{\arraystretch}{1.5}
\scalebox{0.745}{$\begin{array}{c||c|c|c|c|c||c|c|c|c||c|c|c||c|c||c}
    \alpha_{[x,y]} & [0,0] & [0,1] & [0,2] & [0,3] & [0,4] & [1,1] & [1,2] & [1,3] & [1,4] & [2,2] & [2,3] & [2,4] & [3,3] & [3,4] & [4,4] 
\\[0.1cm]    \hline \hline 

\neg \psi_1  & \abox & \abox & \abox & \abox & \abox & \abox & \abox & \abox & \abox & \areq & \areq & \asat & \abox & \abox & \abox
    
\end{array}$}
}
};

\end{tikzpicture}

\caption{\label{fig:reqExplanationABD} 
A graphical (above) and tabular (below)  account of the behaviour of $\reqR(F)$, $\obsR(F)$, and $\boxR(F)$, with $F \in \atoms$ and $R \in \mathset{A,B,D}$, for the formula $\varphi = \hsAA( \hsEB\hsEB q \rightarrow \hsED p)$.}


\end{figure}

Let us consider Figure~\ref{fig:reqExplanationABD}, which provides an example of a consistent 
atom labelling of a model of a $\mathsf{BDA}_{hom}$ formula $\varphi$.
For what concerns $\reqR(\cdot), \boxR(\cdot),$ and $\obsR(\cdot)$, with 
$R \in \mathset{B,D}$, the same considerations we made in describing the example of 
Figure~\ref{fig:reqExplanationBD} in Section~\ref{sec:compass} apply. 
We now explain how the behaviour of sets $\reqA(\cdot), \boxA(\cdot),$ and $\obsA(\cdot)$ differ from
their counterparts $\reqR(\cdot), \boxR(\cdot),$ and $\obsR(\cdot)$, with 
$R \in \mathset{B,D}$, and on giving an initial account of the behaviour of  
the marking functions $\alpha_{[x,y]}$.

We first observe that, while $\reqR$, with $R \in \mathset{B,D}$,  is ``monotone'' 
for atoms labelling intervals which are in the same $R$-relation, this is is not true when $R=A$.
This is a direct consequence of the fact that Allen's relations $\allenB$ and $\allenD$ are transitive, while
relation $\allenA$ is not. As an example, in Figure \ref{fig:reqExplanationABD}, we have that $[0,1] \allenA [1,2] \allenA [2,3]$, but $\reqA(F^{[0,1]})= \reqA(F^{[2,3]}) =\emptyset$  and $\reqA(F^{[0,1]})= \{\neg \psi_1\}$.

Let us now focus on the newly introduced second component $\alpha_{[x,y]}$ of each atom 
which is reported on the very bottom of Figure~\ref{fig:reqExplanationABD}.
In the example of Figure~\ref{fig:reqExplanationABD}, it holds that $\TFA = \{\neg \psi_1\}$
and thus $\alpha_{[x,y]}$ assigns to the interval $[x,y]$ the ``status'' of $\neg \psi_1$
on it. In particular, if $\neg \psi_1 \not\in \reqA(F^{[x,x]})$, then $\hsAA \psi_1 \in F^{[x,x]}$, which forces the formula
$\psi_1$ to belong to $F^{[x,x']}$, for all the intervals $[x,x']$, with $x \leq x'$, according to Lemma~\ref{lem:aprop}.

In general, it holds that $\alpha_{[x,y]}(\neg \psi_1) = \abox$ if and only if $\neg \psi_1 \notin \reqA(F^{[x,y]})$,
which means that $\neg \psi_1$ \emph{ is not requested by } $F^{[x,x]}$ and thus $\psi_1$ must be satisfied 
on all the intervals $[x,y]$. This is the case, for instance, with intervals $[0,0], [1,1], [3,3]$, and $[4,4]$ in Figure~\ref{fig:reqExplanationABD}, which impose that  $\alpha_{[x,y]}(\neg \psi_1) = \abox$, and, consequently, $\psi_1 \in F^{[x,y]}$, for all $[x,y] \in \{ [x,y]: 0 \leq x \leq y \leq 4, x \neq 2 \}$. 
If $\alpha_{[x,x]}(\neg \psi_1) \neq \abox$, then $\alpha_{[x,y]}(\neg \psi_1) \in \{\areq, \asat\}$,
for every $y \geq x$, which means that the request $\hsEA \neg \psi_1$ is pending on $[x,x]$, that is, $\neg \psi_1 \in \reqA(F^{[x,y]})$, and must be satisfied by some interval of the form $[x,y]$,
for some $y \geq x$. If we take the minimum  $y$ such that $\neg \psi_1 \in \obsA(F^{[x,y]})$,
we have that:
\begin{compactitem}
\item $\alpha_{[x, y']}(\neg \psi_1) = \areq$, for every $x\leq y' < y$, which means that 
the pending $\hsEA$-request $\neg \psi_1$ is not fulfilled by the intervals ending in $x$,
if we consider the model up to $y'$; 
\item $\alpha_{[x, y']}(\neg \psi_1) = \asat$, for every $x \leq y \leq y'$, which means that 
the pending-$\hsEA$ request $\neg \psi_1$ is fulfilled for the intervals ending in $x$,
if we consider the model up to $y'$, and, obviously, it stays fulfilled for such intervals ever after. 
\end{compactitem}
In Figure~\ref{fig:reqExplanationABD}, this is the case with interval $[2,2]$ for which $\neg \psi_1 \in \reqA(F^{[2,2]}) $ holds. 
However, since we have $\neg\psi_1 \notin \obsA(F^{[2,2]})$ and $\neg\psi_1 \notin \obsA(F^{[2,3]})$,
it turns out that $\alpha_{[2,2]}(\neg \psi_1)=\alpha_{[2,3]}(\neg \psi_1)= \areq$.
On the other hand, $\neg \psi_1$ appears ``for the first time'' in $\obsA(F^{[2,y]})$
when $y= 4$, and thus $\alpha_{[2,4]}(\neg \psi_1)= \asat$.


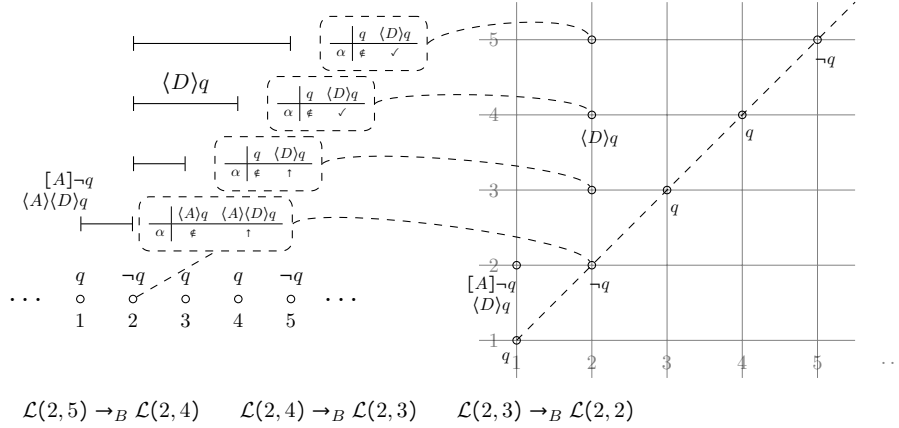
\begin{figure}
\centering
\scalebox{1.1}{\begin{tikzpicture}

\begin{scope}[node distance=0.7cm]

\node(0) {$\dots$};

\node[draw, circle, inner sep=1, right of=0, 
label={[]270:$\scalebox{0.7}{$1$}$},
label={[]90:$\scalebox{0.7}{$q$}$}](1) {};

\node[draw, circle, inner sep=1, right of=1, 
label={[]270:$\scalebox{0.7}{$2$}$},
label={[]90:$\scalebox{0.7}{$\neg q$}$}](2) {};

\node[shape=rectangle,rounded corners,draw,dashed,above of=2, xshift=1.1cm, node distance = 1cm](2alpha) {
    \scalebox{0.5}{$\begin{array}{c|cc}
        & \hsEA q  & \hsEA\hsED q \\
        \hline
        \alpha &  \abox & \areq 
    \end{array}$}
};

\draw[dashed] (2) -- (2alpha.south);

\node[draw, circle, inner sep=1, right of=2, label={[]270:$\scalebox{0.7}{$3$}$},
label={[]90:$\scalebox{0.7}{$q$}$}](3) {};

\node[draw, circle, inner sep=1, right of=3, label={[]270:$\scalebox{0.7}{$4$}$},
label={[]90:$\scalebox{0.7}{$q$}$}](4) {};

\node[draw, circle, inner sep=1, right of=4, label={[]270:$\scalebox{0.7}{$5$}$},
label={[]90:$\scalebox{0.7}{$\neg q$}$}](5) {};

\node[right of=5](6) {$\dots$};

\draw[|-|] ($(1.center) + (0,1)$) -- ($(2.center)+ (0,1)$)
node[pos=0.5, above, label={[xshift=-1.5cm, yshift=0.3cm]0:\scalebox{0.6}{$
    \begin{array}{r}
        \hsAA\neg q \\ \hsEA\hsED q \ 
    \end{array}
    $}}](12) {};


\draw[|-|] ($(2.center) + (0,1.8)$) -- ($(3.center)+ (0,1.8)$)
node[pos=1, xshift=1.1cm, shape=rectangle,rounded corners,draw,dashed](3alpha) {
    \scalebox{0.5}{$\begin{array}{c|cc}
        & q  & \hsED q \\
        \hline
        \alpha &  \abox & \areq
    \end{array}$}
};

\draw[|-|] ($(2.center) + (0,2.6)$) -- ($(4.center)+ (0,2.6)$)
node[pos=1, xshift=1.1cm, shape=rectangle,rounded corners,draw,dashed](4alpha) {
    \scalebox{0.5}{$\begin{array}{c|cc}
        &  q  & \hsED q \\
        \hline
        \alpha &  \abox & \asat 
    \end{array}$}
}
node[pos=0.5, above] {\scalebox{0.75}{$\hsED q$}}
;

\draw[|-|] ($(2.center) + (0,3.4)$) -- ($(5.center)+ (0,3.4)$)
node[pos=1, xshift=1.1cm, shape=rectangle,rounded corners,draw,dashed](5alpha) {
    \scalebox{0.5}{$\begin{array}{c|cc}
        &  q  & \hsED q \\
        \hline
        \alpha &  \abox & \asat 
    \end{array}$}
};

\end{scope}

\node[ yshift=-1.5cm, xshift=4cm]{\scalebox{0.75}{
    \begin{tabular}{ccccc}
        $\cL(2,5) \thenB \cL(2,4)$ & &
        $\cL(2,4)\thenB \cL(2,3)$ &  &
        $\cL(2,3)\thenB \cL(2,2)$       
\end{tabular}

}};

\pgftransformshift{\pgfpoint{6.5cm}{-0.55cm}}

\draw[dashed] (0,0) -- (4.5,4.5);
\draw[step=1.0,black, opacity=0.5, very thin,xshift=-1cm,yshift=-1cm] (0.5,0.5)
grid (5.5,5.5);

\begin{scope}[opacity=0.5]0

\node[yshift=-0.3cm](0) {$\scalebox{0.7}{$1$}$};
\node[ right of=0](0) {$\scalebox{0.7}{$2$}$};
\node[ right of=0](0) {$\scalebox{0.7}{$3$}$};
\node[ right of=0](0) {$\scalebox{0.7}{$4$}$};
\node[ right of=0](0) {$\scalebox{0.7}{$5$}$};
\node[ right of=0](0) {$\scalebox{0.7}{$\ldots$}$};

\node[xshift=-0.3cm, yshift=-0.0cm](0) {$\scalebox{0.7}{$1$}$};
\node[ above of=0](0) {$\scalebox{0.7}{$2$}$};
\node[ above of=0](0) {$\scalebox{0.7}{$3$}$};
\node[ above of=0](0) {$\scalebox{0.7}{$4$}$};
\node[ above of=0](0) {$\scalebox{0.7}{$5$}$};

\end{scope}

\node[draw, circle, inner sep=1, label={[xshift=-0.95cm, yshift=-0.4cm]0:\scalebox{0.6}{$
    \begin{array}{r}
        \hsAA\neg q \\ \hsED q \ 
    \end{array}
    $}}](2) at (0, 1) {};

\node[draw, circle, inner sep=1, label={[xshift=-0.1cm]-90:\scalebox{0.6}{$ $}}](3) at (1, 2) {};
\node[draw, circle, inner sep=1, label={[xshift=0.1cm]-90:\scalebox{0.6}{$\hsED q$}}](4) at (1, 3) {};
\node[draw, circle, inner sep=1, label={[xshift=0.1cm]-90:\scalebox{0.6}{$ $}}](5) at (1, 4) {};

\node[draw, circle, inner sep=1, label={[xshift=0.1cm]-150:\scalebox{0.6}{$ q$}}](1) at (0, 0) {};


\node[draw, circle, inner sep=1, label={[xshift=0.1cm]-90:\scalebox{0.6}{$\neg q$}}](22) at (1, 1) {};

\node[draw, circle, inner sep=1, label={[xshift=0.1cm]-90:\scalebox{0.6}{$q$}}](33) at (2, 2) {};

\node[draw, circle, inner sep=1, label={[xshift=0.1cm]-90:\scalebox{0.6}{$q$}}](44) at (3, 3) {};

\node[draw, circle, inner sep=1, label={[xshift=0.1cm]-90:\scalebox{0.6}{$\neg  q$}}](55) at (4, 4) {};


 

\draw[dashed] (2alpha.east) to[in=120,looseness=0.25] (22);
\draw[dashed] (3alpha.east) to[in=120,looseness=0.25] (3);
\draw[dashed] (4alpha.east) to[in=120,looseness=0.25] (4);
\draw[dashed] (5alpha.east) to[in=120,looseness=0.45] (5);

\end{tikzpicture}}

\caption{\label{fig:arelation} A graphical account of the extension of the  $\thenB$ relation to $\mathsf{A}$-marked atoms
both from the interval point of view (left) and the spatial one (right). }
\vspace{-0.5cm}

\end{figure}

To conclude, in Figure~\ref{fig:arelation}, we give an intuitive account of how the second component of an atom behaves with respect to the relations $\thenB$ and $\thenD$. Intuitively, it is associated with an interval $[x,y]$ to keep track of the $\mathsf{A}$-requests featured by $[x,x]$ which have been satisfied by intervals $[x, y']$,
with $y'\leq y$, that is, the ones marked with $\asat$, against those which are still pending, that is, those marked with $\areq$.


\section{Adding modality \texorpdfstring{$\mathtt{A}$}{A} to \texorpdfstring{$\mathsf{BD}_{hom}$}{BD_hom}: the logic \texorpdfstring{$\mathsf{BDA}_{hom}$}{BDA_hom}}
\label{appendix:aregexp}




In this appendix, in analogy to what we did for modalities $\hsEB$ and $\hsED$ in Section~\ref{sec:logic}, we investigate the counterpart of modality $\hsEA$ in terms of a suitable extension of generalized $*$-free regular expressions. Basically, we enrich the semantics of generalized $*$-free regular expressions with what we call a ``right context''. We will prove that the resulting semantics subsumes the original one, that is, the notion of generalized $*$-free regular expression given in Section~\ref{sec:logic} is just a specialization of it. In particular, the encoding of both $\Pre(e)$ and $\Sub(e)$ directly transfers to this new semantics without any modification. We will conclude the appendix by giving an example that shows how the operator corresponding  to modality $\hsEA$ has an explicit counterpart in the generalized $*$-free regular expressions used 
for real-world programming languages.  

As a preliminary remark, we would like to observe that one may be tempted to interpret modality $\hsEA$ as a logical counterpart of the concatenation operator. This is wrong. Intuitively, modality $\hsEA$ characterizes words with a specific ``right context''.
Such an idea can be formalized as follows.

In order to identify the right generalized $*$-free regular 
expression for modality $\hsEA$, we provide an alternative, yet 
equivalent, semantics for these expressions. In such a semantics, the 
language $\Rlang(e)$ of a generalized $*$-free regular expression $e$ is interpreted over pairs 
of finite words, that is, $\Rlang(e) \subseteq \Sigma^+ \times \Sigma^*$.
A pair $(w,w') \in \Rlang(e)$ represents the word $w$ belonging to the language $\Lang(e)$, according to the semantics given in  Section~\ref{sec:logic}, together with its ``right context'' word $w'$, which is the word that must appear immediately after $w$. 

\smallskip

Formally, generalized $*$-free regular expressions of Section~\ref{sec:logic} are extended as follows:
\[ e ::= \emptyset \ |\ a\ |\ \neg e\ |\ e + e \ |\ \ \Pre(e)\ 
|\ \Sub(e)\ |\  \Rcon(e),
   \mbox{ for any $a \in \Sigma$}
\]
Their semantics is defined as follows: 

\smallskip

\begin{compactenum}[(i)]
\item $\Rlang(\emptyset) = \emptyset$;
\item $\Rlang(a) = \{ (a, w): w \in \Sigma^* \}$;
\item $\Rlang(\neg e) =  \Sigma^+ \times \Sigma^* \setminus  \Rlang(e)$;
\item $\Rlang(e + e') =  \Lang(e) \cup \Lang(e')$;
\item $\Rlang(\Pre(e)) =  \{(wv, u): v \in \Sigma^+, (w, vu) \in \Rlang(e)  \}$;
\item $\Rlang(\Sub(e)) =  \{(uwv, z): u, v \in \Sigma^+, (w, vz) \in \Rlang(e) \}$;
\item $\Rlang(\Rcon(e)) =  \{(w, u): u \in  \Rlang(e) \}$.
\end{compactenum} 

Let us denote the empty word by $\epsilon$. With a little abuse of
notation, we say that, for every $w \in \Sigma^+$, $w \in \Lang(e)$  if and only if $(w,\epsilon) \in \Rlang(e)$. Then, it is easy to prove that, for any expression $e ::= \emptyset \ |\ a\ |\ \neg e\ |\ e + e \ |\ \ \Pre(e)\ |\ \Sub(e)$, $w \in \Lang(e)$ if and only if $(w, \epsilon) \in \Rlang(e)$.
In such a way, the original (restricted) semantics turns out to be a specialization of the extended one. 

It can be easily shown that the extended semantics preserves the 
mapping from a restricted expression $e$ to an equivalent $\mathsf{BD}_{hom}$ formula $\varphi_e$ given in Section~\ref{sec:logic}. In order to capture the language $\Rlang(\Rcon(e))$ in $\mathsf{BDA}_{hom}$, we extend the mapping with the rule:
$$
\varphi_{\Rcon(e)} = \hsEA(\hsEB \top \wedge 
\hsAB\hsAB \bot \wedge \hsEA \psi_e ).
$$
Let us assume that $\varphi_{\Rcon(e)}$ holds over an interval $[x,y]$. Then, it predicates over ``the right context'' of $[x,y]$ by stating that there exists an interval $[y,y+1]$ (the constraint on the length of such an interval is imposed by the first two conjuncts $\hsEB \top \wedge \hsAB\hsAB \bot$) which has an adjacent-to-the-right interval $[y+1, y']$ where $\psi_e$ holds
(third conjunct $\hsEA \psi_e$).

\smallskip


In order to show the significance of the proposed extension of generalized $*$-free regular expression, we explore an interesting correspondence between the operator $\Rcon$ (and thus, indirectly, modality $\hsEA$) and an operator of the regular expressions typically used in popular programming languages like, for instance, Python \cite{van1995python}. 
It is easy to see that the $\Rcon$ operator corresponds to the \emph{lookahed operation}. Such an operation is usually 
implemented as \emph{positive lookahead}, whose syntax is 
$\mathtt{(?= \mathbf{e} )}$, and \emph{negative lookahed},
whose syntax is $\mathtt{(?\ !\ \mathbf{e} )}$, where $e$ is a regular 
expression. In many real-world applications, regular expressions are used to execute \emph{pattern matching} inside a long text as an effective alternative to the task of checking whether such a long text belongs to a certain language. This is 
the case especially in the domain of natural language processing from 
which the following toy example is borrowed. Let us suppose that we 
want to capture a pattern that consists of an English word followed 
by a list of words in English separated by commas and whose last word is 
prefixed by the word \emph{``and''}. An example of a sentence containing such a pattern is the following:
\emph{``This paper deals with $\LogicHS$ \textbf{operators}
\myUnderline{meets}, \myUnderline{begins}, and \myUnderline{during} 
under homogeneity assumption.''}

In such a toy example, a motivation for matching the word 
\textbf{operators} may be related to the fact that the noun preceding a natural language description of items may represent their type. In the above sentence, ``meets'', ``begins'', and ``during'' are indeed of type ``operators''. In such an interpretation, we are assuming that the  word denoting the type is put immediately before the list of words and thus conjunctions like, e.g., ``such as'' or ``like'' are not contemplated. However, they may be captured by longer, but not much more complex than the one we are going to show, regular expressions. For the sake of simplicity, we assume that the number of words in 
the list is greater than or equal to $3$ and each word is a single one.
As an example, the pattern \emph{``Concepts such \textbf{as} \myUnderline{atoms}, \myUnderline{compass structures}, and \myUnderline{requests} will be introduced in this section''} is not captured. A regular expression 
$\mathtt{re}$, which works in any modern programming language, 
that captures such a pattern is:
\[
   \mathtt{re} = \mathtt{(\escapebackslash w+)(?=\respace(?:\escapebackslash w+,\respace)\{2,\}}and\mathtt{\respace\escapebackslash w+)}
\]
\noindent where $\respace$ is used to highlight the single white space `` ''. Since it is outside the scope of this paper, we will not delve too much into the syntax of this kind of regular expressions. For that matter, wonderful websites, such as \cite{Regexp101}, exist (they provide a quick reference for syntax and semantics together with examples and, more importantly, a full on-line environment for testing and debugging regular expressions). 

Let us briefly explain how  $\mathtt{re}$ captures the desired pattern.
First of all, we have that $\mathtt{(e)}$ is used to capture any pattern in $e$. The $\mathtt{(?= e )}$ operator checks whether the current position is followed by a pattern belonging to the language of $e$. The $\escapebackslash w$ variable represents any word-character, both lower and upper case. The operator $+$ is analogous to the operator $e^+ = ee^*$ in standard regular expressions. Thus, $\escapebackslash w+$ means any single word. The operator $(e)\{n,\}$,
with $n \geq 0$, captures a sequence of $n$ or more occurrences of pattern $e$. Finally, the operator $(?: e)$ represents just standard parentheses. 
A graphical account of the various parts of
regular expression $re$ is shown in Figure~\ref{fig:re}.

\begin{figure}
\begin{tikzpicture}[anchor=west]

\tikzset{explain/.style = {opacity=0.7} }
\tikzset{block/.style = {draw, rectangle} }

\node(RE0){$\mathtt{(\escapebackslash w+)}$};
\node(RE1) at (RE0.east) {$\mathtt{(?=}$};
\node(RE2) at (RE1.east) {$\mathtt{\respace}$};
\node(RE3) at (RE2.east) {$\mathtt{(?:\escapebackslash w+,\respace)\{2,\}}$};
\node(RE4) at (RE3.east) {$\mathtt{and\mathtt{\respace\escapebackslash w+}}$};
\node(RE5) at (RE4.east) {$\mathtt{)}$};

\draw [decorate,decoration={brace,amplitude=5pt,aspect=0.5}]
   ($(RE0.north west)$) -- node[above=4.5pt, inner sep=0](B1) {}
   ($(RE0.north east)$) ;

\draw [decorate,decoration={brace,amplitude=5pt,aspect=0.5}]
   ($(RE1.north west)$) -- node[above=4.5pt, inner sep=0](B2) {}
   ($(RE5.north east)$) ;

\draw [decorate,decoration={brace,amplitude=5pt,aspect=0.5, mirror, raise=0.1cm}]
   ($(RE2.south west)$) -- node[below=4.5pt, inner sep=0](B3) {}
   ($(RE2.south east)$) ;   

\draw [decorate,decoration={brace,amplitude=5pt,aspect=0.5, mirror}]
   ($(RE3.south west)$) -- node[below=4.5pt, inner sep=0](B4) {}
   ($(RE3.south east)$) ;

\draw [decorate,decoration={brace,amplitude=5pt,aspect=0.5, mirror}]
   ($(RE4.south west)$) -- node[below=4.5pt, inner sep=0](B5) {}
   ($(RE4.south east)$) ;   

\node[explain](EX0) at (8.25,2) {A };
\node[block](EX1) at (EX0.east){ single word is matched };
\node[explain, anchor=north west](EX2) at (EX0.south west) {if and only if it is};
\node[block, anchor=north west](EX3) at (EX2.south west) {immediately followed by};
\node[explain, anchor=north west](EX4) at (EX3.south west) {a concatenation of the following};
\node[explain, anchor=north west](EX5) at (EX4.south west) {three elements:};
\node[block, anchor=north west](EX6) at (EX5.south west) {
   (i) \begin{tabular}{l}a whitespace;\end{tabular}
};
\node[block, anchor=north west](EX7) at (EX6.south west) {
   (ii) \begin{tabular}{l}
   a sequence of two or more\\ concatenations of a single\\ word, a comma, and\\ a whitespace; 
   \end{tabular}
};
\node[block, anchor=north west](EX8) at (EX7.south west) {
   (iii) \begin{tabular}{l}the concatenation of  \\
    the word ``and'', a whitespace,\\ and a single word.\end{tabular}
};

\draw(B1) edge[looseness=0.6, out=90] (EX1.west);
\draw(B2) edge[looseness=0.7, out=90] (EX3.west);

\draw(B3) edge[looseness=0.7, out=-90] (EX6.west);
\draw(B4) edge[looseness=0.7, out=-90] (EX7.west);
\draw(B5) edge[looseness=0.7, out=-90] (EX8.west);

\end{tikzpicture}
\caption{\label{fig:re}
A graphical account of $re$ and its sub-expressions.
}
\end{figure}
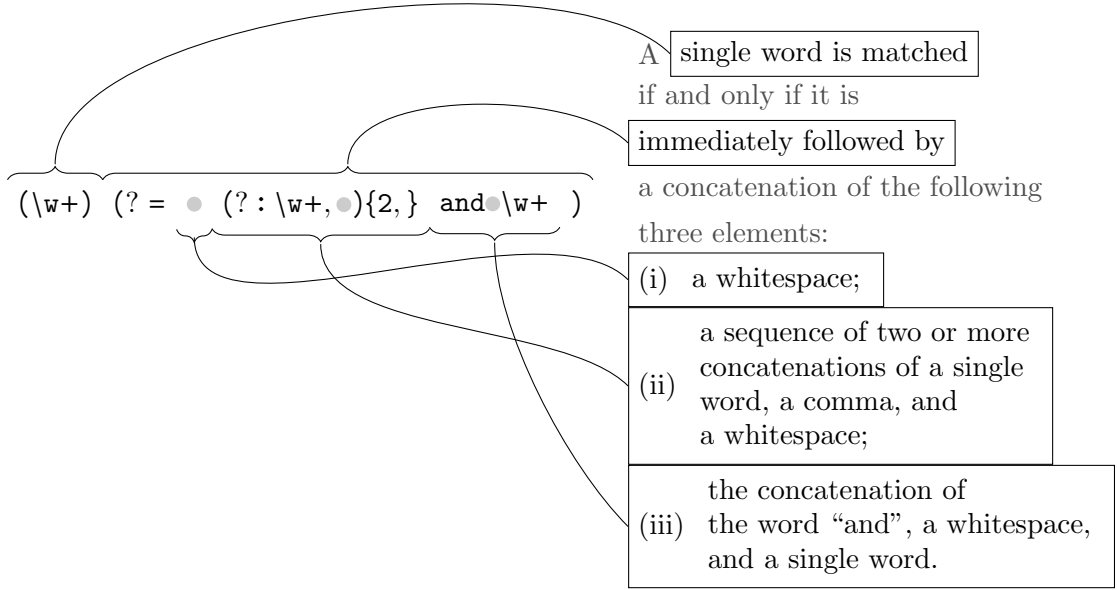

Let $\Sigma = W \cup S$, where $W =\{a, \ldots, z, A, \ldots, Z \}$ (word symbols) and $S = \{\respace, \mathtt{.}, \mbox{`} \mathtt{,} \mbox{'} \}$ (separator symbols).
For the sake of brevity, we omit the intermediate phase of translating $re$ into our $*$-free restricted fragment and we jump directly to the translation into $\mathsf{BDA}_{hom}$. For the sake of simplicity, we do not apply the literal translation here; instead, we make use of a shorter, more understandable encoding which is tailored to the structure of the
specific regular expression $re$. As a preliminary step, we provide some shorthands and assumptions that make the encoding formulas more compact. 
In the encoding, we will make use of the 
shorthands $len_{\geq n}$ and $len_n$  for any $n \in \bN$, that constrain the length of the interval on which they hold to be greater than or equal to $n$ and exactly equal to $n$, respectively.
More precisely, given a model $\bfM=(\bI_N, \cV)$, we have that $[x,y] \models len_{\geq n}$ if and only if $y - x \geq n$,
and $\bfM, [x,y] \models len_n$ if and only if $y - x = n$.
In $\LogicABDhom$, we may capture the semantics of
$len_{\geq n}$ and $len_n$ by means of the formulas $\hsEB^n \pi$ and $len_{\geq n} \wedge \hsAB^{n + 1} \bot$, respectively.\footnote{Notice that we provide  a unary encoding of the length constraints. It is possible to make a binary encoding analogous to the one proposed in \cite {DBLP:journals/lmcs/BozzelliMMPS22}.}
Since in the proposed encoding we will make use of
proposition letters in $\Sigma$ to represent words as points of an interval model (Figure~\ref{fig:reinterval}), we need to force each point to hold \emph{exactly} one symbol
$\sigma \in \Sigma$. Such a constraint is imposed
by putting the formula $[G]( \pi \rightarrow 
\bigvee_{\sigma \in \Sigma}
(\sigma \wedge \bigwedge_{\sigma' \in \Sigma \setminus \{\sigma'\}  } \neg \sigma' ) )$ in conjunction with the encoding of $re$. For the sake of brevity, we will tacitly assume that this is the case. Finally, with a little abuse of notation, in the encoding of $re$ we will make use of $W$ as a shorthand for $\bigvee_{\sigma \in W} \sigma$, which basically allows us to state that a certain (point-)interval
holds a word symbol. 

Now, we are ready to encode $re$ by a formula $\psi_{re}$. More precisely, we will make
use of $\hsED\psi_{re}$ as the main formula, where $\psi_{re}$ just encodes the matching part. Thus, by 
``reading'' a model $\bfM=(\bI_N, \cV)$ for $\hsED\psi_{re},$ we can easily retrieve every
matching by taking all and only those intervals  $[x,y]$ such that $\bfM,[x,y]\models \psi_{re}$.
As an example, in Figure~\ref{fig:reinterval} we have that $\bfM, [0,90] \models \hsED \psi_{re}$, while $\
[24,34] \models \psi_{re}$. In fact, $[24,34]$ is the only interval that satisfies $\psi_{re}$ in the
model of Figure~\ref{fig:reinterval} and, as we will see when we will discuss $\psi_{re}$ in more detail, this is determined both by the points belonging to $[24,34]$ and by the formulas that hold in its ``right context'', that is, the intervals $[x,y]$, with $34\leq x \leq 90$. 

Let $\psi_{re} = \psi^{gm}_{w^+} \wedge
\hsEA(\psi_{\respace(w^,\respace)^{2+}} \wedge  \hsEA \psi^{gm}_{and\respace w^+} )$. Intuitively,
$\psi_{re}$ requires the presence of three adjacent intervals $[x,y], [y,z],$ and $[z, w]$ such that $\bfM,[x,y]\models \psi^{gm}_{w^+}$, $\bfM,[y,z]\models \psi_{\respace(w^,\respace)^{2+}}$, and $\bfM,[z,w]\models 
\psi^{gm}_{and\respace w^+}$. These sub-formulas constrain the three regular expressions whose concatenation forms $re$ as follows: 

\smallskip

\begin{compactitem}

\item $\psi^{gm}_{w^+}=\hsEB\neg W \wedge \hsAB(len_{\geq 1} \rightarrow \hsEA W)
\wedge \hsEB\hsEA W \wedge \hsEA \neg W$. This formula holds over an interval $[x,y]$ if and only if point-intervals $[x,x]$ and $[y,y]$ do not hold a word symbol (conjuncts $\hsEB\neg W$ and $\hsEA \neg
W$, respectively), but a word symbol holds at all the internal point-intervals $[x',x']$, with $x<x'<y$  (conjunct $\hsAB(len_{\geq 1} \rightarrow \hsEA W)$). Finally, it constrains the interval $[x,y]$ to contain at least one word symbol (conjunct $\hsEB\hsEA W$). Intuitively,  $\psi^{gm}_{w^+}$ encodes the greedy match (gm) of a single non-empty word preceeded and followed by two singleton separator symbols. As an example, in Figure \ref{fig:reinterval}, we have that $\psi^{gm}_{w^+}$ holds over the interval $[24,34]$.

\item 
$\psi_{\respace(w^+,\respace)^{2+}} = \hsEA \respace \wedge \hsEB (len_1 \wedge \hsEB \respace \wedge \hsEA W)
\wedge \hsAD(len_1\!\rightarrow\!(\hsEB \respace \wedge \hsEA W) \vee (\hsEB , \wedge \hsEA \respace) \vee (\hsEB
W \wedge \hsEA W) \vee (\hsEB W \wedge \hsEA ,)) \wedge \hsEB(\hsEA ,\wedge  \hsEB\hsEA ,)$. This formula holds over an interval $[x,y]$ if and only if the following conditions hold:

\smallskip

\begin{compactenum}

\item the symbol $\respace$ holds at point-interval $[x,x]$ and a word symbol holds at point-interval $[x+1,x+1]$ (conjunct $\hsEB (len_1 \wedge \hsEB \respace \wedge \hsEA W)$); 

\item the symbol $\respace$ holds at point-interval $[y,y]$  (conjunct $\hsEA \respace$);
\item for every strict sub-interval of $[x,y]$ of the form $[x', x'+1]$, we have that either $[x',x']$ is labelled with $\respace$ and $[x'+1, x'+1]$ 
by a word symbol (disjunct $\hsEB \respace \wedge \hsEA W$), or
$[x',x']$ is labelled by ``$,$'' and $[x'+1, x'+1]$ 
by $\respace$  (disjunct $\hsEB , \wedge \hsEA \respace$), or
both $[x',x']$ and $[x'+1, x'+1]$ 
are labelled by a word symbol (disjunct $\hsEB W \wedge \hsEA W$),  or
$[x',x']$ is labelled by a word symbol  and $[x'+1, x'+1]$ 
with ``$,$'' (disjunct $\hsEB W \wedge \hsEA ,$); 

\item the symbol ``$,$'' appears as a label of  \emph{at least} two distinct point-intervals $[x',x']$ and $[x'', x'']$ in $[x,y]$, i.e., with $x< x'< x''<y$  (conjunct $\hsEB(\hsEA ,\wedge \hsEB\hsEA ,)$). In Figure~\ref{fig:reinterval}, such a condition is satisfied by point-intervals $[40,40]$ and $[48,48]$, which are included in the interval $[34,49]$.

\end{compactenum}

\smallskip

Intuitively, the conjunct $\hsAD(len_1 \rightarrow (\hsEB \respace \wedge \hsEA W) \vee (\hsEB , \wedge \hsEA \respace) \vee (\hsEB W \wedge \hsEA W) \vee (\hsEB W \wedge \hsEA ,))$ constrains the word underlying $[x+1, y+1]$ to belong to the language of $((w)^+, \respace)^*$, while the conjunct $\hsEB(\hsEA , \hsEB\hsEA ,)$ forces at least two iterations of the $*$ operation in such a language. Thus, together they force  such a word to belong to $((w)^+, \respace)^{2+}$;

\item $\psi^{gm}_{and\respace w^+} =  \hsEB a
\wedge \hsEB(len_1 \wedge \hsEA n)
\wedge \hsEB(len_2 \wedge \hsEA d)
\wedge \hsEB(len_3 \wedge \hsEA \respace)
 \wedge \hsEB(len_{\geq 4} \rightarrow \hsEA W) \wedge \hsEA \neg W$. This formula holds over an interval $[x,y]$ if and only if the word underlying the interval $[x, x + 3]$ is exactly ``$and\respace$'' (conjuncts $\hsEB a$,
$\hsEB(len_1 \wedge \hsEA n)$,
$\wedge \hsEB(len_2 \wedge \hsEA d)$, and 
$\wedge \hsEB(len_3 \wedge \hsEA \respace)$) followed by an uninterrupted sequence of word symbols 
underlying the interval $[x+4, y-1]$ (conjunct $\hsEB(len_{\geq 4} \rightarrow \hsEA W)$).
In addition, it imposes the word underlying the interval $[x+4, y-1]$ to be a greedy match, that is, 
an entire word is captured,  since we force a separator symbol on $[y+1, y+1]$ by means of the conjunct $\hsEA \neg W$.

\end{compactitem}

\begin{figure}
\begin{tikzpicture}
   \begin{scope}[node distance=0.3cm]

\node[opacity=0.5,draw, circle, inner sep=1, 
label={[scale=1, opacity=0.5,]180:$\scalebox{0.525}{$\ldots$}$},
label={[scale=1, opacity=0.5,]270:$\scalebox{0.525}{$ 22 $}$},
      label={[scale=1, opacity=0.5,]90:$\scalebox{0.525}{$ H $}$}](22-22) [] {};

\node[opacity=0.5,draw, circle, inner sep=1, label={[scale=1, opacity=0.5,]270:$\scalebox{0.525}{$ 23 $}$},
      label={[scale=1, opacity=0.5,]90:$\scalebox{0.525}{$ S $}$}
      ](23-23) [right of=22-22] {};

\node[opacity=0.5,draw, circle, inner sep=1, label={[scale=1, opacity=0.5,]270:$\scalebox{0.525}{$ 24 $}$},
      label={[scale=1, opacity=0.5,]90:$\scalebox{0.525}{$ \respace $}$}](24-24) [right of=23-23] {};

\node[opacity=0.5,draw, circle, inner sep=1, label={[scale=1, opacity=0.5,]270:$\scalebox{0.525}{$ 25 $}$},
      label={[scale=1, opacity=0.5,]90:$\scalebox{0.525}{$ o $}$}](25-25) [right of=24-24] {};

\node[opacity=0.5,draw, circle, inner sep=1, label={[scale=1, opacity=0.5,]270:$\scalebox{0.525}{$ 26 $}$},
      label={[scale=1, opacity=0.5,]90:$\scalebox{0.525}{$ p $}$}](26-26) [right of=25-25] {};

\node[opacity=0.5,draw, circle, inner sep=1, label={[scale=1, opacity=0.5,]270:$\scalebox{0.525}{$ 27 $}$},
      label={[scale=1, opacity=0.5,]90:$\scalebox{0.525}{$ e $}$}](27-27) [right of=26-26] {};

\node[opacity=0.5,draw, circle, inner sep=1, label={[scale=1, opacity=0.5,]270:$\scalebox{0.525}{$ 28 $}$},
      label={[scale=1, opacity=0.5,]90:$\scalebox{0.525}{$ r $}$}](28-28) [right of=27-27] {};

\node[opacity=0.5,draw, circle, inner sep=1, label={[scale=1, opacity=0.5,]270:$\scalebox{0.525}{$ 29 $}$},
      label={[scale=1, opacity=0.5,]90:$\scalebox{0.525}{$ a $}$}](29-29) [right of=28-28] {};

\node[opacity=0.5,draw, circle, inner sep=1, label={[scale=1, opacity=0.5,]270:$\scalebox{0.525}{$ 30 $}$},
      label={[scale=1, opacity=0.5,]90:$\scalebox{0.525}{$ t $}$}](30-30) [right of=29-29] {};

\node[opacity=0.5,draw, circle, inner sep=1, label={[scale=1, opacity=0.5,]270:$\scalebox{0.525}{$ 31 $}$},
      label={[scale=1, opacity=0.5,]90:$\scalebox{0.525}{$ o $}$}](31-31) [right of=30-30] {};

\node[opacity=0.5,draw, circle, inner sep=1, label={[scale=1, opacity=0.5,]270:$\scalebox{0.525}{$ 32 $}$},
      label={[scale=1, opacity=0.5,]90:$\scalebox{0.525}{$ r $}$}](32-32) [right of=31-31] {};

\node[opacity=0.5,draw, circle, inner sep=1, label={[scale=1, opacity=0.5,]270:$\scalebox{0.525}{$ 33 $}$},
      label={[scale=1, opacity=0.5,]90:$\scalebox{0.525}{$ s $}$}](33-33) [right of=32-32] {};

\node[opacity=0.5,draw, circle, inner sep=1, label={[scale=1, opacity=0.5,]270:$\scalebox{0.525}{$ 34 $}$},
      label={[scale=1, opacity=0.5,]90:$\scalebox{0.525}{$ \respace $}$}](34-34) [right of=33-33] {};

\node[opacity=0.5,draw, circle, inner sep=1, label={[scale=1, opacity=0.5,]270:$\scalebox{0.525}{$ 35 $}$},
      label={[scale=1, opacity=0.5,]90:$\scalebox{0.525}{$ m $}$}](35-35) [right of=34-34] {};

\node[opacity=0.5,draw, circle, inner sep=1, label={[scale=1, opacity=0.5,]270:$\scalebox{0.525}{$ 36 $}$},
      label={[scale=1, opacity=0.5,]90:$\scalebox{0.525}{$ e $}$}](36-36) [right of=35-35] {};

\node[opacity=0.5,draw, circle, inner sep=1, label={[scale=1, opacity=0.5,]270:$\scalebox{0.525}{$ 37 $}$},
      label={[scale=1, opacity=0.5,]90:$\scalebox{0.525}{$ e $}$}](37-37) [right of=36-36] {};

\node[opacity=0.5,draw, circle, inner sep=1, label={[scale=1, opacity=0.5,]270:$\scalebox{0.525}{$ 38 $}$},
      label={[scale=1, opacity=0.5,]90:$\scalebox{0.525}{$ t $}$}](38-38) [right of=37-37] {};

\node[opacity=0.5,draw, circle, inner sep=1, label={[scale=1, opacity=0.5,]270:$\scalebox{0.525}{$ 39 $}$},
      label={[scale=1, opacity=0.5,]90:$\scalebox{0.525}{$ s $}$}](39-39) [right of=38-38] {};

\node[opacity=0.5,draw, circle, inner sep=1, label={[scale=1, opacity=0.5,]270:$\scalebox{0.525}{$ 40 $}$},
      label={[scale=1, opacity=0.5,]90:$\scalebox{0.525}{$ , $}$}](40-40) [right of=39-39] {};

\node[opacity=0.5,draw, circle, inner sep=1, label={[scale=1, opacity=0.5,]270:$\scalebox{0.525}{$ 41 $}$},
      label={[scale=1, opacity=0.5,]90:$\scalebox{0.525}{$ \respace $}$}](41-41) [right of=40-40] {};

\node[opacity=0.5,draw, circle, inner sep=1, label={[scale=1, opacity=0.5,]270:$\scalebox{0.525}{$ 42 $}$},
      label={[scale=1, opacity=0.5,]90:$\scalebox{0.525}{$ b $}$}](42-42) [right of=41-41] {};

\node[opacity=0.5,draw, circle, inner sep=1, label={[scale=1, opacity=0.5,]270:$\scalebox{0.525}{$ 43 $}$},
      label={[scale=1, opacity=0.5,]90:$\scalebox{0.525}{$ e $}$}](43-43) [right of=42-42] {};

\node[opacity=0.5,draw, circle, inner sep=1, label={[scale=1, opacity=0.5,]270:$\scalebox{0.525}{$ 44 $}$},
      label={[scale=1, opacity=0.5,]90:$\scalebox{0.525}{$ g $}$}](44-44) [right of=43-43] {};

\node[opacity=0.5,draw, circle, inner sep=1, label={[scale=1, opacity=0.5,]270:$\scalebox{0.525}{$ 45 $}$},
      label={[scale=1, opacity=0.5,]90:$\scalebox{0.525}{$ i $}$}](45-45) [right of=44-44] {};

\node[opacity=0.5,draw, circle, inner sep=1, label={[scale=1, opacity=0.5,]270:$\scalebox{0.525}{$ 46 $}$},
      label={[scale=1, opacity=0.5,]90:$\scalebox{0.525}{$ n $}$}](46-46) [right of=45-45] {};

\node[opacity=0.5,draw, circle, inner sep=1, label={[scale=1, opacity=0.5,]270:$\scalebox{0.525}{$ 47 $}$},
      label={[scale=1, opacity=0.5,]90:$\scalebox{0.525}{$ s $}$}](47-47) [right of=46-46] {};

\node[opacity=0.5,draw, circle, inner sep=1, label={[scale=1, opacity=0.5,]270:$\scalebox{0.525}{$ 48 $}$},
      label={[scale=1, opacity=0.5,]90:$\scalebox{0.525}{$ , $}$}](48-48) [right of=47-47] {};

\node[opacity=0.5,draw, circle, inner sep=1, label={[scale=1, opacity=0.5,]270:$\scalebox{0.525}{$ 49 $}$},
      label={[scale=1, opacity=0.5,]90:$\scalebox{0.525}{$ \respace $}$}](49-49) [right of=48-48] {};

\node[opacity=0.5,draw, circle, inner sep=1, label={[scale=1, opacity=0.5,]270:$\scalebox{0.525}{$ 50 $}$},
      label={[scale=1, opacity=0.5,]90:$\scalebox{0.525}{$ a $}$}](50-50) [right of=49-49] {};

\node[opacity=0.5,draw, circle, inner sep=1, label={[scale=1, opacity=0.5,]270:$\scalebox{0.525}{$ 51 $}$},
      label={[scale=1, opacity=0.5,]90:$\scalebox{0.525}{$ n $}$}](51-51) [right of=50-50] {};

\node[opacity=0.5,draw, circle, inner sep=1, label={[scale=1, opacity=0.5,]270:$\scalebox{0.525}{$ 52 $}$},
      label={[scale=1, opacity=0.5,]90:$\scalebox{0.525}{$ d $}$}](52-52) [right of=51-51] {};

\node[opacity=0.5,draw, circle, inner sep=1, label={[scale=1, opacity=0.5,]270:$\scalebox{0.525}{$ 53 $}$},
      label={[scale=1, opacity=0.5,]90:$\scalebox{0.525}{$ \respace $}$}](53-53) [right of=52-52] {};

\node[opacity=0.5,draw, circle, inner sep=1, label={[scale=1, opacity=0.5,]270:$\scalebox{0.525}{$ 54 $}$},
      label={[scale=1, opacity=0.5,]90:$\scalebox{0.525}{$ d $}$}](54-54) [right of=53-53] {};

\node[opacity=0.5,draw, circle, inner sep=1, label={[scale=1, opacity=0.5,]270:$\scalebox{0.525}{$ 55 $}$},
      label={[scale=1, opacity=0.5,]90:$\scalebox{0.525}{$ u $}$}](55-55) [right of=54-54] {};

\node[opacity=0.5,draw, circle, inner sep=1, label={[scale=1, opacity=0.5,]270:$\scalebox{0.525}{$ 56 $}$},
      label={[scale=1, opacity=0.5,]90:$\scalebox{0.525}{$ r $}$}](56-56) [right of=55-55] {};

\node[opacity=0.5,draw, circle, inner sep=1, label={[scale=1, opacity=0.5,]270:$\scalebox{0.525}{$ 57 $}$},
      label={[scale=1, opacity=0.5,]90:$\scalebox{0.525}{$ i $}$}](57-57) [right of=56-56] {};

\node[opacity=0.5,draw, circle, inner sep=1, label={[scale=1, opacity=0.5,]270:$\scalebox{0.525}{$ 58 $}$},
      label={[scale=1, opacity=0.5,]90:$\scalebox{0.525}{$ n $}$}](58-58) [right of=57-57] {};

\node[opacity=0.5,draw, circle, inner sep=1, label={[scale=1, opacity=0.5,]270:$\scalebox{0.525}{$ 59 $}$},
      label={[scale=1, opacity=0.5,]90:$\scalebox{0.525}{$ g $}$}](59-59) [right of=58-58] {};

\node[opacity=0.5,draw, circle, inner sep=1, label={[scale=1, opacity=0.5,]270:$\scalebox{0.525}{$ 60 $}$},
      label={[scale=1, opacity=0.5,]90:$\scalebox{0.525}{$ \respace $}$}](60-60) [right of=59-59] {};

\node[opacity=0.5,draw, circle, inner sep=1, label={[scale=1, opacity=0.5,]270:$\scalebox{0.525}{$ 61 $}$},
      label={[scale=1, opacity=0.5,]90:$\scalebox{0.525}{$ u $}$}](61-61) [right of=60-60] {};

\node[opacity=0.5,draw, circle, inner sep=1, label={[scale=1, opacity=0.5,]270:$\scalebox{0.525}{$ 62 $}$},
      label={[scale=1, opacity=0.5,]90:$\scalebox{0.525}{$ n $}$}](62-62) [right of=61-61] {};

\node[opacity=0.5,draw, circle, inner sep=1, label={[scale=1, opacity=0.5,]270:$\scalebox{0.525}{$ 63 $}$},
      label={[scale=1, opacity=0.5,]90:$\scalebox{0.525}{$ d $}$}](63-63) [right of=62-62] {};

\node[opacity=0.5,draw, circle, inner sep=1, label={[scale=1, opacity=0.5,]270:$\scalebox{0.525}{$ 64 $}$},
      label={[scale=1, opacity=0.5,]90:$\scalebox{0.525}{$ e $}$}](64-64) [right of=63-63] {};

\node[opacity=0.5,draw, circle, inner sep=1, label={[scale=1, opacity=0.5,]270:$\scalebox{0.525}{$ 65 $}$},
label={[scale=1, opacity=0.5,]0:$\scalebox{0.525}{$ \ldots $}$},
      label={[scale=1, opacity=0.5,]90:$\scalebox{0.525}{$ r $}$}](65-65) [right of=64-64] {};
      
\def\y{2cm}
\draw[-] ($(25-25)+(0, \y)$) -- ($(63-63)+ (0, \y)$)
node[fill=white, pos=0.5] {\scalebox{0.75}{$\hsED \psi_{re}$}}; 
\draw[dashed] ($(25-25)+(0, \y)$) -- ($(25-25)+(-1, \y)$);    
\draw[dashed] ($(63-63)+(0, \y)$) -- ($(63-63)+(1, \y)$);

\def\y{1cm}
\def\b{24}
\def\e{34}
\draw[|-|] ($(\b-\b)+(0, \y)$) -- ($(\e-\e)+ (0, \y)$)
node[fill=white, pos=0.5] {\scalebox{0.75}{$\psi_{re}, \psi^{gm}_{w^+}$}};

\def\y{1cm}
\def\b{34}
\def\e{49}
\draw[|-|] ($(\b-\b)+(0, \y)$) -- ($(\e-\e)+ (0, \y)$)
node[fill=white, pos=0.5] {\scalebox{0.75}{$\psi_{\respace(w^+,\respace)^{2+}}$}};

\def\y{1cm}
\def\b{49}
\def\e{60}
\draw[|-|] ($(\b-\b)+(0, \y)$) -- ($(\e-\e)+ (0, \y)$)
node[fill=white, pos=0.5] {\scalebox{0.75}{$\psi_{and\respace w^+}$}};

  \end{scope}

\end{tikzpicture}
\caption{\label{fig:reinterval} a graphical account of how a $\hsED \psi_{re}$
 holds over an interval model representing a text.}
\end{figure}
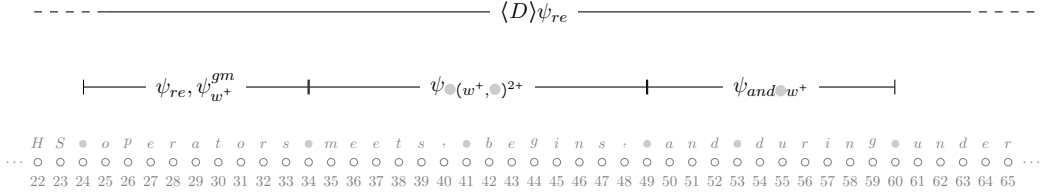

\smallskip

We conclude this appendix with some remarks about the practical use of regular expressions. To the best of our knowledge, in their implementation the majority of existing programming languages do not support the free use of negation in regular expressions, but they allow for positive/negative lookahead/lookbehind. In this section, we showed how to deal with positive/negative lookahead by means of modality $\hsEA$. Moreover, we argued that positive/negative lookbehind may be captured by adding  modality $\hsEAbar$, which is the converse of modality $\hsEA$, to $\mathsf{BDA}_{hom}$, thus obtaining the logic $\mathsf{BDA\overline{A}}_{hom}$. 
For the sake of simplicity, we did not take modality $\hsEAbar$ into consideration
in this work, as its introduction involves a number of  technicalities. However, in view of the results 
established in the paper, we conjecture that, under the homogeneity assumption, the satisfiability problem for $\mathsf{BDA\overline{A}}_{hom}$ belongs to the same complexity class as its proper fragment $\mathsf{BDA}_{hom}$.

}

\end{document}